\documentclass[rmp,aps,nofootinbib,showpacs,floatfix]{revtex4}

\usepackage{graphicx}
\usepackage{latexsym} 
\usepackage{epstopdf}
\usepackage{amsmath,amssymb,slashed}
\usepackage{dsfont}
\usepackage{natbib}
\usepackage[colorlinks=false,linktocpage=true]{hyperref}

\def\lg{{\mathchoice{~\raise.58ex\hbox{$<$}\mkern-14.8mu\lower.52ex\hbox{$>$}~}
                    {~\raise.58ex\hbox{$<$}\mkern-14.8mu\lower.52ex\hbox{$>$}~}
                    {\raise.59ex\hbox{{$\scriptscriptstyle <$}}\mkern-12.8mu%
                     \lower.01ex\hbox{{$\scriptscriptstyle >$}}}   {}   }} 
\def\gl{{\mathchoice{~\raise.58ex\hbox{$>$}\mkern-12.8mu\lower.52ex\hbox{$<$}~}
                    {~\raise.58ex\hbox{$>$}\mkern-12.8mu\lower.52ex\hbox{$<$}~}
                    {\raise.62ex\hbox{{$\scriptscriptstyle >$}}\mkern-12.0mu%
                     \lower.05ex\hbox{{$\scriptscriptstyle <$}}}  {}    }}   

\def\ca{{\mathchoice{~\raise.58ex\hbox{$c$}\mkern-9.0mu\lower.52ex\hbox{$a$}~}
                    {~\raise.58ex\hbox{$c$}\mkern-9.0mu\lower.52ex\hbox{$a$}~}
                    {\raise.59ex\hbox{{$\scriptscriptstyle c$}}\mkern-7.0mu%
		     \lower.01ex\hbox{{$\scriptscriptstyle a$}}}   {} 	}} 
\def\ac{{\mathchoice{~\raise.58ex\hbox{$a$}\mkern-10.0mu\lower.52ex\hbox{$c$}~}
                    {~\raise.58ex\hbox{$a$}\mkern-10.0mu\lower.52ex\hbox{$c$}~}
		    {\raise.62ex\hbox{{$\scriptscriptstyle a$}}\mkern-9.0mu%
		     \lower.05ex\hbox{{$\scriptscriptstyle c$}}}  {} 	}} 
		       
\newcommand{\be}{\begin{equation}}
\newcommand{\ee}{\end{equation}}
\newcommand{\ba}{\begin{eqnarray}}
\newcommand{\ea}{\end{eqnarray}}
\newcommand{\ban}{\begin{eqnarray*}}
\newcommand{\ean}{\end{eqnarray*}}
\newcommand{\non}{\nonumber}

\newcommand{\Tr}{\mbox{${\rm Tr}$}}
\newcommand{\sla}{\!\!\!/ \,}

\newcommand{\subthreesection}[1]{
\vspace{5mm}
\noindent
{\em #1}
\vspace{5mm}
}

\newcommand{\tr}{\,{\rm Tr}\,}
\newcommand{\6}{\partial }
\newcommand{\beq}{\begin{equation}}
\newcommand{\eeq}{\end{equation}}
\newcommand{\bqa}{\begin{eqnarray}}
\newcommand{\eqa}{\end{eqnarray}}
\newcommand{\bea}{\begin{eqnarray}}
\newcommand{\eea}{\end{eqnarray}}
\newcommand{\nn}{\nonumber}
\newcommand{\0}{\over }

\newcommand{\9}{}

\usepackage[usenames]{color}

\begin{document}

\title{Color Instabilities in the Quark-Gluon Plasma}

\author{Stanis\l aw Mr\' owczy\' nski}

\affiliation{Institute of Physics, Jan Kochanowski University, 
ul. \'Swi\c etokrzyska 15, PL - 25-406 Kielce, Poland \\
and National Centre for Nuclear Research, Warsaw, Poland, 
ul. Ho\.za 69, PL - 00-681 Warsaw, Poland}

\author{Bj\"orn Schenke}

\affiliation{Physics Department, Bldg. 510A,
Brookhaven National Laboratory, Upton, NY 11973, USA}

\author{Michael Strickland}

\affiliation{Kent State University, Kent, OH 44240, USA\\
and Frankfurt Institute for Advanced Studies, 
Goethe-Universit\"at Frankfurt\\
Max-von-Laue-Stra\ss{}e~1, D-60438 Frankfurt am Main, Germany}

\begin{abstract}

When the quark-gluon plasma (QGP) - a system of deconfined quarks and gluons - is in a nonequilibrium state, it is usually unstable with respect to color collective modes. The instabilities, which are expected to strongly influence dynamics of the QGP produced in relativistic heavy-ion collisions, are extensively discussed under the assumption that the plasma is weakly coupled. We begin by presenting the theoretical approaches to study the QGP, which include: field theory methods based on the Keldysh-Schwinger formalism, classical and quantum kinetic theories, and fluid techniques. The dispersion equations, which give the spectrum of plasma collective excitations, are analyzed in detail. Particular attention is paid to a momentum distribution of plasma constituents which is obtained by deforming an isotropic momentum distribution. Mechanisms of chromoelectric and chromomagnetic instabilities are explained in terms of elementary physics. The Nyquist analysis, which allows one to determine the number of solutions of a dispersion equation without explicitly solving it, and stability criteria are also discussed. We then review various numerical approaches - purely classical or quantum - to simulate the temporal evolution of an unstable quark-gluon plasma. The dynamical role of instabilities in the processes of plasma equilibration is analyzed.

\end{abstract}

\date{May 26, 2017}

\pacs{11.15.Bt, 11.10.Wx, 12.38.Mh, 25.75.-q, 52.27.Ny, 52.35.-g}

\maketitle

\tableofcontents

\section{Introduction}
\label{sec-intro} 

The quark-gluon plasma (QGP) is a system of quarks and gluons which are not confined in hadrons' interiors but can move in the entire volume occupied by the system. The dynamics of the plasma is governed by Quantum Chromodynamics (QCD) which is the non-Abelian quantum field theory of quarks and gluons. In the domain of asymptotic freedom, where the coupling constant is small, QCD strongly resembles Quantum Electrodynamics (QED) and thus the QGP reveals some similarities to an electromagnetic plasma (EMP). An important common feature of EMP and QGP is the collective character of the dynamics. The range of electrostatic interaction is, in spite of the screening, usually much larger than the inter-particle spacing in EMP. There are many particles in the Debye sphere - the sphere of the radius equal to the screening length - and these particles are highly correlated. There is a similar situation in the deconfined perturbative phase of  QCD, see {\it e.g.} \cite{Blaizot:2001nr}. In an equilibrium plasma, the Debye mass is of order $gT$, where $g$ is the QCD  coupling constant and $T$ is the temperature. Since the particle density in the QGP is of order $T^3$, the number of partons\footnote{The term `parton' is used to denote a quasiparticle fermionic (quark) or bosonic (gluon) excitation of the quark-gluon plasma.} in  the Debye sphere, which is roughly $1/g^3$, is large in the weakly coupled ($1/g \gg 1$) QGP. 

There is a very rich spectrum of collective excitations of both EMP and QGP. When the plasma is out of equilibrium, there are various unstable modes which can strongly influence the temporal evolution of the system. The problem is of experimental relevance since the QGP is produced in relativistic heavy-ion  collisions.

\subsection{Quark-Gluon Plasma from Relativistic Heavy-Ion Collisions}
\label{subsec-QGP-RHIC}

Active experimental studies of the QGP started in the mid 1980s when atomic nuclei were accelerated to sufficiently high energies in the hopes of creating a drop of QGP in the laboratory. The large scale programs at European Organization for Nuclear Research (CERN) and Brookhaven National Laboratory (BNL) provided evidence of QGP production in the early stage of nucleus-nucleus collisions, when the system is extremely hot and dense, but properties of the QGP remain somewhat enigmatic. The key question is whether the system generated in relativistic heavy ion collisions is truly strongly coupled and has a large coupling constant, $\alpha_s \equiv g^2/4\pi$, or whether the coupling constant is not asymptotically large but the system behaves like a strongly coupled system because of its high density and through complex dynamical effects, like the plasma instabilities that are the subject of this review. The experiments at CERN Large Hadron Collider (LHC) aim to clarify the situation.

The QGP as created in relativistic heavy-ion collisions manifests a fluid-like behavior which is particularly evident in studies of the collective flow, see the reviews \cite{Voloshin:2008dg,Heinz:2013th,Gale:2013da}. A hydrodynamic description requires the system to be close to local thermal equilibrium and experimental data on the particle spectra and collective flow suggest, when analyzed within the hydrodynamic model, that an equilibration time of the parton system produced does not much exceed 1 ${\rm fm}/c$ \cite{Heinz:2004pj,Bozek:2010aj}\footnote{Note, however, that viscous hydrodynamical models used to describe plasma dynamics possess sizable momentum space anisotropies \cite{Strickland:2013uga}. As a consequence, many authors now refer to the {\it hydronization} of the plasma rather than the equilibration.}. Such a fast equilibration can be explained assuming that the quark-gluon plasma is strongly coupled \cite{Shuryak:2008eq}. However, it is plausible that, due to the high-energy density generated during the early stage of the collision when the flow is generated \cite{Sorge:1998mk}, the plasma is weakly coupled because of asymptotic freedom. Thus, the question arises whether the weakly interacting plasma can be equilibrated on timescales of 1 ${\rm fm}/c$. 

Models that assume that parton-parton collisions are responsible for the thermalization of weakly coupled plasmas usually obtain a significantly longer equilibration time. To thermalize the system one needs either a few hard collisions of momentum transfer of order of the characteristic parton momentum\footnote{Although anisotropic systems are considered, the characteristic momentum in all directions is assumed for the purpose of qualitative discussion to be of the same order.}, which is denoted here as $T$ (the temperature of the equilibrium system), or many collisions of smaller transfer. As discussed in e.g. \cite{Baier:2000sb}, the inverse equilibration time is on the order of $\tau_{\rm eq}^{-1} \sim g^{26/5} Q_s$ 
where $Q_s \sim 1 \div 3$ GeV is the so-called QCD saturation scale. Recent detailed calculations of the role of instabilities suggest that instabilities work to reduce the equilibrium time to $\tau_{\rm eq}^{-1} \sim g^5 Q_s$ \cite{Kurkela:2011ti}.  This represents only a short reduction in the equilibration time, however, one must keep in mind that this is an estimate for when the system is fully equilibrated, with isotropic momentum-space distributions in the local rest frame.  Going beyond such parametric estimates, numerical studies of QGP plasma instability evolution have demonstrated that the turbulent cascade induced by the chromo-Weibel instability, in particular, results in fast thermalization of the QGP, in the sense that an anisotropic but thermal gluon distribution function is established, within 1 fm/c after the collision \cite{Attems:2012js}.  Such QGP plasma instability studies have dramatically improved our understanding of the early-time dynamics of gauge fields generated in heavy-ion collisions, demonstrating that instabilities drive the system to maintain large gluon occupancies which result in a fundamental change to our understanding of perturbative approaches to QGP dynamics.  As a result, the turbulent cascade induced by QGP plasma instabilities now forms a starting point for studies of QGP thermalization using weak-coupling methods \cite{Berges:2013fga,Gelis:2013rba}.  In addition, the existence of a high-occupancy turbulent field background causes one to revisit the calculations of QGP transport coefficients in order to take into account the effect of strong-field domains which induce a so-called anomalous shear viscosity \cite{Asakawa:2006tc,Asakawa:2006jn}. In this review, we aim to provide an accessible entry point to understanding the role of plasma instabilities for the equilibration and various signatures of a QGP.

\subsection{Plasma Instabilities}
\label{subsec-insta}

If one considers the early-time dynamics of a QGP as created in heavy ion collisions, one finds that at very early times after the nuclear impact, large momentum-space anisotropies are generated.  Studies using both weak and strong coupling methods, see \cite{Gelis:2009wh,Gelis:2013rba,Li:2016rcr} and \cite{Chesler:2008hg,Heller:2011ju,vanderSchee:2012qj}, respectively, find that at early-times after the collisions $\tau \sim 0.2-0.3$ fm/c the QGP possesses an oblate momentum-space anisotropy, with the transverse pressure in the local rest frame $({\cal P}_T)$ greatly exceeding the longitudinal pressure  $({\cal P}_L)$.  The initial degree of momentum-space anisotropy increases as one increases the shear viscosity of the QGP $(\eta)$.  In the strong coupling limit, one finds early-time pressure anisotropies with ${\cal P}_L/{\cal P}_T \sim 0.3$ at $\tau \simeq 0.2$ fm/c.   Since the strong coupling limit places a putative lower bound on the shear viscosity over entropy density ratio $(\eta/s)$, this means that the strong coupling thermalization studies mentioned above place an upper bound on the ratio of the longitudinal and transverse pressures at early times, {\it i.e.} ${\cal P}_L/{\cal P}_T \lesssim 0.3$. A similar picture emerges in the weak coupling limit, where one finds once again that the transverse pressure in the local rest frame greatly exceeds the longitudinal pressure at early times after the nuclear impact, however, the oblate momentum-space anisotropy predicted is more extreme \cite{Gelis:2009wh,Gelis:2013rba,Li:2016rcr,Kurkela:2015qoa,Keegan:2015avk}.  

As mentioned above, QGP equilibration can be accelerated by instabilities which turn on in an anisotropic weakly coupled quark-gluon plasma \cite{Mrowczynski:2005ki,Kurkela:2011ti}. The characteristic inverse time for instability growth is of order $\tau_{\rm unstable}^{-1} \sim g T$ for a sufficiently anisotropic momentum distribution. Thus, the instabilities are much `faster' than the collisions which have an inverse time scale on the order of $\tau_{\rm coll}^{-1} \sim g^n T$ with $ 2\le n \le 4$ in the weak coupling regime \cite{Arnold:1998cy}. The index $n=4$ for hard collisions and it goes down when the momentum transfer decreases. 

The QGP is expected to possess an array of instabilities just like an EMP does. The history of plasma physics is said to be the successive discovery of new types of instabilities. Plasma instabilities can be divided into two general groups: (1) {\it hydrodynamic} instabilities, which are caused by coordinate space inhomogeneities, and (2) {\it kinetic} instabilities, which are caused by a non-equilibrium momentum distribution of plasma particles. Hydrodynamic instabilities are usually associated with phenomena occurring at the plasma boundaries. In the case of QGP, this is the domain of highly non-perturbative QCD where the non-Abelian nature of the theory cannot be ignored. Then, the behavior of QGP is presumably very different from that of an EMP, and thus, we will not speculate about possible analogies.  

The kinetic instabilities are simply collective modes with positive imaginary part of the mode frequency $\Im\omega$. As a result, the mode amplitude exponentially grows in time $t$ as $e^{\Im\omega t}$. There are longitudinal (electrostatic or electric) and transverse (magnetic) instabilities with the wave vector parallel and perpendicular, respectively, to the generated electric field. The former corresponds to a growing charge density and the latter to a growing current. In a non-relativistic plasma, the electric instabilities are usually much more important than the magnetic ones, since the magnetic effects are suppressed by a factor of $(v/c)^2$, where $v$ is the typical velocity of plasma constituents \footnote{This factor can be derived, for example, by considering the relative strength of magnetic to electric forces acting between two parallel infinite wires which are charged and the charges flow with the velocity $v$. }. In a relativistic plasma, both types of instabilities are of similar strength. As will be discussed later on, electric instabilities occur when the momentum distribution of plasma particles has more than one maximum, as in {\it e.g.} the two-stream system. In the weak-coupling limit, a sufficient condition for magnetic instabilities is the  momentum-space anisotropy of the one-particle distribution function. We will pay particular attention to (chromo-)magnetic instabilities, which are relevant for the QGP produced in relativistic heavy-ion collisions.

\subsection{Outline of the paper}
\label{subsec-outline}

The goal of this review is to present results on the weakly-coupled unstable QGP and to make a pedagogical introduction to the theoretical tools which are used to derive the results.  In Section \ref{sec-field-theory} we discuss field theory methods and, after a brief presentation of QCD,  we discuss the real-time (or Keldysh-Schwinger) formalism for non-equilibrium systems. The real-time formalism is then used to obtain the gluon polarization tensor and quark self-energy which, in particular, provide dispersion relations for collective excitations in the QGP.  In this Section we also present the so-called hard-loop effective action describing long wavelength phenomena in QGP in a gauge-invariant manner.

The two subsequent sections are devoted to kinetic theory. Sec.~\ref{sec-kinetic-theory-class} presents classical transport theory where a color charge is represented by a continuous variable. In Section \ref{sec-kinetic-theory-quant} the transport theory of quantum color is introduced. In both cases, a derivation of the kinetic equations of quarks and gluons is given. Using the equations, a linear response analysis is performed. In this way, we obtain the gluon polarization tensor, which was earlier computed diagrammatically. Since EMPs are often described in terms of fluid dynamics, an analogous approach to the QGP is presented in Sec.~\ref{sec-chromo-fluid}. We also perform a linear response analysis, showing how the gluon polarization tensor can be found with the help of chromohydrodynamics. 

Having obtained the gluon polarization tensor and quark self energy, we are then ready to discuss the dispersion relations of collective excitations in quark-gluon plasma. We start Sec.~\ref{sec-collective} with a discussion of collective modes in an isotropic QGP which includes the case of an equilibrium QGP. We then proceed to the case of a QGP with a non-equilibrium momentum distribution, starting with the two-stream system, and we derive the spectrum of collective excitations. We find several unstable modes and we then explain the mechanisms of the instabilities. The Nyquist analysis, which allows one to determine the number of solutions of a dispersion equation without solving the equation, and the stability criteria are also discussed.  We then focus on a particular momentum distribution which has proven to be very useful in studying various problems of an unstable QGP.  The distribution is obtained by deforming an isotropic momentum distribution. We present a detailed analysis of the collective modes in such a deformed system in Section~\ref{subsec-deformed}.  We also discuss the limiting cases of extremely prolate or oblate distributions. 

The analytic results presented are very instructive, however, they are usually limited to rather simplified situations. To study phenomena in their full complexity, one has to resort to numerical methods. In Sec.~\ref{sec-simulate} we discuss various numerical methods which have been used to study the temporal evolution of an unstable QGP. In the beginning we present the hard-loop simulations in the static and expanding geometry. Then, we move to purely classical frameworks, which include the classical statistical lattice gauge theory and the color glass condensate approach, where the system's dynamics is fully represented by that of classical non-Abelian fields. Finally, we present the so-called Wong-Yang-Mills simulations where, apart from the classical chromodynamic fields, particles with classical color charges play an important role. The various numerical simulations naturally reproduce the semi-analytic linear response of an unstable QGP, however, the main focus of numerical studies is to determine the late-time non-linear evolution of the QGP after the instability growth saturates.
The review is summarized in Sec.~\ref{sec-conclusions} where our final remarks are also collected. The appendices \ref{app-eps-vs-pi} and \ref{app-abelianization} deal with two rather technical issues.  

We do not discuss in this review a strongly coupled QGP studied in the framework of AdS/CFT duality. Instead, we refer the reader to the review articles \cite{Son:2007vk,CasalderreySolana:2011us} where the framework is introduced and numerous results are presented. 

Throughout the article we use natural units with $\hbar = c = k_{\rm B} = 1$. Our convention of the metric tensor is $g^{\mu\nu} = {\rm diag}(1,-1,-1,-1)$.  The strong coupling constant $\alpha_s \equiv g^2/4\pi$ is assumed to be small and in most places we assume that quarks are massless.


\section{Field-theory methods}
\label{sec-field-theory}  

Properties of the QGP are, in principle, all encoded in the structure of Quantum Chromodynamics (QCD). We therefore open the review with a brief presentation of QCD as a classical field theory.  The quantized theory is discussed in terms of the Keldysh-Schwinger approach which is designed to study quantum fields in and out of equilibrium. The quark and gluon self-energies, which are further used to derive QGP collective modes, are computed perturbatively at one-loop level. Finally, we consider the effective action of soft excitations in the QGP.

\subsection{QCD as a classical field theory}                                                                            
\label{subsec-QCD-class}   

QCD is a non-Abelian gauge theory which governs the propagation and interactions of quark and gluon fields. In Nature there are three colors and the gauge group is ${\rm SU}(3)$. For generality, we consider QCD with $N_c$ colors in which case the gauge group is ${\rm SU}(N_c)$. The quark fields $\psi_q^i(x)$ are Dirac bispinors which additionally carry a flavor index $q =1, \, 2, \, \ldots N_f$ and a fundamental color index  $i =1, \, 2, \, \ldots N_c$.  In what follows $x\equiv (t, {\bf x})$ is the position four-vector. There are in Nature six quark flavors ($u,d,c,s,t,b$) with the masses ranging from 2 MeV for the up $(u)$ quark to 170 GeV for the top $(t)$ quark. Since the temperature of QGP is usually in the interval 200-600 MeV, the plasma is mostly composed of light quarks ($u,d$) with an admixture of strange quarks ($s$). So, we typically have $N_f = 2$ or $N_f = 3$ in QGP studies. 

The quark fields transform under local gauge transformation as
\be
\label{quark-gauge-transform}
\psi_q(x) \rightarrow U(x) \psi_q(x) , 
\;\;\;\;\;\;\;
\bar{\psi}_q(x) \rightarrow \bar{\psi}_q(x) U^\dagger(x) ,
\ee
where $U(x)$ is a unitary $N_c \times N_c$ matrix belonging to the fundamental representation of the ${\rm SU}(N_c)$ group. The transformation matrix $U(x)$ is usually parametrized as
\be
\label{U-gauge}
U(x) = e^{i \omega^a(x) \, \tau^a} ,
\ee
where $\omega^a(x) $ with $a = 1,\,2, \, \dots N_c^2-1$ are real functions of $x$ and $\tau^a$ are the generators of the  fundamental representation of ${\rm SU}(N_c)$. The generators obey the commutation relations
\be
\label{SUN-algebra}
[\tau^a, \tau^b] = i f^{abc} \tau^c ,
\ee
where $f^{abc}$ are totally antisymmetric ${\rm SU}(N_c)$ structure constants. The generators are hermitian traceless matrices and, as a result, the matrix (\ref{U-gauge}) is automatically unitary and its determinant equals unity.  The generators are normalized in the canonical way
\be
\label{tr-tau-tau}
\Tr [\tau^a \tau^b] =\frac{1}{2}\delta^{ab} .
\ee

The gauge field $A^\mu$, which describes gluons, is a four-vector with the Lorentz index $\mu = 0, \,1, \, 2, \,3$. It is usually written either in the fundamental representation, in which case it is an $N_c \times N_c$ hermitian traceless matrix, or in the adjoint representation in which case one has $N_c^2 -1$ real functions $A^\mu_a$. The fields in the two representations are related as $A^\mu = A^\mu_a \tau^a$. The transformation law of the gauge field is deduced from the requirement of gauge invariance of the QCD Lagrangian (see below).

The Lagrangian density of QCD in Minkowski space equals
\be
\label{lagrangian-QCD}
{\cal L}_{\rm QCD} = -\frac{1}{2}
\Tr[F^{\mu \nu} F_{\mu \nu}]
                     + \sum_{q=1}^{N_f} {\bar \psi}_q  
                                \left( i \gamma_\mu D^\mu - m_q \right) 
                                \psi_q  + {\cal L}_{\rm gf},
\ee
where $D^\mu$ is the gauge-covariant derivative in the fundamental representation 
\be
\label{covariant-derivative}
D^{\mu} \equiv \partial^\mu \, \mathds{1} - i g A^\mu ,
\ee
which acts on a color vector such as $\psi_q$; $g$ is the QCD coupling constant. Above, $F^{\mu \nu}$ is the chromodynamic field strength tensor, which in the fundamental representation is
\be
\label{strength-tensor}
F^{\mu \nu} \equiv \frac{i}{g} \,
               [D^\mu,D^\nu] = \partial^\mu A^\nu - \partial^\nu A^\mu 
                                - i g [A^\mu, A^\nu] .
\ee
In the Lagrangian, $\gamma_\mu$ are the Dirac matrices, $m_q$ are bare quark masses, and ${\cal L}_{\rm gf}$ is the general gauge fixing term which is usually introduced when QCD is quantized.

Demanding gauge invariance of the fermion term in the Lagrangian (\ref{lagrangian-QCD}), one finds that, when the quark field transforms according to Eq.~(\ref{quark-gauge-transform}), the chromodynamic field in the fundamental representation has to transform as
\be
A^\mu (x) \rightarrow U(x) \, A^\mu (x) U^\dagger (x) 
+ \frac{i}{g} U(x) \partial^\mu U^\dagger (x) .
\ee
As a consequence, the field strength tensor transforms covariantly 
\be
F^{\mu \nu}(x) \rightarrow U(x) \, F^{\mu \nu} (x) U^\dagger (x) ,
\ee
and the first term in the Lagrangian (\ref{lagrangian-QCD}) is automatically gauge invariant. 

When the chromodynamic field and field strength tensor belong to the adjoint representation of the ${\rm SU}(N_c)$ group, the latter can be expressed through the former one as
\be
F^{\mu \nu}_a = \partial^\mu A^\nu_a - \partial^\nu A^\mu_a 
+ g f^{abc}A^\mu_b A^\nu_c .
\ee
The gauge transformation laws in the adjoint representation read
\be
\label{gauge-trans-adjoint}
A^\mu_a (x) \rightarrow {\cal U}_{ab}(x) \, A^\mu_b (x) + \frac{i}{g} C^\mu_a(x),
\;\;\;\;\;\;\;\;
F^{\mu \nu}_a (x) \rightarrow {\cal U}_{ab}(x) \, F^{\mu \nu}_b (x) ,
\ee
where
\be
\label{M-C-def}
{\cal U}_{ab}(x) \equiv 2\Tr [\tau^a U(x) \, \tau^b U^\dagger (x) ] ,
\;\;\;\;\;\;\;\;
 C^\mu_a(x) \equiv 2\Tr[ \tau^a  U(x) \partial^\mu U^\dagger (x) ].
\ee
The matrix ${\cal U}_{ab}$ and the vector $C^\mu_a$ acquire a simple form for infinitesimally small gauge transformations. 
Substituting the parametrization (\ref{U-gauge}) into the definitions (\ref{M-C-def}) and keeping only the terms linear in $\omega^a$, one obtains
\be
\label{gauge-trans-adjoint-small}
{\cal U}_{ab}(x) \approx \delta^{ab} + f^{abc} \omega^c(x) ,
\;\;\;\;\;\;\;\;
 C^\mu_a(x) \approx -i \partial^\mu \omega^a(x) .
\ee

The  Lagrangian (\ref{lagrangian-QCD}) leads to the equations of motion of quark and gluon fields
\ba
\label{Dirac-eq}
&&\left( i \gamma_\mu D^\mu - m_q \right)  \psi_q  = 0,
\\
 \label{YM-eq}
&&D_\mu F^{\mu \nu}  \equiv \partial_\mu F^{\mu \nu} -ig [A_\mu, F^{\mu \nu}] =  j^\nu ,
\ea
where the chromodynamic gauge field is in the fundamental representation and the color current is defined as $ j^\mu = j^\mu_a \tau^a$ with 
\be
\label{quark-current}
j^\mu_a \equiv - g \sum_{q=1}^{N_f} {\bar \psi}_q  \tau^a \gamma^\mu \psi_q .
\ee
The sign of the current (\ref{quark-current}) is such that the Lagrangian of quark-gluon interaction is ${\cal L}_{\rm int}^{qg} \equiv -  j^\mu_a A^a_\mu$.  We note that the form of the covariant derivative depends on whether it acts on a color vector as $\psi_q$ or a color tensor as $F^{\mu \nu}$. We also note that the current is not conserved but it is covariantly conserved {\it i.e.} $D_\mu j^\mu = 0$.

The equation of motion of the chromodynamic field in the adjoint representation is
\be
\label{YM-eq-adjoint}
{\cal D}_\mu^{ab} F^{\mu \nu}_b  =  j^\nu_a \, ,
\ee
where the covariant derivative is given by
\be 
\label{cov-D-adjoint}
{\cal D}^\mu_{ab} \equiv \partial^\mu \delta^{ab} - g f^{abc} A^\mu_c.
\ee
The generators of the ${\rm SU}(N_c)$ group in the adjoint representation are $(N_c^2 -1) \times (N_c^2 -1)$ hermitian traceless matrices of the form
\be
\label{ad-gen}
(T^a)_{bc} \equiv - if^{abc}.
\ee
Therefore, the covariant derivative can be written down analogously to the fundamental representation, {\em i.e.} 
\be
\label{cov-D-adjoint-tensor}
{\cal D}^\mu \equiv \partial^\mu \mathds{1} - ig T^a A^\mu_a = \partial^\mu \mathds{1} - ig {\cal A}^\mu,
\ee
where ${\cal A}^\mu \equiv  T^a A^\mu_a$ is a color matrix of dimension $(N_c^2 - 1) \times (N_c^2 - 1)$.  We note that, as in the fundamental representation, the covariant derivative in the adjoint representation depends on whether it acts on a color vector with $(N_c^2 - 1)$ components  or on a color tensor which is an $(N_c^2 - 1) \times (N_c^2 - 1)$ matrix. In the former case, the covariant derivative is given by the formula (\ref{cov-D-adjoint}) or (\ref{cov-D-adjoint-tensor}). When the covariant derivative acts on a color tensor, say the strength tensor  ${\cal F}^{\mu \nu} \equiv T^a F^{\mu \nu}_a$,  we have
\be
{\cal D}_\mu {\cal F}^{\mu \nu}  \equiv \partial_\mu {\cal F}^{\mu \nu} -ig [{\cal A}_\mu, {\cal F}^{\mu \nu}].
\ee

We also write down here the Yang-Mills equation (\ref{YM-eq-adjoint}) in the form which will be applied to derive a gluon transport equation. Since 
\be
{\cal F}^{\mu \nu} \equiv T^a F^{\mu \nu}_a 
= \partial^\mu {\cal A}^\nu - \partial^\nu {\cal A}^\mu - ig [{\cal A}^\mu, {\cal A}^\nu ] 
=\frac{i}{g} [{\cal D}^\mu, {\cal D}^\nu ] ,
\ee
Eq.~(\ref{YM-eq-adjoint}) reads
\be
\label{YM-eq-D2-1}
\Big[g^{\mu \nu} {\cal D}^2 -  {\cal D}^\mu {\cal D}^\nu  + i g {\cal F}^{\mu \nu} \Big] A_\mu = j^\nu .
\ee
Using the relation 
\be
{\cal D}^\mu {\cal D}^\nu = {\cal D}^\nu {\cal D}^\mu + [{\cal D}^\mu ,{\cal D}^\nu]  
= {\cal D}^\nu {\cal D}^\mu - ig {\cal F}^{\mu \nu} ,
\ee
Eq.~(\ref{YM-eq-D2-1}) can be also written in the form
\be
\label{YM-eq-D2}
\Big[g^{\mu \nu} {\cal D}^2 -  {\cal D}^\mu {\cal D}^\nu  - 2 i g {\cal F}^{\mu \nu} \Big] A_\nu = j^\mu ,
\ee
which will be used in Sec.~\ref{sec-glue-trans-eq}.

\subsection{Keldysh-Schwinger Formalism for Non-Equilibrium Field Theories}                                                                            
\label{subsec-Keldysh-Schwinger}   

The Keldysh-Schwinger formulation of  quantum field  theory \cite{Schwinger:1960qe,Keldysh:1964ud}  provides a natural framework to study equilibrium and non-equilibrium many body systems. We will present here a simplified version of the formalism, assuming that the plasma's momentum distribution is, in general, anisotropic but the plasma is homogeneous (translationally invariant) in coordinate space. Our presentation of the Keldysh-Schwinger formalism follows closely the study  \cite{Mrowczynski:1992hq}.

\begin{figure}[t]
\centerline{\includegraphics[width=8cm]{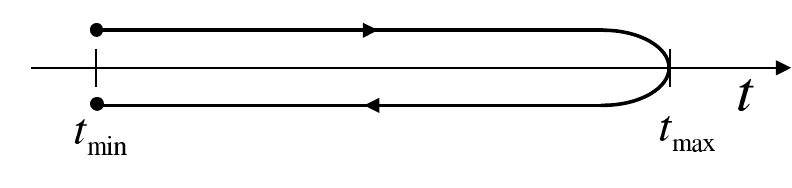}}
\caption{The contour in the complex time plane for evaluation of operator 
expectation values.} 
\label{fig-contour}
\end{figure}

The function which plays a central role in the Keldysh-Schwinger formalism is the so-called contour Green's function. For a spinor field $\psi(x)$ and for a vector field $A^\mu(x)$  the contour Green's functions are defined as
\ba
\label{contour-S}
i\big( S(x,y) \big)_{\alpha \beta}^{ij} \buildrel \rm def \over 
= \langle  \tilde T [\psi_\alpha^i (x) \bar\psi_\beta^j (y) ]\rangle ,
\\ 
\label{contour D}
i\big( D (x,y) \big)_{\mu \nu}^{ab} \buildrel \rm def \over 
= \langle  \tilde T [A_\mu^a (x) A_\nu^b (y)] \rangle \, ,
\ea
where $\alpha, \beta = 1,\,2,\,3,\,4$ and $\mu, \nu= 0,\,1,\,2,\,3$ are spinor and Lorentz indices, respectively; $i, j = 1,\,2, \dots N_c$ and $a,b = 1,\,2, \dots N_c^2-1$ are color indices of the fundamental and adjoint representation of the ${\rm SU}(N_c)$ group, respectively; $\langle ...\rangle$ denotes the ensemble average at time $t_0$ (usually  taken to be $-\infty $) which should be understood as $\langle \dots \rangle \equiv \Tr[\rho(t_0) \dots ]$ with $\rho(t_0)$ being the density matrix at the time $t_0$; $\tilde T$ is the time ordering operation along the directed contour shown in Fig.~\ref{fig-contour}. The parameter $t_{\rm max}$ is taken to $+\infty$ in calculations.  The time arguments have an infinitesimal positive or negative imaginary part which places them on the upper or lower branch of the contour. The operation of ordering the operators $X(x)$ and $Y(y)$ along the contour is defined as
\ba
\label{c-ordering}
\tilde T [X (x) Y (y)] \buildrel \rm def \over = 
\tilde\Theta (x_0,y_0) X (x) Y (y)  \pm \tilde\Theta (y_0,x_0) Y(y) X (x) ,
\ea
where $\tilde\Theta (x_0,y_0)$ equals 1 if $x_0$ succeeds $y_0$ on the contour, and  0 if $x_0$  precedes $y_0$; the upper sign is for bosonic operators and the lower one applies when $X$ and $Y$ are fermionic operators.

The contour propagator involves four functions, which can be thought of as corresponding to propagation  along the top branch of the contour, from the top branch to the bottom branch, from the bottom branch to the top branch, and along the bottom branch.  We first define four Green's functions with real time arguments and then discuss their relationship to the contour Green's function.  These are
\ba 
\label{contour-parts}
i\big(S^> (x,y)  \big)_{\alpha \beta}^{ij} &\buildrel \rm def \over =&
 \langle  \psi_\alpha^i (x) \bar\psi_\beta^j (y) \rangle , 
\\ [2mm] 
i\big(S^<  (x,y) \big)_{\alpha \beta}^{ij} &\buildrel \rm def \over =&
- \langle \bar\psi_\beta^j (y) \psi_\alpha^i (x) \rangle , 
\\ [2mm] 
i\big(S^c (x,y) \big)_{\alpha \beta}^{ij} &\buildrel \rm def \over =& 
\langle  T^c [\psi_\alpha^i (x) \bar\psi_\beta^j (y) ]\rangle , 
\\ [2mm] 
i\big(S^a (x,y) \big)_{\alpha \beta}^{ij} &\buildrel \rm def \over = &
\langle  T^a [\psi_\alpha^i (x) \bar\psi_\beta^j (y) ]\rangle , 
\ea
where $T^c\,(T^a)$ indicates chronological (anti-chronological) time ordering
\ban
T^c  [X (x) Y (y)] &\buildrel \rm def \over =& 
\Theta (x_0-y_0) X (x) Y (y)  \pm \Theta (y_0-x_0) Y(y) X (x) ,
\\
T^a [X (x) Y (y)] &\buildrel \rm def \over = &
\Theta (y_0-x_0) X (x) Y (y)  \pm \Theta (x_0-x_0) Y(y) X (x) .
\ean
The upper and lower signs are again for bosonic and fermionic operators, respectively.

The functions $D^>, \; D^<, \; D^c$, and $D^a$ are defined analogously to $S^>, \; S^<, \; S^c$, and $S^a$ with the only difference being that there is no minus sign in the definition of $D^<$, {\em i.e.} $iD^<_{\mu \nu}  (x,y) \buildrel \rm def \over =  \langle A_\nu (y) A_\mu (x) \rangle$. The functions carrying indices $>, \; <, \; c$, and $a$ are related to the contour Green's function in the following manner:
\ba 
\label{ca><}
S^c(x,y) \equiv S (x,y) && \text{for $x_0$ and $y_0$ on the upper branch,} 
\\ [2mm] \nonumber
S^a(x,y) \equiv S (x,y) && \text{for $x_0$ and $y_0$ on the lower branch,} 
\\ [2mm] \nonumber 
S^<(x,y) \equiv S (x,y) && \text{for $x_0$ on the upper branch and}  
\\ \nonumber
&& \text{$y_0$ on the lower branch,}  
\\ [2mm] \nonumber 
S^>(x,y) \equiv S (x,y) && \text{for $x_0$ on the lower branch and}  
\\ \nonumber
&& \text{$y_0$ on the upper branch.} 
\ea

Instead of the contour Green's function, one can equivalently use the matrix formulation where
\be
\label{matrix-form}
S  = \left(  
   \begin{matrix} 
   S_{11} & S_{12} \cr
   S_{21} & S_{22} \cr
   \end{matrix}  
   \right) 
= \left(  
   \begin{matrix} 
   S^c & S^< \cr
   S^> & S^a \cr
   \end{matrix}  
   \right) .
\ee
Directly from the definitions, one finds the following identities 
\ba
\label{ident-c}
&& S^{\ca}(x,y) = \Theta (x_0 - y_0) S^{\gl}(x,y) +
\Theta (y_0 - x_0) S^{\lg}(x,y) , 
\\ [2mm] 
&&~~~~~~~~~~~~ 
(i S^{\ca}(x,y) \gamma^0 )^\dagger = iS^{\ac}(x,y) \gamma^0, 
\\ [2mm] 
&&~~~~~~~~~~~~
(i S^{\gl}(x,y)\gamma^0)^\dagger = i S^{\gl}(x,y)  \gamma^0,
\ea
where a $\dagger$ denotes hermitian conjugation which involves an interchange of the arguments of the Green's function. In the analogous relations for the Green's functions of gauge fields, the gamma matrices $\gamma^0$ are absent. The relation (\ref{ident-c}) shows that the four components of the contour Green's function are not independent from each other, but instead they obey the identity
\ba
\label{circling}
S^c(x,y) + S^a(x,y) - S^<(x,y)- S^>(x,y)=0 .
\ea

One often needs to work with the retarded $(+)$, advanced $(-)$, and symmetric Green's functions which are defined as
\ba
\label{ret-xspace}
S^\pm (x,y) &\buildrel \rm def \over=& 
= \pm \Theta(\pm x_0 \mp y_0) \big(S^>(x,y) - S^<(x,y) \big) , 
\\ [2mm] 
\label{sym-xspace}
S^{\rm sym}(x,y) &\buildrel \rm def \over=& 
S^>(x,y) + S^<(x,y) .
\ea

Let us briefly discuss the physical interpretation of the Green's functions we have introduced above. The functions labeled with the indices $c, ~a,~ +$, and $-$ describe the propagation of a disturbance in which a single particle or antiparticle is added to a many-particle system at space-time point $y$ and then is removed from it at a space-time  point $x$. The function labeled with $c$ describes a particle disturbance propagating forward in time and an antiparticle disturbance propagating backward in time. The meaning of the function labeled with $a$ is analogous but particles are propagated backward in time and antiparticles forward in time. In the zero density limit, the function labeled with $c$ coincides with the Feynman propagator. In the case of the retarded (advanced) Green functions, both particles and antiparticles evolve forward (backward) in time. 

The physical meaning of the functions labeled with $\lg$ is more transparent when one considers their Wigner transforms, which are given by
\be
\label{Wigner-trans-def}
S^{\lg}(X,p) \buildrel \rm def \over = \int d^4u \; e^{ip \cdot u}
S^{\lg}(X+{1 \over 2}u,X-{1 \over 2}u) .
\ee
Then, the functions $S^{\lg}(X,p)$ appear to be the phase-space densities. For example, the current $\langle \bar\psi (X) \gamma^\mu \psi (X) \rangle$ is expressed as
\be
\langle \bar\psi (X) \gamma^\mu \psi (X) \rangle 
= i \int {d^4p \over (2\pi)^{4}} \Tr[ \gamma^\mu S^>(X,p) ] .
\ee
The functions labeled with $\lg$ are the phase-space densities and, as a result, are the quantum analog of classical distribution  functions. The functions $iS^\lg$ and $iD^\lg$ are hermitian, but they are not positive definite, and thus the probabilistic interpretation is only approximately valid. One should also note that, in contrast to the classical distribution function, the functions labeled with $\lg$ are in general nonzero for off-mass-shell four-momenta. 

The evolution along the contour illustrated in Fig.~\ref{fig-contour} is formally very similar to evolution along the real time axis, and consequently it is possible to define a perturbation expansion on such a contour.  As a consequence, one can perturbatively compute the contour Green's functions or the contour self-energies. The calculations, however, involve summations over contributions from both branches of the contour and, as a result, are more difficult than their vacuum counterparts to compute.  The self-energies with real-time arguments can be extracted from the contour self-energy. 

Let us discuss the free Green's functions which are needed for perturbative calculations presented in the subsequent sections. We adopt three important simplifications. 
\begin{itemize}
\item The system under study  is assumed to be {\em on average homogeneous} and therefore is {\em translationally invariant}. Consequently, the functions $S(x,y)$ and $D(x,y)$ are independent of $X \equiv (x+y)/2$.  In this case, the Wigner transform defined via 
Eq.~(\ref{Wigner-trans-def}) is simply the Fourier transform with respect to $x-y$. 

\item The QGP is assumed to be {\em locally colorless}. The functions  $S(x,y)$ and $D(x,y)$ are then unit matrices in color space; $S(x,y)$  is the matrix  $N_c \times N_c$ and $D(x,y)$ is  the matrix  $(N_c^2 -1) \times (N_c^2 -1)$. 

\item The QGP is assumed to be unpolarized, {\it i.e.} the spin states of quarks, antiquarks and gluons are uniformly distributed.
\end{itemize}

Taking into account these simplifying assumptions, the Green's functions of the free massless quark field are found to be
\ba
\label{S-pm}
\big(S^{\pm}(p) \big)_{\alpha \beta}^{ij} 
&=&  \frac{\delta^{ij} {p\sla}_{\!\alpha \beta}}{p^2\pm i\, {\rm sgn}(p_0)0^+},
\\
\label{S->}
\big(S^>(p) \big)_{\alpha \beta}^{ij} 
&=& \delta^{ij} \frac{i\pi}{E_{\bf p}} {p\sla}_{\!\alpha \beta} 
\Big( \delta (E_{\bf p} - p_0)  \big[ n_q ({\bf p}) -1\big]
+ \delta (E_{\bf p} + p_0) \bar n_q (-{\bf p}) \Big),
\\
\label{S-<}
\big(S^<(p) \big)_{\alpha \beta}^{ij}
&=& \delta^{ij} \frac{i\pi}{E_{\bf p}} {p\sla}_{\!\alpha \beta} \Big( \delta (E_{\bf p} - p_0)  n_q({\bf p})
+ \delta (E_{\bf p} + p_0) \big[ \bar n_q (-{\bf p}) - 1\big] \Big),
\ea
where $0^+$ is an infinitesimally small positive number, $E_{\bf p} \equiv |{\bf p}|$, $n_q({\bf p})$ and $\bar n_q({\bf p})$ are the distribution functions of quarks and of antiquarks, respectively.  The functions are normalized in such a way that the quark density of a given flavor equals
\be
\label{norm-quark-n}
\rho_q = 2N_c \int \frac{d^3p}{(2\pi)^3}\, n_q({\bf p}) ,
\ee
where the factor of 2 takes into account the two spin states of each quark. The functions (\ref{S-pm}), (\ref{S->}), and (\ref{S-<}) obey the identity $S^>(p) - S^< (p) = S^+(p) - S^-(p)$. The contributions to $S^\gl(p)$ proportional to  $n_q({\bf p})$ or $\bar n_q({\bf p})$ represent effects of the plasma, the remaining part of $S^\gl(p)$ is the vacuum contribution.

The Green's functions of the free gluon field in Feynman gauge are \cite{Czajka:2014eha}
\ba
\label{D-pm}
\big(D^\pm \big)^{ab}_{\mu\nu}(p) &=&- \frac{g_{\mu\nu} \delta^{ab}}{p^2 \pm i\textrm{sgn}(p_0)0^+},
\\ [2mm]
\label{D->}
\big(D^> \big)^{ab}_{\mu\nu}(p) &=& \frac{i\pi}{E_{\bf p}} g_{\mu\nu} \delta^{ab} 
\Big[\delta(E_{\bf p}-p_0)\big(n_g({\bf p})+1\big) + \delta(E_{\bf p}+p_0)n_g(-{\bf p})\Big], 
\\[2mm]
\label{D-<}
\big(D^< \big)^{ab}_{\mu\nu}(p) &=& \frac{i\pi}{E_{\bf p}} g_{\mu\nu} \delta^{ab} 
\Big[\delta(E_{\bf p}-p_0)n_g({\bf p}) + \delta(E_{\bf p}+p_0)\big(n_g(-{\bf p})+1\big)\Big],
\ea
where $n_g({\bf p})$ is a distribution function of gluons which is normalized in such a way that the gluon density is given as
\be
\label{norm-gluon-n}
\rho_g = 2 (N_c^2 -1)  \int \frac{d^3p}{(2\pi)^3}\, n_g ({\bf p}) ,
\ee
where the factor of 2 takes into account the two gluon spin states. So, the function $n_g({\bf p})$  takes into account only the {\it physical} transverse gluons.

Since there are unphysical gluon degrees of freedom in the Feynman gauge, one needs the Faddeev-Popov ghosts to eliminate them. The Green's functions of the free ghost field, which are obtained from the gluon ones by means of the Slavnov-Taylor identity, are \cite{Czajka:2014eha}
\ba
\label{G-pm}
\Delta_{ab}^\pm (p)&=&\frac{\delta_{ab}}{p^2 \pm i \textrm{sgn}(p_0)0^+}, 
\\[2mm]
\label{G->}
\Delta_{ab}^>(p)&=& - \delta^{ab}  \frac{i\pi}{E_{\bf p}}
\Big[ \delta(E_{\bf p}-p_0)\big(n_g({\bf p})+1\big) + \delta(E_{\bf p}+p_0)n_g(-{\bf p}) \Big], 
\\[2mm]
\label{G-<}
\Delta_{ab}^<(p)&=& - \delta^{ab}  \frac{i\pi}{E_{\bf p}}
\Big[ \delta(E_{\bf p}-p_0)n_g({\bf p}) + \delta(E_{\bf p}+p_0)\big(n_g(-{\bf p})+1\big) \Big] .
\ea
As can be seen, the distribution function $n_g({\bf p})$ of {\it physical} gluons enters the ghost Green's functions.

\subsection{Self-energies}
\label{subsec-self-energies}                                                                                

In this section we compute the one-loop retarded self-energies of gluons and quarks in the Hard Loop Approximation which were first computed for an anisotropic QGP in \cite{Mrowczynski:2000ed}. However, the first computation of the gluon polarization tensor was not quite complete and we follow here the detailed analysis presented in \cite{Czajka:2014eha}. The self-energies will be used in Sec.~\ref{sec-collective} to discuss collective modes 
in QGP.  

\begin{figure}[t]
\begin{center}
\includegraphics[width=11cm]{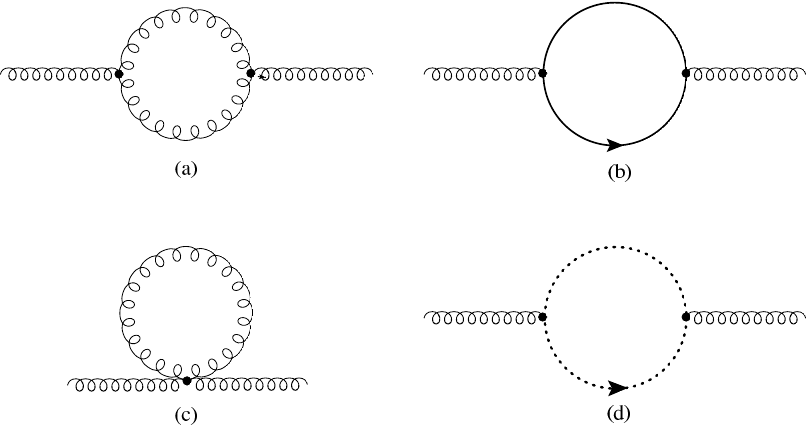}
\end{center}
\caption{Feynman graphs which contribute to the one-loop gluon 
polarization tensor. The curly, plain, and dotted lines denote, respectively, gluon, quark, and ghost fields.}
\label{fig-glue-self-energy}
\end{figure}

As already noted, the Green's functions of a many-body system with the field operators ordered along the contour shown in Fig.~\ref{fig-contour} have a perturbative expansion which is very similar to that of the time-ordered vacuum Green's functions. However, the integrals along the real time axis are replaced by the integrals along the contour.  Special treatment is also required to deal with the so-called tadpole diagrams, where a loop is attached to the remaining part of the diagram through a single vertex. An example of a tadpole is shown in Fig.~\ref{fig-glue-self-energy}c.  In the perturbative expansion of the time-ordered vacuum Green's function, the tadpole diagrams are usually eliminated due to the normal ordering of field operators. In the many-body perturbative expansion the tadpole diagrams represent effects of a finite density of real quanta such as quarks, antiquarks, and gluons which are constituents of QGP. 

After the interacting contour Green's function is diagrammatically expressed through the free Green's functions, one infers its components with the indices $>, <, c$, and $a$. In general it is a difficult technical task which is, however, quite simple in the case of one-loop diagrams considered here.

\subsubsection{Polarization tensor}
\label{subsubsec-polarization-tensor}                                                                                

The polarization tensor $\Pi$ can be defined by means of the Dyson-Schwinger equation
\be
i{\cal D} = i D + i D \, (i\Pi) \, i{\cal D} ,
\ee
where ${\cal D}$ and $D$ are the interacting and free gluon propagators, respectively. The lowest order contributions to $\Pi^{\mu \nu}$ are given by the four diagrams shown in Fig.~\ref{fig-glue-self-energy}. 

Applying the Feynman rules, the contribution to the contour $\Pi$ coming from the quark loop corresponding to the graph from Fig.~\ref{fig-glue-self-energy}b is immediately written down in coordinate space as
\be
 i \big( \Pi_q (x,y) \big)^{\mu \nu}_{ab} = (-1)(-ig)^2 
{\rm Tr} [\gamma ^\mu \tau^a iS (x,y) \gamma ^\nu \tau^b iS(y,x)] ,
\ee
where the factor $(-1)$ occurs due to the fermion loop. The trace is taken over the spinor and fundamental  color indices. Using the relation (\ref{tr-tau-tau}), one obtains
\be
\label{contour-Pi}
\big( \Pi_q (x,y) \big)^{\mu \nu}_{ab} = - \frac{i}{2} g^2 \delta^{ab}
{\rm Tr} [\gamma ^\mu S (x,y) \gamma ^\nu S(y,x)] ,
\ee
where the trace is now taken only over the spinor indices. We see that the polarization tensor is a unit matrix in color space which results from the fact that the QGP is assumed to be locally colorless. 

We are actually interested in the retarded polarization tensor. The retarded $(+)$ and advanced $(-)$ polarization tensors are given by
\be
\label{Pi-plus-minus}
\Pi^{\pm} (x,y) = \Pi_\delta (x) \,\delta^{(4)}(x-y)  \pm \Theta(\pm x_0 \mp y_0)
\Big( \Pi^> (x,y) - \Pi^< (x,y) \Big).
\ee
where the first term represents a contribution due to one-point tadpole diagrams such as that shown in Fig.~\ref{fig-glue-self-energy}c. The polarization tensors $\Pi^\lg(x,y)$ are found from the contour tensor (\ref{contour-Pi}) by locating the argument $x_0$ on the upper (lower) and $y_0$ on the lower (upper) branch of the contour. Consequently, one obtains
\be
\label{Pi->-<-x-y}
\big( \Pi_q^{\lg} (x,y) \big)^{\mu \nu}_{ab}
= - \frac{i}{2} g^2  \delta^{ab}
{\rm Tr} [\gamma^\mu S^\lg (x,y) \gamma^\nu S^\gl (y,x)] .
\ee

As already mentioned, the system under consideration is assumed to be translationally invariant. As a consequence, the two-point functions as $S(x,y)$ effectively depend on $x$ and $y$ only through $x-y$. Therefore, we can set $y=0$ and write $S(x,y)$ as $S(x)$ and $S(y,x)$ as $S(-x)$. Equation (\ref{Pi->-<-x-y}) then becomes
\be
\label{Pi->-<-x-y-1}
\big( \Pi_q^{\lg} (x) \big)^{\mu \nu}_{ab}
= - \frac{i}{2} g^2 \delta^{ab}
{\rm Tr} [\gamma^\mu S^\lg (x) \gamma^\nu S^\gl (-x)] .
\ee
Since
\be
S^{\pm} (x) = \pm \Theta(\pm x_0)
\Big(S^> (x) - S^< (x) \Big) ,
\ee
the retarded polarization tensor $\Pi_q^+ (x)$ is found to be 
\be
\label{Pi-x-1}
\big( \Pi_q^+ (x) \big)^{\mu \nu}_{ab}
 = - i \frac{g^2}{4}  \delta^{ab}
{\rm Tr}\big[\gamma _\mu S^+(x) \gamma _\nu S^{\rm sym}(-x)
+ \gamma _\mu S^{\rm sym}(x) \gamma _\nu S^-(-x) \big].
\ee

In momentum space this becomes
\be
\label{Pi-k-e-1}
\big( \Pi_q^+ (k) \big)^{\mu \nu}_{ab}
= -  i\frac{g^2}{4}  \delta^{ab}
 \int \frac{d^4p}{(2\pi )^4}
{\rm Tr} \big[\gamma^\mu S^+(p+k)\gamma ^\nu S^{\rm sym}(p)
+ \gamma ^\mu S^{\rm sym}(p) \gamma ^\nu S^-(p-k) \big].
\ee
Substituting the functions $S^{\pm}$ and $S^{\rm sym} \equiv S^> + S^<$ given by Eqs.~(\ref{S-pm}), (\ref{S->}), and (\ref{S-<}) into the formula (\ref{Pi-k-e-1}), one finds
\ba
\label{Pi-k-e-3}
\big( \Pi_q^+ (k) \big)^{\mu \nu}_{ab}
&=&
- \frac{g^2}{8}  \delta^{ab}
 \int \frac{d^3p}{(2\pi )^3} \, 
\frac{n_q ({\bf p}) + \bar{n}_q ({\bf p}) -1}{E_{\bf p}}
\\ \non
&&\times
{\rm Tr} \bigg[
\bigg(
\frac{ \gamma^\mu (p\sla + k\sla)\gamma ^\nu  p\sla
+ \gamma ^\mu p\sla \gamma ^\nu(p\sla + k\sla)}
{(p+k)^2 + i\, {\rm sgn}\big((p+k)_0\big)0^+}
+  \frac{\gamma ^\mu p\sla \gamma ^\nu(p\sla -k\sla)
+  \gamma^\mu (p\sla - k\sla)\gamma ^\nu  p\sla}
{(p-k)^2 - i\, {\rm sgn}\big((p-k)_0\big)0^+}
\bigg) \bigg],
\ea
where, after performing the integration over $p_0$, the momentum ${\bf p}$ was changed into  $-{\bf p}$ in the negative energy contribution. Computing the traces of gamma matrices and using $p^2 =0$,  one finds
\ba
\label{Pi-k-e-4}
\big( \Pi_q^+ (k) \big)^{\mu \nu}_{ab}
&=& 
 - g^2   \delta^{ab} 
\int \frac{d^3p}{(2\pi )^3} \, \frac{n_q ({\bf p}) + \bar{n}_q ({\bf p}) -1}{E_{\bf p}}
\\ \non
&& \times
\bigg(
\frac{2p^\mu p^\nu  + k^\mu p^\nu + p^\mu k^\nu - g^{\mu \nu} (k \cdot p)  }
{(p+k)^2 + i\, {\rm sgn}\big((p+k)_0\big)0^+}
+ \frac{2p^\mu p^\nu  -  k^\mu p^\nu - p^\mu k^\nu + g^{\mu \nu} (k \cdot p) }
{(p-k)^2 - i\, {\rm sgn}\big((p-k)_0\big)0^+}
\bigg) .
\ea

We are interested in collective modes which occur when the wavelength of a quasi-particle is much larger than the characteristic interparticle distance in the plasma. Thus, we look for the polarization tensor in the limit $k^\mu \ll p^\mu$ which is the condition of the Hard Loop Approximation for anisotropic systems  \cite{Mrowczynski:2000ed,Mrowczynski:2004kv}.  The approximation
is implemented by observing that 
\ba
\label{HL-approx-1}
\frac{1}{(p+k)^2 + i0^+}
+ \frac{1} {(p-k)^2 - i0^+}
= \frac{2k^2}{(k^2)^2 - 4 (k\cdot p)^2 - i {\rm sgn}(k\cdot p) 0^+}
\approx -\frac{1}{2}  \frac{k^2}{(k\cdot p + i 0^+)^2} ,
\\ [2mm]
\label{HL-approx-2}
\frac{1}{(p+k)^2 + i0^+}
- \frac{1} {(p-k)^2 - i0^+}
= \frac{4(k \cdot p)}{(k^2)^2 - 4 (k\cdot p)^2 - i {\rm sgn}(k\cdot p) 0^+}
\approx \frac{k\cdot p}{(k\cdot p + i 0^+)^2}.
\ea
We note that $(p+k)_0 > 0$ and $(p-k)_0 > 0$ for $p^\mu \gg k^\mu$. With the above formulas, Eq.~(\ref{Pi-k-e-4}) becomes
\ba
\label{Pi-k-e-final}
\big( \Pi_q^+ (k) \big)^{\mu \nu}_{ab} &=&
g^2    \delta^{ab}
 \int \frac{d^3p}{(2\pi )^3} \, \frac{n_q ({\bf p}) + \bar{n}_q ({\bf p}) -1}{E_{\bf p}} \,
\frac{k^2 p^\mu p^\nu - \big(k^\mu p^\nu + p^\mu k^\nu - g^{\mu \nu} (k \cdot p) \big) (k \cdot p)}
{(k\cdot p + i 0^+)^2} ,
\ea
which has the well-known structure of the polarization tensor of photons in ultrarelativistic QED plasmas, see {\em e.g.} the reviews 
\cite{Mrowczynski:2007hb,Blaizot:2001nr}. As can be seen from this expression, $\Pi_q^+(k)$ is symmetric with respect to Lorentz indices 
\be
\big( \Pi_q^+ (k) \big)^{\mu \nu}_{ab} = \big( \Pi_q^+ (k) \big)^{\nu \mu}_{ab},
\ee 
and transverse 
\be
k_\mu \big( \Pi_q^+ (k) \big)^{\mu \nu}_{ab} = 0,
\ee
as required by gauge invariance.

When $n_q$ and $\bar{n}_q$ both vanish, the polarization tensor (\ref{Pi-k-e-final}) is still nonzero. It is actually infinite and represents the vacuum contribution. Since we are interested in medium effects, the vacuum contribution should be subtracted from the formula (\ref{Pi-k-e-final}). Then, $n_q + \bar{n}_q  -1$ is replaced by $n_q + \bar{n}_q $. Eq.~(\ref{Pi-k-e-final}) gives the contribution of quarks of one flavor. The integral should be multiplied by $N_f$ to obtain the contribution of all quark flavors. 

In analogy to the quark-loop expression (\ref{Pi-k-e-1}), one finds the gluon-loop contribution to the retarded polarization tensor shown in Fig.~\ref{fig-glue-self-energy}a to be
\ba
\nonumber
\big( \Pi_g^+ (k) \big)^{\mu \nu}_{ab}   &=& -i\frac{g^2}{4} N_c \delta_{ab}
 \int \frac{d^4p}{(2\pi )^4}  \int \frac{d^4q}{(2\pi )^4} D^{\rm sym}_0(p)
\Big[ (2\pi)^4 \delta^{(4)}(k+p-q)
M^{\mu \nu} (k,q,p) D^+_0(q)
\\ [2mm]
\label{Pi-gluon-loop-2}
&& \;\;\;\;\;\;\;\;\;\;\;\;\;\;\;\;\;\;\;\;\;\;\;\;\;\;\;\;\;\;\;\;\;\;\;\;\;\;\;\;\;\;\;\;\;\;\;\;\;\;\;\;\;\;\;\;\;\;
+ (2\pi)^4 \delta^{(4)}(k-p+q)
M^{\mu \nu} (k,-q,-p) D^-_0(q)
\Big],
\ea
where the functions $D^\pm_0$ and $D^{\rm sym}_0\equiv D^>_0 + D^<_0$  are the free gluon functions $D^\pm$ and $D^{\rm sym} \equiv D^> + D^<$ given by Eqs.~(\ref{D-pm}), (\ref{D->}), and (\ref{D-<}) stripped of the color and Lorentz factors, {\it i.e.}, $D^{\mu \nu}_{ab} \equiv \delta_{ab} g^{\mu \nu} D_0$. The combinatorial factor $1/2$ is included in Eq.~(\ref{Pi-gluon-loop-2}) and 
\be
\label{tensor-M-def}
M^{\mu \nu} (k,q,p) \equiv
\Gamma^{\mu \sigma \rho} (k,-q,p)
\Gamma^{\;\;\,\nu}_{\sigma \;\; \rho} (q,-k,-p) 
\ee
with the three-gluon coupling
\be
\label{3-g-vertex-2}
\Gamma^{\mu \nu \rho} (k,p,q) \equiv
g^{\mu \nu }(k-p)^\rho
+g^{\nu \rho}(p-q)^\mu +g^{\rho \mu}(q-k)^\nu .
\ee

Within the hard loop approximation, the tensor (\ref{tensor-M-def}) is computed as
\be
\label{tensor-M-F-HL}
M^{\mu \nu} (k,p \pm k, \pm p) \approx \pm 2 g^{\mu \nu} (k\cdot p)
+ 10 p^\mu p^\nu
\pm 5(k^\mu p^\nu + p^\mu k^\nu),
\ee
where we have used $p^2=0$. Substituting the expressions (\ref{tensor-M-F-HL}) into Eq.~(\ref{Pi-gluon-loop-2}) and using  the explicit form of the functions $D^\pm$ and $D^{\rm sym}$, one obtains
\ba
\label{Pi-gluon-loop-5}
\big( \Pi_g^+ (k) \big)^{\mu \nu}_{ab} =
\frac{g^2}{4} N_c \delta_{ab}
 \int \frac{d^3p}{(2\pi )^3} \frac{2n_g({\bf p})+1}{E_{\bf p}}
\frac{5k^2 p^\mu p^\nu - 2 g^{\mu \nu} (k\cdot p)^2
- 5(k^\mu p^\nu + p^\mu k^\nu)(k\cdot p)}{(k\cdot p + i 0^+)^2}.
\ea

The gluon tadpole contribution to the contour $\Pi (x,y)$, which is shown in Fig.~\ref{fig-glue-self-energy}c, is proportional to $\delta^{(4)}(x-y)$ and thus it does not contribute to $\Pi^{\gl}(x,y)$ because it vanishes when the points $x_0$ and $y_0$ are located on different branches of the time contour. The tadpole represents the first term in the formula (\ref{Pi-plus-minus}) and its contribution to the retarded polarization tensor equals
\be
\label{Pi-gluon-tadpole-1}
\big( \Pi_t^+ (k) \big)^{\mu \nu}_{ab} = - \frac{g^2}{2}
 \int \frac{d^4p}{(2\pi )^4}
\Gamma^{\mu \nu \rho}_{abcc, \rho} D^<_0(p)  ,
\ee
where the combinatorial factor $1/2$ is included and the four-gluon coupling $\Gamma^{\mu \nu \rho \sigma }_{abcd}$ equals
\be
\label{4-g-vertex}
\Gamma^{\mu \nu \rho \sigma }_{abcd} \equiv
f_{abe}f_{ecd}(g^{\mu \sigma} g^{\nu \rho} - g^{\mu \rho} g^{\nu \sigma})
+ f_{ace}f_{edb}(g^{\mu \rho} g^{\nu \sigma} - g^{\mu \nu} g^{\rho \sigma})
+ f_{ade}f_{ebc}(g^{\mu \nu} g^{\rho \sigma} - g^{\mu \sigma} g^{\nu \rho}).
\ee
With the explicit form of the function $D^<(p)$ given by Eq.~(\ref{D-<}), the formula (\ref{Pi-gluon-tadpole-1}) gives
\be
\big( \Pi_t^+ (k) \big)^{\mu \nu}_{ab} = \frac{3}{2} g^2 N_c \, \delta_{ab} g^{\mu \nu}
\int \frac{d^3p}{(2\pi )^3} \frac{2 n_g({\bf p}) +1}{E_{\bf p}} .
\ee

The ghost-loop contribution to the retarded polarization tensor, which is shown in Fig.~\ref{fig-glue-self-energy}d, equals
\ba
\label{Pi-ghost-loop-2}
\big( \Pi_{gh}^+ (k) \big)^{\mu \nu}_{ab} &=& i\frac{g^2}{2} N_c \delta_{ab}
 \int \frac{d^4p}{(2\pi )^4} \;\Delta^{\rm sym}_0(p)
\Big[ (p+k)^\mu p^{\nu} \Delta^+_0(p+k)
+ p^\mu (p-k)^\nu \Delta^-_0(p-k) \Big].
\ea
where the factor $(-1)$  is included since we deal with a fermionic loop  and the functions $\Delta^\pm_0$ and $\Delta^{\rm sym}_0 \equiv \Delta^>_0 + \Delta^<_0$ are the ghost functions  $\Delta^\pm$ and $\Delta^{\rm sym} \equiv \Delta^> + \Delta^<$ given by Eqs.~(\ref{G-pm}), (\ref{G->}), and (\ref{G-<}) stripped of the color factor $\delta^{ab}$. Using the explicit forms of the ghost functions (\ref{G-pm}), (\ref{G->}), and (\ref{G-<}), the formula (\ref{Pi-ghost-loop-2}) becomes
\be
\label{Pi-ghost-loop-4}
 \big( \Pi_{gh}^+ (k) \big)^{\mu \nu}_{ab}= - \frac{g^2}{4} N_c \delta_{ab}
\int \frac{d^3p}{(2\pi )^3} \; \frac{2n_g({\bf p})+1}{E_{\bf p}}
\frac{k^2 p^\mu p^\nu - (k^\mu p^\nu + p^\mu k^\nu) (k\cdot p)}{(k\cdot p + i 0^+)^2},
\ee
which holds in the hard loop approximation.

As already mentioned, the quark-loop contribution to the retarded polarization tensor (\ref{Pi-k-e-final}) is symmetric and transverse with respect to Lorentz indices. The same holds for the sum of the contributions of pure gluodynamics: gluon-loop, gluon-tadpole, and ghost-loop. The complete QCD result is obtained by summing all four contributions and subtracting the vacuum term. Then, one obtains the final formula 
\ba
\label{Pi-k-final}
\big( \Pi^+ (k) \big)^{\mu \nu}_{ab} &=&
\frac{g^2}{2}  \delta^{ab} \int \frac{d^3p}{(2\pi )^3} \, 
\frac{f ({\bf p})}{E_{\bf p}} \,
\frac{k^2 p^\mu p^\nu -\big(k^\mu p^\nu + p^\mu k^\nu - g^{\mu \nu} (k \cdot p) \big) (k \cdot p)}
{(k\cdot p + i 0^+)^2} ,
\ea
where $f ({\bf p}) \equiv 2 N_f \big(n_q ({\bf p}) + \bar{n}_q ({\bf p})\big) + 4N_c n_g({\bf p})$. The normalization of the effective distribution function $f({\bf p})$ is such that in an equilibrium QGP with temperature $T = \beta^{-1}$ and quark chemical potential $\mu$ one has
\be
\label{f-equilibrium}
f^{\rm eq}({\bf p}) = \frac{2 N_f}{e^{\beta (E_{\bf p} -\mu)}+1} + \frac{2N_f}{e^{\beta (E_{\bf p} +\mu)}+1} +  \frac{4N_c}{e^{\beta E_{\bf p}}-1},
\ee
where the spin factors are included.

The polarization tensor (\ref{Pi-k-final}) is obviously symmetric and transverse. Performing an integration by parts and requiring that $\lim_{|{\bf p}| \rightarrow \infty} f({\bf p}) = 0$, we find the following expression for the retarded polarization tensor
\be
\label{Pi-k-final-2}
\big( \Pi^+ (k) \big)^{\mu \nu}_{ab} =  
- \frac{g^2}{2}  \delta^{ab} \int \frac{d^3p}{(2\pi )^3} \, 
\frac{p^\mu}{E_{\bf p}} \,
\frac{\partial f({\bf p})}{\partial p^\rho} \left( g^{\nu \rho} - 
\frac{p^\nu k^\rho}{k\cdot p + i 0^+}\right) .
\ee
The integration by parts, which leads from the formula (\ref{Pi-k-final-2}) to (\ref{Pi-k-final}), is easily performed if the integration measure $d^3p/E_{\bf p}$ is replaced by $2 d^4p \, \delta(p^2) \, \Theta(p_0)$, as then all components of $p^\mu$ can be treated as independent from each other. 

It is sometimes useful to use the chromodielectric tensor (chromoelectric permittivity) $\varepsilon^{ij}(k)$ instead of the polarization tensor $\Pi^{\mu \nu}(k)$.  We show in Appendix~\ref{app-eps-vs-pi} that the two quantities are related to each other as 
\be
\label{epsilon-Pi}
\varepsilon^{ij}(k) = \delta^{ij} + {1 \over \omega^2} \Pi^{ij}(k) ,
\ee
where the indices $i,j,k = 1,2,3$ label the Cartesian coordinates of three-vectors. The dielectric tensor corresponding to the expressions 
(\ref{Pi-k-final}) and (\ref{Pi-k-final-2}) read
\ba
\label{epsilon-noderiv}
\varepsilon^{ij} (k) &=& 
 \delta^{ij} + 
{g^2 \over 2\omega^2}  \int \frac{d^3p}{(2\pi )^3} \, 
\frac{f ({\bf p})}{E_{\bf p}} \,
\frac{(\omega^2 - {\bf k}^2) v^i v^j  -  \big(k^i v^j + v^i k^j 
+ \delta^{ij} (\omega - {\bf k} \cdot {\bf v}) \big) 
(\omega - {\bf k} \cdot {\bf v})}
{(\omega - {\bf k} \cdot {\bf v} + i 0^+)^2}
\\ [2mm] 
\label{epsilon-deriv}
&=&  \delta^{ij} + 
{g^2 \over 2\omega} \int {d^3 p \over (2\pi )^3}
{ v^i \over \omega - {\bf k} \cdot {\bf v} + i0^+} 
{\partial f({\bf p}) \over \partial p^k} 
\bigg[ \Big( 1 - {{\bf k} \cdot {\bf v} \over \omega} \Big) \delta^{kj}
+ {k^k v^j \over \omega} \bigg] ,
\ea
where the color indices $a,b$ are dropped since $\varepsilon^{ij}(k)$ is a unit matrix in color space.

\begin{figure}[t]
\begin{center}
\includegraphics[width=5cm]{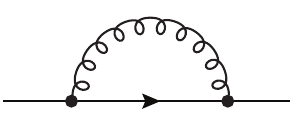}
\end{center}
\caption{Feynman graph corresponding to the leading order contribution to the quark self-energy. }
\label{fig-quark-self-energy}
\end{figure}

\subsubsection{Quark self-energy}
\label{subsubsec-quark-self-energy}                                                                                

The quark self-energy $\Sigma$ can be defined by means of the Dyson-Schwinger equation 
\be 
i{\cal S} = i S  - i S \, i\Sigma \, i{\cal S} , 
\ee 
where ${\cal S}$ and $S$ are the interacting and free propagators, respectively. The lowest order contribution to $\Sigma$ is given by the diagram shown in Fig.~\ref{fig-quark-self-energy}. 

Appropriately modifying Eq.~(\ref{Pi-k-e-1}), one finds
\be
\label{Si-k-a-1}
\Sigma^+ (k) =  \frac{i}{2} g^2
 \int \frac{d^4p}{(2\pi )^4} \big[
\gamma_\mu \tau^a S^+(p+k)  \Big(D^{\rm sym}(p) \Big)^{\mu \nu}_{ab} \tau^b \gamma_\nu
+ \gamma_\mu  \tau^a S^{\rm sym} (p) \Big(D^-(p-k) \Big)^{\mu \nu}_{ab} \tau^b \gamma_\nu 
\big].
\ee
The self-energy carries spinor and color indices (in the fundamental representation) which are not shown. Substituting the functions $D^{\pm}_{\mu \nu}$, $D^{\rm sym}_{\mu \nu}$ and $S^{\pm}$, $S^{\rm sym}$ given by Eqs.~(\ref{D-pm}) - (\ref{D-<}) and  (\ref{S-pm}) - (\ref{S-<}) into Eq.~(\ref{Si-k-a-1}) and using the identity $\tau^a\tau^a = C_F \, \mathbf{1}$ with $C_F \equiv (N_c^2-1)/(2N_c)$ being the quadratic Casimir invariant of the fundamental representation, one finds
\ba
\label{Si-k-a-3}
\Sigma^+ (k) &=&  g^2 \frac{C_F}{2}
 \int \frac{d^3p}{(2\pi )^3 E_{\bf p}}
\bigg\{ \bigg[
  \frac{p\sla+k\sla }{(p+k)^2 + i\, {\rm sgn}\big((p+k)_0\big)0^+}
- \frac{p\sla - k\sla}{(p-k)^2 - i\, {\rm sgn}\big((p-k)_0\big)0^+}
\bigg]  \big[2 n_g ({\bf p}) +1\big]
\\ \non
&& \;\;\;\;\;\;\;\;\;\;\;\;\;\;\;\;\;\;\;\;\;\;\;\;\;
- \bigg[
\frac{p\sla }{(p-k)^2 - i\, {\rm sgn}\big((p-k)_0\big)0^+}
- \frac{p\sla }{(p+k)^2 + i\, {\rm sgn}\big((p+k)_0\big)0^+}
  \bigg] \big[n_q({\bf p}) + \bar{n}_q({\bf p}) - 1\big]  \bigg\},
\ea
where the change of variables ${\bf p} \rightarrow -{\bf p}$ was made in the negative energy terms.  Applying the Hard Loop Approximation by using the formulas (\ref{HL-approx-1}) and (\ref{HL-approx-2}), one obtains
\ba
\label{Si-k-final}
\Sigma^+ (k) &=& g^2 \frac{C_F}{4}
\int \frac{d^3p}{(2\pi )^3} \,
\frac{ \tilde f ({\bf p}) }{E_{\bf p}}  \,
\frac{p\sla}{k\cdot p + i 0^+} ,
\ea
where $ \tilde f ({\bf p}) \equiv  2\big(n_q ({\bf p}) +  \bar{n}_q ({\bf p})\big) + 4 n_g ({\bf p}) $. The formula (\ref{Si-k-final}) has the well-known form of the electron self-energy in an ultrarelativistic QED plasma, see {\em e.g.} the review \cite{Blaizot:2001nr}.

The self-energies (\ref{Pi-k-final}) and  (\ref{Si-k-final}) have been obtained in the hard-loop approximation, that is for the external momentum $k$ being much smaller than the internal momentum $p$ which is carried by a plasma constituent. However, it appears that the results (\ref{Pi-k-final}) and (\ref{Si-k-final}) are only valid when the external momentum $k$ is not too small. This is most easily seen in case of the fermion self-energy (\ref{Si-k-final}) which diverges as $k \rightarrow 0$. When we deal with an equilibrium (isotropic) plasma of the temperature $T$, the characteristic momentum of (massless) plasma constituents is of the order $T$. One observes that if the external momentum $k$ is of the order $g^2 T$, which is the so-called {\it magnetic} or  {\it ultrasoft} scale, the self-energy (\ref{Si-k-final}) is not perturbatively small as it is of the order ${\cal O}(g^0)$. Therefore, the expression (\ref{Si-k-final}) is meaningless for $k \le g^2 T$.  Since $k$ must be much smaller than $p \sim T$, one arrives at the well-known conclusion that the self-energy (\ref{Si-k-final}) is valid at the {\it soft} scale, that is when $k$ is of the order $gT$.  Analyzing higher order corrections to the self-energies (\ref{Pi-k-final}) and (\ref{Si-k-final}) one finds that they are indeed valid for $k \sim gT$ and they break down at the magnetic scale because of the infrared problem of gauge theories, see {\it e.g.} \cite{Lebedev:1989ev} or the review \cite{Kraemmer:2003gd}. When the momentum distribution of plasma particles is anisotropic, instead of the temperature $T$, we have a characteristic four-momentum ${\cal P}^\mu$ of plasma constituents and the hard-loop approximation requires that ${\cal P}^\mu \gg k^\mu$ which should be understood as a set of four conditions for each component of the four-momentum $k^\mu$. The validity of the self-energies (\ref{Pi-k-final}) and (\ref{Si-k-final}) is then limited to $k^\mu \sim g {\cal P}^\mu$.  

\subsection{The Hard-Loop Effective Action}
\label{subsec-actions}                                                                                

The two-point functions at soft momentum, which were derived above, allow one to construct an effective action for soft excitations in an anisotropic QGP~\cite{Mrowczynski:2004kv}. The action not only generates the gluon polarization tensor (\ref{Pi-k-final}) and quark self-energy (\ref{Si-k-final}), but also the vertex functions and all $n-$point functions. 

The relevant terms in the action which generate (\ref{Pi-k-final}) and (\ref{Si-k-final}) read
\ba \label{g-2-action1}
{\cal L}^{(A)}_2(x) &=&  
{1\over2} \int d^4y \, A^a_\mu(x) \Pi^{\mu \nu}_{ab}(x-y) A^b_\nu(y) \;, 
\\ [2mm]  \label{q-2-action1}
{\cal L}^{(\Psi)}_2(x) &=&  
\int d^4y \, \bar{\Psi}(x) \Sigma (x-y) \Psi (y) \; ,
\ea
where the subscript `2' indicates that the effective actions above only generate two-point functions. 

\subthreesection{Quark effective action}

Using the explicit form of the quark self-energy (\ref{Si-k-final}), one immediately rewrites the action (\ref{q-2-action1}) as
\ba 
\label{q-2-action2}
{\cal L}^{(\Psi)}_2(x) =  
- i {C_F \over 4}  g^2 \int  \frac{d^3p}{(2\pi )^3} \, { \tilde f ({\bf p}) \over E_{\bf p}}
\; \bar{\Psi}(x) {p\sla \over p \cdot \partial}
\Psi (x) ,
\ea
where
$$
{1 \over p\cdot \partial} \Psi (x) \buildrel \rm def \over =
i \int \frac{d^4k}{(2\pi)^4} {e^{-i k \cdot x} \over p\cdot k} \: \Psi (k) .
$$ 
Following \cite{Braaten:1991gm}, we modify the action (\ref{q-2-action2}) to comply with the requirement of gauge invariance. We simply replace the derivative $\partial^\mu$ by the covariant derivative $D^\mu = \partial^\mu - ig A^\mu$ in the fundamental representation. Thus, we obtain
\ba 
\label{q-action}
{\cal L}^{(\Psi)}(x) = - i {C_F \over 4} g^2 
\int  \frac{d^3p}{(2\pi )^3} \, { \tilde f ({\bf p}) \over E_{\bf p}}
\; \bar{\Psi}(x) {p\sla \over p\cdot D} \Psi (x) ,
\ea
where 
\be 
{1 \over p\cdot D}\: \Psi (x)  \buildrel \rm def \over =
{1\over p\cdot \partial}
\sum_{n=0}^{\infty}\bigg(ig \:  p\cdot A(x) {1 \over p \cdot \partial} \bigg)^n
\: \Psi (x) .
\ee

In equilibrium, the integrals over the momentum absolute value and the solid-angle factorize, and the quark action (\ref{q-action}) reduces to the Braaten-Pisarski result
\be
{\cal L}^{(\Psi)}_{\rm HTL}(x) =  
- i m_q^2  \, \Big\langle \bar{\Psi}(x) {\hat p \cdot \gamma \over \hat p\cdot D}
\Psi (x) \Big\rangle_{\hat{\bf p}} ,
\ee 
where
\be
m_q^2 = {C_F \over 4} g^2 
\int \frac{d^3p}{(2\pi )^3} \, {\tilde f ^{\rm eq}({\bf p}) \over E_{\bf p}}
= {C_F \over 8} g^2 \bigg(T^2 + {\mu^2 \over \pi^2} \bigg) ,
\ee
and $\langle \cdots \rangle_{\hat{\bf p}} \equiv \int d^2\Omega/4\pi \cdots$ denotes an average over the orientation of the unit vector $\hat{\bf p} = {\bf p}/ |{\bf p}|$ which defines the light-like four-vector $\hat p \equiv (1, \hat{\bf p})$ and $\mu$ is the quark chemical potential.

\subthreesection{Gluonic effective action}

Let us now derive the gluon effective action which is expected to be local and quadratic in the field strength tensor $F^{\mu \nu}(x)$. Therefore, we first look for an operator ${\cal M}^{\mu \nu}(x)_{ab}$ that satisfies the equation
\be
{1 \over 2} \int d^4y A^a_\mu(x) \Pi^{\mu \nu}_{ab}(x-y) A^b_\nu(y) 
= {1\over 4} (\partial_\mu A_\nu^a (x) - \partial_\nu A_\mu^a (x))
{\cal M}^{\nu \rho}_{ab}(x) 
(\partial_\rho A^{b \,\mu} (x) - \partial^\mu A_\rho^b (x)) ,
\ee
giving
\ba 
\label{Pi-M}
\Pi^{\mu \nu}_{ab}(k) = 2 k^2 {\cal M}^{\sigma \rho}_{ab}(k)
\; P_{\rho \sigma}^{\;\;\;\;\mu \nu}(k) \;,
\ea
where
\be
P^{\rho \sigma \mu \nu}(k) =
{1 \over k^2} \Big[ k^2 g^{\rho \nu} g^{\sigma \mu} 
+ k^\rho k^\sigma g^{\mu \nu} 
- k^\rho k^\nu g^{\sigma \mu}
- k^\sigma k^\mu  g^{\rho \nu} \Big] .
\ee
Since $P$ is the projection operator ($P^{\rho \sigma \mu \nu}(k) P_{\nu \mu}^{\;\;\;\; \delta \lambda}(k) 
= - P^{\rho \sigma  \delta \lambda}(k)$), $P^{-1}$ does not exist. Therefore, there is no unique solution of Eq.~(\ref{Pi-M}); various  solutions differ from each other by the components parallel to $k$. Because $k_\mu P^{\rho \sigma \mu \nu}(k) = k_\nu P^{\rho \sigma \mu \nu}(k) = 0$, Eq.~(\ref{Pi-M}) complies with the transversality of $\Pi^{\mu \nu}(k)$.

Substituting the explicit form of the gluon self-energy (\ref{Pi-k-final}) into Eq.~(\ref{Pi-M}), one finds that the equation is satisfied by
\be
{\cal M}^{\mu \nu}_{ab}(k) = \delta_{ab} \: \frac{g^2}{2}
 \int \frac{d^3p}{(2\pi )^3} \, { f({\bf p}) \over E_{\bf p} } \; 
{p^\mu p^\nu \over (p \cdot k)^2} ,
\ee
which gives
\ba \label{g-2-action2}
{\cal L}^{(A)}_2(x) =
\frac{g^2}{2}  \int \frac{d^3p}{(2\pi )^3} \, { f({\bf p}) \over E_{\bf p} } \; 
(\partial_\mu A_\nu^a (x) - \partial_\nu A_\mu^a (x))
{p^\nu p^\rho \over (p \cdot \partial)^2} \;
(\partial_\rho A^{a\,\mu} (x) - \partial^\mu A_\rho^a (x))  .
\ea
In order to generate the higher-order vertices we invoke the requirement of gauge invariance, replacing
$\partial^\mu A^\nu_a - \partial^\nu A^\mu_a $ by the field strength tensor $F^{\mu \nu}_a \equiv \partial^\mu A^\nu_a - \partial^\nu A^\mu_a + g f_{abc} A^\mu_b A^\nu_c$, and $\partial^\mu$ by the covariant derivative in the adjoint representation $D^\mu_{ab} \equiv \partial^\mu \delta_{ab} + g f_{acb} A^\mu_c$. Thus, we obtain the effective action
\ba \label{g-action}
{\cal L}^{(A)}(x) =
\frac{g^2}{2}  \int \frac{d^3p}{(2\pi )^3} \, { f({\bf p}) \over E_{\bf p} } \; 
F_{\mu \nu}^a (x)
\bigg({p^\nu p^\rho \over (p \cdot D)^2} \bigg)_{ab} \;
F_\rho^{\;\;b \,\mu} (x) .
\ea

In equilibrium, the gluon action (\ref{g-action}) reduces, as the quark action, to the respective Braaten-Pisarski result
\be
{\cal L}^{(A)}_{\rm HTL}(x) = 
\frac{1}{2}\, m_D^2
\Big\langle 
F_{\mu \nu}^a (x)
\bigg({\hat p^\nu \hat p^\rho \over (\hat p \cdot D)^2} \bigg)_{ab} \;
F_\rho^{\;\;b \,\mu} (x) 
\Big\rangle_{\hat{\bf p}} ,
\ee
where $m_D$ is the Debye mass given as
\be
\label{eq:minfty}
m_D^2 \equiv g^2  \int \frac{d^3p}{(2\pi )^3} \, 
{ f^{\rm eq}({\bf p}) \over E_{\bf p} } \;  
= {N_c \over 3} g^2 T^2 
+ {N_f \over 6} g^2 \bigg(T^2 + {3 \over \pi^2} \mu^2 \bigg) .
\ee
%

\subthreesection{Complete hard-loop effective action}

Combining the gluonic (\ref{g-action}) and fermionic (\ref{q-action}) contributions to the effective action one obtains the full hard-loop effective action
\begin{equation}
\label{Saniso}
  S_{\rm HL}[A, \bar\psi, \psi ] =
\frac{g^2}{2} \int d^4x \int \frac{d^3p}{(2\pi )^3} \,  
\left[ { f(\mathbf p) \over E_{\bf p} }\;
F_{\mu \nu} (x)
{p^\nu p^\rho \over (p \cdot D)^2} \;
F_\rho^{\mu} (x)
- i {C_F \over 2}
{ \tilde f ({\bf p}) \over E_{\bf p} }\;
 \bar{\Psi}(x) {p \cdot \gamma \over p\cdot D}
\Psi (x) \right].
\end{equation}
We note that in the case of pure-glue theory we can express the hard-loop effective
action in terms of an auxiliary field $W$ defined via
\be
(p \cdot D) W^\mu(x,{\bf \hat p}) = p^\rho F_\rho^{\mu} (x),
\ee 
\be
\label{Saniso-glue}
S_{\rm HL}^{\rm pure \; glue}[W] =  \frac{g^2}{2} \int d^4x \int \frac{d^3p}{(2\pi )^3} \,  
{ f(\mathbf p) \over E_{\bf p} }\; W^\mu (x,{\bf \hat p}) \; W_\mu (x,{\bf \hat p}) .
\ee
Using the hard-loop action (\ref{Saniso})  one can derive the quark-gluon, triple-gluon, and four-gluon vertex functions for an anisotropic QGP \cite{Mrowczynski:2004kv}. These vertex functions satisfy the appropriate Ward-Takahashi identities. Actually, these identities are guaranteed to be satisfied since the effective action (\ref{Saniso}) is gauge invariant by construction. Let us finally note that, as discussed in the end of Sec.~\ref{subsubsec-quark-self-energy}, the Hard Loop Approach is valid at the soft scale of the order of $g {\cal P}$ where ${\cal P}$ is, as previously, the characteristic momentum of plasma constituents.


\section{Classical Kinetic theory}
\label{sec-kinetic-theory-class}

A natural theoretical tool to describe non-equilibrium plasmas is kinetic theory. Such a theory for the QGP has been formulated in two ways.  Within the first approach the color degrees of freedom are treated quantum mechanically and the distribution function of plasma constituents is a matrix in color space \cite{Heinz:1983nx,Winter:1984,Elze:1986qd,Elze:1986hq,Elze:1987yb,Mrowczynski:1989np}, see also the reviews \cite{Elze:1989un, Blaizot:2001nr}.  In the second approach, one assumes that the color is a continuous classical variable \cite{Heinz:1983nx,Heinz:1984yq,Heinz:1985qe,Kelly:1994ig,Kelly:1994dh}, see also the review \cite{Litim:2001db}.  We start with the latter approach which is much simpler than the first. We will set up the system of Wong equations \cite{Wong:1970fu} that describe a classical particle carrying a classical color charge interacting with the chromodynamic field.  Using the resulting Wong equations, we then derive a classical transport equation of the Vlasov type.

\subsection{Wong equations}

The Wong equations are a set of classical equations of motion for a point-like particle interacting with a chromodynamic field. We start with the electromagnetic case, as the Wong equations are just the extension to chromodynamics. Ignoring the dynamical effects of spin degrees of freedom, which are not of our interest here, the equations of motion of a point-like particle with charge $q$ and mass $m$ are 
\ba
\label{easyeqm}
\frac{d \mathbf{x} (t)}{dt} &=& \mathbf{v}(t) = \frac{\mathbf{p}(t)}{E_{\bf p}(t)}, 
\\ [2mm]
\frac{d\mathbf{p}(t)}{dt} &=& - q \Big(\mathbf{E}\big(t,\mathbf{x} (t)\big) 
+\mathbf{v}(t)\times\mathbf{B}\big(t,\mathbf{x} (t)\big)  \Big),
\ea
where $\mathbf{x} (t)$, $\mathbf{v} (t)$, $\mathbf{p} (t)$, and $E_{\bf p}(t)$ are the particle's position, velocity, momentum, and energy, respectively; $\mathbf{E}\big(t,\mathbf{x} (t)\big)$ and $\mathbf{B}\big(t,\mathbf{x} (t)\big) $ are the electric and magnetic fields along the particle's trajectory. The sign of the charge $q$ is chosen to agree with the color current defined by Eq.~(\ref{quark-current}). 

When we move from electrodynamics to chromodynamics with an ${\rm SU}(N_c)$ gauge group, instead of one electric field and one magnetic field we have instead $N_c^2 -1$ chromoelectric and $N_c^2 -1$ chromomagnetic fields labeled with the indices $a,b,c$. We also have $N_c^2 -1$ charges $q^a(t)$ which, in contrast to electrodynamics, evolve in time by rotating in color space. This happens because chromodynamic interactions are associated with a color exchange. The equation describing the time dependence of $q^a(t)$ can be obtained in the following way. We first define the color charge density $\rho^a$ and current ${\bf j}^a$ of a single particle as
\ba
\label{j1y}
\rho^a (t,\mathbf{x}) &=& - g q^a(t) \, 
\delta^{(3)}\big(\mathbf{x}-\mathbf{x}(t)\big),
\\
\label{j2y}
\mathbf{j}^a (t,\mathbf{x})  &=& - g q^a (t) \, \mathbf{v}(t) \, 
\delta^{(3)}\big(\mathbf{x}-\mathbf{x}(t)\big),
\ea
and we compute
\ba
\label{rhodot}
\frac{d \rho^a (t,\mathbf{x})}{dt} 
&=&  - g \frac{d q^a(t) }{dt} \, 
\delta^{(3)}\big(\mathbf{x}-\mathbf{x}(t)\big) 
+ g q^a (t) \, \frac{\partial \, \delta^{(3)}\big(\mathbf{x}-\mathbf{x}(t)\big)}{\partial \mathbf{x}} \,
\mathbf{v}(t) \nonumber ,
\\
&=&  - g \frac{d q^a(t) }{dt} \, 
\delta^{(3)}\big(\mathbf{x}-\mathbf{x}(t)\big) 
- \nabla \cdot \mathbf{j}^a (t,\mathbf{x}) ,
\ea
which can be written as
\be
\label{eq-rho-dot-7}
g \frac{d q^a(t) }{dt} \, \delta^{(3)}\big(\mathbf{x}-\mathbf{x} (t)\big)
= - \frac{d \rho^a (t,\mathbf{x})}{dt} - \nabla \cdot \mathbf{j}^a (t,\mathbf{x}) 
= - \partial^\mu j_\mu^a(x),
\ee
with the four-current $j^\mu_a=(\rho^a,\mathbf{j}^a)$. In QCD the four-current is not conserved 
but it is covariantly conserved
\be
\label{cov-conserve-j}
D^\mu_{ab} \, j_\mu^b(x) =
\big(\partial^\mu \delta^{ab} - gf^{abc} A^\mu_c(x) \big) \, j_\mu^b(x) = 0,
\ee
and consequently $\partial^\mu  j_\mu^b(x) = gf^{abc} A^\mu_c(x) \, j_\mu^b(x)$. Eq.~(\ref{eq-rho-dot-7}) thus
provides the equation we seek. The complete set of Wong equations thus reads
\ba
\label{wong1}
\frac{d \mathbf{x} (t)}{dt} &=& \mathbf{v}(t) = \frac{\mathbf{p}(t)}{E_{\bf p}(t)}, 
\\ [2mm]
\label{wong2}
\frac{d\mathbf{p}(t)}{dt} &=& - g q^a \Big(\mathbf{E}_a\big(t,\mathbf{x} (t)\big) 
+\mathbf{v}(t)\times\mathbf{B}_a\big(t,\mathbf{x} (t)\big)  \Big),
\\ [2mm]
\label{wong3}
\frac{d q^a(t) }{dt} &=&  - g f^{abc} q^b (t) \, A^\mu_c\big(x(t)\big)  \, \frac{p_\mu (t)}{E_{\bf p}(t)} .
\ea

Let us write down the Wong equations (\ref{wong1}), (\ref{wong2}), and (\ref{wong3}) in Lorentz-covariant form. For this purpose, the time $t$ is replaced by the particle's proper time $\tau$ such that $d\tau = \gamma^{-1} (t) \, dt$ with $\gamma (t) = \frac{E_{\bf p}(t)}{m}$ being the Lorentz factor of the particle's motion. Instead of the particle's position $\mathbf{x}$, velocity $\mathbf{v}$, and momentum $\mathbf{p}$, one introduces the corresponding four-vectors $x^\mu$, $u^\mu \equiv \frac{p^\mu}{m}$, and $p^\mu$. As a result the Wong equations (\ref{wong1}), (\ref{wong2}), and (\ref{wong3}) become
\ba
\label{EOM-1a}
\frac{d x^\mu(\tau)}{d \tau} &=& u^\mu(\tau ) ,
\\ [2mm]
\label{EOM-1b}
\frac{d p^\mu(\tau)}{d \tau} &=& - g q^a(\tau ) \, F_a^{\mu \nu}\big(x(\tau )\big) 
\, u_\nu(\tau ) ,
\\ [2mm]
\label{EOM-1c}
\frac{d q_a(\tau)}{d \tau} &=& - g f^{abc} u_\mu (\tau ) \, q_b(\tau) \, A^\mu _c \big(x(\tau )\big)  .
\ea
The chromoelectric and chromomagnetic fields are expressed here through the field strength tensor as
\be
E^i_a = F_a^{i0}, \;\;\;\;\; B^i_a = - \frac{1}{2} \,\epsilon^{ijk} F_a^{jk},
\ee
where the indices $i, j, k = 1, 2, 3$ label Cartesian coordinates and $ \epsilon^{ijk}$ is the totally antisymmetric Levi-Civita tensor.

If the classical color charge $q^a$ is assumed to transform under local gauge transformation covariantly (as the field strength tensor $F_a^{\mu \nu}$), the form of the Wong equations is gauge independent. Equations (\ref{EOM-1a}) and (\ref{EOM-1b}) are trivially gauge invariant. Equation (\ref{EOM-1c}) is evidently gauge covariant when written as
\be
u_\mu(\tau ) \, D^\mu_{ab} q^b(\tau ) = 0,
\ee
where $D^\mu_{ab}$ is the covariant derivative in the adjoint representation (\ref{cov-D-adjoint}).

The Wong equations have the same form for both quarks and gluons. The difference appears when averaging over color charges, which is needed to obtain gauge-independent quantities, is performed according to the rules
\ba
\label{int-q}
\int Dq \,q^a &=& 0,
\\
\label{int-qq}
\int Dq \,q^a q^b &=& C_2 \delta^{ab},
\ea
where $Dq$ is the integration measure over the space of classical color charges, $C_2 = 1/2$ for a quark (in the fundamental representation),  and $C_2 = N_c$ for a gluon (in the adjoint representation). The explicit form of the integration measure $Dq$, which is gauge invariant, is given in \cite{Litim:2001db}. For us it will be sufficient to use the rules (\ref{int-q}) and (\ref{int-qq}). 

\subsection{Liouville equation}

Now we are going to derive the classical transport equation in the Vlasov or mean-field approximation. The equation is obtained from the Liouville equation 
\be
\label{Liouville-eq}
\frac{d}{dt} {\cal F} (\mathbf{x},\mathbf{p},\vec{q}\,) = 0 ,
\ee
which is obeyed by the microscopic distribution function ${\cal F}$, which for $N$ colored particles is defined as 
\be
\label{micro-f}
 {\cal F} (\mathbf{x},\mathbf{p},\vec{q}\,)
= \sum_{i=1}^N \delta^{(3)}\big(\mathbf{x}-\mathbf{x}_i(t)\big) \; (2\pi)^3\delta^{(3)}\big(\mathbf{p}-\mathbf{p}_i(t)\big)
\delta^{(N_c^2-1)}\big(\vec{q}-\vec{q}_i(t)\big),
\ee
where the vector $\vec{q} = (q^1,q^2, \dots q^{N_c^2-1})$ represents $N_c^2-1$ color charges. We note that the function ${\cal F} (\mathbf{x},\mathbf{p},\vec{q}\,)$ does not explicitly depend on time and the phase space includes the classical color charge vector $\vec{q}$ in addition to the vectors $\mathbf{x}$ and $\mathbf{p}$.

The Liouville equation (\ref{Liouville-eq}) gives
\ba
\frac{d}{dt} {\cal F} (\mathbf{x},\mathbf{p},\vec{q}\,)
&=&  \sum_{i=1}^N \bigg[ \,
\frac{d\mathbf{x}_i(t)}{dt} \;
\frac{d \delta^{(3)}\big(\mathbf{x}-\mathbf{x}_i(t) \big)}{d\mathbf{x}} \;
\delta^{(3)}\big(\mathbf{p}-\mathbf{p}_i(t)\big) \;
\delta^{(N_c^2-1)}\big(\vec{q} - \vec{q}_i(t)\big)
\\ \nonumber
&& \;\;\;\;\;\;\;
+ \frac{d\mathbf{p}_i(t)}{dt} \;
\delta^{(3)}\big(\mathbf{x}-\mathbf{x}_i(t)\big) \;
\frac{d \delta^{(3)}\big(\mathbf{p}-\mathbf{p}_i(t)\big)}{d\mathbf{p}} \;
\delta^{(N_c^2-1)}\big(\vec{q} - \vec{q}_i(t)\big)
\\ \nonumber
&& \;\;\;\;\;\;\;
+ \frac{d \vec{q}_i(t)}{dt} \;
\delta^{(3)}\big(\mathbf{x}-\mathbf{x}_i(t)\big) \;
\delta^{(3)}\big((\mathbf{p}-\mathbf{p}_i(t)\big) \;
\frac{d \delta^{(N_c^2-1)}\big(\vec{q} -\vec{q}_i(t)\big)}{d\vec{q}} \, \bigg] = 0.
\ea
Using the Wong equations (\ref{wong1})-(\ref{wong3}), one obtains
\ba
\label{wongvlasov}
\frac{d}{dt} {\cal F} (\mathbf{x},\mathbf{p},\vec{q}\,)
= - \bigg[ {\bf v} \cdot \nabla 
- g q^a \big(\mathbf{E}_a (t,\mathbf{x}) 
               + \mathbf{v}\times\mathbf{B}_a (t,\mathbf{x}) \big) \cdot \nabla_{\bf p} 
- g f^{abc} q^b A^\mu_c(x) \, \frac{p_\mu}{E_{\bf p}} \frac{\partial}{\partial q^a}
\bigg] {\cal F} (\mathbf{x},\mathbf{p},\vec{q}\,) = 0 .
\ea

Since the distribution function ${\cal F}$ does not explicitly depend on time, that is $\frac{\partial }{\partial t} {\cal F} (\mathbf{x},\mathbf{p},\vec{q}\,) = 0$, we can write $E_{\bf p} {\bf v} \cdot  \nabla {\cal F} = p^\mu \partial_\mu   {\cal F}$. Using the fact that $\frac{\partial }{\partial p_0} {\cal F} (\mathbf{x},\mathbf{p},\vec{q}\,) = 0$ and expressing the chromoelectric and chromomagnetic fields through the strength tensor, one obtains
\be
\label{Liouville-eq-final}
p^\mu  \Big(\partial_\mu + g q^a F^a_{\mu \nu}(x) \frac{\partial}{\partial p_\nu}
- g f^{abc} q^b A_\mu^c(x)  \,  \frac{\partial}{\partial q^a}
\Big) {\cal F} (\mathbf{x},\mathbf{p},\vec{q}\,) = 0 .
\ee 
Equation (\ref{Liouville-eq-final}) needs to be supplemented by the Yang-Mills equations which describe the self-consistent generation of chromodynamic fields due to the particle current
\be
\label{currentwym}
D^\mu_{ab} F_{\mu \nu}^b (x) = j_\nu^a(x)  
= - g \int \frac{d^3p}{(2\pi)^3} \int Dq \; 
\frac{p_\nu}{E_{\bf p}} \, q^a {\cal F} (\mathbf{x},\mathbf{p},\vec{q}\,) .
\ee

Equations (\ref{Liouville-eq-final}) and (\ref{currentwym}) are exact but they require a precise knowledge of all particles' trajectories. This information is usually not available, nevertheless the equations are very useful for numerical simulations where the quark-gluon plasma is represented by a relatively small set of test particles. Such classical simulations are discussed in Sec.~\ref{Wong-Yang-Mills}. 

\subsection{Vlasov equation}

Equations (\ref{Liouville-eq-final}) and (\ref{currentwym}) are the starting point for a derivation of a transport equation for the macroscopic distribution function $f(x,\mathbf{p},\vec{q}\,)$ which is the microscopic distribution function (\ref{micro-f}) averaged over a statistical ensemble 
\be
f(x,\mathbf{p},\vec{q}\,) \equiv \langle {\cal F}(\mathbf{x},\mathbf{p},\vec{q}\,)\rangle ,
\ee
where $x \equiv (t, {\bf x})$ and thus an explicit time dependence of  $f$ is included. The transport equation of $f(x,\mathbf{p},\vec{q})$ should be deduced from Eq.~(\ref{Liouville-eq-final}) averaged over the statistical ensemble.  The Vlasov or mean-field equation is obtained by assuming that the chromodynamic field is dominated by the macroscopic (ensemble averaged)  field which slowly varies in space-time. This field is generated by the current associated with $f(x,\mathbf{p},q)$. Thus, we have
\ba
\label{Vlasov-classical}
p^\mu  \Big(\partial_\mu + g q^a F^a_{\mu \nu}(x) \frac{\partial}{\partial p_\nu}
- g f^{abc} q^b A_\mu^c(x)  \,  \frac{\partial}{\partial q^a} \Big) 
f(x,\mathbf{p},\vec{q}\,) 
&=& 0 , \\
\label{field-generation}
D^\mu F_{\mu \nu}^a(x) = j_\nu^a(x)  
= - g   \int Dq \int \frac{d^3p}{(2\pi)^3}
\frac{p_\nu}{E_{\bf p}} \, q^a f (x,\mathbf{p},\vec{q}\,) ,
\ea
where, to simplify our notation, we use the same symbols $F^{\mu\nu}$ and $A^\mu$ to denote the macroscopic and microscopic chromodynamic fields. 

\subsection{Gauge symmetry}

The microscopic and macroscopic distribution functions both transform under local gauge transformations as a scalar field
$f(x,\mathbf{p},\vec{q}\,) \rightarrow f'(x,\mathbf{p},\vec{q}\,')=f(x,\mathbf{p},\vec{q}\,)$ when $\vec{q} \rightarrow \vec{q}\,'$. Consequently, the transport equations (\ref{Liouville-eq-final}) and  (\ref{Vlasov-classical}) appear to be gauge invariant.  The Vlasov terms (those with $F_{\mu \nu}$) in these equations are trivially gauge invariant, but one has to show that
\be
\label{gauge-inv-eq}
 \Big(\partial_\mu - g f^{abc} q^b A_\mu^c   \frac{\partial}{\partial q^a}  \Big)
f(x,\mathbf{p},\vec{q}\,) 
=  \Big(\partial_\mu - g f^{abc} {q'}^b {A'}_\mu^c  \frac{\partial}{\partial {q'}^a} \Big)
f(x,\mathbf{p},\vec{q}\,(\vec{q}\,')) 
\ee
to prove the gauge invariance of Eqs.~(\ref{Liouville-eq-final}) and (\ref{Vlasov-classical}). We consider an infinitesimal gauge transformation of the form 
\be
{q'}^a = q^a + f^{abc} q^b \, \omega^c,
\;\;\;\;\;\;\;\;\;\;
{q}^a = {q'}^a - f^{abc} {q'}^b \, \omega^c,
\;\;\;\;\;\;\;\;\;\;
\frac{\partial}{\partial {q'}^a} 
= \frac{\partial q^b}{\partial {q'}^a} \frac{\partial}{\partial q^b}
= -  f^{abc} \omega^c \frac{\partial}{\partial q^b} ,
\ee
and
\be
{A'}_\mu^a = A_\mu^a+ f^{abc} A_\mu^b \, \omega^c  + \frac{1}{g}  \partial_\mu \omega^a  .
\ee
Since ${q'}^a$ depends on $x$ through $\omega^a(x)$, we have
\be
\label{eq-q-x-7}
\partial_\mu f(x,\mathbf{p},\vec{q}\,(\vec{q}\,'))  
= \partial_\mu f(x,\mathbf{p},\vec{q}\,)
+ 
\big(\partial_\mu {q}^a(\vec{q}\,') \big) 
\frac{\partial}{\partial q^a} f(x,\mathbf{p},\vec{q}\,) 
= \partial^\mu f(x,\mathbf{p},\vec{q}\,)
- f^{abc} q^b (\partial_\mu \omega_c) 
\frac{\partial}{\partial q^a} f(x,\mathbf{p},\vec{q}\,) .
\ee
A somewhat cumbersome calculation gives
\be
g f^{abc} {q'}^b {A'}_\mu^c  \,  \frac{\partial}{\partial {q'}^a} 
f(x,\mathbf{p},\vec{q}\,(\vec{q}\,')) = 
f^{abc} q^b (\partial_\mu \omega^c)  \,  \frac{\partial}{\partial q^a} 
f(x,\mathbf{p},\vec{q}) ,
\ee
which combined with Eq.~(\ref{eq-q-x-7}) leads to the equality (\ref{gauge-inv-eq}). In this way, the gauge invariance of the transport equations (\ref{Liouville-eq-final}) and (\ref{Vlasov-classical}) is proven. The gauge covariance of the field generation equations (\ref{currentwym}) and (\ref{field-generation}) is evident provided the integration measure $Dq$ is gauge invariant. 

\subsection{Linear response analysis}

In this section we study how the QGP, which is (on average) colorless, homogeneous, and stationary, responds to color fluctuations. Such an analysis was originally performed in \cite{Kelly:1994ig,Kelly:1994dh}. The distribution function is assumed to be of the form
\be
\label{f+delta-f}
f(x,{\bf p},\vec{q}\,) = f^0({\bf p}) + \delta f(x,{\bf p},\vec{q}\,) ,
\ee
where the function $f^0({\bf p})$, which is independent of $x$ and $\vec{q}$, represents a space-time homogeneous and colorless state of the plasma, while $\delta f(x,{\bf p},\vec{q}\,)$, describes a colorful fluctuation. It is assumed that
\be
f^0({\bf p}) \gg |\delta f(x,{\bf p},\vec{q}\,)|,
\;\;\;\;\;\;
|\partial_p^i f^0({\bf p})|  \gg |\partial_p^i \delta f(x,{\bf p},\vec{q}\,)| .
\ee
                                                                             
Substituting the distribution function in the form (\ref{f+delta-f}) into the current (\ref{field-generation}) and using Eq.~(\ref{int-q}), one obtains
\be
\label{j-delta-f}
j^{\mu}_a(x) = - g  \int Dq \int \frac{d^3p}{(2\pi)^3}\; 
\frac{p^\mu}{E_{\bf p}} \, q^a \delta f (x,\mathbf{p},\vec{q}\,) .
\ee
As can be seen from the above expression, the color-current results from deviations from the colorless state. 

To proceed, we substitute the distribution function (\ref{f+delta-f}) into the transport equation (\ref{Vlasov-classical}) and we assume that $A^\mu$ and $F^{\mu \nu}$ are of the same order as  $\delta f$. Then, keeping only the terms linear in $\delta f$, $A^\mu$, and $F^{\mu \nu}$, the transport equation (\ref{Vlasov-classical}) gives
\be
\label{lin-trans-eq}
p^{\mu} \partial_\mu \delta f(x,{\bf p},\vec{q}\,)  
= - g q_a p_{\mu}F^{\mu \nu}_a(x){\partial f^0({\bf p}) \over \partial p^{\nu}} .
\ee
We note that the left-hand side of the linearized equation is not gauge invariant, but the invariance can be restored by including the quadratically small term from Eq.~(\ref{Vlasov-classical}) with the derivative with respect to $\vec{q}$.

The linearized equation (\ref{lin-trans-eq}) is easily solved by means of the retarded Green's function $G_{\bf p}(x)$ which satisfies
\be
\label{ret-Green-eq}
p_{\mu}\partial^{\mu} G_{\bf p}(x) = \delta^{(4)}(x)
\ee
and equals
\be
\label{ret-Green}
G_{\bf p}(x) = E_{\bf p}^{-1} \Theta (t) \,  \delta^{(3)} ({\bf x} - {\bf v}t) \;,
\ee
with ${\bf v} \equiv  {\bf p}/E_{\bf p} $.  One finds the Green's function (\ref{ret-Green}) by performing a Fourier transform of Eq.~(\ref{ret-Green-eq}), solving the algebraic equation, and finally performing the inverse Fourier transform. 

The solution of  Eq.~(\ref{lin-trans-eq}) is
\be
\label{sol-non-cov}
\delta f(x,{\bf p},\vec{q}\,)  = - g \int d^4 y \: 
G_{\bf p} (x - y) \;q_a  p_{\mu}F^{\mu \nu}_a(y) 
{\partial f^0({\bf p}) \over \partial p^{\nu}} ,
\ee
and it gives the current (\ref{j-delta-f}), which is
\be
\label{j-F-1}
j^{\rho}_a(x) = g^2  C_2 
\int \frac{d^3p}{(2\pi)^3}\; \frac{p^\rho}{E_{\bf p}}
\int d^4 y \; G_{\bf p}(x - y) \;  p_{\mu} F^{\mu \nu}_a(y) 
{\partial f^0({\bf p}) \over \partial p^{\nu}} ,
\ee
where color averaging has been performed according to Eq.~(\ref{int-qq}). 

As can be seen, the right-hand side of Eq.~(\ref{j-F-1}) transforms differently under local gauge transformations than the left-hand side. To cure the problem, one introduces a link operator, which is sometimes called the gauge parallel transporter, 
defined in the fundamental representation as 
\be
\label{link-def-fund}
\Omega(x,y) = {\cal P} \exp\Big[ ig \int_y^x dz_\mu A^\mu(z) \Big],
\ee
with ${\cal P}$ denoting ordering along the path which is chosen to be a straight line connecting the points $x$ and $y$. $\Omega(x,y)$ transforms as
\be
\label{link-trans-fund}
\Omega(x,y) \rightarrow U(x) \; \Omega(x,y)\; U^\dagger(y) .
\ee
In the adjoint representation, the link operator is given by
\be
\label{link-def-adjoint}
\Omega(x,y) = {\cal P} \exp\Big[ ig \int_y^x dz_\mu {\cal A}^\mu(z) \Big] ,
\ee
where ${\cal A}^\mu = A^\mu_a T_a$, and it transforms as
\be
\label{link-trans-adjoint}
\Omega(x,y) \rightarrow {\cal U}(x) \; \Omega(x,y)\; {\cal U}^\dagger(y) .
\ee
The matrix ${\cal U}(x)$ is defined in Eq.~(\ref{M-C-def}). We also note that $\Omega(x,y)$ obeys the equation
\be
\label{link-EOM}
D^\mu_x \Omega(x,y) =0 .
\ee

Using the link operator (\ref{link-def-adjoint}), one modifies the current (\ref{j-F-1}) to comply with the gauge covariance. The modified current is
\be
\label{j-F-2}
j^{\mu}_a(x) = g^2  C_2 
\int \frac{d^3p}{(2\pi)^3}\; \frac{p_\nu}{E_{\bf p}}
\int d^4 y \; G_{\bf p}(x - y) \;  p_{\mu} \Omega_{ab}(x,y)\; F^{\mu \nu}_b(y) 
{\partial f^0({\bf p}) \over \partial p^{\nu}} .
\ee
One verifies with the help of Eq.~(\ref{link-EOM}) that the current (\ref{j-F-2}) obeys $D^\mu j_\mu = 0$. 
                                                       
Finally, we are going to perform the Fourier transform of the induced current (\ref{j-F-2}). Before this step, however, we can neglect the terms which are not of leading order in $g$. The parallel-transporter $\Omega$ is approximated by unity and the stress tensor $F_{\mu \nu}$ by $\partial_{\mu} A_{\nu} - \partial_{\nu} A_{\mu}$.  Within such an approximation, the Fourier-transformed induced current (\ref{j-F-2}) is given by
\be
\label{j-F-k}
j^{\mu}_a(k) = g^2 C_2
\int {d^3p \over (2\pi )^3} \: 
\frac{p^{\mu}}{E_{\bf p}}
{\partial f^0({\bf p}) \over \partial p_{\lambda}} \bigg[ g^{\lambda \nu}
- { k^{\lambda} p^{\nu} \over p^{\sigma}k_{\sigma} + i0^+} \bigg] \;
A_{\nu}^a(k) ,
\ee
where $C_2$ equals 1/2 for quarks and antiquarks and $C_2 = N_c$ for gluons. To obtain the complete induced current, one sums up the contributions coming from quarks, antiquarks, and gluons present in the plasma. Since the current (\ref{j-F-k}) depends on $q^a$ quadratically, the quark and antiquark contributions have the same sign. The gluon contribution is enhanced by the factor $2N_c$ when compared to the quark one. The complete induced current thus reads
\be
\label{j-F-k-final}
j^{\mu}_a(k) = \frac{g^2}{2}
\int {d^3p \over (2\pi )^3} \: 
\frac{p^{\mu}}{E_{\bf p}}
{\partial f({\bf p}) \over \partial p_{\lambda}} \bigg[ g^{\lambda \nu}
- { k^{\lambda} p^{\nu} \over p^{\sigma}k_{\sigma} + i0^+} \bigg] \;
A_{\nu}^a(k) ,
\ee
where $f({\bf p}) \equiv 2 N_f n_q({\bf p}) + 2N_f \bar{n}_q({\bf p}) + 4 N_c n_g({\bf p})$ with  $n_q({\bf p})$, $\bar{n}_q({\bf p})$, and $n_g({\bf p})$ denoting the distribution functions of quarks, antiquarks, and gluons which are normalized according to Eqs.~(\ref{norm-quark-n}) and (\ref{norm-gluon-n}).

Since the current can be obtained from the action $S$ as 
\be
j^{\mu}_a(x) = - \frac{\delta S}{\delta A^a_\mu(x)} ,
\ee
and the gluon polarization tensor is
\be
\Pi^{\mu \nu}_{ab}(x,y) = \frac{\delta^2 S}{\delta A^b_\nu(y) \, \delta A^a_\mu(x)},
\ee
the Fourier transformed induced current $j^{\mu}(k)$ in a translationally invariant system can be expressed as
\be
\label{j-Pi-A}
j^{\mu}_a(k) = - \Pi^{\mu \nu}_{ab}(k) A_{\nu}^{b}(k) ,
\ee
which holds in the linearized case. Comparing the relations (\ref{j-F-k-final}) and (\ref{j-Pi-A}), the retarded polarization tensor is found to be
\be
\label{Pi-class}
\Pi^{\mu \nu}_{ab}(k) = - \frac{g^2}{2} \delta^{ab} 
\int {d^3p \over (2\pi )^3} \: 
\frac{p^{\mu}}{E_{\bf p}}
{\partial f({\bf p}) \over \partial p_{\lambda}} \Bigg[ g^{\lambda \nu}
- { k^{\lambda} p^{\nu} \over p^{\sigma}k_{\sigma} + i0^+} \Bigg] .
\ee
In this way we have rederived the result (\ref{Pi-k-final-2}) which was obtained in Sec.~\ref{sec-field-theory} diagrammatically.


\section{Kinetic theory with quantum color}
\label{sec-kinetic-theory-quant}

In this section we present a transport theory for the QGP where the color charge is not a classical variable but is instead encoded in a matrix structure of the theory. Such a formulation, which was initiated in \cite{Heinz:1983nx} and further developed in \cite{Winter:1984,Elze:1986qd,Elze:1986hq,Elze:1987yb,Elze:1989gm,Mrowczynski:1989np,Blaizot:1999xk}, naturally emerges from QCD.  The problem of derivation of QGP transport equations from QCD is not fully resolved, but the Vlasov or collisionless limit, which is of our main interest, is rather well understood. In the subsequent subsections we show how to obtain the transport equations for quarks and gluons in the mean-field approximation and then the resulting transport equations are used to compute the polarization tensor of an anisotropic QGP. Those readers who are not much interested in the rather technical and complex problem of derivation of transport equations from an underlying quantum field theory can skip Sec.~\ref{sec-trans-derivation} and go directly to Sec.~\ref{sec-trans-equations}.

\subsection{Derivation}
\label{sec-trans-derivation}

The transport equations in the Vlasov or mean-field limit were first derived in \cite{Winter:1984,Elze:1986qd,Elze:1986hq,Elze:1987yb,Elze:1989gm} and the approach was put on a more solid ground when the Vlasov limit of kinetic theory was shown to be equivalent to diagrammatic QCD in the Hard Loop Approximation. It was first demonstrated for a quasi-equilibrium plasma \cite{Blaizot:1993zk,Blaizot:1993be,Kelly:1994ig,Kelly:1994dh}, see also the reviews \cite{Blaizot:2001nr,Litim:2001db}, and later on the equivalence was extended to an anisotropic QGP \cite{Mrowczynski:2000ed,Mrowczynski:2004kv}. 

In our derivation of the QGP Vlasov equations we essentially follow the original works \cite{Elze:1986qd,Elze:1986hq,Elze:1987yb,Elze:1989gm}. However, there is an important difference. The transport equations were found in \cite{Elze:1986qd,Elze:1986hq,Elze:1987yb} for an 
arbitrary mean field and the semiclassical Vlasov limit was obtained by performing a gradient expansion. Our derivation assumes from the very beginning that the mean field weakly varies in space-time.

\subsubsection{Quark transport equation}

The derivation of the quark transport equation is rather straightforward. It was worked out in \cite{Winter:1984,Elze:1986qd}. One starts with the definition of the Wigner function, which is the quantum analog of the classical distribution function. For a quark field $\psi$ the Wigner function can be defined as
\be
\label{Wigner-noncov-def-1}
W^{ij}_{\alpha \beta}(X,p) \equiv \int d^4u \, e^{ip\cdot u} 
\langle \psi_\alpha^i (X+u/2) \: {\bar \psi}_\beta^j (X- u/2) \rangle ,
\ee
where $i,j = 1,\,2, \dots N_c$ are color indices and $\alpha, \beta = 1,\,2,\,3,\,4$ are spinor indices; $\langle \dots \rangle$ denotes, as previously, averaging over a statistical ensemble. As in Sec.~\ref{subsec-Keldysh-Schwinger}, we use here the coordinates $X \equiv (x+y)/2$ and $u=x-y$. To simplify the notation, we write the Wigner function as
\be
\label{Wigner-noncov-def-2}
W(X,p) \equiv \int d^4u \, e^{ip \cdot u} \langle \psi (X+u/2) \otimes \bar \psi (X-u/2)  \rangle,
\ee
where the indices are not shown. It is sometimes useful to express the definition (\ref{Wigner-noncov-def-2}) in the following way
\be
\label{Wigner-noncov-def-3}
W(X,p) \equiv \int d^4u \, e^{ip\cdot u} 
\langle e^{\frac{u}{2}\cdot \partial} \psi (X) 
\otimes
\bar \psi (X) 
e^{-\frac{u}{2} \cdot \buildrel\leftarrow\over{\partial}} \rangle,
\ee
where $ e^{\frac{u}{2}\cdot \partial}$ is the operator of space-time translation by the four-vector $u/2$. 

The Wigner function defined by Eq.~(\ref{Wigner-noncov-def-1}), (\ref{Wigner-noncov-def-2}) or (\ref{Wigner-noncov-def-3}) does not transform covariantly with respect to gauge  transformations. The problem is cured by modifying the definition (\ref{Wigner-noncov-def-2}) as
\be
\label{Wigner-def-1}
W(X,p) \equiv \int d^4u \, e^{ip\cdot u} 
\langle \Omega(X,X+u/2) \; \psi (X + u/2) 
\otimes
\bar \psi (X-u/2) \; \Omega(X-u/2,X) 
\rangle,
\ee
where  $\Omega(x,y)$ is the link operator defined by Eq.~(\ref{link-def-fund}). Since the fields $A^\mu(x)$ and $\psi(x)$ transform under a gauge transformation $U(x)$ as
\be
A^{\mu}(x) \rightarrow U(x) A^{\mu}(x) U^\dagger (x) 
+ U(x) \partial^{\mu} U^\dagger (x) , \;\;\;\;\;\;\;\;
\psi(x) \rightarrow U(x)  \psi(x) ,
\ee
the Wigner function (\ref{Wigner-def-1}) transforms covariantly, {\em i.e.}
\be
\label{W-transform}
W(X,p) \rightarrow U(X) \, W(X,p) \, U^{\dag }(X) .
\ee

In analogy to Eq.~(\ref{Wigner-noncov-def-3}) the covariant Wigner function can be expressed as 
\be
\label{Wigner-def-2}
W(X,p) \equiv \int d^4u \, e^{ip \cdot u} 
\langle 
e^{\frac{u}{2}\cdot D } \; \psi (X) 
\otimes
\bar \psi (X) \; e^{-\frac{u}{2}\cdot \buildrel\leftarrow\over{D^\dagger}}
\rangle,
\ee
where $D^\mu \equiv \partial^\mu -igA^\mu(X)$ is the covariant derivative.

The definition (\ref{Wigner-def-2}) can be used to derive the equations of motion of $W(X,p)$ which come from the Dirac equations (\ref{Dirac-eq}) obeyed by the  fields $\psi$ and $\bar{\psi}$. The derivation starts with the equations
\ba 
\label{EOM1}
\int d^4u \, e^{ip\cdot u} 
\langle
e^{\frac{u}{2}\cdot D} \; [\gamma^\mu D_\mu +im] \psi (X) 
\otimes 
\bar \psi (X) \; e^{-\frac{u}{2}\cdot \buildrel\leftarrow\over{D^\dagger}}
\rangle =0 ,
\\ [2mm]
\label{EOM2}
\int d^4u \, e^{ip\cdot u} 
\langle
e^{\frac{u}{2}\cdot D} \; \psi (X) 
\otimes 
\bar \psi (X) \; [\gamma^\mu \buildrel\leftarrow\over{D_\mu^\dagger}  - im]
e^{-\frac{u}{2}\cdot \buildrel\leftarrow\over{D^\dagger}}
\rangle
=0 .
\ea
The transport equation of $W(X,p)$ in the mean-field or Vlasov limit  is obtained by assuming that
\begin{itemize}
\item the chromodynamic field is dominated by the mean field:  
$|\langle A^\mu (X) \rangle | \gg | A^\mu (X) - \langle A^\mu (X) \rangle |$,
and consequently $A^\mu (X) \approx \langle A^\mu (X) \rangle$;
\item the Wigner function weakly varies in coordinate and momentum
spaces that is $| W(X,p) | \gg | \partial_p^\mu \partial^\nu W(X,p) |$;
\item the mean field is sufficiently weak and thus
$| W(X,p) | \gg | \partial_p^\mu (\partial^\nu - ig \langle A^\nu (X) \rangle \big) W(X,p) |$.
\end{itemize}

Keeping in mind these simplifying assumptions, the operator $e^{\pm \frac{u}{2}\cdot D}$, which enters the Wigner function definition,
can be approximated as 
\be
e^{\pm \frac{u}{2}\cdot D} = 1\pm \frac{1}{2} u_\nu D^\nu + {\cal O}(D^2) .
\ee
Then, one derives the following commutation relations
\ba
\label{formula11}
e^{\frac{u}{2}\cdot D} \; D^\mu  
&=& \Big(D^\mu -  \frac{i}{2} g u_\nu F^{\nu \mu} \Big) e^{\frac{u}{2}\cdot D} 
+ {\cal O}(D^3),
\\ [2mm]
\label{formula22}
\buildrel\leftarrow\over{D_\mu^\dagger}
e^{-\frac{u}{2}\cdot \buildrel\leftarrow\over{D^\dagger}}
&=&
e^{-\frac{u}{2}\cdot \buildrel\leftarrow\over{D^\dagger}}
\Big(\buildrel\leftarrow\over{D_\mu^\dagger} 
- \frac{i}{2} g u^\nu F_{\nu \mu}  \Big) + {\cal O}(D^3) ,
\ea
where the field strength tensor $F^{\mu \nu}$ is defined by Eq.~(\ref{strength-tensor}). Because $\delta A^\mu (X) \equiv A^\mu (X) - \langle A^\mu (X) \rangle$ is neglected in the consideration presented in this section, the mean field $\langle A^\mu (X) \rangle$ is denoted as $A^\mu (X)$ to simplify the notation. 

With the help of Eqs.~(\ref{formula11}) and (\ref{formula22}), equations (\ref{EOM1}) and (\ref{EOM2}) are transformed to 
\ba 
\label{EOM11}
\int d^4u \, e^{ip\cdot u}
\langle 
\big[\gamma_\mu 
\big(D^\mu -  \frac{i}{2} g u_\nu F^{\nu \mu} \big) +im \big]
e^{\frac{u}{2}\cdot D} \;  \psi (X) 
\otimes  
\bar \psi (X) \;  e^{- \frac{u}{2}\cdot \buildrel\leftarrow\over{D^\dagger}}
\rangle =0 ,
\\ [2mm]
\label{EOM22}
\int d^4u \, e^{ip\cdot u} 
\langle 
e^{\frac{u}{2}\cdot D} \; \psi (X) 
\otimes
\bar \psi (X) \; 
e^{- \frac{u}{2}\cdot \buildrel\leftarrow\over{D^\dagger}}
\big[\gamma^\mu 
\big(\buildrel\leftarrow\over{D_\mu^\dagger} 
- \frac{i}{2} g u^\nu F_{\nu \mu}  \big) - im\big]
\rangle =0 .
\ea

Since we are not interested in the spin degrees of freedom, we take the trace of Eqs.~(\ref{EOM11}) and (\ref{EOM22}) with respect to the spinor indices and sum the equations. As a result, we obtain the transport equation of the Wigner function $W(X,p)$
\be
\label{trans-W}
D^\mu \Tr [\gamma_\mu W(X,p)] - \frac{g}{2} \frac{\partial}{\partial p^\nu}
\big\{F^{\nu \mu}(X), \Tr [\gamma_\mu W(X,p)]\big\} =0 ,
\ee
where the covariant derivative $D^\mu$ acts on the color tensor and thus $D^\mu \equiv \partial^\mu -ig [ A^\mu, \dots]$; $\{ \dots , \dots\}$ denotes an anticommutator. 

Keeping in mind the form of the quark unordered Green's functions (\ref{S->}) and (\ref{S-<}),  one introduces the distribution function for quarks $Q (X,{\bf p})$ and antiquarks $\bar Q(X,{\bf p})$ as
\be
\label{distri-f-def}
W(X,p) = \frac{\pi}{E_{\bf p}} p^\mu \gamma_\mu 
\Big( \delta (E_{\bf p} - p_0)  \big[Q (X,{\bf p}) -1\big]
+ \delta (E_{\bf p} + p_0) \bar Q(X,-{\bf p}) \Big).
\ee 
In contrast to the Wigner function $W(X,p)$,  the distribution functions $Q (X,{\bf p})$ and $\bar Q(X,{\bf p})$ are nonzero only for the four-momenta obeying the mass-shell constraint $p^2=0$. We also note that the definition (\ref{distri-f-def}) assumes that the spin degrees of freedom are uniformly populated, and hence that the system described by $W(X,p)$ is unpolarized. 

Substituting Eq.~(\ref{distri-f-def}) into Eq.~(\ref{trans-W}) and integrating over $p_0$, one obtains two equations corresponding to positive and negative energies.  These are the transport equations for the quark and antiquark distribution functions which read
\ba
\label{trans-q}
p^{\mu} D_\mu Q(X,{\bf p}) + {g \over 2}\, p^{\mu}  \frac{\partial}{\partial p_\nu}
\left\{ F_{\mu \nu}(X), Q(X,{\bf p}) \right\}
&=& 0 ,
 \\ [2mm]
\label{trans-barq} 
p^{\mu} D_\mu \bar Q(X,{\bf p}) - {g \over 2}\, p^{\mu}  \frac{\partial}{\partial p_\nu}
\left\{ F_{\mu \nu}(X),  \bar Q(X,{\bf p}) \right\}
&=& 0 .
\ea
We note that the term from Eq.~(\ref{distri-f-def}), which corresponds to $-1$, vanishes identically when the Eq.~(\ref{distri-f-def}) is substituted into Eq.~(\ref{trans-W}) and the integration over $p_0$ is performed. This happens because the covariant derivative $D_\mu$ cancels the term and so does the momentum derivative contracted with the antisymmetric tensor $F_{\mu \nu}$. 

\subsubsection{Gluon transport equation}
\label{sec-glue-trans-eq}

The derivation of the gluon transport equation is a much more complex problem. The equation was derived in \cite{Elze:1986hq}, see also \cite{Elze:1987yb}, using the fundamental representation with the gluon Wigner function being a tensor with four color indices. The final transport equation had a rather complicated structure. However, it was soon observed \cite{Mrowczynski:1989np} that the gluon Vlasov equation has the same structure as the corresponding quark equation when a specific adjoint representation is used. Although the transport equation appeared to be correct, the derivation was not really convincing because of a rather {\it ad hoc} treatment of the mean-field contribution to the gluon Wigner function.  A reliable derivation using the background field method was worked out in \cite{Elze:1989gm}, see also \cite{Blaizot:2001nr}. We mostly follow here the original derivation \cite{Elze:1989gm} simplifying it noticeably by adopting the mean-field approximation and taking into account only the lowest nontrivial order of the gradient expansion. 

Gluons play a double role in the quark-gluon plasma: transverse gluons, which are on mass-shell, are plasma constituents similarly to quarks and antiquarks while virtual gluons are carriers of color forces of two types. The smooth classical gluon field is responsible for the mean-field or Vlasov dynamics and the quantum field of off-mass-shell gluons controls collisions of plasma constituents. The latter effect is neglected and the gluon field $A^\mu$ is thus split into the classical mean field $\bar{A}^\mu$ and the quantum fluctuations $a^\mu$ corresponding to plasma constituents.  So, we write $A^\mu = \bar{A}^\mu + a^\mu$. The splitting suggests application of the background field method to derive the transport equation of gluons. The gauge transformation of the fields $\bar{A}^\mu_a$ and $a^\mu_a$ in the fundamental representation is 
\ba
\label{trans-barA}
\bar{A}^\mu (x) &\rightarrow& U(x) \, \bar{A}^\mu (x) U^\dagger (x) 
+ \frac{i}{g} U(x) \partial^\mu U^\dagger (x) ,
\\
\label{trans-a}
a^\mu (x) &\rightarrow& U(x) \, a^\mu (x) U^\dagger (x) ,
\ea
where the derivative term is included in the transformation of $\bar{A}^\mu$, as the transformation of $a^\mu$ is then homogeneous. 

Let us discuss the equations of motion of the fields $\bar{A}^\mu_a$ and $a^\mu_a$ in the adjoint representation which will be used later on.  The equation of motion of classical background field $\bar{A}^\mu$ is found by substituting $A^\mu = \bar{A}^\mu + a^\mu$ into  the Yang-Mills equation (\ref{YM-eq-adjoint}). Then, one takes the average of the resulting equation and requires that $\langle a^\mu \rangle = 0$ and $\langle \bar{A}^\mu \rangle = \bar{A}^\mu $.  In this way one finds
\be
\label{YM-eq-adjoint-barF}
\bar{\cal D}_\mu^{ab} \bar{F}^{\mu \nu}_b  =  \langle j^\nu_a \rangle + \langle J^\nu_a \rangle ,
\ee
where $\bar{\cal D}_\mu$ and $\bar{F}^{\mu \nu}$ are defined as ${\cal D}_\mu$ and $F^{\mu \nu}$ but with 
$\bar{A}^\mu$ instead of $A^\mu$; $ J^\nu_a$ is the gluon current
\be
\label{g-current}
J^\nu_a \equiv  g f^{abc} \Big[  a_\mu^b \partial^\nu a^\mu_c  
-  2 a^\mu_b  \partial_\mu a^\nu_c  -  (\partial_\mu a^\mu_b  )a^\nu_c   \Big],
\ee
where the terms of order $g^2$, or equivalently cubic in $a^\mu$, are neglected.

The equation of motion of $a^\mu$  can be found by substituting $A^\mu = \bar{A}^\mu + a^\mu$ into Eq.~(\ref{YM-eq-adjoint}) and subtracting Eq.~(\ref{YM-eq-adjoint-barF}). However, we rather start with Eq.~(\ref{YM-eq-D2}) and subtract the analogous equation for $\bar{A}^\mu$. Thus one finds
\be
\label{YM-eq-a-D2}
\Big[g^{\mu \nu} \bar{\cal D}^2 -  \bar{\cal D}^\mu \bar{\cal D}^\nu  - 2 i g \bar{\cal F}^{\mu \nu} \Big] a_\nu 
=  j^\mu  + \delta J^\mu ,
\ee
where 
\be
\delta J^\nu_a \equiv 
 g f^{abc} \Big[ 
\Big( a_\mu^b \partial^\nu a^\mu_c  - \langle a_\mu^b \partial^\nu a^\mu_c  \rangle \Big) 
-  2 \Big( a^\mu_b  \partial_\mu a^\nu_c - \langle a^\mu_b  \partial_\mu a^\nu_c \rangle \Big) 
-  \Big(  (\partial_\mu a^\mu_b  )a^\nu_c  -  \langle (\partial_\mu a^\mu_b  )a^\nu_c \rangle  \Big)  \Big] .
\ee 

We are now ready to start a derivation of the transport equation of gluons. The Wigner function for gluons is defined as
\be
\label{Gamma-def}
\Gamma^{\mu \nu}(X,p) \equiv \int d^4u  \, e^{ip \cdot u} 
\langle 
e^{\frac{u}{2}\cdot \bar{\cal D}} a^\mu (X) 
\otimes 
a^\nu (X)  e^{- \frac{u}{2}\cdot  \buildrel\leftarrow\over{\bar{\cal D}}} 
\rangle ,
\ee
where the potential $a^\mu$ is in the adjoint representation and consequently, the Wigner function is an $(N_c^2-1) \times (N_c^2-1)$ matrix. The covariant derivative $\bar{\cal D}^\mu$ involves only the background field that is $\bar{\cal D}^\mu \equiv \partial^\mu - ig T^a \bar{A}^\mu_a = \partial^\mu - ig \bar{\cal A}^\mu$. We note that the definition of the Wigner function (\ref{Gamma-def}) can be rewritten as
\be
\label{Gamma-def-2}
\Gamma^{\mu \nu}(X,p) \equiv 
\int d^4u  \, e^{ip \cdot u} 
\langle 
\bar\Omega(X,X+u/2) \, a^\mu (X+u/2) 
\otimes 
a^\nu (X-u/2) \, \bar\Omega(X-u/2,X)
\rangle ,
\ee
where $\bar\Omega(x,y)$ is the link operator in the adjoint representation 
\be
\bar\Omega(x,y) \equiv {\cal P}e^{ig \int_y^x d_\mu z\, \bar{\cal A}^\mu(z)},
\ee
which transforms as
\be
\bar\Omega(x,y) \rightarrow {\cal U}(x) \, \bar\Omega(x,y) \, {\cal U}^\dagger (y) ,
\ee
and consequently the gluon Wigner function transforms covariantly 
\be
\Gamma^{\mu \nu}(X,p)  \rightarrow {\cal U}(X) \, \Gamma^{\mu \nu}(X,p) \, {\cal U}^\dagger (X) .
\ee
The matrix ${\cal U}(X)$ is defined in Eq.~(\ref{M-C-def}). 

Let us compute the covariant drift term of the transport equation which is
\ba
\label{trans-eq-0}
p\cdot \bar{\cal D} \, \Gamma^{\mu \nu}(X,p) =
 \int d^4u  \, e^{ip \cdot u} p_\rho 
\Big\langle \big(
\bar{\cal D}^\rho e^{\frac{u}{2}\cdot \bar{\cal D}} a^\mu (X) \big)
\otimes 
a^\nu (X)  e^{- \frac{u}{2}\cdot  \buildrel\leftarrow\over{\bar{\cal D}}} 
+ 
e^{\frac{u}{2}\cdot \bar{\cal D}} a^\mu (X) 
\otimes 
\big( a^\nu (X)  e^{- \frac{u}{2}\cdot  \buildrel\leftarrow\over{\bar{\cal D}}} 
\buildrel\leftarrow\over{\bar{\cal D}^\rho} \big)
\Big\rangle .
\ea
Using the approximate commutation relations
\ba
\label{formula01}
\bar{\cal D}^\mu e^{\frac{u}{2} \bar{\cal D}}  
&=& e^{\frac{u}{2}\cdot \bar{\cal D}} \Big(\bar{\cal D}^\mu -  \frac{i}{2} g u_\nu {\cal F}^{\mu \nu} \Big) 
+ {\cal O}(\bar{\cal D}^3),
\\[4mm]
\label{formula02}
e^{-\frac{u}{2}\cdot \buildrel\leftarrow\over{\bar{\cal D}}}  \buildrel\leftarrow\over{\bar{\cal D}^\mu} 
&=& 
\Big(\buildrel\leftarrow\over{\bar{\cal D}^\mu} +  \frac{i}{2} g u_\nu {\cal F}^{\nu \mu} \Big) 
e^{-\frac{u}{2}\cdot \buildrel\leftarrow\over{\bar{\cal D}}} 
+ {\cal O}(\bar{\cal D}^3), 
\ea
one manipulates Eq.~(\ref{trans-eq-0}) to the form
\ba
\label{trans-eq-1}
p\cdot \bar{\cal D} \, \Gamma^{\mu \nu}(X,p) &=&
 \int d^4u  \, e^{ip \cdot u} p_\rho 
\Big\langle 
e^{\frac{u}{2}\cdot \bar{\cal D}} 
\Big[ \Big(\bar{\cal D}^\rho -  \frac{i}{2} g u_\sigma \bar{\cal F}^{\rho \sigma} \Big)
a^\mu (X) \Big]
\otimes 
a^\nu (X)  e^{- \frac{u}{2}\cdot  \buildrel\leftarrow\over{\bar{\cal D}}} 
\\[2mm] \non
&&  ~~~~~~~~~~~~~~~~~~~~~+ ~
e^{\frac{u}{2}\cdot \bar{\cal D}} a^\mu (X) 
\otimes 
\Big[ a^\nu (X)  
\Big( \buildrel\leftarrow\over{\bar{\cal D}^\rho} +  \frac{i}{2} g u_\sigma \bar{\cal F}^{\sigma \rho} \Big) \Big]
e^{- \frac{u}{2}\cdot  \buildrel\leftarrow\over{\bar{\cal D}}}  
\Big\rangle .
\ea
We note here that $u_\nu {\cal F}^{\nu \mu} \sim {\cal O}(\bar{\cal D}^2)$ because  ${\cal F}^{\nu \mu} \sim {\cal O}(\bar{\cal D})$ and $u_\nu$ corresponds to the derivative with respect to momentum.   Consequently, $\big[ e^{\pm \frac{u}{2}\cdot \bar{\cal D}} ,  {\cal F}^{\mu \nu} \big] = {\cal O}(\bar{\cal D}^3)$ and $\bar{\cal F}^{\mu \nu}$, which enter Eq.~(\ref{trans-eq-1}), can be interchanged with $e^{\pm \frac{u}{2}\cdot \bar{\cal D}}$ within the adopted accuracy. We additionally replace $i u^\nu e^{ip \cdot u}$ by $\frac{\partial }{\partial p_\nu} e^{ip \cdot u}$. Thus, we obtain
\ba
\label{trans-eq-2}
&& p\cdot \bar{\cal D} \, \Gamma^{\mu \nu}(X,p)  
+ \frac{g}{2} p_\rho \frac{\partial }{\partial p_\sigma} 
\Big\{ \bar{\cal F}^{\rho \sigma}(X), \Gamma^{\mu \nu}(X,p) \Big\} 
\\[2mm]\non
&& ~~~~~~~~~~~ 
= \int d^4u  \, e^{ip \cdot u} p_\rho 
\Big\langle 
e^{\frac{u}{2}\cdot \bar{\cal D}} \Big(\bar{\cal D}^\rho a^\mu (X) \Big)
\otimes 
a^\nu (X)  e^{- \frac{u}{2}\cdot  \buildrel\leftarrow\over{\bar{\cal D}}} 
+
e^{\frac{u}{2}\cdot \bar{\cal D}} a^\mu (X) 
\otimes 
\Big( a^\nu (X)  \buildrel\leftarrow\over{\bar{\cal D}^\rho} \Big)
e^{- \frac{u}{2}\cdot  \buildrel\leftarrow\over{\bar{\cal D}}}  
\Big\rangle  ,
\ea
where two terms have been moved from the right-hand-side to the left-hand-side of Eq.~(\ref{trans-eq-2}). 

In the next step, one replaces $p_\rho$ by $-i \frac{\partial}{\partial u^\rho} \, e^{i p \cdot u}$ on the right-hand-side of Eq.~(\ref{trans-eq-2}) and then performs the partial integration. Consequently, $\frac{\partial}{\partial u^\rho}$ acts on $e^{\pm \frac{u}{2}\cdot \bar{\cal D}}$ producing $\pm \frac{1}{2}\bar{\cal D}_\rho$. So, we obtain
\ba
\label{trans-eq-4}
p\cdot \bar{\cal D} \Gamma^{\mu \nu}(X,p) 
&+& \frac{g}{2} p_\rho \frac{\partial }{\partial p_\sigma} 
\Big\{ \bar{\cal F}^{\rho \sigma}(X), \Gamma^{\mu \nu}(X,p) \Big\} 
\\[2mm] \non
&=&
\frac{i}{2} \int d^4u  \, e^{ip \cdot u} 
\Big\langle 
e^{\frac{u}{2}\cdot \bar{\cal D}} 
\Big( \bar{\cal D}^2 a^\mu (X)  \Big)
\otimes 
a^\nu (X)  e^{- \frac{u}{2}\cdot  \buildrel\leftarrow\over{\bar{\cal D}}} 
-
e^{\frac{u}{2}\cdot \bar{\cal D}} a^\mu (X) 
\otimes  
\Big( a^\nu (X)  \buildrel\leftarrow\over{\bar{\cal D}^2} \Big)
e^{- \frac{u}{2}\cdot  \buildrel\leftarrow\over{\bar{\cal D}}}  
\Big\rangle .
\ea

Now we use the equation of motion (\ref{YM-eq-a-D2}) which allows one to convert  Eq.~(\ref{trans-eq-4}) into
\ba
\label{trans-eq-5}
p\cdot \bar{\cal D} \Gamma^{\mu \nu}(X,p) 
&+& \frac{g}{2} p_\rho \frac{\partial }{\partial p_\sigma} 
\Big\{ \bar{\cal F}^{\rho \sigma}(X), \Gamma^{\mu \nu}(X,p) \Big\} 
\\[2mm] \non
&=&
\frac{i}{2} \int d^4u  \, e^{ip \cdot u} 
\Big\langle 
e^{\frac{u}{2}\cdot \bar{\cal D}} 
\Big( 
\big( \bar{\cal D}^\mu \bar{\cal D}^\rho  + 2 i g \bar{\cal F}^{\mu \rho} \big) a_\rho(X) 
+  j^\mu(X)  + \delta J^\mu(X)
 \Big)
\otimes 
a^\nu (X)  e^{- \frac{u}{2}\cdot  \buildrel\leftarrow\over{\bar{\cal D}}} 
\\[2mm] \non
&& ~~~~~~~~~~~~~~~~~~~~ -
e^{\frac{u}{2}\cdot \bar{\cal D}} a^\mu (X) 
\otimes  
\Big( a_\sigma(X)
\big( \buildrel\leftarrow\over{\bar{\cal D}^\sigma \bar{\cal D}^\nu}  + 2 i g \bar{\cal F}^{\nu \sigma} \big) 
+   j^\nu(X)  + \delta J^\nu(X)
 \Big)
e^{- \frac{u}{2}\cdot  \buildrel\leftarrow\over{\bar{\cal D}}}  
\Big\rangle .
\ea
One observes that the terms, where the quark current $j^\mu$ enters, vanish because the terms are linear in $a^\mu$.  The terms with $\delta J^\mu$ also vanish as these terms are proportional to either $\langle a^\mu \rangle$, which vanish, or to $\langle a^\mu a^\nu a^\rho \rangle$, which are neglected. Thus, we have 
\ba
\label{trans-eq-6}
p\cdot \bar{\cal D} \Gamma^{\mu \nu}(X,p) 
&+& \frac{g}{2} p_\rho \frac{\partial }{\partial p_\sigma} 
\Big\{ \bar{\cal F}^{\rho \sigma}(X), \Gamma^{\mu \nu}(X,p) \Big\} 
\\[2mm] \non
&=&
\frac{i}{2} \int d^4u  \, e^{ip \cdot u} 
\Big\langle 
e^{\frac{u}{2}\cdot \bar{\cal D}} 
\Big( \bar{\cal D}^\mu \bar{\cal D}^\rho  a_\rho(X) \Big)
\otimes 
a^\nu (X)  e^{- \frac{u}{2}\cdot  \buildrel\leftarrow\over{\bar{\cal D}}} 
-
e^{\frac{u}{2}\cdot \bar{\cal D}} a^\mu (X) 
\otimes  
\Big( a_\sigma(X) \buildrel\leftarrow\over{\bar{\cal D}^\sigma \bar{\cal D}^\nu}  \Big)
e^{- \frac{u}{2}\cdot  \buildrel\leftarrow\over{\bar{\cal D}}}  
\Big\rangle 
\\[2mm] \non
&& ~~~~~~~~~~~~~~~~~~~~~~~~~~~~~~~~~~~~~~~~~~~~~~~~~~~~~~~~~~~~~~~~ -
g \bar{\cal F}^{\mu \rho} (X)\Gamma^{\;\;\nu}_\rho(X,p) 
- g \Gamma^\mu_{\;\;\sigma}(X,p) \bar{\cal F}^{\sigma \nu}(X) ,
\ea
where additionally  $\bar{\cal F}^{\mu \rho}$ and $\bar{\cal F}^{\sigma \nu}$ have been pulled out of the average $\langle \dots \rangle$.

Until now we have not applied any gauge condition which is done right now by requiring
\be
\label{gauge-cond}
\bar{\cal D}^\mu a_\mu(x) = 0 .
\ee
Because the covariant derivative transforms as
\be
\bar{\cal D}^\mu \rightarrow {\cal U}(x) \, \bar{\cal D}^\mu \, {\cal U}^\dagger (x) ,
\ee
the gauge condition (\ref{gauge-cond}) is invariant under the gauge transformation (\ref{trans-a}). Applying the gauge condition  (\ref{gauge-cond}), the two terms on the right-hand-side of Eq.~(\ref{trans-eq-6}) vanish and the transport equation of the Wigner function obtains its final form
\ba
\label{trans-eq-final}
p\cdot \bar{\cal D} \Gamma^{\mu \nu}(X,p) 
+ \frac{g}{2} p_\rho \frac{\partial }{\partial p^\sigma} 
\Big\{ \bar{\cal F}^{\rho \sigma}(X), \Gamma^{\mu \nu}(X,p) \Big\} 
+ g \bar{\cal F}^{\mu \rho} (X)\Gamma^{\;\;\nu}_\rho(X,p) 
+ g \Gamma^\mu_{\;\;\sigma}(X,p) \bar{\cal F}^{\sigma \nu}(X) = 0 ,
\ea
which is the main result of this section.

We assume that the Lorentz structure of the gluon Wigner function is the same as that of the gluon propagator in the general covariant gauge, that is
\be
\label{Lorentz-structure}
\Gamma^{\mu \nu}(X,p) = \Big( g^{\mu \nu} - (1- \xi) \frac{p^\mu p^\nu}{p^2} \Big) {\cal G}(X,p) ,
\ee
where $\xi$ is the gauge parameter. Substituting the Wigner function in the form (\ref{Lorentz-structure}) in the transport equation (\ref{trans-eq-final}) and taking the trace with respect to the Lorentz indices, one finds
\ba
\label{trans-eq-cal-G}
p\cdot \bar{\cal D} {\cal G} (X,p) 
+ \frac{g}{2} p_\rho \frac{\partial }{\partial p^\sigma} 
\Big\{ \bar{\cal F}^{\rho \sigma}(X), {\cal G} (X,p)  \Big\} =0  ,
\ea
where an overall factor of $(3+\xi)$ factors out and thus it has been eliminated. The terms from the right-hand-side of Eq.~(\ref{trans-eq-final}) have vanished because of the antisymmetry of $\bar{\cal F}^{\mu \nu} = - \bar{\cal F}^{\nu \mu}$ and the symmetry of $\Gamma^{\mu \nu} = \Gamma^{\nu \mu}$.

The on-mass-shell distribution function for gluons $G^{ab} (X,{\bf p})$, which is an $(N_c^2 -1) \times (N_c^2 -1)$ matrix,  is defined via the relation
\be 
\label{distri-fun-G}
{\cal G}^{ab}(X,p) = - \frac{\pi}{E_{\bf p}} 
\Big( \delta (E_{\bf p} - p_0)  \big[G^{ab} (X,{\bf p}) +1\big]
+ \delta (E_{\bf p} + p_0) G^{ba} (X,-{\bf p}) \Big).
\ee
We note that the color indices are transposed in the second term. Substituting the Wigner function in the form (\ref{distri-fun-G}) into Eq.~(\ref{trans-eq-cal-G}) and integrating over $p_0$, we obtain the transport equation of interest
\be
\label{eq-trans-G}
p_\mu \bar{\cal D}^\mu G(X,{\bf p})
+ \frac{g}{2} p_\mu \{ \bar{\cal F}^{\mu \nu}(X), \partial^p_\nu G(X,{\bf p})\} =0 .
\ee
In contrast to the quark case, the positive and negative energy parts of the gluon Wigner function (\ref{distri-fun-G}) give the same equation (\ref{eq-trans-G}). One should remember that the ${\rm SU}(N_c)$ generators in the adjoint representation of the form (\ref{ad-gen}) obey $(T^a)^T = - T^a$ and consequently ${\cal F}\cdot G^T = - (G \cdot {\cal F})^T$ where $T$ denotes the transposition of color indices.

\subsubsection{Generation of the mean field}

The set of transport equations (\ref{trans-q}), (\ref{trans-barq}), and (\ref{eq-trans-G}) needs to be supplemented by the Yang-Mills equation (\ref{YM-eq-adjoint-barF}) describing the self-consistent generation of the chromodynamic mean field $ \bar{\cal F}^{\mu \nu}.$ Needless to say, the chromodynamic field $A^\mu$ should be replaced by $\bar{A}^\mu$ in the transport equations of quarks and antiquarks  (\ref{trans-q}, \ref{trans-barq}). Our goal here is to express the quark and gluon contributions to the color current, which enters the Yang-Mills equation (\ref{YM-eq-adjoint-barF}), through the quark, antiquark and gluon distribution functions. 

Using the quark Wigner function (\ref{Wigner-def-1}), the ensemble averaged quark contribution to the color current is
\be
\label{color-current-W}
\langle j^\mu_a (X) \rangle \equiv
- g\langle \psi(X) \gamma^\mu \tau^a {\bar \psi}(X) \rangle 
= - g\int \frac{d^4p}{(2\pi)^4} \Tr[ \gamma^\mu \tau^a W (X,p) ],
\ee
where the trace is taken over spinor and color indices in the fundamental representation. We note that one gets the same current with the noncovariant Wigner function (\ref{Wigner-noncov-def-1}) as the current is local in coordinate space. Expressing the Wigner function (\ref{Wigner-def-1}) through the quark and antiquark distribution functions according to Eq.~(\ref{distri-f-def}), the quark current (\ref{color-current-W}) equals
\be
\label{quark-current-ave}
\langle j^\mu_a (X) \rangle 
= - 2g \int \frac{d^3p}{(2\pi)^3} \; 
\frac{p^\mu}{E_{\bf p}} \; 
\Tr \big[ \tau^a \big(Q(X,{\bf p}) - \bar Q (X,{\bf p}) \big)\big] .
\ee

Let us now discuss the gluon contribution to the color current. To express the gluon current (\ref{g-current}) through the gluon Wigner function (\ref{Gamma-def}), we first define the gluon propagator
\be
\label{gluon-prop}
i\,d^{\mu \nu}_{ab}(x,y) \equiv \langle a^\mu_a (x) a^\nu_b (y) \rangle ,
\ee
which allows to write down the gluon current (\ref{g-current}) as
\ba
\label{j-g-2}
\langle J^\nu_a (x) \rangle
= i g f^{abc} 
\lim_{y \rightarrow x} 
\big[
\partial^\nu_y d_{bc \mu}^{\;\;\;\;\; \mu}(x,y) 
- 2 \partial_\mu^y d^{\mu \nu}_{bc}(x,y) 
- \partial_\mu^x d^{\mu \nu}_{bc}(x,y)  \big] .
\ea

Using the Wigner transform $d^{\mu \nu}_{ab}(X,p)$, one computes
\ba
\partial^\rho_x d^{\mu \nu} (x,y) 
=\int \frac{d^4p}{(2\pi)^4}  \, e^{- i p \cdot u} \Big(\frac{1}{2} \frac{\partial}{\partial X_\rho}  - i p^\rho  \Big) d^{\mu \nu} (X,p)  ,
\\[2mm]
\partial^\rho_y d^{\mu \nu}(x,y) 
=\int \frac{d^4p}{(2\pi)^4}  \, e^{- i p \cdot u} \Big(\frac{1}{2} \frac{\partial}{\partial X_\rho}  + i p^\rho  \Big) d^{\mu \nu} (X,p)  .
\ea
Neglecting the gradients of $d^{\mu \nu} (X,p)$ with respect to $X$, the current (\ref{j-g-2}) reads
\ba
\label{j-g-3}
\langle J^\nu_a (X) \rangle 
= - g f^{abc} 
\int \frac{d^4p}{(2\pi)^4} \Big[ 
p^\nu d_{bc \mu}^{\;\;\;\;\; \mu}(X,p) 
- p_\mu d^{\mu \nu}_{bc}(X,p)  \Big] .
\ea

The gluon Wigner function  (\ref{Gamma-def}) differs from the gluon propagator (\ref{gluon-prop}) because of the link operators $\bar\Omega(X,X+u/2)$ and $\bar\Omega(X-u/2,X)$ present in the former one, see Eq.~(\ref{Gamma-def-2}). However, the gluon current (\ref{j-g-2}) is effectively a one- not two-point function, because of the limit $y \rightarrow x$. Since the links $\bar\Omega(X,X+u/2)$ and $\bar\Omega(X-u/2,X)$ smoothly tend to unity as $u \rightarrow 0$, we simply replace $i d^{\mu \nu}_{bc}(X,p)$ by $\Gamma^{\mu \nu}_{bc}(X,p)$ in Eq.~(\ref{j-g-3}) to obtain 
\ba
\label{j-g-Gamma}
\langle J^\nu_a (X) \rangle 
= i g f^{abc} 
\int \frac{d^4p}{(2\pi)^4} \Big[ 
p^\nu \Gamma_{bc \mu}^{\;\;\;\;\; \mu}(X,p) 
- p_\mu \Gamma^{\mu \nu}_{bc}(X,p)  \Big] .
\ea

Assuming again that the Lorentz structure of the gluon Wigner function is given by Eq.~(\ref{Lorentz-structure}), the gluon current (\ref{j-g-Gamma}) reads
\ba
\label{j-g-cal-G}
\langle J^\nu_a (X) \rangle 
= - 3i g f^{abc}  \int \frac{d^4p}{(2\pi)^4} \,p^\nu {\cal G}_{bc} (X,p) .
\ea
The gauge parameter $\xi$ has dropped out but expression (\ref{j-g-cal-G}) is still gauge dependent! One observes that in the physical {\it radiation gauge}, when $a^0(x) = 0$ and $\nabla \cdot {\bf a}(x)=0$, the Lorentz structure of the Wigner function is expected to be  
\ba
\Gamma^{00}(X,p) &=& \Gamma^{0i}(X,p) =  \Gamma^{i0}(X,p) = 0 ,
\\
\Gamma^{ij}(X,p) &=& \Big(g^{ij} + \frac{p^i p^j}{{\bf p}^2}\Big) {\cal G}(X,p) ,
\ea
and the gluon current (\ref{j-g-Gamma}) equals
\be
\label{j-g-cal-G-radiation}
\langle J^\nu_a (X) \rangle 
=  - 2 i g f^{abc}  \int \frac{d^4p}{(2\pi)^4} \,p^\nu {\cal G}_{bc} (X,p) .
\ee
The coefficients 3 and 2 in Eq.~(\ref{j-g-cal-G}) and (\ref{j-g-cal-G-radiation}), respectively, equal the number of degrees of freedom of gluons in the covariant and radiation gauge. The right number is obviously 2, as in the physical radiation gauge where there are only transverse gluons of two polarizations.  As well known, one should introduce the Faddeev-Popov ghosts to cancel the unphysical degree of freedom in the covariant gauge. Here we simply use the formula (\ref{j-g-cal-G-radiation}) as the gluon current. 

Substituting the definition of the gluon distribution function (\ref{distri-fun-G}) into Eq.~(\ref{j-g-cal-G-radiation}), the gluon current is
\ba
\label{j-gl-G-0}
\langle J^\nu_a (X) \rangle 
= - 2 i g f^{abc}  \int \frac{d^3p}{(2\pi)^3} \, \frac{p^\nu}{E_{\bf p}} 
\Big[G^{bc} (X,{\bf p}) +\frac{1}{2}\Big] .
\ea
As seen, in the vacuum limit, when $G \rightarrow 0$, there remains a nonzero, actually divergent, contribution to the current (\ref{j-gl-G-0}). 
After subtraction of the vacuum effect, the final formula of the gluon current reads
\ba
\label{j-gl-G}
\langle J^\nu_a (X) \rangle = - 2  g  \int \frac{d^3p}{(2\pi)^3} \, \frac{p^\nu}{E_{\bf p}} 
{\rm Tr}\big[T^a G(X,{\bf p}) \big],
\ea
where the trace is taken over the color indices.

With the currents (\ref{quark-current-ave}) and (\ref{j-gl-G}) the Yang-Mills equation (\ref{YM-eq-adjoint-barF}) describing the self-consistent generation of the chromodynamic mean field $ \bar{A}^\mu$ is
\be
\label{YM-eq-adjoint-barF-final}
\bar{\cal D}_\mu^{ab} \bar{F}^{\mu \nu}_b  =  
 - 2g \int \frac{d^3p}{(2\pi)^3} \; 
\frac{p^\nu}{E_{\bf p}} \Big\{
\Tr \big[ \tau^a \big(Q(X,{\bf p}) - \bar Q (X,{\bf p}) \big)\big] 
+ {\rm Tr}\big[T^a G(X,{\bf p}) \big] \Big\}.
\ee
Using the identity
\be
\tau^a_{ij} \tau^a_{kl} = \frac{1}{2} \delta^{il}\delta^{jk} - \frac{1}{2N_c} \delta^{ij}\delta^{kl},
\ee
one rewrites Eq.~(\ref{YM-eq-adjoint-barF-final}) to the fundamental representation as
\be
\label{YM-eq-fun-barF-final}
\bar{D}_\mu \bar{F}^{\mu \nu}  =  
 - g \int \frac{d^3p}{(2\pi)^3} \; 
\frac{p^\nu}{E_{\bf p}} \Big\{
Q(x,{\bf p}) - \bar Q (x,{\bf p})
-{1 \over N_c}{\rm Tr}\big[Q(x,{\bf p}) - \bar Q (x,{\bf p})\big]
+ 2 \tau^a {\rm Tr}\big[T^a G(x,{\bf p}) \big]
 \Big\}.
\ee
\subsection{Transport equations}
\label{sec-trans-equations}

We recapitulate here the main facts of the transport theory of quark-gluon plasma. We also simplify the rather complicated notation which was used to derive the transport equations.

The transport theory of a quark-gluon plasma is formulated in terms of particles and classical fields. The particles - quarks, antiquarks and gluons - should be understood as sufficiently hard quasiparticle excitations of quantum fields of QCD while the classical fields are highly populated soft gluonic modes. An excitation is called `hard', when its momentum in the equilibrium rest frame is of order of the temperature $T$, and it is called `soft' when the momentum is of order $gT$.

The distribution function of quarks $Q(x,{\bf p})$ is a hermitian $N_c\times N_c$ matrix in  color space. The distribution function is gauge dependent and it transforms under a local gauge transformation $U(x)$ as
\be
\label{Q-transform}
Q(x,{\bf p}) \rightarrow U(x) \, Q(x,{\bf p}) \, U^{\dag }(x) .
\ee
Here and in most cases below the color indices are suppressed. The distribution function of antiquarks, which we denote by $\bar Q(x,{\bf p})$, is also a hermitian $N_c\times N_c$ matrix and it transforms according to Eq.~(\ref{Q-transform}). The distribution function of gluons is a hermitian $(N_c^2-1)\times (N_c^2-1)$ matrix which transforms as
\be
\label{G-transform}
G(x,{\bf p}) \rightarrow {\cal U}(x) \: G(x,{\bf p})
\:{\cal U}^{\dag }(x) ,
\ee
where we recall the definition
\be
{\cal U}_{ab}(x) = 2{\rm Tr}\bigr[\tau^a U(x)
\tau^b U^{\dag }(x)] .
\ee
                                                                                
The color current in the fundamental representation equals
\ba
\label{col-current}
j^{\mu }(x) = 
- g \int \frac{d^3p}{(2\pi)^3} \; 
\frac{p^\mu}{E_{\bf p}} \;
\Big[ Q(x,{\bf p}) - \bar Q (x,{\bf p})
-{1 \over N_c}{\rm Tr}\big[Q(x,{\bf p}) - \bar Q (x,{\bf p})\big]
+ 2 \tau^a {\rm Tr}\big[T^a G(x,{\bf p}) \big]\Big] .
\ea
A sum over spin states, two per particle, and over quark flavors $N_f$ is understood in Eq.~(\ref{col-current}), even though it is not explicitly written. The current can be decomposed as $j^\mu (x) = j^\mu_a (x) \tau^a$ with $j^\mu_a (x) = 2 {\rm Tr} (\tau_a j^\mu (x))$.  Since not only quarks and antiquarks but also gluons contribute to the current (\ref{col-current}), it is clearly of a non-Abelian nature, as gluon-gluon coupling is taken into account. The distribution functions, which are proportional to the unit matrix in color space, are gauge independent and contribute a vanishing current.

Gauge invariant quantities, such as the baryon current $(b^\mu )$ and the energy-momentum tensor $(t^{\mu \nu})$, are given by the traces of the distribution functions: 
\ban
b^{\mu }(x) &=& {1 \over 3} \int \frac{d^3p}{(2\pi)^3}\; p^{\mu} \;
{\rm Tr}\Big[ Q(x,{\bf p}) - \bar Q (x,{\bf p}) \Big] \; ,
\\[2mm]
t^{\mu \nu}(x) &=&
\int \frac{d^3p}{(2\pi)^3}\; p^{\mu} p^{\nu} \;
{\rm Tr}\Big[ Q(x,{\bf p})
+ \bar Q(x,{\bf p}) + G(x,{\bf p}) \Big] ,
\ean
where we use the same symbol ${\rm Tr}[\cdots]$ for the trace in both the fundamental and adjoint representations.
                                                                                
The distribution functions of quarks, antiquarks, and gluons satisfy the transport equations:
\ba
\label{transport-q} 
p^{\mu} D_{\mu}Q(x,{\bf p}) + {g \over 2}\: p^{\mu}
\left\{ F_{\mu \nu}(x), \partial^\nu_p Q(x,{\bf p}) \right\}
&=& C[Q,\bar Q,G] ,
\\ [2mm]
\label{transport-barq} 
p^{\mu} D_{\mu}\bar Q(x,{\bf p}) - {g \over 2} \: p^{\mu}
\left\{ F_{\mu \nu}(x), \partial^\nu_p \bar Q(x,{\bf p})\right\}
&=& \bar C[Q,\bar Q,G],
\\ [2mm]
\label{transport-gluon}
p^{\mu} {\cal D}_{\mu}G(x,{\bf p}) + {g \over 2} \: p^{\mu}
\left\{ {\cal F}_{\mu \nu}(x), \partial^\nu_p G(x,{\bf p}) \right\}
&=& C_g[Q,\bar Q,G],
\ea
where the covariant derivatives $D_{\mu}$ and ${\cal D}_{\mu}$ act as
\be
D_{\mu} = \partial_{\mu} - ig[A_{\mu}(x),...\; ]\;,\;\;\;\;\;\;\;
{\cal D}_{\mu} = \partial_{\mu} - ig[{\cal A}_{\mu}(x),...\;]\;,
\ee
with $A_{\mu }$ and ${\cal A}_{\mu }$ being four-potentials of the mean chromodynamic field in the fundamental and adjoint representations, respectively:
\be
A^{\mu }(x) = A^{\mu }_a (x) \tau^a , \;\;\;\;\;
{\cal A}^{\mu }(x) = A^{\mu }_a (x) T^a ;
\ee
$\{...,...\}$ denotes the anticommutator and $\partial^\nu_p$ the four-momentum derivative. Since the distribution functions do not depend on $p_0$, the derivative with respect to $p_0$ is identically zero.  The field strength tensors $F^{\mu \nu}$ and ${\cal F}^{\mu \nu}$, which belong to the fundamental and adjoint representation, respectively, are defined through $A^{\mu }$ and ${\cal A}^{\mu }$ in a standard way, that is as
$F^{\mu \nu} \equiv \partial^\mu A^\nu - \partial^\nu A^\mu - i g [A^\mu, A^\nu]$ and ${\cal F}^{\mu \nu} \equiv \partial^\mu {\cal A}^\nu - \partial^\nu {\cal A}^\mu - i g [{\cal A}^\mu, {\cal A}^\nu]$.

On the right-hand side, $C, \bar C$, and $C_g$ represent collision terms which, in spite of some efforts \cite{Selikhov:1991br,Selikhov:1994xn,Blaizot:1999xk,Mrowczynski:2009gf},  have not been satisfactorily  derived for an arbitrary nonequilibrium QGP state. The collisions can be easily taken into account using the approximate BGK collision terms \cite{Manuel:2004gk,Schenke:2006xu}. Within a more realistic approach color charges are treated in a similar way as spin degrees of freedom, and one uses the so-called Waldmann-Snider collision terms \cite{Arnold:1998cy,Manuel:2003zr} which are usually applied to study spin transport. In our further considerations we will use the collisionless transport equations where $C = \bar C = C_g = 0$.

The chromodynamic mean field, which enters the transport equations (\ref{transport-q}), (\ref{transport-barq}), and (\ref{transport-gluon}), is generated self-consistently by quarks, antiquarks, and gluons present in the plasma. The transport equations are thus supplemented by the Yang-Mills equation 
\be
\label{yang-mills}
D_{\mu} F^{\mu \nu}(x) = j^{\nu}(x) ,
\ee
where the current is given by Eq.~(\ref{col-current}).

\subsection{Linear response analysis }

We study here how the quark-gluon plasma, which is (on average) colorless, homogeneous, and stationary, responds to color fluctuations. The distribution functions are assumed to be of the form
\ba
\label{5.1a}
Q_{ij}(x,{\bf p}) &=& f_q({\bf p})\delta_{ij} + \delta Q_{ij}(x,{\bf p}) ,
\\ 
 \label{5.1b}
\bar Q_{ij}(x,{\bf p}) &=& \bar{f}_q({\bf p})\delta_{ij}
+ \delta \bar Q_{ij}(x,{\bf p}) ,
\\ 
\label{5.1c}
G_{ab}(x,{\bf p}) &=& f_g({\bf p})\delta_{ab} + \delta G_{ab}(x,{\bf p}),
\ea
where the functions $f_q({\bf p})\delta_{ij}$, $\bar{f}_q({\bf p})\delta_{ij}$, and $f_g({\bf p})\delta_{ab}$ represent the colorless, homogeneous, and stationary state of the plasma. The functions are related to the distributions $n_q({\bf p})$, $\bar{n}_q ({\bf p})$, and $n_g({\bf p})$ from Sec.~\ref{sec-field-theory}  as $f_q({\bf p}) = N_f n_q({\bf p})$, $\bar{f}_q({\bf p}) = N_f \bar{n}_q ({\bf p})$, and $f_g({\bf p}) = n_g({\bf p})$.  The functions  $\delta Q_{ij}(x,{\bf p})$, $\delta \bar Q_{ij}(x,{\bf p})$, and $\delta G_{ab}(x,{\bf p})$, which describe colorful fluctuations, are assumed to be much smaller than the respective colorless functions. The same is assumed in the momentum gradients of these functions. 
                                                                             
Substituting (\ref{5.1a}), (\ref{5.1b}), and (\ref{5.1c}) in (\ref{col-current}), one obtains
\begin{eqnarray}
\label{5.2}
j^{\mu}(x) = -\frac{g}{2} \int {d^3p \over (2\pi )^3}\, 
\frac{p^{\mu}}{E_{\bf p}}
\Bigr[ \delta Q(x,{\bf p}) - \delta \bar Q (x,{\bf p})
-{1 \over N_c}{\rm Tr}\bigr[\delta Q(x,{\bf p})
- \delta \bar Q (x,{\bf p})\bigl]
+ 2\tau^a {\rm Tr}\big[T^a\delta G(x,{\bf p})\big] \Bigl] .
\end{eqnarray}
As can be seen, the current occurs due to deviations from the colorless state. 

We are going to derive $\delta Q(x,{\bf p})$, $\delta \bar Q(x,{\bf p})$, and $\delta G(x,{\bf p})$ as solutions of the transport equations. For this purpose we substitute the distribution functions  (\ref{5.1a}), (\ref{5.1b}), and (\ref{5.1c}) into Eqs.~(\ref{transport-q}), (\ref{transport-barq}), and (\ref{transport-gluon}). Assuming that $A^\mu$ and $F^{\mu \nu}$ are of the same order as  $\delta Q$, $\delta \bar Q$, and $\delta G$ and linearizing the equations with respect to $\delta Q$, $\delta \bar Q$, and $\delta G$ one finds
\ba
\label{5.4a}
p^{\mu} \partial_\mu \delta Q(x,{\bf p})
&=& - g p^{\mu}F_{\mu \nu}(x){\partial f_q({\bf p}) \over \partial p_{\nu}} ,
\\ 
\label{5.4b}
p^{\mu} \partial_\mu \delta \bar Q(x,{\bf p})
&=& \;\;\;
g p^{\mu}F_{\mu \nu}(x){\partial \bar{f}_q({\bf p}) \over\partial p_{\nu}} ,
\\ 
\label{5.4c}
p^{\mu} \partial_\mu \delta G(x,{\bf p})
&=&-
g p^{\mu}{\cal F}_{\mu \nu}(x){\partial f_g({\bf p}) \over \partial p_{\nu}} .
\ea
The gauge covariance of the transport equation is now lost. To restore it, the derivatives $\partial_\mu$ should be replaced by the covariant derivatives $D_\mu$ and ${\cal D}_\mu$, respectively, in the drift terms.

The linearized equations (\ref{5.4a}), (\ref{5.4b}), and (\ref{5.4c}) are easily solved by means of the retarded Green's function (\ref{ret-Green}). The solutions read
\ba
\label{5.5.1a}
\delta Q(x,{\bf p}) &=& - g \int d^4 y \:
G_{\bf p}(x - y) \; p^{\mu}F_{\mu \nu}(y) \;
{\partial f_q({\bf p}) \over \partial p_{\nu}} ,
\\
\label{5.5.1b}
\delta \bar Q(x,{\bf p}) &=& \;\;\; g \int d^4 y \:
G_{\bf p}(x - y) \; p^{\mu}F_{\mu \nu}(y)\;
{\partial \bar{f}_q({\bf p}) \over \partial p_{\nu}} ,
\\ 
\label{5.5.1c}
\delta G(x,{\bf p}) &=& - g \int d^4 y \:
G_{\bf p}(x - y) \; p^{\mu}{\cal F}_{\mu \nu}(y) \;
{\partial f_g({\bf p}) \over \partial p_{\nu}}.
\ea
To cast the solutions into a gauge covariant form, the gauge parallel transporter (\ref{link-def-fund}) is inserted 
\ba
\label{5.5a}
\delta Q(x,{\bf p}) &=& - g \int d^4 y \:
G_{\bf p}(x - y) \; \Omega(x,y)\;
p^{\mu}F_{\mu \nu}(y) \;
\Omega(y,x)\;{\partial f_q ({\bf p}) \over \partial p_{\nu}} ,
\\
\label{5.5b}
\delta \bar Q(x,{\bf p}) &=& \;\;\; g \int d^4 y \:
G_{\bf p}(x - y) \; \Omega(x,y)\;
p^{\mu}F_{\mu \nu}(y)\;
\Omega(y,x)\;{\partial \bar{f}_q({\bf p}) \over \partial p_{\nu}} ,
\\ 
\label{5.5c}
\delta G(x,{\bf p}) &=& - g \int d^4 y \:
G_{\bf p}(x - y) \; \Omega(x,y)\;
p^{\mu}{\cal F}_{\mu \nu}(y) \; \Omega(y,x)
\;{\partial f_g({\bf p}) \over \partial p_{\nu}} .
\ea
The parallel transporter $\Omega$ belongs to the fundamental representation when it acts on $F^{\mu \nu}$ and to the adjoint in case of ${\cal F}^{\mu \nu}$. One checks, with the help of Eq.~(\ref{link-EOM}), that the expressions (\ref{5.5a}), (\ref{5.5b}), and (\ref{5.5c}) indeed solve the linearized equations (\ref{5.4a}), (\ref{5.4b}), and (\ref{5.4c}) with the covariant derivatives in the drift terms. 

Substituting the solutions (\ref{5.5a}), (\ref{5.5b}), and (\ref{5.5c}) in Eq. (\ref{5.2}), one finds the gauge covariant color current
\be
\label{ind-current}
j^{\mu }(x) =  g^2 
\int {d^3p \over (2\pi )^3}\, 
\frac{p^{\mu}p^{\lambda}}{E_{\bf p}}
\int d^4y \: G_{\bf p}(x - y) \; \Omega(x,y)\;
F_{\lambda \nu}(y) \; \Omega(y,x)\;
{\partial f({\bf p})\over \partial p_{\nu}} ,
\ee
where $f({\bf p}) \equiv f({\bf p}) + \bar{f}_q({\bf p}) + 2N_c f_g({\bf p}) = N_f n({\bf p}) + N_f\bar{n}_q({\bf p}) + 2N_c n_g({\bf p})$.
                                                                 
To proceed, we perform a Fourier transform of the induced current (\ref{ind-current}). Before this step, however, we neglect the terms which are not of leading order in $g$. As a result, the parallel-transporters $\Omega$ are approximated by unity and the stress tensor $F_{\mu \nu}$ by $\partial_{\mu} A_{\nu} - \partial_{\nu} A_{\mu} $. Within such an approximation, the Fourier transformed induced current (\ref{ind-current}), which is no longer gauge covariant, equals 
\be
\label{ind-current-k}
j^{\mu}(k) = g^2
\int {d^3p \over (2\pi )^3} \: 
\frac{p^{\mu}}{E_{\bf p}}
{\partial f ({\bf p}) \over \partial p_{\lambda}} \Bigg[ g^{\lambda \nu}
- { k^{\lambda} p^{\nu} \over p^{\sigma}k_{\sigma} + i0^+} \Bigg] \;
A_{\nu}(k) .
\ee
We have thus rederived the current  (\ref{j-F-k-final}) which provides - via the relation (\ref{j-Pi-A}) - the gluon polarization tensor (\ref{Pi-class}) obtained in Sec.~\ref{sec-field-theory} diagrammatically.


\section{Chromohydrodynamics}
\label{sec-chromo-fluid}

Electromagnetic plasmas are often described by means of  fluid equations \cite{Kra73} which result from the conservation laws combined with Maxwell equations.  The approach is less detailed than that based on kinetic theory but the fluid equations are noticeably simpler to solve and the hydrodynamic approach proved useful in numerical simulations of plasma evolution, studies of  nonlinear dynamics, etc. \cite{Kra73}.  It is thus not surprising that chromohydrodynamics was discussed by several authors over a long period of time \cite{Kajantie:1980jj,Holm:1984hg,Holm:1985yh,Mrowczynski:1987ch,Mrowczynski:1989bv,Bhatt:1988tq,Jackiw:2000cd,Bistrovic:2002jx,Manuel:2003zr,Dai:2006te,Bambah:2006yg,PeraltaRamos:2012er,Calzetta:2013nqa}. However,  the equations of color fluids have not helped much to achieve a deeper insight into the dynamics of the quark-gluon plasma. Recently, a new form of chromohydrodynamics has been proposed \cite{Manuel:2006hg} which has appeared useful in studies of  color instabilities \cite{Mannarelli:2007gi}.  Another hydrodynamic approach to color instabilities is presented in \cite{Calzetta:2013nqa}. We discuss here the chromohydrodynamics as formulated in \cite{Manuel:2006hg} to cover a whole spectrum of theoretical tools applicable to the unstable QGP. 

Before proceeding to the main subject of this section, an important point has to be clarified.  Real hydrodynamics deals with systems that are in local equilibrium, and thus it is applicable only at sufficiently long time scales. The continuity and the Euler or Navier-Stokes equations are supplemented by the equation of state to form a complete set of equations. The equations can be derived from kinetic theory, using a local equilibrium distribution function, which by definition maximizes the entropy density, and thus the function cancels the collision terms of the transport equations. Such an approach to electromagnetic plasmas is known as magnetohydrodynamics (MHD) and it is frequently used to solve various problems of electromagnetic plasmas. An analogous approach to QGP -- chromohydrodynamics -- was derived in \cite{Manuel:2003zr} where the local equilibrium state was found using a collision term of the Waldman-Snider form. The resulting chromohydrodynamics was trivial in the sense that, although the local equilibrium can be colorful, all color components of the plasma move with the same hydrodynamic velocity. Therefore, chromodynamic effects disappear entirely once the system is neutralized. It actually occurs even before local equilibrium is achieved \cite{Manuel:2004gk}, since the process of plasma whitening is faster than momentum equilibration. Thus, there is no QCD analog of magnetohydrodynamics which appears due to a large difference of the electron and ion masses which effectively slows down mutual equilibration of the electrons and ions.  Therefore, at a relatively long time scale, one deals with a charged electron fluid in local equilibrium which moves in a passive background of positive ions.

Since the hydrodynamic equations express macroscopic conservation laws, the equations hold not only for systems in local equilibrium but also for non-equilibrium systems. In particular, the equations can be applied at time scales significantly shorter than that for local equilibration. At such a short time scale the collision terms of the transport equations can be neglected. However, extra assumptions are then needed to close the set of equations, since the (equilibrium) equation of state cannot be used.  In the next two subsections we derive the fluid equations relevant for this regime and show how to make use of them to derive the gluon polarization tensor.

\subsection{Fluid equations}
\label{subsec-fluid-eq}

We assume that there are several streams in the plasma system under consideration and that the distribution functions of quarks, antiquarks, and gluons of each stream satisfy the collisionless transport equation. The streams are labeled with the index $\alpha$.
                                                                                
Further analysis is limited to quarks, but inclusion of anti-quarks and gluons is straightforward. The distribution function of quarks belonging to the stream $\alpha$ is denoted as $Q_\alpha(x,{\bf p})$. Integrating the collisionless transport Eq.~(\ref{transport-q}) satisfied by $Q_\alpha$ over momentum, one finds the covariant continuity equation
\be
\label{cont-eq-qgp}
D_\mu n^\mu_\alpha = 0 ,
\ee
where $n^\mu_\alpha$ is an $N_c\times N_c$ matrix defined as
\be
\label{flow-qgp}
n^\mu_\alpha (x) \equiv \int \frac{d^3p}{(2\pi)^3}\;
p^\mu Q_\alpha(x,{\bf p}) .
\ee
The four-flow $n^\mu_\alpha$ transforms under gauge transformations as the quark distribution function, {\it i.e.} according to Eq.~(\ref{Q-transform}).
                                                                                
Multiplying the transport equations (\ref{transport-q}) by the four-momentum and integrating the product over momentum, one obtains
\be
\label{en-mom-eq-qgp}
D_\mu T^{\mu \nu}_\alpha
- {g \over 2}\{F_{\mu}^{\;\; \nu}, n^\mu_\alpha \}= 0 ,
\ee
where the energy-momentum tensor is
\be
\label{en-mom-q-qgp}
T^{\mu \nu}_\alpha (x) \equiv \int \frac{d^3p}{(2\pi)^3}\;
p^\mu p^\nu  Q_\alpha(x,{\bf p}) .
\ee
                                                                                
We further assume that the structure of $n^\mu_\alpha$ and $T^{\mu \nu}_\alpha$ is
\ba
\label{flow-id-qgp}
n^\mu_\alpha(x)  &=& n_\alpha (x) \, u_\alpha^\mu(x) ,
\\[2mm]
\label{en-mom-id-qgp}
T^{\mu \nu}_\alpha(x)  &=& {1 \over 2}
\big(\epsilon_\alpha (x) + p_\alpha (x)\big)
\big\{u^\mu_\alpha (x), u^\nu_\alpha (x) \big\}
- p_\alpha (x) \, g^{\mu \nu},
\ea
where the hydrodynamic velocity $u^\mu_\alpha$ is, as $n_\alpha$, $\epsilon_\alpha$, and $p_\alpha$, an $N_c\times N_c$ matrix. The anticommutator of $u^\mu_\alpha$ and $u^\nu_\alpha$ is present in Eq.~(\ref{en-mom-id-qgp}) to guarantee the symmetry of $T^{\mu \nu}_\alpha$ with respect to $\mu \leftrightarrow \nu$ which is evident in Eq.~(\ref{en-mom-q-qgp}).
                                                                                
The relativistic version of the Euler equation can be obtained from the Abelian version of Eq.~(\ref{en-mom-eq-qgp}) by removing from it the part which is parallel to $u^\mu_\alpha$. Such a procedure is not possible for a non-Abelian plasma because the matrices $n_\alpha$, $u^\mu_\alpha$, and $u^\nu_\alpha$, in general, do not commute with one other. Therefore, one has to work directly with Eqs.~(\ref{cont-eq-qgp}) and (\ref{en-mom-eq-qgp}) with $n^\mu_\alpha$ and $T^{\mu \nu}_\alpha$ defined by Eqs.~(\ref{flow-id-qgp}) and (\ref{en-mom-id-qgp}). The equations have to be supplemented by the Yang-Mills Eq.~(\ref{yang-mills}) with the color current of the form
\be
\label{hydro-current-qgp}
j^\mu(x) = -\frac{g}{2} \sum_\alpha \Big(n_\alpha u^\mu_\alpha
- {1 \over N_c}{\rm Tr}\big[n_\alpha u^\mu_\alpha  \big]\Big) ,
\ee
where only the quark contribution is taken into account for now.

The fluid Eqs.~(\ref{cont-eq-qgp}) and (\ref{en-mom-eq-qgp}) do not form a closed set of equations but can be closed with an equation analogous to the equation of state. Since quarks and gluons are (approximately) massless, $p^2 = 0$, the energy-momentum tensor (\ref{en-mom-q-qgp}) is traceless ($T^\mu_{\mu \, \alpha}(x) = 0$). Then, Eq.~(\ref{en-mom-id-qgp}) combined with the constraint $u_\alpha^\mu(x)u_{\alpha \,\mu}(x) = 1$ provides the desired equation
\be
\label{EoS}
\epsilon_\alpha (x) = 3 p_\alpha (x) ,
\ee 
which relates the matrix-valued functions $\epsilon_\alpha$ and $p_\alpha$. It formally coincides with the equation of state of an ideal gas of massless particles. 

\subsection{Linear response analysis}
\label{subsec-lin-res-chromohydro}

In this section the hydrodynamic equations (\ref{cont-eq-qgp}) and (\ref{en-mom-eq-qgp}) are linearized around a stationary, homogeneous, and colorless state described by $\bar n$, $\bar \epsilon$, $\bar p$, and $\bar u^\mu$. We mostly omit here the index $\alpha$ in order to simplify the notation. The index is restored in the very final formulae.

Since every stream is assumed to be colorless, the matrices $\bar n$, $\bar \epsilon$, $\bar p$, and $\bar u^\mu$ are proportional to a unit matrix in color space. Therefore,
\be
\label{neutral-cov}
\bar n  \, \bar u^\mu
- {1 \over N_c}{\rm Tr}\big[\bar n \, \bar u^\mu  \big]
= 0 ,
\ee
which means that the color four-current of every stream vanishes. The quantities of interest are decomposed as
\be
n (x) = \bar n  + \delta n(x)
, \;\;\;\;\;\;\;
\epsilon (x) = \bar \epsilon + \delta \epsilon (x)
,
\ee
\be
p (x) = \bar p  + \delta p(x)
, \;\;\;\;\;\;\;
u^\mu (x) = \bar u^\mu + \delta u^\mu (x)
.
\ee
Since the state described by $\bar n$, $\bar \epsilon$, $\bar p$ and $\bar u^\mu$ is assumed to be stationary, homogeneous, and colorless, we have
\be
D^\mu \bar n  = 0 , \;\;\;\;\;\;\;
D^\mu \bar \epsilon  = 0 , \;\;\;\;\;\;\;
D^\mu \bar p  = 0 , \;\;\;\;\;\;\;
D^\mu \bar u^\nu  = 0 . \;\;\;\;\;\;\;
\ee
Because we consider only small deviations from the stationary, homogeneous, and colorless state, the following conditions are obeyed
\be
\bar n \gg \delta n , \;\;\;\;\;\;\;
\bar \epsilon \gg \delta \epsilon , \;\;\;\;\;\;\;
\bar p \gg \delta p , \;\;\;\;\;\;\;
\bar u^\mu \gg \delta u^\mu .
\ee
Note that $\delta n$, $\delta \epsilon$, $\delta p$, and $\delta u^\mu$ should be diagonalized in order to be compared to $\bar n$, $\bar \epsilon$, $\bar p$, and $\bar u^\mu$.

Substituting the linearized $n^\mu$ and $T^{\mu \nu}$, which are
\be
\label{flow-lin}
n^\mu  = \bar n \, \bar u^\mu + \bar n \, \delta u^\mu
+ \delta n \,\bar u^\mu
,
\ee
\be
\label{en-mom-lin}
T^{\mu \nu} = (\bar \epsilon + \bar p )
\bar u^\mu \, \bar u^\nu - \bar p  \, g^{\mu \nu}
+ (\delta \epsilon + \delta p )
\bar u^\mu \, \bar u^\nu
+ (\bar\epsilon + \bar p )
(\bar u^\mu \, \delta u^\nu + \delta u^\mu \, \bar u^\nu )
- \delta p  \, g^{\mu \nu}
,
\ee
into Eqs.~(\ref{cont-eq-qgp}) and (\ref{en-mom-eq-qgp}), one finds
\be
\label{cont-lin-eq}
\bar n \, D_\mu \delta u^\mu
+ (D_\mu \delta n ) \, \bar u^\mu = 0 ,
\ee
\be
\label{en-mom-lin-eq}
\big(D_\mu (\delta \epsilon + \delta p )\big)
\bar u^\mu \, \bar u^\nu
+ (\bar\epsilon + \bar p )
\big(\bar u^\mu \, (D_\mu \delta u^\nu)
+ (D_\mu \delta u^\mu) \, \bar u^\nu \big)
- D^\nu \delta p
- g F^{\mu \nu} \bar n \bar u_\mu = 0 .
\ee
Projecting Eq.~(\ref{en-mom-lin-eq}) with $\bar u^\nu$, one finds
\be
\label{en-den-lin-eq}
\bar u^\mu D_\mu \delta \epsilon
+ (\bar\epsilon + \bar p ) D_\mu \delta u^\mu = 0 .
\ee
To derive Eq.~(\ref{en-den-lin-eq}) one uses the fact that $\bar u^\mu \delta u_\mu = {\cal O}\big((\delta u)^2\big)$ and $\bar u^\mu D^\nu \delta u_\mu = {\cal O}\big((\delta u)^2\big)$. This happens because $u^\mu u_\mu = 1 = \bar u^\mu  \bar u_\mu$, and consequently $2 \bar u^\mu \delta u_\mu + \delta u^\mu\delta u_\mu = 0$.

Acting on Eq.~(\ref{en-mom-lin-eq}) with the projection operator $(g_{\sigma \nu} - \bar u_\sigma \bar u_\nu)$, one obtains the linearized relativistic Euler equation
\be
\label{Euler-lin}
(\bar \epsilon + \bar p ) \bar u_\mu D^\mu  \delta u^\nu
- (D^\nu - \bar u^\nu \bar u_\mu D^\mu ) \delta p
- g \bar n  \bar u_\mu F^{\mu \nu}= 0.
\ee
As already mentioned, Eqs.~(\ref{cont-lin-eq}), (\ref{en-den-lin-eq}), and (\ref{Euler-lin}) do not form a closed set of equations. In the following, we solve the equations, utilizing two different methods for closing the system.

\subsubsection{Pressure gradients neglected}

The system is closed when the pressure gradients are neglected, which physically means that the system's dynamics is dominated by the mean field. When the term with the pressure gradient is dropped in Eq.~(\ref{Euler-lin}), the equation 
(\ref{en-den-lin-eq}) effectively decouples from the remaining equations, and one has to solve only two equations, (\ref{cont-lin-eq}) and (\ref{Euler-lin}), where $\delta \epsilon$ is absent.

In principle, Eqs.~(\ref{cont-lin-eq}) and (\ref{Euler-lin}) with the pressure term neglected can be formally solved in a gauge covariant manner, using the link operator (\ref{link-def-fund}). However, we are interested here only in computing the 
polarization tensor. Thus, we replace the covariant derivatives by the normal ones in order to fully linearize the equations, and the gauge independence of the result is checked {\it a posteriori}.  After performing a Fourier transform, Eqs.~(\ref{cont-lin-eq}) and (\ref{Euler-lin}) become of the form
\be
\label{cont-eq3}
\bar u^\mu k_\mu \delta n
+ \bar n k_\mu \delta u^\mu = 0 ,
\ee
\be
\label{cov-Euler3}
i(\bar \epsilon + \bar p )
\bar u_\mu k^\mu  \delta u^\nu
+ g \bar n  \bar u_\mu F^{\mu \nu}= 0.
\ee
Eqs.~(\ref{cont-eq3}) and (\ref{cov-Euler3}) are easily solved, providing
\be
\delta n = -i g \;
\frac{\bar n^2}{\bar \epsilon + \bar p} \;
\frac{\bar u_\mu k_\nu}{(\bar u \cdot k)^2} \;
F^{\mu \nu} ,
\ee
\be
\delta u^\nu = i g \;
\frac{\bar n}{\bar \epsilon + \bar p} \;
\frac{\bar u_\mu}{\bar u \cdot k} \;
F^{\mu \nu} .
\ee

Since ${\rm Tr}[ F^{\mu \nu}] = 0$, the induced current equals 
\be
\label{curr-lin} \delta j^\mu = -\frac{g}{2}\sum_\alpha (\bar
n_\alpha \delta u^\mu_\alpha + \delta n_\alpha \, \bar
u^\mu_\alpha ),
\ee
where the index labeling the streams present in the plasma system is restored. Since the linearized field strength tensor equals 
$F^{\mu \nu}(k) = - ik^\mu A^\nu(k) + ik^\nu A^\mu(k)$, one finally finds
\be 
\delta j^\mu(k) = -
\Pi^{\mu \nu}(k) A_\nu(k) , 
\ee 
with
\be 
\label{Pi-hydro}
\Pi^{\mu \nu}(k) = \frac{g^2}{2} \sum_\alpha \frac{\bar
n_\alpha^2}{\bar \epsilon_\alpha + \bar p_\alpha} 
\frac{k^2  \bar u^\mu_\alpha \bar u^\nu_\alpha  
- \big(k^\mu \bar u^\nu_\alpha + \bar u^\mu_\alpha k^\nu - (\bar u_\alpha \cdot k) g^{\mu \nu} \big) (\bar u_\alpha \cdot k)} {(\bar u_\alpha \cdot k)^2} .
\ee 
The polarization tensor $\Pi^{\mu \nu}(k)$ is proportional to a unit matrix in color space and it is symmetric. The tensor is also transverse ($k_\mu \Pi^{\mu \nu}(k) = 0$), and thus it is gauge independent. 

The polarization tensor (\ref{Pi-hydro}) can be obtained from the formula (\ref{Pi-k-final}), which was derived diagrammatically and within the kinetic theory,  when the distribution function is of the form
\be
\label{tsunami}
f({\bf p}) = \sum_\alpha \bar n_\alpha \, \bar u^0_\alpha \;
\delta^{(3)}\Big({\bf p}
- \frac{\bar \epsilon_\alpha + \bar p_\alpha}
{\bar n_\alpha} \, \bar{\bf u}_\alpha \Big) .
\ee

The delta-like distribution function (\ref{tsunami}), and consequently the approximation, which neglects the pressure gradients, is applicable for systems where the thermal momentum ($p_{\rm thermal}$) is much smaller than the collective momentum ($p_{\rm collec}$) of the hydrodynamic flow. In an electromagnetic plasma of electrons and ions, such a situation occurs for sufficiently low temperatures when the effects of pressure are indeed expected to be small. For a system of massless partons, the condition $p_{\rm thermal} \ll p_{\rm collec}$ is achieved by requiring that $p_{\rm collec}$ is large, rather than $p_{\rm thermal}$ is small. For massless particles in local equilibrium $p_{\rm thermal} \sim T$, where $T$ is the local temperature, while the formula (\ref{tsunami}) clearly shows that $p_{\rm collec} \sim T \bar\gamma_\alpha \bar v_\alpha$ where $\bar v_\alpha $ and $\bar\gamma_\alpha$ are the velocity and Lorentz factor of the collective flow. Therefore, the condition $p_{\rm collec} \gg p_{\rm thermal}$ requires $\bar\gamma_\alpha \gg 1$.

\subsubsection{Pressure gradients included}

Equation (\ref{EoS}) allows one to close the system of fluid equations, not neglecting the pressure gradients. Using the relation (\ref{EoS}), one has to solve three equations (\ref{cont-lin-eq}), (\ref{en-den-lin-eq}), and (\ref{Euler-lin}). Performing the linearization analogous to that from the previous section and the Fourier transformation, Eqs.~(\ref{cont-lin-eq}), (\ref{en-den-lin-eq}), and (\ref{Euler-lin}) become of the 
form
\be
\label{cont-eq6}
\bar u^\mu k_\mu \delta n
+ \bar n k_\mu \delta u^\mu = 0 ,
\ee
\be
\label{ener-den-eq6}
\bar u^\mu k_\mu \delta \epsilon +
\frac{4}{3}\bar \epsilon k_\mu  \delta u^\mu
= 0,
\ee
\be
\label{cov-Euler6}
i\frac{4}{3}\bar\epsilon
\bar u_\mu k^\mu  \delta u^\nu
+ i\frac{1}{3}
(\bar u^\mu \bar u^\nu k_\mu - k^\nu )
\delta \epsilon
+ g\bar n \bar u_\mu F^{\mu \nu}= 0.
\ee

Substituting $\delta \epsilon$ obtained from Eq.~(\ref{ener-den-eq6}) into the Euler equation (\ref{cov-Euler6}), one finds
\be
\label{cov-Euler7}
\Big[g^{\nu \mu} + \frac{1}{3(\bar u \cdot k)^2}
\big(k^\nu k^\mu - \bar u^\nu k^\mu (\bar u \cdot k) \big) \Big]
\delta u_\mu =
i\frac{3}{4} g \,
\frac{\bar n}{\bar\epsilon \,(\bar u \cdot k)} \,
\bar u_\mu F^{\mu \nu} .
\ee
Observing that
\be
\big(k_\sigma k_\nu - \bar u_\sigma k_\nu(\bar u \cdot k) \big)
\big(k^\nu k_\mu - \bar u^\nu k_\mu (\bar u \cdot k) \big)
\propto (k_\sigma k_\mu - \bar u_\sigma k_\mu(\bar u \cdot k)\big) ,
\ee
the operator on the left-hand-side of Eq.~(\ref{cov-Euler7}) can be inverted as
\be
\Big[g_{\sigma \nu} - \frac{1}{k^2 + 2(\bar u \cdot k)^2}
\big(k_\sigma k_\nu - \bar u_\sigma k_\nu
(\bar u \cdot k) \big) \Big] \;
\Big[g^{\nu \mu} + \frac{1}{3(\bar u \cdot k)^2}
\big(k^\nu k^\mu - \bar u^\nu k^\mu (\bar u \cdot k) \big) \Big]
= g_\sigma^{\;\;\mu} ,
\ee
and Eq.~(\ref{cov-Euler7}) is exactly solved by
\be
\delta u_\sigma = i\frac{3}{4} g \,
\frac{\bar n}{\bar\epsilon \,(\bar u \cdot k)}
\Big[g_{\sigma \nu} - \frac{1}{k^2 + 2(\bar u \cdot k)^2}
\big(k_\sigma k_\nu - \bar u_\sigma k_\nu
(\bar u \cdot k) \big) \Big]
\bar u_\mu F^{\mu \nu} .
\ee

Substituting $\delta n$ and $\delta u^\mu$ into the current (\ref{curr-lin}), one finds the polarization tensor to be
\ba
\label{Pi2}
\nonumber
\Pi^{\mu \nu}(k) = \frac{g^2}{2} \sum_\alpha
\frac{3 \bar n_\alpha^2}{4 \bar\epsilon_\alpha}
\frac{1}{(\bar u_\alpha \cdot k)^2}
&\Big[&
k^2  \bar u^\mu_\alpha \bar u^\nu_\alpha  
- \big(k^\mu \bar u^\nu_\alpha + \bar u^\mu_\alpha k^\nu - (\bar u_\alpha \cdot k) g^{\mu \nu} \big) (\bar u_\alpha \cdot k)
\\[2mm]
&+&
\frac{(\bar u_\alpha \cdot k)k^2
(k^\mu \bar u^\nu_\alpha + k^\nu \bar u^\mu_\alpha )
- (\bar u_\alpha \cdot k)^2 k^\mu k^\nu
- k^4  \bar u^\mu_\alpha \bar u^\nu_\alpha}
{k^2 + 2(\bar u_\alpha \cdot k)^2} \: \Big],
\ea
where the stream index $\alpha$ has been restored. The first term in Eq.~(\ref{Pi2}) corresponds to the polarization tensor (\ref{Pi-hydro}) found when the pressure gradients are neglected while the second term gives the effect of the pressure gradients. The first term is, as already mentioned, symmetric and transverse. The second term is symmetric and transverse as well. Thus, the whole polarization tensor (\ref{Pi2}) is symmetric and transverse.

As explained below Eq.~(\ref{tsunami}), the pressure gradients can be neglected and the distribution function can be approximated by the delta-like form (\ref{tsunami}) when $\bar \gamma_\alpha \gg 1$. Since the four-velocity is expressed as $ \bar u^\mu_\alpha = (\bar\gamma_\alpha ,\bar\gamma_\alpha \bar{\bf  v}_\alpha)$, it is easy to check that the second term in Eq.~(\ref{Pi2}) is small, when compared to the first one, for $\bar \gamma_\alpha \gg 1$. However, one has to be careful here: when the wave vector ${\bf k}$ is exactly parallel to $\bar{\bf v}_\alpha$ and both are, say, in the $z$ direction, then the two terms of the non-zero components of the polarization tensor (\ref{Pi2}) ($\Pi^{00}$, $\Pi^{0z}$, $\Pi^{zz}$) are comparable and vanish as $1/\bar\gamma_\alpha^2$. For the remaining non-zero components of (\ref{Pi2}) ($\Pi^{xx}$, $\Pi^{yy}$), the second term is identically zero.


\section{Collective Modes}
\label{sec-collective}                                                                                

Let us consider the quark-gluon plasma in a homogeneous and stationary state in which case there are no net local color charges and no net currents. At a moment the state is perturbed by either a random fluctuation or external field. As a result, there appear in the plasma local charges or currents which generate chromoelectric and chromomagnetic fields. The fields in turn interact with colored partons which contribute to the local charges and currents.  If the wavelength of the perturbation exceeds the typical inter-particle spacing, the plasma undergoes a collective motion involving many partons which are present within the interaction range. The collective motions caused by imbalanced charges or currents are classically called the {\it plasma oscillations} or {\it plasma waves}. Quantum-mechanically we deal with quasi-particle collective excitations of the plasma.

The spectrum of collective excitations is a fundamental characteristic of any many-body system. It carries a great deal of information on the thermodynamic and transport properties of an equilibrium system, it also controls to a large extent the temporal evolution of a non-equilibrium one. In the quark-gluon plasma there are collective modes corresponding to plasma particles, that is quarks and (transverse) gluons, there are also collective excitations being genuine many-body phenomena like longitudinal gluon modes (longitudinal plasmons), plasminos, and phonons which are density fluctuations. We discuss here the longitudinal and transverse gluon, as well as the quark collective modes.

\subthreesection{Gluon modes}

The dispersion relations for gluonic collective modes naturally appear when one considers the equation of motion of the Fourier-transformed gauge potential $A^{\mu}(k)$ which represents the chromodynamic mean field. Starting with the Yang-Mills equation  (\ref{YM-eq}) and using the relation (\ref{j-Pi-A}), the equation of motion of $A^{\mu}(k)$ in the absence of external sources is found to be
\be
\label{eq-A}
\Big[ k^2 g^{\mu \nu} -k^{\mu} k^{\nu} - \Pi^{\mu \nu}(k) \Big]
A_{\nu}(k) = 0 ,
\ee
where $k \equiv (\omega, {\bf k})$ and $\Pi^{\mu \nu}(k)$ is the retarded polarization tensor which contains dynamical information about the system. Specifically, it describes how the plasma responds to chromodynamic perturbations. The tensor $\Pi^{\mu \nu}(k)$ must obey the Taylor-Slavnov identity to get gauge independent results. 

In QED with gauge fixing condition which is linear in gauge potential, the Ward-Takahashi identity -- the Taylor-Slavnov indentity for the case of Abelian fields -- implies transversality of the polarization tensor, $k_\mu \Pi^{\mu \nu}(k) = 0$. In non-Abelian gauge theories transversality holds only in certain gauges, in particular, in the Feynman gauge used in Sec.~\ref{sec-field-theory}. In general the problem of getting gauge independent results is rather complex and different gauges require different treatments, see \cite{Kobes:1990xf, Kobes:1990dc} and the review article \cite{Kraemmer:2003gd}. The situation is greatly simplified in the hard-loop approximation which was applied to obtain the polarization tensor (\ref{Pi-k-final}). As we haven seen in Sec.~\ref{subsec-actions}, the hard-loop polarization tensor, which is transverse, leads to the gauge invariant hard-loop effective action. Therefore, the transversality of $\Pi^{\mu \nu}(k)$ guarantees its gauge independence within the hard-loop approximation.

A dispersion equation of gluon collective modes is obtained as a condition that the {\em homogeneous} equation of motion (\ref{eq-A}) has non-trivial solutions. One is tempted to require 
\be
\label{dispersion-pi}
{\rm det}\Big[ k^2 g^{\mu \nu} -k^{\mu} k^{\nu} - \Pi^{\mu \nu}(k) \Big] = 0 .
\ee
However, due to the transversality of $\Pi^{\mu \nu}(k)$, the determinant from Eq.~(\ref{dispersion-pi}) vanishes identically for any wave four-vector $k$. This is certainly not the right dispersion equation but one should remember that not all four components of the gauge potential $A^{\mu}$, which enter the equation of motion (\ref{eq-A}), are independent from each other when a gauge condition is imposed.  A component of $A^{\mu}$, which can be expressed through other components, must be eliminated. This is easily done in the temporal axial gauge where $A^0=0$. Then, the dispersion equation reads 
\be
\label{dispersion-g}
{\rm det}\big[\Xi (\omega,{\bf k}) \big]  = 0 ,
\ee
and the $3\times3$ matrix $\Xi (\omega,{\bf k})$, which is the inverse gluon propagator in the temporal axial gauge, equals
\be
\label{Xi-def}
\Xi^{ij} (\omega,{\bf k}) \equiv {\bf k}^2 \delta^{ij} -k^i  k^j
- \omega^2 \varepsilon^{ij}(\omega,{\bf k}) .
\ee
We have introduced here the chromodielectric tensor $\varepsilon^{ij}(\omega,{\bf k})$ which is related to the polarization tensor by Eq.~(\ref{epsilon-Pi}). 

The dispersion equation (\ref{dispersion-g}) with the matrix (\ref{Xi-def}), which is written in the from known from electromagnetic plasmas  \cite{Rukhadze-Silin-1961}, is solved by functions $\omega ({\bf k})$ giving the dispersion relations of gluon collective modes. The gauge potential $A^{\mu}$ from Eq.~(\ref{eq-A}) can be understood as a classical field or expectation value of the gauge field operator. This is justified due to the long wavelengths of collective excitations. However, we obtain the same dispersion equation (\ref{dispersion-g}) when, instead of the classical field $A^{\mu}$, one starts with the equation of motion of a retarded propagator of the gauge field. Then, the dispersion relations correspond to poles of the propagator. 

Solutions to the dispersion equations $\omega ({\bf k})$, which are, in general, complex-valued, represent plasma modes. If the imaginary part of a mode's frequency $\Im\omega({\bf k})$ is negative,  the mode is damped - its amplitude exponentially decays in time as $e^{\Im\omega({\bf k}) \, t}$. The mode is {\it over-damped} when additionally $- \Im\omega({\bf k}) \ge |\Re \omega ({\bf k})|$. When $\Im\omega({\bf k}) =0$, we have a stable mode with a constant amplitude. Finally, if $\Im\omega({\bf k}) > 0$, the mode's amplitude exponentially grows in time, {\em i.e.} there is an instability. 

When the electric field of a mode is parallel to its wave vector ${\bf k}$, the mode is called {\it longitudinal}. There is no magnetic field associated with such a mode. A mode is called {\it transverse} when the electric field is transverse to the wave vector. The Maxwell equations show that the longitudinal modes, also known as {\it electric} or {\it electrostatic} modes, are associated with electric charge oscillations.  The transverse modes, which are also known as {\it magnetic}, correspond to oscillations of electric current. The term `magnetic mode' might be somewhat misleading, as there is never a purely magnetic field at nonzero frequency and the transverse modes are rather like electromagnetic waves in vacuum. 

\subthreesection{Quark modes}

Since the quark field $\psi (k)$ obeys the equation 
\be 
\Big[ k\sla  - \Sigma (k)  \Big] \psi (k) =0 , 
\ee
where $\Sigma (k)$ is the retarded quark self-energy, the dispersion equation of quark modes is
\be
\label{dis-quark-1}
 {\rm det}\Big[ k\sla  - \Sigma (k) \Big]  = 0 . 
\ee
The quark self-energy (\ref{Si-k-final}) has the following spinor structure 
\be 
\label{structure-quark} 
\Sigma (k) = \gamma^{\mu} \Sigma_{\mu}(k) ,
\ee 
with
\be
\label{Si-vec}
\Sigma^\mu (k) = g^2 \frac{C_F}{4} \int \frac{d^3p}{(2\pi )^3} \,
\frac{ \tilde f ({\bf p}) }{E_p}  \, 
\frac{p^\mu}{k\cdot p + i 0^+} .
\ee
Substituting the expression (\ref{structure-quark}) into Eq.~(\ref{dis-quark-1}) and computing the determinant as explained in Appendix 1 of \cite{Mrowczynski:1992hq}, one obtains
\be
\label{dis-quark-2} 
\Big[\big( k^{\mu} - \Sigma^{\mu}(k)
\big) \big(k_{\mu} - \Sigma_{\mu}(k) \big)\Big]^2  = 0 ,
\ee
which is the dispersion equation to be discussed later on.

\subsection{Collective modes in an isotropic QGP}
\label{subsec-collective-iso}

Although we are mostly interested in an unstable anisotropic plasma, we start our discussion of collective modes with the case of an isotropic QGP which, in particular, includes the case of an equilibrium plasma. Isotropic plasmas are discussed in textbooks, see {\it e.g.} \cite{Landau-Lifshitz-1981,lebellac}, and are included here for the sake of completeness, and as a reference for  our analysis of anisotropic plasmas. 

\subsubsection{Gluon modes}
\label{sec-gluon-modes-iso}

If the plasma is isotropic, the dielectric tensor can be expressed as
\be
\label{e-L-T}
\varepsilon^{ij}(\omega,{\bf k}) = \varepsilon_T (\omega,{\bf k})
 \Big(\delta^{ij} - \frac{k^i k^j}{{\bf k}^2} \Big)
+ \varepsilon_L (\omega,{\bf k}) \; \frac{k^i k^j}{{\bf k}^2} ,
\ee
where the longitudinal ($\varepsilon_L (\omega,{\bf k})$) and transverse 
($\varepsilon_T (\omega,{\bf k})$) components equal
\be
\varepsilon_L (\omega,{\bf k}) = \frac{k^i k^j}{{\bf k}^2} \, \varepsilon^{ij}(\omega,{\bf k}) ,
\;\;\;\;\;\;\;\;\;\;
\varepsilon_T (\omega,{\bf k}) = \frac{1}{2} 
\big( \varepsilon^{ii}(\omega,{\bf k}) - \varepsilon_L (\omega,{\bf k}) \big) .
\ee
Then, the dispersion Eq.~(\ref{dispersion-g}) splits into two equations
\be
\label{dis-eq-iso}
\varepsilon_L (\omega,{\bf k}) = 0 , \;\;\;\;\;\;\;\;\;\;\;
\omega^2\varepsilon_T (\omega,{\bf k}) - {\bf k}^2 = 0 .
\ee
We note that in vacuum, where $\varepsilon_L=\varepsilon_T = 1$, Eqs.~(\ref{dis-eq-iso}) give two transverse modes $\omega = \pm |{\bf k}|$ and no longitudinal one. 

For further analysis, one needs explicit expressions for $\varepsilon_L (\omega,{\bf k})$ and $\varepsilon_T (\omega,{\bf k})$. Using Eq.~(\ref{epsilon-deriv}), the  longitudinal and transverse chromodielectric functions are found to be
\ba
\label{epsilon-L-1}
\varepsilon_L (\omega,{\bf k}) &=& 
 1 + 
{g^2 \over 2 \omega \, {\bf k}^2} \int {d^3 p \over (2\pi )^3}
{ ({\bf k} \cdot {\bf v})^2 \over \omega - {\bf k} \cdot {\bf v} + i0^+} 
\; {\partial f({\bf p}) \over \partial E_{\bf p}} ,
\\ [2mm] 
\label{epsilon-T-1}
\varepsilon_T (\omega,{\bf k}) &=&  1+ 
{g^2 \over 4 \omega \, {\bf k}^2} \int {d^3 p \over (2\pi )^3}
{ ({\bf k} \times {\bf v})^2 \over \omega - {\bf k} \cdot {\bf v} + i0^+} 
{\partial f({\bf p}) \over \partial E_{\bf p}} ,
\ea
where we have used the fact that in an isotropic plasma the distribution function $f({\bf p})$ depends on a particle's momentum only through $E_{\bf p}$ and thus
\be
{\partial f({\bf p}) \over \partial {\bf p}} = {\bf v}  \, {\partial f({\bf p}) \over \partial E_{\bf p}} .
\ee

One computes $\varepsilon_L (\omega,{\bf k})$ and $\varepsilon_T (\omega,{\bf k})$ in spherical coordinates with the $z$-axis chosen to lie along the vector ${\bf k}$. As a result, the integral over the azimuthal angle is trivial while the integrals over $E_{\bf p} = |{\bf p}|$ and $\theta$ factorize when plasma particles are massless and $|{\bf v}| =1$. In this way one finds
\ba
\label{epsilon-L-final}
\epsilon_L (\omega,{\bf k}) &=& 
 1 + \frac{m_D^2}{{\bf k}^2} \bigg( 1 -  \frac{\omega}{2|{\bf k}|} \ln 
\Big| \frac{ \omega + |{\bf k}|} { \omega - |{\bf k}|} \Big| \bigg)
+ i \frac{\pi}{2} \, \Theta ( {\bf k}^2 - \omega^2 )  \frac{m_D^2\omega }{|{\bf k}|^3} ,
\\ [2mm] 
\label{epsilon-T-final}
\varepsilon_T (\omega,{\bf k}) &=&  1 - 
\frac{m_D^2}{2{\bf k}^2} 
\bigg( 1 -  \frac{\omega^2 - {\bf k}^2}{2 \omega |{\bf k}|} \ln 
\Big| \frac{ \omega + |{\bf k}|} { \omega - |{\bf k}|} \Big| \bigg)
+ i \frac{\pi}{4} \, \Theta ( {\bf k}^2 - \omega^2 )  
\frac{m_D^2(\omega^2 - {\bf k}^2)}{\omega |{\bf k}|^3} ,
\ea
where $m_D$ is the Debye mass given as 
\be
\label{m_D^2}
m_D^2 \equiv - \frac{g^2}{4\pi^2} \int_0^\infty dE_{\bf p} E_{\bf p}^2 
\frac{d f({\bf p})}{d E_{\bf p}} = \frac{g^2}{2\pi^2} \int_0^\infty dE_{\bf p} E_{\bf p}  f({\bf p}) 
= g^2\int {d^3 p \over (2\pi )^3} \frac{f({\bf p})}{E_{\bf p}}.
\ee

As can be seen from Eqs.~(\ref{epsilon-L-1}) and (\ref{epsilon-T-1}), the denominators of the integrands vanish for $\omega = {\bf k} \cdot {\bf v}$, giving rise to $\Im\epsilon_L$ and $\Im\epsilon_T$ through the relation
\be
\frac{1}{x \pm i0^+} = {\rm P}\bigg[ \frac{1}{x} \bigg] \mp i\pi \delta(x) ,
\ee
where ${\rm P}$ denotes the principal value of the integral. As a consequence, the imaginary contributions to $\varepsilon_L (\omega,{\bf k})$ and $\varepsilon_T (\omega,{\bf k})$ appear when $\omega^2 < {\bf k}^2$, as seen in Eqs.~(\ref{epsilon-L-final}) and (\ref{epsilon-T-final}). The dielectric functions (\ref{epsilon-L-final}) and (\ref{epsilon-T-final}) of an ultrarelativistic plasma were derived for the first time in \cite{Silin:1960-Russian}.

\begin{figure}[t]
\vspace{-3mm}
\begin{minipage}{8.5cm}
\begin{center}
\includegraphics*[width=1.05\textwidth]{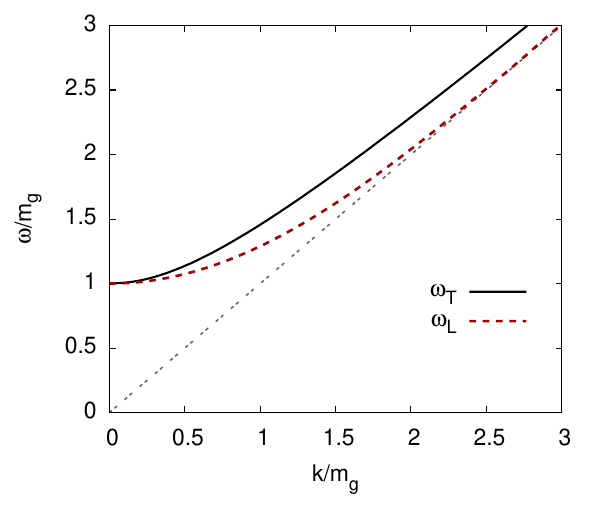}
\vspace{-10mm}
\caption{Gluon collective modes (plasmons) in an isotropic plasma.  Solid line represents the transverse mode and the dashed line the longitudinal one. The gray double-dashed line denotes the light cone $\omega = |{\bf k}|$.} 
\label{fig-iso-modes-L-T}
\end{center}
\end{minipage}
\hspace{5mm}
\begin{minipage}{8.5cm}
\begin{center}
\includegraphics*[width=1.05\textwidth]{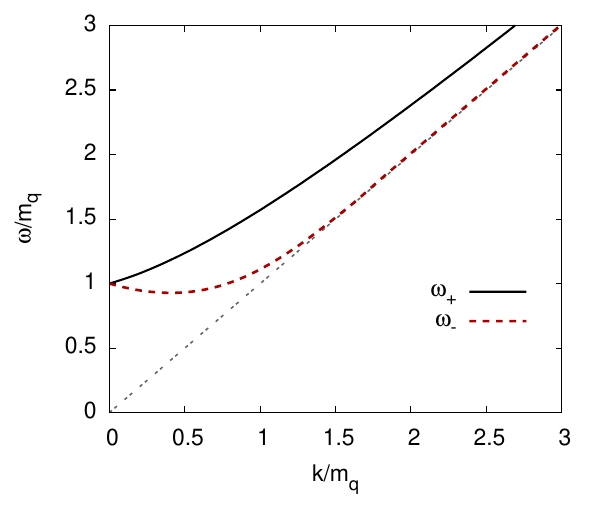}
\vspace{-10mm}
\caption{Quark collective modes in an isotropic plasma. Solid black line represents the normal mode and the red dashed line the plasmino.  The grey double-dashed line denotes the light cone $\omega = |{\bf k}|$.} 
\label{fig-iso-modes-quark}
\end{center}
\end{minipage}
\end{figure}

Substituting the dielectric functions (\ref{epsilon-L-final}) and (\ref{epsilon-T-final}) into Eq.~(\ref{dis-eq-iso}) one obtains explicit dispersion equations which must be solved numerically.  However, they can easily be solved in the long wavelength limit $\omega \gg |{\bf k}|$. Then, the logarithms in the formulas (\ref{epsilon-L-final}) and (\ref{epsilon-T-final}) can be expanded in powers of $|{\bf k}|/\omega$ and 
\ba
\label{eL-iso-small-k}
\varepsilon_L (\omega,{\bf k})  &=& 1 - \frac{m^2_D}{3\omega^2} \Big[ 1 + \frac{3{\bf k}^2}{5 \omega^2}
+ {\cal O}\Big( \frac{{\bf k}^4}{ \omega^4}\Big) \Big]  ,
\\[2mm] 
\label{eT-iso-small-k}
\varepsilon_T (\omega,{\bf k})    &=& 1 - \frac{m^2_D}{3 \omega^2} \Big[ 1 + \frac{{\bf k}^2}{ 5\omega^2}
+ {\cal O}\Big( \frac{{\bf k}^4}{ \omega^4}\Big) \Big] .
\ea
With the approximate expressions of $\varepsilon_L$ and $\varepsilon_T$, one easily obtains the following dispersion relations for long-wavelength longitudinal and transverse modes 
\ba
\label{omega-long}
\omega_L^2({\bf k}) = m_g^2 + \frac{3}{5} {\bf k}^2 + {\cal O}\Big( \frac{{\bf k}^4}{m_g^2}\Big) ,
\\
\label{omega-trans}
\omega_T^2({\bf k}) = m_g^2 + \frac{6}{5} {\bf k}^2 + {\cal O}\Big( \frac{{\bf k}^4}{m_g^2}\Big),
\ea
where  $m_g \equiv m_D/\sqrt{3}$ is the plasmon mass. Classically this is the lowest frequency of plasma oscillations which is the same for transverse and longitudinal modes. 

The dispersion equations (\ref{dis-eq-iso}) with the dielectric functions  (\ref{epsilon-L-final}) and (\ref{epsilon-T-final}) can also be solved analytically in the short wavelength limit (${\bf k}^2 \gg m^2_D$) and the dispersion relations are
\ba
\label{T-plasmon-iso}
\omega^2_T({\bf k}) &\approx& \frac{1}{2}m^2_D + {\bf k}^2 ,
\\[2mm]
\label{omega-long-large-k}
\omega^2_L({\bf k}) &\approx& {\bf k}^2  \Big[ 1 + 4 \exp\Big({-2\frac{{\bf k}^2}{m^2_D} -2} \Big)\Big].
\ea

Numerical solutions of the dispersion equations (\ref{dis-eq-iso}) are shown in Fig.~\ref{fig-iso-modes-L-T}.  The transverse and longitudinal modes are both above the light cone $\omega = \pm |{\bf k}|$, even so the longitudinal mode approaches the light cone as ${\bf k}^2 \rightarrow \infty$, in agreement with the formula (\ref{omega-long-large-k}).  Since the dispersion curves stay above the light cone, the phase velocity of the plasma waves exceeds the speed of light. Consequently, there is no Landau damping which occurs when the velocity of a plasma particle equals the phase velocity of the wave.  A detailed explanation of the mechanism of Landau damping is given in Sec.~\ref{subsec-mechanism-long}.

\subsubsection{Quark modes}

As already mentioned, when the plasma is isotropic, the distribution function $\tilde f({\bf p})$,  which enters the quark self-energy (\ref{Si-vec}),  depends on the particle's momentum only through $E_{\bf p}$. Computing the self-energy (\ref{Si-vec}) in spherical coordinates with the $z$-axis  chosen along the vector ${\bf k}$, the integral over the azimuthal angle is trivial while the integrals over $E_{\bf p} = |{\bf p}|$ and $\theta$ factorize when plasma particles are massless and $|{\bf v}| =1$. As a result, one finds
\ba
\label{self-quark-0}
\Sigma^0(k) &=& { m_q^2 \over 2|{\bf k}|} 
\Big[
\ln \Big| \frac{ \omega + |{\bf k}|} { \omega - |{\bf k}|} \Big|
-i \pi \Theta ({\bf k}^2 - \omega^2 ) \Big] , 
\\ [3mm]
\label{self-quark-vec}
{\bf \Sigma} (k) &=& { m_q^2 \over {\bf k}^2} \; {\bf k}
- { m_q^2 \omega \over 2|{\bf k}|^3 }
\Big[
\ln \Big| \frac{ \omega + |{\bf k}|} { \omega - |{\bf k}|} \Big|
-i \pi \Theta ({\bf k}^2 - \omega^2 ) \Big] \; {\bf k}  ,
\ea
where 
\be
m_q^2 \equiv  \frac{g^2}{8 \pi^2}C_F \int_0^{\infty} dE_{\bf p} E_{\bf p} \, \tilde f({\bf p}) .
\ee
For the case of an equilibrium plasma, the self-energies (\ref{self-quark-0}) and (\ref{self-quark-vec}) were found long ago by  \cite{Weldon:1982bn,Blaizot:1993bb}.

With the self-energies (\ref{self-quark-0}) and (\ref{self-quark-vec}), the dispersion equation (\ref{dis-quark-2}), which can be rewritten as 
\be
\label{dis-quark-final} 
\big(\omega - \Sigma^0(k) \big)^2 - \big({\bf k} - {\bf \Sigma}(k) \big)^2  = 0 ,
\ee
cannot be solved analytically. However, in the long wavelength limit ($ \omega^2 \gg {\bf k}^2$) the dispersion equation (\ref{dis-quark-final}) simplifies to
\be
\Big( 1 - {m_q^2 \over \omega^2} \Big)^2 \omega^2 - {\bf k}^2 = 0,
\ee
which is solved by
\be
\label{modes-eq}
\omega_{\pm}^2({\bf k}) \approx m_q^2  + {1 \over 2} {\bf k}^2  \pm m_q |{\bf k}|.
\ee

Numerical solutions of dispersion equation (\ref{dis-quark-final}) with the self-energies (\ref{self-quark-0}) and (\ref{self-quark-vec}) are shown for a broad range of momenta in Fig.~\ref{fig-iso-modes-quark}. As seen, $\omega_+({\bf k})$ monotonically grows  as a function of $|{\bf k}|$. In the case of the $\omega_-$ mode, its frequency initially decreases with $|{\bf k}|$ and there is a shallow minimum at $|{\bf k}| \approx m_q$ and further $\omega_-({\bf k})$ monotonically grows.  As discussed in the textbook \cite{lebellac}, the modes $\omega_\pm({\bf k})$ have opposite helicity over chirality ratio.  The $\omega_+({\bf k})$ mode corresponds to the positive energy fermion, the $\omega_-({\bf k})$, which is sometimes called a plasmino, is a specific medium effect. 

The gluon and quark modes, which are discussed above, are purely real that is they have infinite lifetimes. However, one expects that all excitations of weakly coupled equilibrium plasmas can have long but finite lifetimes. Damping rates have been indeed obtained in perturbation theory in next-to-leading order, see  {\it e.g.} the review \cite{Blaizot:2001nr}, and the ratio of imaginary over real part of mode frequency is generically of order $g \log (1/g)$. Unfortunately, there is no known way of carrying out next-to-leading order calculations for anisotropic systems, which are of our main interest, because unstable modes lead to non-integrable singularities.

\subsection{Two-stream system}
\label{subsec-2-stream}

After the case of the isotropic plasma we discuss a system consisting of two interpenetrating streams of partons. The spectrum of collective excitations of such a system appears to be rich and some solutions to the dispersion equations can be easily obtained analytically. Since the two-stream system is not very similar to the actual state of the QGP created in relativistic heavy-ion collisions, we will not attempt to discuss collective modes in the system in full generality. This is done in \cite{Deja:2015qsa}. Instead, we consider here two special cases where we can straightforwardly identify the unstable modes which are our primary interest. We can then analyze the underlying mechanisms responsible for the instabilities. 

\subsubsection{Gluon modes}

The two-stream system is naturally described in terms of chromohydrodynamics presented in Sec.~\ref{sec-chromo-fluid}. Then, the polarization tensor is immediately given by Eq.~(\ref{Pi-hydro}) or Eq.~(\ref{Pi2}). Equivalently,  one can use a more general expression (\ref{Pi-k-final}) with the distribution function of the form
\be
\label{f-2-streams}
f({\bf p}) = (2\pi )^3 n 
\Big[\delta^{(3)}({\bf p} - {\bf q}) + \delta^{(3)}({\bf p} + {\bf q}) \Big] \;,
\ee
where $n$ is the effective parton density in a single stream. The distribution function (\ref{f-2-streams}) should be treated as an idealization of the two-peak distribution where the particles have momenta close to ${\bf q}$ or $-{\bf q}$, but it is not required that the momenta are exactly ${\bf q}$ or $-{\bf q}$. 

The distribution function (\ref{f-2-streams}) substituted into Eq.~(\ref{epsilon-noderiv}) gives a dielectric tensor of the form
\ba
\label{epsilon-2-stream}
\varepsilon^{ij} (\omega, {\bf k}) =
\delta^{ij} \big( 1 - \frac{\mu^2}{\omega^2} \Big)
+ u^i u^j  \frac{\mu^2}{\omega^2} 
\frac{(\omega^2 - {\bf k}^2) \big(\omega^2 + ({\bf k} \cdot {\bf u})^2\big)}
{\big(\omega^2 - ({\bf k} \cdot {\bf u})^2 \big)^2}
- (k^i u^j + u^i k^j) \, \frac{\mu^2}{\omega^2}
\frac{{\bf k} \cdot {\bf u}}{\omega^2 - ({\bf k} \cdot {\bf u})^2} ,
\ea
where $\mu^2 \equiv g^2n/E_{\bf q}$ and ${\bf u} \equiv {\bf q}/E_{\bf q}$ is the stream velocity. Since quarks and gluons are massless, $E_{\bf q} = |{\bf q}|$ and ${\bf u}^2 =1$. However, when the distribution (\ref{f-2-streams}) is treated as an approximation of the two-peak structure and partons have non-vanishing momenta perpendicular to the stream velocity, $E_{\bf q} \ge |{\bf q}|$ and ${\bf u}^2 \le1$. In the subsequent sections, where the cases  ${\bf k}\cdot {\bf u} = 0$ and  $({\bf k}\cdot {\bf u})^2 = {\bf k}^2 {\bf u}^2$ are discussed, we assume that ${\bf u}^2 \le1$. 

With the dielectric tensor (\ref{epsilon-2-stream}) the complete spectrum of plasmons can be found in a closed form \cite{Deja:2015qsa} but the solutions are of a rather complex form. Therefore, we discuss below two special cases when the solutions are relatively simple and their physical content is more transparent. 

\subthreesection{The case ${\bf k} \perp {\bf u}$}

We start with the situation when the wave vector ${\bf k}$ is perpendicular to the stream velocity ${\bf u}$. The easy way to find the collective modes in such a special case is to explicitly compute the matrix $\Xi(\omega, {\bf k}) \equiv {\bf k}^2 \delta^{ij} -k^i  k^j - \omega^2 \varepsilon^{ij}(\omega, {\bf k})$ assuming that ${\bf u} = (0,0,u)$ and ${\bf k} = (k,0,0)$. Then, Eq.~(\ref{dispersion-g}) reads
\ba
\label{dis-eq-2-stream-k-perp-u}
{\rm det} \left[
\begin{array}{ccc}
\omega^2 - \mu^2 & 0 & 0 
\\
0 & \omega^2 - \mu^2 -k^2 & 0
\\
0 & 0 & 
\omega^2 - \mu^2 - k^2 
+ \frac{m^2(\omega^2 - k^2)u^2}{\omega^2}
\end{array}
\right] =0,
\ea
which gives 
\be
(\omega^2 - \mu^2)(\omega^2 - \mu^2 - k^2)
\Big(\omega^2 - \mu^2 - k^2 
+ \frac{\mu^2(\omega^2 - k^2)u^2}{\omega^2}\Big) = 0.
\ee
The solutions are 
\be
\label{solutions-k-perp-n}
\omega_1^2(k) = \mu^2 \,,\;\;\;\;\;\;\;
\omega_2^2(k) = \mu^2 + k^2 \,,\;\;\;\;\;\;\;
\omega_{\pm}^2(k) = 
\frac{1}{2}\Big( \mu^2(1 - u^2) + k^2 
\pm \sqrt{ \big(\mu^2(1 - u^2) + k^2\big)^2+4 \mu^2 u^2 k^2}\;\Big) .
\ee
The solutions $\omega_1^2$, $\omega_2^2$, and $\omega_+^2$, which are all positive, correspond to stable modes, while the solution $\omega_-^2$, which is negative, describes an unstable mode and an overdamped mode. The instability is known as the filamentation or Weibel instability. The mode is transverse: the wave vector is along the axis $x$ and the chromoelectric field is parallel to the axis $z$. As explained below Eq.~(\ref{Xi-def}), the dispersion equation (\ref{dis-eq-2-stream-k-perp-u}) can be treated as resulting from the equation of motion of the chromoelectric field  $[{\bf k}^2 \delta^{ij} -k^i  k^j - \omega^2 \varepsilon^{ij}(\omega, {\bf k})]E^j(\omega, {\bf k}) =0$ which allows one to identify the components of the field associated with a given mode. The mechanism responsible for the Weibel instability is discussed in Sec.~\ref{subsec-mechanism-trans}.

\subthreesection{The case ${\bf k} \parallel {\bf u}$}

Assuming now that ${\bf u} = (0,0,u)$ and ${\bf k} = (0,0,k)$, one computes the matrix ${\bf k}^2 \delta^{ij} -k^i  k^j - \omega^2 \varepsilon^{ij}(\omega, {\bf k})$. The resulting dispersion equation (\ref{dispersion-g}) has the form
\begin{displaymath}
{\rm det}\left[
\begin{array}{ccc}
\omega^2 -  \mu^2 - k^2 & 0 & 0 
\\
0 & \omega^2 - \mu^2 - k^2 & 0
\\
0 & 0 & \omega^2 - \mu^2 - \frac{2 \mu^2 k^2 u^2}{ \omega^2 -  k^2 u^2}
-  \frac{\mu^2 (\omega^2 +  k^2 u^2) (k^2 - \omega^2)  u^2}{ (\omega^2 -  k^2 u^2)^2}
\end{array}
\right] = 0 ,
\end{displaymath}
which gives
\be
(\omega^2 - \mu^2 - k^2)^2 
\left( \omega^2 - \mu^2 - \frac{2 \mu^2 k^2 u^2}{ \omega^2 -  k^2 u^2}
-  \frac{m^2 (\omega^2 +  k^2 u^2) (k^2 - \omega^2)  u^2}{ (\omega^2 -  k^2 u^2)^2} 
\right) =0.
\ee
There are two simple solutions representing the transverse modes $\omega_2^2(k) = \mu^2 + k^2 $ and two solutions corresponding to the longitudinal modes
\be
\label{solution-L-parallel-u}
\omega_\pm^2(k) = k^2 u^2 + \frac{\lambda^2}{2} 
\pm \frac{\lambda}{2} \sqrt{8 k^2 u^2 + \lambda^2},
\ee
where $\lambda \equiv \mu \sqrt{1 - u^2}$. As can be seen from this expression, $\omega_+^2(k)$ is always positive
but $\omega_-^2(k)$ is negative for $ k^2 < \lambda^2/u^2$.  For momenta below this bound, the system is unstable. 
In plasma physics, this instability is called the two-stream electrostatic instability. It is interesting to note that the longitudinal mode becomes stable as $u \rightarrow 1$. Then, $\lambda = 0$ and $\omega_\pm^2(k) = k^2$. The mechanism responsible for the instability is discussed in Sec.~\ref{subsec-mechanism-long}.

\subsubsection{Quark modes}

The distribution function which enters the quark self-energy $\tilde f({\bf p})$ is of the form (\ref{f-2-streams}), but the parton density parameter $n$ is replaced by $\rho$. Substituting this distribution into Eq.~(\ref{Si-vec}) one finds
\ba
\label{Si-0-2streams}
\Sigma^0 (k) &=& M^2
\Big( \frac{1}{\omega  - {\bf k}\cdot{\bf u}} +  \frac{1}{\omega + {\bf k}\cdot{\bf u}}\Big) 
= \frac{2 M^2 \omega}{\omega^2  - ({\bf k}\cdot{\bf u})^2},
\\ [2mm]
\label{Si-vec-2streams}
{\bf \Sigma} (k) &=& M^2 
\Big( \frac{1}{\omega  - {\bf k}\cdot{\bf u}} 
- \frac{1}{\omega + {\bf k}\cdot{\bf u}}\Big)  {\bf u}
=  \frac{2 M^2 ({\bf k}\cdot{\bf u}) }{\omega^2  - ({\bf k}\cdot{\bf u})^2} {\bf u} ,
\ea
where $M^2 \equiv  g^2 C_F\rho/4E_q$ and ${\bf u} \equiv {\bf q}/E_{\bf q}$ is the stream velocity. As in the case of the gluon modes in the two-stream system,  the distribution (\ref{f-2-streams}) is treated as an approximation of the two-peak structure and partons can have non-vanishing momenta perpendicular to the stream velocity. Then, $E_{\bf q} \ge |{\bf q}|$ and ${\bf u}^2 \le1$. 

With the self-energies (\ref{Si-0-2streams}) and (\ref{Si-vec-2streams}), the dispersion equation (\ref{dis-quark-2}) becomes
\be
\label{dis-quark-2-stream} 
\bigg( \omega - \frac{2 M^2 \omega}{\omega^2  - ({\bf k}\cdot{\bf u})^2} \bigg)^2
- \bigg({\bf k}  - \frac{2 M^2 ({\bf k}\cdot{\bf u}) }{\omega^2  - ({\bf k}\cdot{\bf u})^2} {\bf u} \bigg)^2 
 = 0 ,
\ee
which gives
\be
\label{dis-quark-2-stream-4} 
(\omega^2 - {\bf k}^2 - 4 M^2) \big(\omega^2  - ({\bf k}\cdot{\bf u})^2\big)^2 
+ 4 M^4 \big(\omega^2 - ({\bf k}\cdot{\bf u})^2 {\bf u}^2 \big)
= 0 , 
\ee
provided $\omega^2 \not= ({\bf k}\cdot{\bf u})^2$.

If ${\bf u}^2=1$ the equation simplifies to 
\ba
\label{dis-quark-2-stream-5} 
(\omega^2 - {\bf k}^2 - 4 M^2) \big(\omega^2  - ({\bf k}\cdot{\bf u})^2\big)
+ 4 M^4 = 0 
\ea
and the solutions read
\be
\omega_\pm^2({\bf k}) = \frac{1}{2}
\Big( {\bf k}^2 + ({\bf k}\cdot {\bf u})^2 + 2M^2 
\pm 
\sqrt{\big({\bf k}^2 - ({\bf k}\cdot {\bf u})^2 + 8M^2\big)\big({\bf k}^2 - ({\bf k}\cdot {\bf u})^2\big)} \; \Big) .
\ee
One can verify that $\omega_\pm^2({\bf k})$ are real and positive. So, there is no instability. When ${\bf u}^2 <1$, the dispersion equation (\ref{dis-quark-2-stream}) is cubic in $\omega^2$. There are thus explicit solutions which, however, are rather complicated. Nevertheless, one can show that for all three solutions  $\omega^2({\bf k})$ is real and positive. 

As it was shown for the first time in \cite{Mrowczynski:2001az}, there are no fermionic unstable modes even in the extremely anisotropic configuration such as the two-stream system. Qualitative features of the modes are also rather independent of the momentum distribution of plasma constituents \cite{Schenke:2006fz}. Therefore, we will not discuss the quark collective excitations further but we will focus on the plasmons.

\subsection{Mechanisms of instabilities}
\label{subsec-mechanism}

Studying the collective excitations in the two-stream system, we have found both transverse and longitudinal unstable modes. Now we are going to explain mechanisms of the instabilities in terms of elementary physics. 

\subsubsection{Unstable Transverse Modes}
\label{subsec-mechanism-trans}

Let us first discuss how the unstable transverse modes are triggered by fluctuations. We will then show how the unstable modes are amplified. 

\subthreesection{Seeds of filamentation}

Since the system of partons under consideration is on average locally colorless, the average color current vanishes. Color fluctuations are possible, however. The current-current correlator for a classical system of non-interacting massless partons 
equals \cite{Mrowczynski:1996vh,Mrowczynski:2008ae}
\ba
\label{cur-cor-x}
M^{\mu \nu}_{ab} (t,{\bf x}) &\buildrel \rm def \over =& 
\langle j^{\mu}_a (t_1,{\bf x}_1) j^{\nu}_b (t_2,{\bf x}_2) \rangle 
= {1 \over 2} \,g^2\; \delta^{ab} 
\int {d^3p \over (2\pi )^3} \; {p^{\mu} p^{\nu} \over E_{\bf p}^2} 
f({\bf p}) \; \delta^{(3)} ({\bf x} -{\bf v} t) ,
\ea
where ${\bf v} \equiv {\bf p}/ E_{\bf p}$ and $(t,{\bf x}) \equiv (t_2-t_1,{\bf x}_2-{\bf x}_1)$. Due to the average space-time homogeneity, the correlation tensor (\ref{cur-cor-x}) depends only on the difference $(t_2-t_1,{\bf x}_2-{\bf x}_1)$. The space-time points $(t_1,{\bf x}_1)$ and $(t_2,{\bf x}_2)$ are correlated in the system of non-interacting particles if there is a particle which travels from $(t_1,{\bf x}_1)$ to $(t_2,{\bf x}_2)$.  For this reason the delta  $\delta^{(3)} ({\bf x} - {\bf v} t)$ is present in the formula (\ref{cur-cor-x}). The momentum integral of the distribution function simply represents the summation over particles. The fluctuation spectrum is found via a Fourier transform of the tensor (\ref{cur-cor-x}), {\it i.e.}, 
\be
\label{cur-cor-k}
M^{\mu \nu}_{ab} (\omega ,{\bf k}) = {1 \over 2} \,g^2\; \delta^{ab} 
\int {d^3p \over (2\pi )^3} \; 
{p^{\mu} p^{\nu} \over E_{\bf p}^2} \; f({\bf p})  \;
2\pi \delta (\omega -{\bf kv}).
\ee

To compute the fluctuation spectrum, the parton momentum distribution has to be specified. Calculations with two forms of the  momentum distribution are presented in \cite{Mrowczynski:1996vh}. Here we only qualitatively discuss Eqs.~(\ref{cur-cor-x}) and (\ref{cur-cor-k}), assuming that the parton momentum distribution is anisotropic.

With the momentum distribution elongated in the  $z$ direction,  Eqs.~(\ref{cur-cor-x}) and (\ref{cur-cor-k}) clearly show that the correlator $M^{zz}$ is larger than $M^{xx}$ or  $M^{yy}$. It is also clear that $M^{zz}$ is the largest when the wave vector ${\bf k}$ is along the direction of the momentum deficit. Then, the delta function $\delta (\omega -{\bf kv})$ does not much constrain the integral in Eq.~(\ref{cur-cor-k}). Since the momentum distribution is elongated in the $z$ direction, the current fluctuations are the largest when the wave vector ${\bf k}$ is in the $x\!-\!y$ plane. Thus, we conclude that some fluctuations in the anisotropic system are large, much larger than in the isotropic one. In other words, an anisotropic system has a natural tendency to split into current filaments parallel to the direction of the momentum surplus. These currents are seeds of the filamentation instability.

\subthreesection{Mechanism of filamentation}

\begin{figure}
\includegraphics[width=9cm]{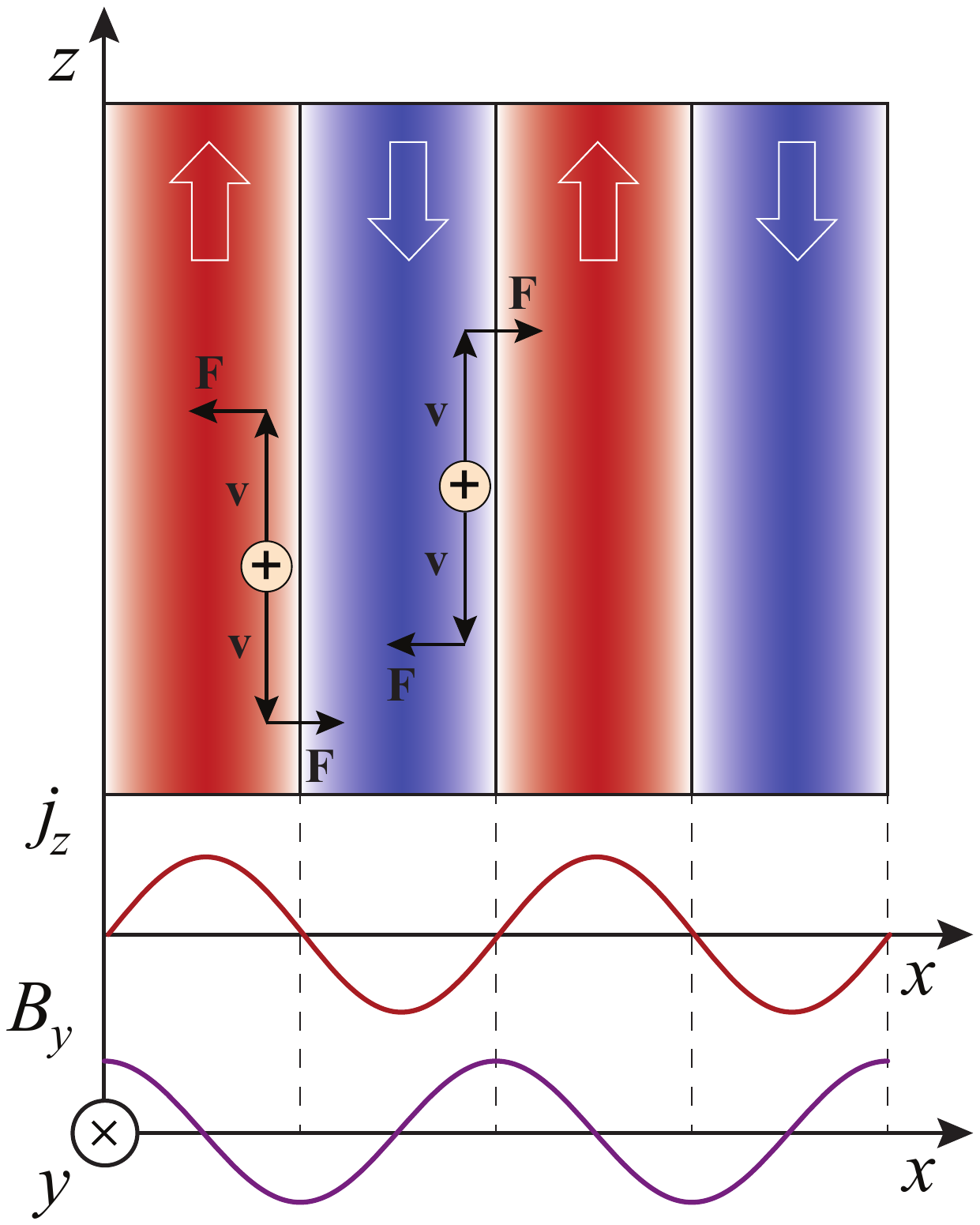}
\caption{The mechanism of filamentation instability, 
see text for a description.}
\label{fig-mechanism} 
\end{figure}

Let us now explain why the fluctuating currents, which flow in the direction of the momentum surplus, can grow in time. To simplify the discussion, which follows \cite{Mrowczynski:1996vh}, we consider an electromagnetic anisotropic system. The form of the fluctuating current is assumed to be
\be
\label{flu-cur}
{\bf j}(x) = j \: \hat {\bf e}_z \: {\rm cos}(k_x x) ,
\ee
where $\hat {\bf e}_z$ is the unit vector in the $z$ direction. According to Eq.~(\ref{flu-cur}), there are current filaments of the thickness $\pi /\vert k_x\vert$ with the current flowing in the opposite directions in the neighboring filaments. Once there is the fluctuation current (\ref{flu-cur}), it generates a magnetic field 
\ban
{\bf B}(x) = {j \over k_x} \: \hat {\bf e}_y \: {\rm sin}(k_x x) .
\ean
Consequently, the partons flying in the $z$-direction experience a Lorentz force
\ban
{\bf F}(x) = q \: {\bf v} \times {\bf B}(x) = 
- q \: v_z \: {j \over k_x} \: \hat {\bf e}_x \: {\rm sin}(k_x x) \;,
\ean
where $q$ is the electric charge. One observes, see Fig.~\ref{fig-mechanism}, that the force distributes the partons in such a way that those, which positively contribute to the current in a given filament, are focused in the filament center while those, which negatively contribute, are moved to the neighboring one. Thus, the initial current is growing and the magnetic field generated by this current is growing as well. The instability is driven by the energy transferred from the particles to fields. More specifically, the kinetic energy related to a motion along the direction of the momentum surplus is used to generate the magnetic field. The mechanism of Weibel instability is explained somewhat differently in \cite{Arnold:2003rq}.

\subsubsection{Unstable Longitudinal Modes}
\label{subsec-mechanism-long}

The unstable longitudinal modes occur due to the same Landau mechanism which is responsible for damping of plasma oscillations in isotropic non-relativistic plasmas. To simplify the terminology we explain here the Landau mechanism in terms of electromagnetism.  Let us consider a plane wave of pure electric field with the wave vector and the electric field both along the $z$ axis. For a charged particle, which moves along the axis $z$ with a velocity $v= p_z/E_{\bf p}$ equal to the phase velocity of the wave $v_\phi= \omega/k$,  the electric field does not oscillate but it is constant. The particle is then either accelerated or decelerated depending on the field's phase. For an electron with $v=v_\phi$ the probability to be accelerated and to be decelerated are equal to each other, as the time intervals spent by the particle in the acceleration zone and in the deceleration zone are equal to each other.

\begin{figure}
\includegraphics[width=12cm]{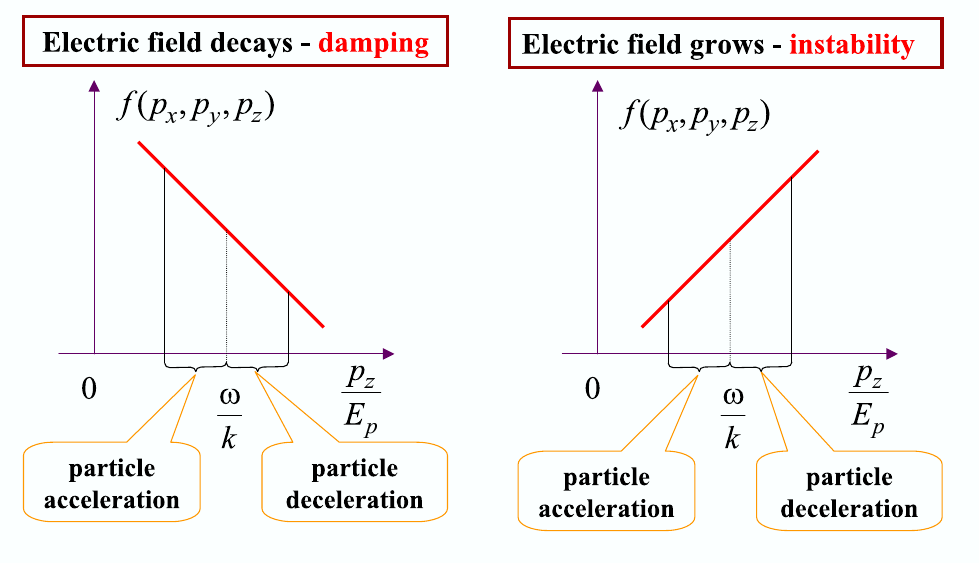}
\caption{The mechanism of energy transfer between particles
and fields.}
\label{fig-elec-mechanism}
\end{figure}

Let us now consider electrons with velocities somewhat smaller than the phase velocity of the wave. Such particles spend more time in the acceleration zone than in the deceleration zone, and the net result is that the particles with $v < v_\phi$ are accelerated. Consequently, the energy is transferred from the electric field to the particles. The particles with  $v > v_\phi$ spend more time in the deceleration zone than in the acceleration zone, and thus they are effectively decelerated - the energy is transferred from the particles to the field. If the momentum distribution is such that there are more electrons in a system with  $v < v_\phi$ than with $v > v_\phi$, the wave looses energy which is gained by the particles, as shown in the left-hand-side of Figure~\ref{fig-elec-mechanism}. This is the mechanism of the famous collisionless Landau damping of the plasma oscillations. If there are more particles with $v > v_\phi$ than with $v < v_\phi$, the particles loose energy which is gained by the wave. Consequently, the wave amplitude grows. This is the mechanism of electric instability. As explained above, it requires the existence of the momentum interval where $f({\bf p})$ grows with ${\bf p}$. Such an interval appears  when the momentum distribution has more than one maximum. This happens in the two-stream system or in the system of a plasma and a beam.  A distribution with more than one maximum is not expected to occur in the quark-gluon plasma from relativistic heavy-ion collisions. Consequently, the longitudinal unstable modes do not seem relevant for such a plasma. 
 


\subsection{Nyquist analysis}
\label{subsec-Nyquist}

A Nyquist analysis allows one to determine the number of solutions of a given equation without solving the equation, see {\it e.g.} the handbook \cite{Kra73}. Knowing the number of solutions is often very important, because it is usually not possible to obtain exact analytic solutions of dispersion equations. Therefore, one typically uses analytic approximations or numerical methods. When an approximation is used, there is a danger to find a solution that is an artifact of the approximation. When numerical methods are used, a solution that is outside the range of the search can be missed. In the next subsection we explain following \cite{Carrington:2014bla} the Nyquist method and in Sec.~\ref{sec-Penrose} we derive the so-called Penrose criterion which signals the existence of an unstable solution. 

\subsubsection{The method}

To explain the idea of a Nyquist analysis, we discuss a generic equation of the form
\be
\label{general-eq}
f(\omega) = 0 ,
\ee
and we define the function 
\be
\label{F-def}
F(\omega) \equiv \frac{f^\prime(\omega)}{f(\omega)} = \frac{d}{d\omega}{\rm ln}f(\omega) .
\ee
We consider the contour integral 
\be
\label{Nyq-int-1}
\oint_C \frac{d\omega}{2\pi i} F(\omega) ,
\ee
where the contour is a positively (counterclockwise) oriented closed loop, which is chosen so that $F(\omega)$ is analytic inside the loop except at isolated points. The integral is equal to the sum of the residues. It is straightforward to show that the residue of $F(\omega)$ at a zero of $f(\omega)$ of order $l$ is $l$, and the residue of $F(\omega)$ at a pole of $f(\omega)$ of order $l$ is $-l$. Thus, we have
\be
\label{Nyq-int-2}
\oint_C \frac{d\omega}{2\pi i} F(\omega) = n_Z - n_P ,
\ee
where $n_Z$ and $n_P$ are the numbers of zeros and poles of $f(\omega)$ inside the contour $C$, taking into account the fact that each zero and pole of order $l$ is counted $l$ times. The aim of the method is to determine $n_Z$.

The first step in the Nyquist analysis is to choose the contour $C$. If $f(\omega)$ has only isolated singular points, then $C$ can be chosen as the big circle which includes the entire plane of complex $\omega$. If $f(\omega)$ has cuts, then the contour must be chosen to exclude these cuts. For example, for an isotropic plasma $f(\omega)$ has a cut for $\omega \in [-|{\bf k}|,|{\bf k}|]$, see the dispersion equations (\ref{dis-eq-iso}) and (\ref{dis-quark-final}) with the functions (\ref{epsilon-L-final}), (\ref{epsilon-T-final}) and (\ref{self-quark-0}), (\ref{self-quark-vec}), respectively. Consequently, the contour $C$ is chosen as in Fig.~\ref{fig-Nyquist-1}. The integrals along the lines connecting the circular contour $C_\infty$ to $C_{\rm cut}$ always compensate each other and therefore the contour integral (\ref{Nyq-int-2}) equals
\be
\label{Nyq-int-LM}
\oint_{C_\infty} \frac{d\omega}{2\pi i} F(\omega) +
\oint_{C_{\rm cut}} \frac{d\omega}{2\pi i} F(\omega) = n_Z - n_P.
\ee

The contribution from the big circle is easy to calculate by writing $\omega = |\omega|e^{i \phi}$ and taking $|\omega|\to\infty$. Using $d\omega = i \omega d\phi$, we have 
\be
\label{Nyq-int-infty}
\oint_{C_\infty} \frac{d\omega}{2\pi i} F(\omega) = \lim_{|\omega| \rightarrow \infty} \omega F(\omega) \equiv n_\infty .
\ee

The integral along the cut can be calculated using the fact that $F(\omega)$, defined by Eq.~(\ref{F-def}), is the logarithmic derivative of $f(\omega)$. Consequently 
\be
\label{Nyq-int-cut}
\oint_{C_{\rm cut}} \frac{d\omega}{2\pi i} F(\omega)  
= \frac{1}{2\pi i}\oint_{C_{\rm cut}}  \frac{d}{d\omega}{\rm ln}f(\omega)
= \frac{1}{2\pi i} \Big( {\rm ln} f(\omega_e) - {\rm ln} f(\omega_s) \Big) \equiv n_W ,
\ee
where $\omega_s$ is the (arbitrarily chosen) starting point of the contour which encloses the cut, and $\omega_e$ is the end point. The points $\omega_s$ and $\omega_e$ have the same modulus, but their phases differ by $2\pi$. The value of the right-hand-side of Eq.~(\ref{Nyq-int-cut}) can be found by mapping the closed contour $C_{\rm cut}$ in the plane of complex $\omega$ onto a path in the plane of complex $f(\omega)$.  Since the logarithm of $f$ has a cut, which runs along the real axis from $f = -\infty$ to $f = 0$, the value of the integral (\ref{Nyq-int-cut}) is a {\it winding number} (denoted $n_W$) which equals the number of times that the curve in the plane of complex $f$, which starts at  $f(\omega_s)$ and ends at $f(\omega_e)$, travels counterclockwise around the point $f=0$.  

Combining the results (\ref{Nyq-int-infty}) and (\ref{Nyq-int-cut}), we rewrite Eq.~(\ref{Nyq-int-LM}) as 
\be
\label{ny-all}
n_Z = n_P + n_\infty + n_W ,
\ee
which tells us that the number of zeros of the function $f(\omega)$ inside the contour $C$ equals the number of poles of $f(\omega)$ inside this contour, plus $n_\infty$ given by the limit (\ref{Nyq-int-infty}), plus the winding number (\ref{Nyq-int-cut}).  The only difficult piece is the calculation of $n_W$, for which one needs to determine the signs of the real and imaginary parts of the function $f(\omega)$ along the contour $C_{\rm cut}$. A detailed discussion of the Nyquist analysis applied to the dispersion equations of QGP with a spheroidal momentum distribution can be found in \cite{Romatschke:2004jh,Schenke:2006fz,Carrington:2014bla}. 

\subsubsection{Penrose criterion}
\label{sec-Penrose}

\begin{figure}[t]
\vspace{-3mm}
\begin{minipage}{8.5cm}
\begin{center}
\includegraphics*[width=0.9\textwidth]{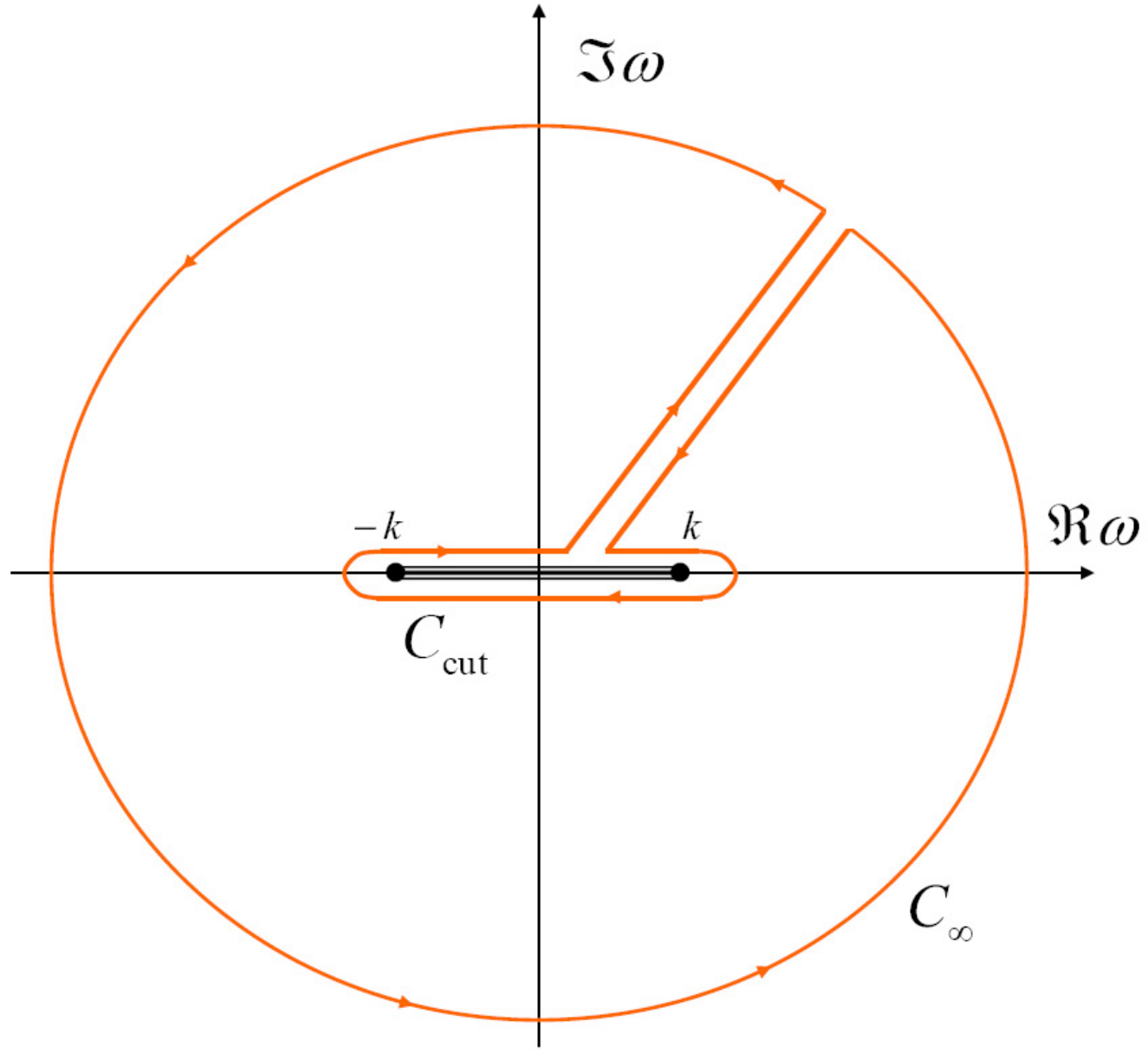}
\vspace{-2mm}
\caption{The contour $C$ in the plane of complex $\omega$ which is used to compute the number of solutions of dispersion equations of an isotropic plasma. Figure from \cite{Carrington:2014bla}.} 
\label{fig-Nyquist-1}
\end{center}
\end{minipage}
\hspace{5mm}
\begin{minipage}{8.5cm}
\begin{center}
\includegraphics*[width=0.9\textwidth]{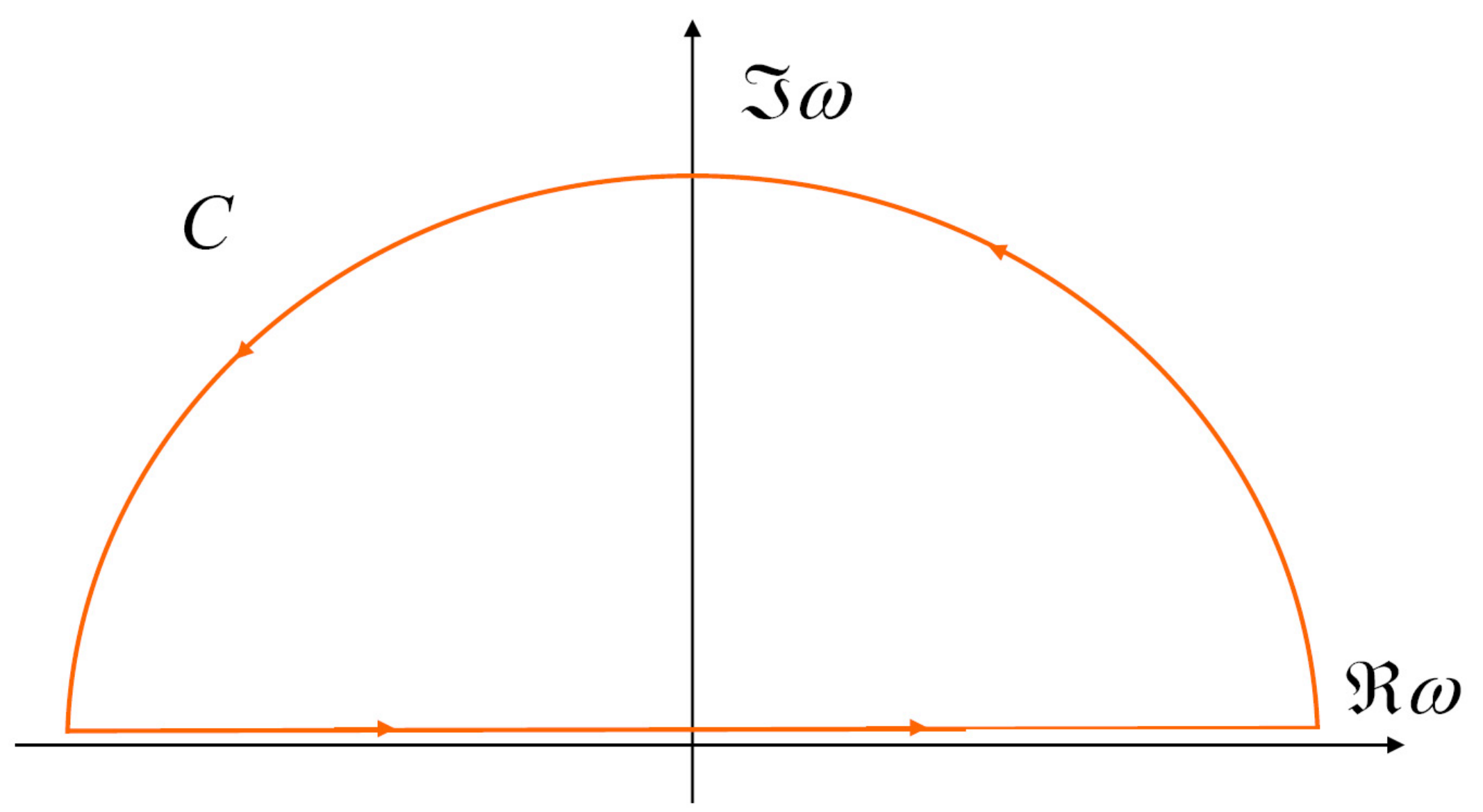}
\vspace{10mm}
\caption{The contour $C$ in the plane of complex $\omega$ which is used to compute the number of unstable solutions of a dispersion equation.} 
\label{fig-contour-Nyquist}
\end{center}
\end{minipage}
\end{figure}

One is often interested not in all solutions of a given dispersion equation but one asks whether the equation has unstable solutions. Then, the Nyquist analysis should be performed with the contour shown in Fig.~\ref{fig-contour-Nyquist}. The contour encircles the upper half-plane of complex $\omega$. If there are zeros of the function $f(\omega)$  located in the  upper half-plane, then there are solutions of Eq.~(\ref{general-eq}) corresponding to instabilities. 

Let us consider the simplest, though typical, case when the function $f(\omega)$ is analytic in the upper-half plane. The cuts, which are usually located at purely real $\omega$, are outside the contour shown in Fig.~\ref{fig-contour-Nyquist}. The integral (\ref{Nyq-int-1}) along the contour then equals the number of zeros $n_Z$ of $f(\omega)$ in the upper half-plane.  The integral equals 
\be
\label{Nyquist-integral-3}
\oint_C \frac{d\omega}{2\pi i} F(\omega)  
= \frac{1}{2\pi i}  {\rm ln}\bigg[\frac{f(\omega_e)}{f(\omega_s)} \bigg] = n_Z ,
\ee
where $\omega_s$ is the (arbitrarily chosen) starting point of the contour $C$, and $\omega_e$ is the end point. The points $\omega_s$ and $\omega_e$ have the same modulus, but their phases differ by $2\pi$. The integral  (\ref{Nyquist-integral-3}) is nonzero when the function $f(\omega)$ acquires a phase along the contour. In case of the dispersion equations under study, it solely depends on the sign of $f(\omega=0)$, which is called the {\it Penrose criterion}, see {\it e.g.} the handbook \cite{Kra73}. It was first applied to the quark-gluon plasma in \cite{Mrowczynski:1994xv}. 

To derive the Penrose criterion and to demonstrate how it works, we consider a specific dispersion equation: the wave vector is chosen along the $x$-axis (${\bf k} = (k,0,0)$) and we assume that the electric field and the induction vector are both along the axis $z$. Then, instead of the general dispersion equation (\ref{dispersion-g}) we have the equation 
\be
\label{dispersion-eq-Nyquist-transverse}
f(\omega) \equiv k^2 - \omega^2 \varepsilon^{zz}(\omega,{\bf k}) = 0,
\ee
where
\be
\epsilon^{zz} (\omega,{\bf k}) = 1 
- \frac{\omega_0^2}{\omega^2} 
+{g^2 \over 2 \omega^2} \int {d^3 p \over (2\pi )^3}
{ k v_z^2 \over \omega - k v_x + i0^+} 
 {\partial f({\bf p}) \over \partial p_x} ,
\ee
with
\be
\omega_0^2 \equiv 
- {g^2 \over 2 } \int {d^3 p \over (2\pi )^3}
v_z {\partial f({\bf p}) \over \partial p_z} .
\ee

\begin{figure}[t]
\begin{center}
\includegraphics*[width=15cm]{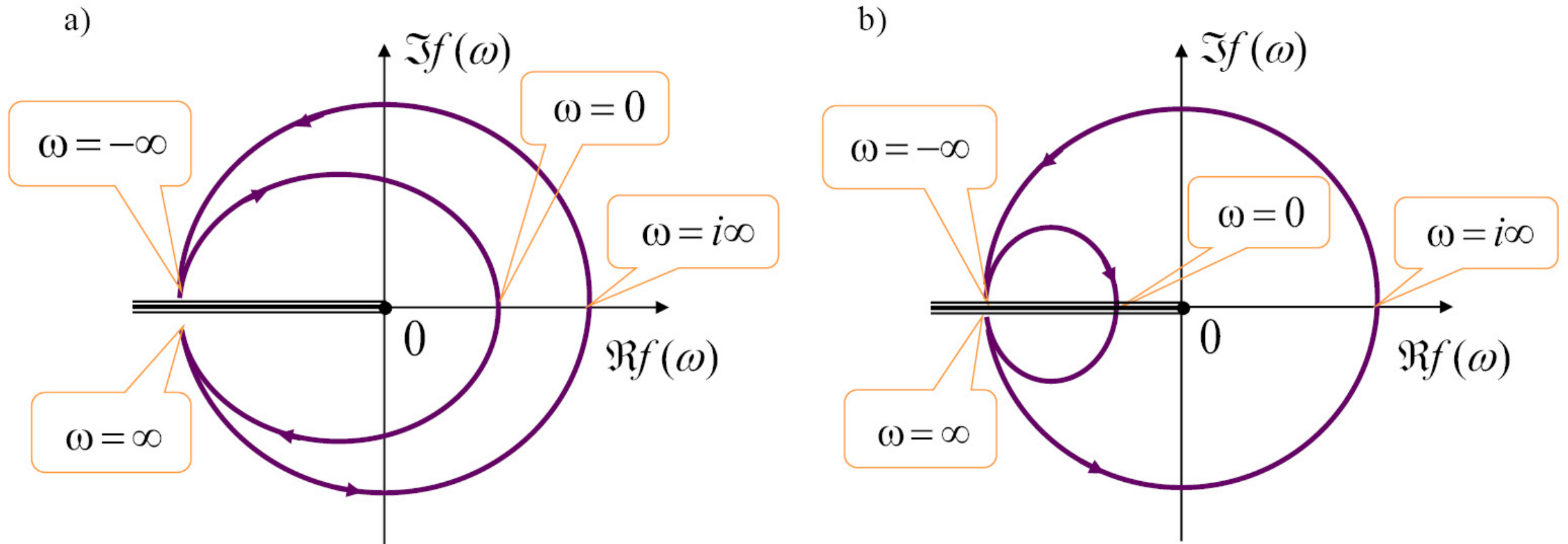}
\end{center}
\vspace{-3mm}
\caption{Mapping of the contour $C$ in the plane of complex $\omega$, which is shown in Fig.~\ref{fig-contour-Nyquist}, into the contour in the plane of complex $f(\omega)$. The plot a) represents a stable system and the plot b) signals presence of an unstable mode. There is indicated the branch cut of the logarithm functions which is along negative real arguments. } 
\label{fig-Penrose}
\end{figure}

To compute the integral (\ref{Nyquist-integral-3}) with the function $f(\omega)$ given by Eq.~(\ref{dispersion-eq-Nyquist-transverse}), one  maps the contour $C$ into the contour in the plane of complex $f(\omega)$ shown in Fig.~\ref{fig-Penrose}. Let us start at $\omega = \infty$ and run along the big half-circle through the point $\omega = i\infty$. Then, $f(\omega) \approx - \omega^2$ and we get a big  circle in the plane of complex $f(\omega)$ which starts and ends at $f(\omega) = - \infty$. We continue the mapping with $\omega$ running along the real axis from $- \infty$ to $\infty$ and we get the contour in the plane of complex $f(\omega)$ which again starts and ends at $f(\omega) = - \infty$. The crucial question is whether the contour crosses the branch cut indicated in Fig.~\ref{fig-Penrose} which goes along the negative real axis. One observes that the branch cut  is not crossed if $f(\omega = 0) > 0$. Then, $\ln f(\omega)$ remains at the same branch of the Riemann surface. Consequently, the integral  (\ref{Nyquist-integral-3}) vanishes and there are no solutions of the equation (\ref{dispersion-eq-Nyquist-transverse}) in the the upper half-plane of complex $\omega$.  The unstable modes appear when $f(\omega = 0) < 0$. Then, the function $f(\omega)$ gains the phase along the contour $C$ and consequently $\ln f(\omega)$ appears to change the Riemann branch once the path along the contour $C$ is closed. Thus, we get the Penrose criterion: {\em there is an unstable mode if $f(\omega = 0) < 0$.} We note that according to Eq.~(\ref{dispersion-eq-Nyquist-transverse}), we have 
\be
\label{D-zero}
f(\omega=0) = k^2 + \omega_p^2
+ {g^2 \over 2} \int {d^3 p \over (2\pi )^3}
{ v_z^2 \over v_x - i{\rm sgn}(k)0^+} {\partial f({\bf p}) \over \partial p_x} .
\ee

Let us consider a simple example of the Gaussian distribution function 
\be
f ({\bf p}) \sim \exp \Big(- \frac{p_x^2}{2\sigma_x^2} 
- \frac{p_y^2}{2\sigma_y^2} - \frac{p_z^2}{2\sigma_z^2} \Big) 
\ee
which substituted into Eq.~(\ref{D-zero}) gives
\be
\label{D-zero-Gauss}
f(\omega=0) = k^2 
+\Big( \frac{1}{\sigma_z^2} - \frac{1}{\sigma_x^2} \Big)
\frac{g^2}{2}  \int \frac{d^3p}{(2\pi )^3} \,
\frac{f ({\bf p})}{E_{\bf p}} \, \frac{p_z^2} {E_{\bf p}}  .
\ee
As seen, $f(\omega = 0) < 0$ and there is an unstable mode for sufficiently small wave vectors if $\sigma_x^2 < \sigma_z^2$. Thus, even infinitely weak anisotropy generates instability. And keeping in mind that ${\bf k} = (k,0,0)$ and ${\bf E} = (0,0,E)$ one observes that the unstable mode occurs when the momentum distribution is narrower in the direction of the wave vector than in the direction of the electric field. 

For a momentum distribution obeying mirror symmetry $f({\bf p}) = f(-{\bf p})$, the stability criterion was discussed in detail in \cite{Arnold:2003rq}. It was found that a sufficient but not necessary condition for the existence of an unstable mode of the wave vector ${\bf k}$ is a negative eigenvalue of the $3 \times 3$ matrix $\Xi^{ij} (\omega =0 , {\bf k})$ with $\Xi^{ij} (\omega , {\bf k})$ given by Eq.~(\ref{Xi-def}).



\subsection{Deformed isotropic system}
\label{subsec-deformed}

In heavy-ion collisions, the existence of momentum-space anisotropy is a generic feature of the parton momentum distribution in the local rest frame. After the first collisions, when the partons are released from the incoming nucleons, the momentum distribution is strongly elongated along the beam - it  is of the {\em prolate} shape with the average transverse momentum being much smaller than the average longitudinal one. The partons include here not only the {\it wee} partons of small $x$ but also the {\it valence} quarks which carry the longitudinal momentum of incoming nuclei. Due to the free streaming, the momentum distribution evolves in the local rest frame to the distribution which is squeezed along the beam - it is of the {\em oblate} shape with the average transverse momentum being much larger than the average longitudinal one. During the evolution, which is quantitatively studied in {\em e.g.} \cite{Jas:2007rw}, the parton momentum distribution can be treated as a deformed isotropic distribution. Romatschke and Strickland \cite{Romatschke:2003ms} introduced an elegant Ansatz to parameterize an anisotropic momentum distribution by deforming an isotropic one. This distribution and its generalizations have appeared very useful to study various aspects of quark-gluon plasma, see {\it e.g.} \cite{Schenke:2006yp,Mauricio:2007vz,Dumitru:2007hy,Martinez:2008di,Martinez:2010sc,Strickland:2011aa,Attems:2012js,Florkowski:2012as,Strickland:2014eua,Nopoush:2015yga,Ryblewski:2015hea,Bhattacharya:2015ada,Krouppa:2015yoa,Alqahtani:2015qja}. In this section we discuss in detail a spectrum of gluon collective excitations - plasmons, following closely a very detailed study \cite{Carrington:2014bla}.  

\subsubsection{Momentum distributions}

The dielectric tensor (\ref{epsilon-noderiv}) and the quark self-energy (\ref{Si-k-final}) are fully determined by the momentum distribution of plasma constituents. As already mentioned, Romatschke and Strickland \cite{Romatschke:2003ms} introduced an elegant Ansatz to model anisotropic distributions by squeezing or stretching isotropic ones. They considered a momentum distribution of the form 
\be
\label{R-S-ansatz}
f_\xi({\bf p}) = C_\xi
f_{\rm iso}\big(\sqrt{{\bf p}^2 +\xi ({\bf p}\cdot{\bf n})^2}\big) ,
\ee
where $ f_{\rm iso}(|{\bf p}|)$ is an isotropic distribution,  $C_\xi$ is a normalization constant, ${\bf n}$ is a unit vector, and the parameter $\xi \in (-1, \infty)$ controls the shape of the distribution. The vector ${\bf n}$ is usually chosen along the beam direction, so that $p_L \equiv {\bf p}\cdot{\bf n}$ and $p_T \equiv |{\bf p} - ({\bf p}\cdot{\bf n}){\bf n}|$. When $\xi =0$ the distribution is isotropic. For $-1<\xi < 0$ the distribution is elongated in the direction of ${\bf n}$ - it is  {\em prolate}. For $\xi > 0$ the distribution is squeezed in the direction of the vector ${\bf n}$ - it is  {\em oblate} - becoming more and more oblate as the parameter $\xi$ increases. 

There is some freedom in choosing the normalization constant $C_\xi$ of the distribution (\ref{R-S-ansatz}).  Initially, Romatschke and Strickland put $C_\xi =1$ \cite{Romatschke:2003ms} but in a later publication \cite{Romatschke:2004jh} they normalized the anisotropic number density to the isotropic one, so that  
\be
\label{norm-xi-1}
\int \frac{d^3p}{(2\pi)^3} \, f_\xi ({\bf p}) = 
\int \frac{d^3p}{(2\pi)^3} \, f_{\rm iso}(|{\bf p}|) ,
\ee
which leads to the normalization constant $C_\xi = \sqrt{1 + \xi}$.

Another normalization condition was proposed in \cite{Carrington:2014bla}. In case of massless plasma constituents, the whole spectrum of collective excitations depends on a single mass parameter. It is usually chosen to be
\be
\label{mass2}
m^2 \equiv g^2 \int {d^3p \over (2\pi)^3} \, 
\frac{f_\xi ({\bf p})}{|{\bf p}|} ,
\ee
which for the isotropic momentum distribution ($\xi = 0$) reduces to the usual Debye mass $m_D$, as seen in Eq.~(\ref{m_D^2}). To compare collective modes at different anisotropies it is natural to use a mass parameter that is independent of $\xi$.  One accomplishes this by demanding the normalization condition 
\be
\label{mass2-iso}
\int {d^3p \over (2\pi)^3} \, 
\frac{f_\xi ({\bf p})}{|{\bf p}|} =\int {d^3p \over (2\pi)^3} \, 
\frac{f_{\rm iso}(|{\bf p}|)}{|{\bf p}|} ,
\ee
which determines the normalization constant as
\ba
\label{norm-constant-xi}
C_\xi  =\left\{ \begin{array}{lll}  
\frac{\sqrt{|\xi|} }{ {\rm Arctanh}\sqrt{|\xi|}}
&& \textrm {for} ~~   -1 \le \xi < 0 ,
\\[3mm] 
\frac{\sqrt{\xi} }{ {\rm Arctan}\sqrt{\xi}}
&& \textrm{for}  ~~~~~~ 0 \le \xi . 
\end{array} \right.
\ea

In addition to Eq.~(\ref{R-S-ansatz}), which we refer to as the $\xi$-distribution, one can also consider a distribution of the form \cite{Carrington:2014bla}
\be
\label{alter-ansatz}
f_\sigma({\bf p}) \equiv 
C_\sigma f_{\rm iso} \big(\sqrt{(\sigma +1) {\bf p}^2 - \sigma ({\bf p}\cdot{\bf n})^2 }\:\big) ,
\ee
where $\sigma\ge - 1$, which we call the $\sigma$-distribution.  For $0 > \sigma \ge - 1$ the distribution (\ref{alter-ansatz}) is oblate, for $\sigma=0$ it is isotropic, for $\sigma > 0$ it is prolate, increasing in prolateness as the parameter $\sigma$ grows. If the normalization constant $C_\sigma$ is determined by requiring that the distributions $f_\sigma ({\bf p})$ and $f_{\rm iso}(|{\bf p}|)$ satisfy the condition analogous to Eq.~(\ref{norm-xi-1}), one finds $C_\sigma = \sigma + 1$.  Requiring the condition analogous to Eq.~(\ref{mass2-iso}), so that the mass parameter (\ref{mass2}) is independent of $\sigma$, one finds
\ba
\label{norm-constant-eta}
C_\sigma  =\left\{ \begin{array}{lll} 
\frac{\sqrt{|\sigma(\sigma + 1)|}}{{\rm Arctan}\sqrt{|\frac{\sigma}{\sigma + 1}|}} 
&& \textrm {for} ~~   -1 \le \sigma < 0 ,
\\[3mm] 
\frac{\sqrt{\sigma(\sigma + 1)} }{{\rm Arctanh}\sqrt{\frac{\sigma}{\sigma + 1}}}
&& \textrm{for}  ~~~~~~ 0 \le \sigma . 
\end{array} \right.
\ea

There are two special cases of particular interest which can be treated analytically: the extremely prolate and extremely oblate distributions. The latter is proportional to $\delta({\bf n} \cdot {\bf p})=\delta(p_L)$ and can be obtained from the $\xi$-distribution (\ref{R-S-ansatz}) by taking the limit $\xi\to \infty$ (it does not correspond to the limit $\sigma\to -1$ of the $\sigma$-distribution (\ref{alter-ansatz})). The extremely prolate distribution is proportional to $\delta({\bf p}^2-({\bf n} \cdot {\bf p})^2) \sim \delta(p_T)$ and corresponds to the limit $\sigma\to \infty$ of the $\sigma$-distribution (but not the limit $\xi\to -1$ of the $\xi$-distribution).

In practice, the simplest way to obtain the extremely oblate and extremely prolate distributions is not to take the limits described above, but to start from the forms
\ba
\label{extreme-oblate} 
f_{\rm ex-oblate}({\bf p}) &=& \delta(p_L) \, h (p_T), 
\\[2mm]
\label{extreme-prolate} 
f_{\rm ex-prolate}({\bf p}) &=& \delta(p_T)\, \frac{|p_L|}{p_T} \, g(p_L) \,,
\ea
and determine the functions $h(p_T)$ and $g(p_L)$ from the normalization conditions analogous to Eq.~(\ref{mass2-iso}): 
\be
m^2 = {g^2 \over 4\pi^2} \int_0^\infty dp_T\, h(p_T) 
= {g^2 \over 4\pi^2} \int_{-\infty}^\infty dp_L \, g(p_L) .
\ee

\subsubsection{Dispersion equations}
\label{sec-dis-eqs-deformed}

To solve the general dispersion equation (\ref{dispersion-g}), one must either find the zeros of the determinant of the matrix $\Xi$ (\ref{Xi-def}), or invert $\Xi$  and find the poles of the inverted matrix. The second strategy, developed in \cite{Romatschke:2003ms,Romatschke:2004jh}, appeared to be very efficient. 

\subthreesection{Decomposition of $\Xi$}

The first step is to decompose the matrix using a complete set of projection operators. In isotropic plasmas, an arbitrary tensor depends only on the wave vector ${\bf k}$, and can be decomposed into two components, which are transverse and longitudinal with respect to ${\bf k}$. In anisotropic plasmas, the number of projection operators that are needed is larger. An important simplifying feature of the distributions (\ref{R-S-ansatz}) and (\ref{alter-ansatz}) is that the momentum distribution is deformed in only one direction, which is given by the vector ${\bf n}$. An arbitrary symmetric tensor which depends on two vectors can be decomposed in terms of four projection operators. Following \cite{Romatschke:2003ms,Kobes:1990dc}, we introduce the vector ${\bf n}_T$ transverse to ${\bf k}$, which equals
\be
\label{nT-def}
n_T^i = \big(\delta^{ij} - \frac{k^i k^j}{{\bf k}^2}\big) \, n^j ,
\ee 
and define four matrices 
\ba
\begin{array}{ccc}
A^{ij}({\bf k}) = \delta^{ij} - \frac{k^i k^j}{{\bf k}^2}, & &
B^{ij}({\bf k}) = \frac{k^i k^j}{{\bf k}^2} ,
\\[2mm] 
C^{ij}({\bf k},{\bf n}) = \frac{n_T^i n_T^j}{{\bf n}_T^2}, & &
D^{ij}({\bf k},{\bf n}) = k^i n_T^j + k^j n_T^i ,
\end{array}
\ea
which obey the following relations 
\ba
\label{ortho}
\begin{array}{cccc}
AA=A, & AB = 0, & AC = C, & (AD)^{ij} = n_T^i k^j,  \\[2mm]
BA =0, & BB = B, & BC = 0, & (BD)^{ij} = k^i n_T^j,  \\[2mm]
CA= C, & CB = 0, & CC =C, & (CD)^{ij} = n_T^i k^j,  \\[2mm]
(DA)^{ij} = k^i n_T^j, & (DB)^{ij} = n_T^i k^j, & (DC)^{ij} = k^i n_T^j, & DD = n_T^2{\bf k}^2 (B+C).
\end{array}
\ea
Using this projector basis, the inverse propagator $\Xi$ given by Eq.~(\ref{Xi-def}) can be decomposed as 
\be
\label{Sigma-A-B-C-D}
\Xi^{ij}(\omega, {\bf k}) = a(\omega, {\bf k})\,A^{ij} +b(\omega, {\bf k})\,B^{ij} +c(\omega, {\bf k})\,C^{ij} +d(\omega, {\bf k})\,D^{ij}\, ,
\ee
and the coefficients $a$, $b$, $c$ and $d$ can be found from the equations
\ba
\label{a-b-c-d}
k^i \Xi^{ij} k^j = {\bf k}^2 b , \;\;\;\;\;
n_T^i \Xi^{ij} n_T^j = {\bf n}_T^2 (a + c) , \;\;\;\;\;
n_T^i \Xi^{ij} k^j = {\bf n}_T^2 {\bf k}^2 d , \;\;\;\;\;
{\rm Tr}\Xi = 2a + b + c .
\ea
We note that the Lorentz covariant version of the decomposition (\ref{Sigma-A-B-C-D}) can be found in \cite{Dumitru:2007hy}. With the help of the relations (\ref{ortho}), we invert the matrix (\ref{Sigma-A-B-C-D}) and obtain 
\ba
\label{dispXX}
(\Xi^{-1})^{ij} = \frac{1}{a} \,A^{ij} 
+ \frac{-a(a+c)\,B^{ij} 
+ (- d^2{\bf k}^2{\bf n}_T^2 +bc)\,C^{ij}
+ad \,D^{ij}}
{a(d^2{\bf k}^2{\bf n}_T^2 - b(a+c))} .
\ea
Keeping in mind that, as already mentioned, $\Xi^{-1}$  equals the gluon propagator in the temporal axial gauge, we rewrite Eq.~(\ref{dispXX}) as
\be 
\label{propagator-meg-notation}
(\Xi^{-1})^{ij} =\Delta^{ij} = \Delta_A (A^{ij}-C^{ij}) 
+ \Delta_G  \big((a + c) B^{ij} + b \, C^{ij} - d \, D^{ij}\big),
\ee
where the functions $\Delta_A$ and $\Delta_G$ are defined as
\ba
\label{Delta-A}
 \Delta_A^{-1}(\omega,{\bf k}) &\equiv& a(\omega,{\bf k})  ,
\\[2mm]
\label{Delta-G}
\Delta^{-1}_G(\omega,{\bf k}) &\equiv& - d^2(\omega,{\bf k}) \, {\bf k}^2 {\bf n}_T^2 + b(\omega,{\bf k}) \big( a(\omega,{\bf k}) + c(\omega,{\bf k}) \big) .
\ea

\subthreesection{Dispersion equations}

The dispersion equations are obtained from the poles of the propagator (\ref{dispXX}) or (\ref{propagator-meg-notation}) and are
\ba
\label{dis-eq-A}
\Delta_A^{-1}(\omega,{\bf k}) = 0 ,
\\[2mm]
\label{dis-eq-G}
\frac{1}{\omega^2}\Delta^{-1}_G(\omega,{\bf k}) = 0 .
\ea
Following the terminology introduced in \cite{Romatschke:2003ms}, we will refer to solutions of the dispersion equation $\Delta^{-1}_A=0$ as $A$-modes, and solutions of the equation $\Delta^{-1}_G/\omega^2=0$ will be called $G$-modes. In the $G$-mode dispersion equation, the factor $1/\omega^2$ is introduced to remove two trivial zero solutions that are of no physical interest. A presence of such solutions of the dispersion equation $\Delta^{-1}_G=0$ is easily seen in the case of an isotropic plasma when $\Delta^{-1}_G(\omega, {\bf k}) = \omega^2 \varepsilon_L (\omega,{\bf k}) $. Then, except the longitudinal plasmons discussed in Sec.~\ref{sec-gluon-modes-iso}, there is the double solution $\omega =0$ of the equation $\Delta^{-1}_G=0$. Removing the zero solutions is important in the context of the Nyquist analysis discussed in Sec.~\ref{subsec-Nyquist} which then provides the number of {\it physical} solutions of a given dispersion equation.

When the anisotropy is weak, the coefficient $d(\omega,{\bf k})$ can be neglected, as shown in Sec.~\ref{sec-weakly-aniso}, and the dispersion equation (\ref{dis-eq-G}) factors into two simpler equations which are
\ba
\label{dis-eq-B}
\frac{1}{\omega^2}\Delta_B^{-1}(\omega, {\bf k})=0,
\\[2mm]
\label{dis-eq-C}
\Delta_C^{-1}(\omega, {\bf k}) = 0 ,
\ea
where
\ba
\label{Delta-B}
\Delta_B^{-1}(\omega, {\bf k}) &\equiv& b(\omega,{\bf k}) ,
\\[2mm]
\label{Delta-C}
\Delta_C^{-1}(\omega, {\bf k}) &\equiv& a(\omega,{\bf k})+c(\omega,{\bf k}) .
\ea
We will refer to the solutions of these equations as $B$-modes and $C$-modes, respectively. In the $B$-mode equation we have again removed two zero solutions. 

For an isotropic plasma, the coefficients $c_{\rm iso}$ and $d_{\rm iso}$ vanish while $a_{\rm iso}$ and $b_{\rm iso}$ are equal to
\ba
\label{a-iso}
a_{\rm iso}(\omega,{\bf k}) &=& \omega^2 - {\bf k}^2 - \omega^2 \varepsilon_T(\omega,{\bf k}) ,
\\
\label{b-iso}
b_{\rm iso}(\omega,{\bf k}) &=& \omega^2 - \omega^2 \varepsilon_L(\omega,{\bf k}) ,
\ea
where the dielectric functions  $\varepsilon_L$ and $\varepsilon_T$ are given by Eq. (\ref{epsilon-L-final}) and (\ref{epsilon-T-final}), respectively. 
The dispersion equations (\ref{dis-eq-A}) and (\ref{dis-eq-B}) are then equivalent to Eqs.~(\ref{dis-eq-iso}).

\subthreesection{Coefficients $a,~b,~c,~d$}

Starting with the decomposition (\ref{Sigma-A-B-C-D}) and solving the set of equations (\ref{a-b-c-d}), one finds the coefficients $a,~b,~c,~d$:
\ba
\label{alpha-gen}
a(\omega,{\bf k}) &=& \omega^2 - {\bf k}^2
- \frac{g^2}{2} \int {d^3p \over (2\pi)^3} \,
\frac{f({\bf p})}{|{\bf p}|}
\bigg[1 + \frac{{\bf k}^2 - \omega^2}{(\omega - {\bf k}\cdot {\bf v} +i0^+)^2}
\Big(1 - \frac{({\bf n}_T \cdot {\bf v})^2}{{\bf n}_T^2} - \frac{({\bf k} \cdot {\bf v})^2}{{\bf k}^2} \Big)\bigg] ,
\\[2mm]
\label{beta-gen}
b(\omega,{\bf k}) &=& \omega^2
-\frac{g^2}{2} \int {d^3p \over (2\pi)^3} \,
\frac{f({\bf p})}{|{\bf p}|}
\bigg[1 + \frac{2({\bf k}\cdot {\bf v})}{\omega - {\bf k}\cdot {\bf v} +i0^+}
 + \frac{({\bf k}^2 - \omega^2)({\bf k}\cdot {\bf v})^2}{{\bf k}^2 (\omega - {\bf k}\cdot {\bf v} +i0^+)^2} \bigg],
\\[2mm]
\label{gamma-gen}
c(\omega,{\bf k}) &=& - 
\frac{g^2}{2} \int {d^3p \over (2\pi)^3} \,
\frac{f({\bf p})}{|{\bf p}|}
\bigg[\frac{{\bf k}^2 - \omega^2}{(\omega - {\bf k}\cdot {\bf v} +i0^+)^2}
\Big(-1 +2 \frac{({\bf n}_T \cdot {\bf v})^2}{{\bf n}_T^2} + \frac{({\bf k} \cdot {\bf v})^2}{{\bf k}^2} \Big) \bigg],
\\[2mm]
\label{delta-gen}
d(\omega,{\bf k}) &=& - 
\frac{g^2}{2} \int {d^3p \over (2\pi)^3} \,
\frac{f({\bf p})}{|{\bf p}|}
\bigg[\frac{1}{\omega - {\bf k}\cdot {\bf v} +i0^+} \frac{{\bf n}_T \cdot {\bf v}}{{\bf n}_T^2}  + \frac{{\bf k}^2 - \omega^2}{(\omega - {\bf k}\cdot {\bf v} +i0^+)^2}
\frac{({\bf n}_T \cdot {\bf v})({\bf k} \cdot {\bf v})}{{\bf n}_T^2 {\bf k}^2} \bigg].
\ea
An important advantage of a momentum distribution in the form (\ref{R-S-ansatz}) or (\ref{alter-ansatz}) is that, for massless plasma constituents, the integral over the magnitude of the momentum and the angular integrals factorize. The momentum distributions (\ref{R-S-ansatz}, \ref{alter-ansatz}) can be written as
\ba
\label{M-defns}
&& f_\xi({\bf p}) = C_\xi f_{\rm iso}(M_\xi |{\bf p}|),
~~~~~M_\xi \equiv \sqrt{ 1+\xi({\bf n}\cdot{\bf v})^2} , 
\\[2mm]
&& f_\sigma({\bf p}) = C_\sigma f_{\rm iso}(M_\sigma |{\bf p}|),
~~~~M_\sigma \equiv \sqrt{1+\sigma + \sigma({\bf n}\cdot{\bf v})^2} ,
\ea
where the functions $M_\xi$ and $M_\sigma$ do not depend on the magnitude $p=|{\bf p}|$. Introducing the variable $\tilde p = M_{\xi/\sigma}|{\bf p}|$, the integrals over $\tilde p$ can be done analytically. The azimuthal integrals can be also done analytically in a straightforward manner. The only difficult integration is over the polar angle which can be done analytically, but the resulting expressions are complicated and not very enlightening. Below, we give analytic expressions (after performing both angular integrations) for $a, \; b, \; c, \; d$  for some special cases where the results are relatively simple. 

\subthreesection{Solutions of dispersion equations}

Solutions $\omega ({\bf k})$ of the dispersion equations (\ref{dis-eq-A}) and (\ref{dis-eq-G}) represent plasmons that are gluon collective modes. With the polarization tensor computed in the leading order, there are no complex solutions, only purely real and purely imaginary ones. The real solutions correspond to undamped propagating modes, and the imaginary ones to unstable or over-damped modes (depending on the sign of the solution). These modes are sometimes called, respectively, stable and unstable solutions, but this terminology is confusing since both the over-damped and propagating modes are stable. We will refer to them as real and imaginary solutions. Every solution has a partner with opposite sign. In the case of imaginary solutions, every unstable mode has a partner over-damped mode. In the case of real solutions, the change of sign corresponds to a phase shift of the plasma wave and is physically unimportant.  

In subsequent sections we present spectra of plasmons for various momentum distributions. We discuss following the study  \cite{Carrington:2014bla} all possible degrees of one dimensional deformation from the extremely prolate case, when the momentum distribution is infinitely elongated in one direction, to the extremely oblate distribution, which is infinitely squeezed in the same direction. For each case, the number of modes is determined using a Nyquist analysis and the complete spectrum of plasmons is found analytically if possible, and numerically when not. Unstable modes exist in all cases except that of an isotropic plasma. The conditions on the wave vectors for the existence of these instabilities are given. 

The extremely prolate system is special in several ways and the notation we use in this case is explained in section \ref{sec-ex-prolate}. In all other cases, we will use the following notation for the dispersion curves: 
\begin{itemize}

\item
red (solid) - real $A$-modes denoted $\omega_\alpha$,

\item
green (dashed) - real $G$-modes which stay above the light cone, denoted $\omega_+$,

\item
blue (dotted) - real $G$-modes which cross the light cone, denoted $\omega_-$,

\item
orange (dashed) - imaginary $A$-modes denoted $\omega_{\alpha i}=i\gamma_\alpha$,

\item
pink (solid) - imaginary $G$-modes  denoted $\omega_{-i}=i\gamma_-$,

\end{itemize}

When plotting real solutions we show only the positive partner, and for imaginary solutions we show the positive imaginary part of the frequency. The curves for the real and imaginary modes are not similar, and therefore there is no ambiguity in plots with two dashed or solid lines. The light cone is always represented as a thin light gray (solid) line. 

We will show that imaginary solutions exist in anisotropic plasmas, when certain conditions on the wave vector are satisfied. These conditions are even functions of $\cos\theta$, where $\theta$ is the angle between the vectors ${\bf k}$ and ${\bf n}$, and therefore when we discuss them we will consider only $0<\theta<90^\circ$.

\subsubsection{Weakly Anisotropic Plasma}
\label{sec-weakly-aniso}

It is interesting to study a weakly anisotropic system because it can be treated analytically to a large extent. The spectrum of plasmons changes qualitatively when an infinitesimal anisotropy is introduced and all qualitative features of the weakly anisotropic plasma survive in case of strong anisotropy.  Such an analysis of a weakly-anisotropic QGP was first presented in \cite{Romatschke:2003ms}. 

To derive the spectrum of collective modes in a weakly anisotropic plasma, one can use the $\xi$-distribution (\ref{R-S-ansatz}) with the assumption $|\xi| \ll 1$ which gives 
\be
\label{R-S-ansatz-weak-aniso}
f_\xi({\bf p}) = \Big(1+\frac{\xi}{3} \Big) f_{\rm iso}(p) 
+ \frac{\xi}{2} \frac{d f_{\rm iso}(p)}{dp} \, p \, ({\bf v}\cdot{\bf n})^2 ,
\ee 
where we have taken into account that the normalization constant (\ref{norm-constant-xi}) equals 
\be
C_\xi = 1 + \frac{\xi}{3} + {\cal O}(\xi^2) .
\ee
The distribution (\ref{R-S-ansatz-weak-aniso}) is weakly prolate for $\xi<0$ and weakly oblate for $\xi>0$. 

Using the formula (\ref{R-S-ansatz-weak-aniso}), the coefficients  $a, \; b, \; c, \; d$ given by Eqs.~(\ref{alpha-gen})-(\ref{delta-gen}) can be computed analytically. For $a$ and $b$ there are contributions of order $\xi^0$ which are just the isotropic results of Sec.~\ref{sec-gluon-modes-iso}. All four functions  $a,~b,~c,~d$ have contributions of order $\xi$. Since the coefficient $d$ enters the dispersion equation (\ref{dis-eq-G}) quadratically, it does not contribute to linear order in $\xi$ and the dispersion equation factorizes into two pieces, so that we have the three dispersion equations of $A$-modes (\ref{dis-eq-A}), $B$-modes (\ref{dis-eq-B}) and $C$-modes (\ref{dis-eq-C}). 

The  coefficients  $a, \; b, \; c$ are computed as
\ba
\non
a(\omega, {\bf k}) &=&  \Big(1 + \frac{\xi}{3}\Big) a_{\rm iso} (\omega, {\bf k}) 
+ \xi  \frac{m^2}{8}
\bigg\{
\frac{8}{3} \cos^2\theta  + \frac{2}{3} \big( 5 - 19 \cos^2\theta \big)\frac{\omega^2}{k^2}
- 2 \big( 1 - 5 \cos^2\theta \big) \frac{\omega^4}{k^4} 
\\[2mm] 
\label{alpha-final}
&& 
+ \bigg[ 1 - 3\cos^2\theta 
-  \Big(2 - 8 \cos^2\theta \Big)  \frac{\omega^2}{k^2} 
+ \Big(1  - 5  \cos^2\theta \Big) \frac{\omega^4 }{k^4}  \bigg] 
\frac{\omega}{k}  \ln\Big( \frac{\omega + k + i0^+}{\omega -k + i0^+} \Big) \bigg\} ,
\\[4mm] 
\non
b (\omega, {\bf k}) &=& \Big(1 + \frac{\xi}{3}\Big) b_{\rm iso} (\omega, {\bf k}) 
+ \xi  m^2 \bigg\{ \Big(- \frac{2}{3} + \cos^2\theta  \Big) \frac{\omega^2}{k^2}
+  (1 - 3 \cos^2\theta )  \frac{\omega^4 }{k^4}
\\[2mm] 
\label{beta-final}
&& 
+ \frac{1}{2} \bigg[  ( 1 - 2 \cos^2\theta ) \frac{\omega^2}{k^2} 
- (1 - 3 \cos^2\theta ) \frac{\omega^4 }{k^4}  \bigg] 
\frac{\omega}{k } \ln\Big( \frac{\omega + k+ i0^+}{\omega -k + i0^+} \Big) \bigg\} ,
\\[4mm]
\label{gamma-final}
c(\omega, {\bf k}) &=& \xi  \frac{m^2}{4}  
 \, \sin^2\theta  \bigg[ - \frac{4}{3}
 +  \frac{10}{3} \frac{\omega^2}{k^2}
- 2 \frac{\omega^4}{k^4}
+ \Big( 1-  2\frac{\omega^2}{k^2} + \frac{\omega^4}{k^4} \Big) 
\frac{\omega}{k}
\ln\Big( \frac{\omega + k + i0^+}{\omega -k + i0^+} \Big) \bigg] ,
\ea
where  $a_{\rm iso}, ~b_{\rm iso}$ are given by Eqs.~(\ref{a-iso}), (\ref{b-iso}) and $\cos\theta = ({\bf k}\cdot {\bf n})/k$.  These results in a different form were derived in \cite{Romatschke:2003ms,Carrington:2014bla}\footnote{The weak-anisotropy results of \cite{Romatschke:2003ms} are slightly different due to the fact that they used $C_\xi=1$.  The results above can be obtained from the original expressions in \cite{Romatschke:2003ms} by a rescaling of the Debye mass.}.

As in the case of the isotropic plasma, the dispersion relations cannot be solved analytically for arbitrary $k$. When $k^2 \ll \omega^2$, the functions $a(\omega,{\bf k})$, $b(\omega,{\bf k})$, $c(\omega,{\bf k})$ are approximated as 
\ba
\label{alpha-small-k}
a(\omega,{\bf k})  
&=& \omega^2 - {\bf k}^2 
- m^2 \bigg\{ \frac{1}{3} \Big(1 - \frac{\xi}{15}\Big)  
+ \frac{1}{5} \Big[ \frac{1}{3} + \frac{\xi}{7}  \Big(\frac{1}{9} + \cos^2\theta \Big) \Big]
\frac{k^2}{\omega^2}
+ {\cal O}\Big( \frac{k^4}{ \omega^4}\Big) \bigg\} ,
\\[4mm]
\label{beta-small-k}
b(\omega,{\bf k})   
&=& \omega^2 
- m^2 \bigg\{ \frac{1}{3} \Big[1 + \frac{\xi}{5}\Big(-\frac{1}{3} + \cos^2\theta \Big) \Big]  
+ \frac{1}{5} \Big[ 1 + \frac{\xi}{7}  \Big(\frac{1}{3} - \cos^2\theta \Big) \Big]
\frac{k^2}{\omega^2}
+ {\cal O}\Big( \frac{k^4}{ \omega^4}\Big) \bigg\} ,
\\[4mm]
\label{gamma-small-k}
c(\omega,{\bf k})  
&=& - \xi  \,m^2 \sin^2\theta \Big[ \frac{1}{15} - \frac{4 \, k^2}{105 \, \omega^2}
+ {\cal O}\Big( \frac{k^4}{ \omega^4}\Big) \Big] .
\ea

In the next three subsections we discuss solutions of the dispersion equations (\ref{dis-eq-A}), (\ref{dis-eq-B}), and (\ref{dis-eq-C}) with the coefficients (\ref{alpha-final}), (\ref{beta-final}), (\ref{gamma-final}) or (\ref{alpha-small-k}), (\ref{beta-small-k}), (\ref{gamma-small-k}). In every case we begin with a discussion of the number of solutions which can be found using a Nyquist analysis described in Sec.~\ref{subsec-Nyquist}. 

\begin{figure}[t]
\center
\includegraphics*[width=9cm]{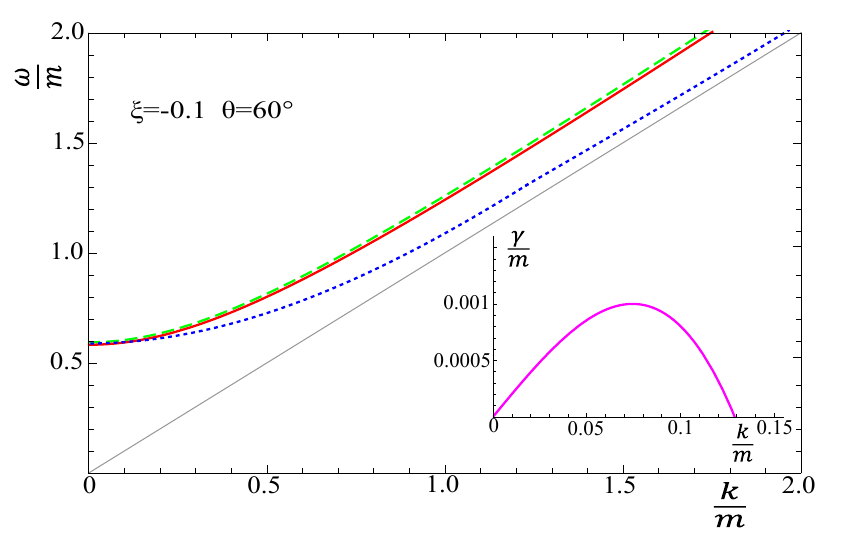}
\caption{(Color online) Dispersion curves of plasmons in a weakly prolate plasma with $\xi = -0.1$ for $\theta = 60^\circ$. The line styles are explained in the text. Figure from \cite{Carrington:2014bla}.}
\label{fig-weak-pro-60}
\end{figure}

\subthreesection{$A$-modes}

The Nyquist analysis shows that the $A$-mode dispersion equation (\ref{dis-eq-A}) has four solutions \cite{Carrington:2014bla} when 
\be
\label{cond-insta-alpha}
k^2 - \xi \, \frac{m^2}{3} \,\cos^2\theta < 0
\ee 
and two solutions otherwise. The condition (\ref{cond-insta-alpha}) is never fulfilled for the prolate plasma ($\xi <0$) and it is fulfilled for any oblate momentum distribution ($\xi >0$) when 
\ba
\label{k-crit-A}
k < k_{\rm A} \equiv \Re\sqrt{\frac{\xi}{3}}\, m|\cos\theta|.
\ea 
Because of the real value in the definition of $k_A$, it vanishes for $\xi < 0$.

The dispersion equation (\ref{dis-eq-A}) cannot be solved analytically. When $\omega^2 \gg k^2$, the coefficient $a(\omega,{\bf k})$ is approximated by the formula (\ref{alpha-small-k}) and Eq.~(\ref{dis-eq-A}) is solved by
\be
\label{sol-weak-aniso-T1}
\omega^2({\bf k}) = \frac{m^2}{3} \Big(1 - \frac{\xi}{15}\Big)  +
\frac{6}{5} \Big[ 1 + \frac{\xi }{14} \Big(\frac{4}{15} + \cos^2\theta \Big) \Big] k^2 
+  {\cal O}\Big(\frac{k^4}{m^2} \Big) ,
\ee
which reduces to the well-known result for the transverse plasmon (\ref{omega-trans}) when $\xi=0$. The plasmon mass, which is given by the first term on the right side of Eq.~(\ref{sol-weak-aniso-T1}), depends on the anisotropy parameter $\xi$ but is independent of the orientation of the wave vector ${\bf k}$. 

One can obtain purely imaginary solutions \cite{Carrington:2014bla} by substituting $\omega = i \gamma$ with $\gamma \in \mathbb{R}$ and assuming $\gamma^2 \ll k^2$. Using the approximate formula
\be
\frac{\omega + k}{\omega - k} = \frac{\gamma^2 - k^2} {\gamma^2 + k^2} -i \frac{2\gamma k} {\gamma^2 + k^2} 
\buildrel{\gamma^2 \ll k^2}\over{\approx} \exp\Big( - i \pi \frac{\gamma}{|\gamma |} \Big) ,
\ee
the coefficient $a(\omega, {\bf k})$ becomes
\be
a(\omega, {\bf k}) =  \omega^2 - {\bf k}^2 + \frac{1}{3}  \xi m^2
 \cos^2\theta + \frac{\pi}{4}\Big[1 -\frac{\xi}{2}\Big(\frac{1}{3} - 3\cos^2\theta \Big)\Big] m^2 \frac{|\gamma |}{k} 
+ {\cal O}\Big(\frac{\gamma^2}{k^2} \Big) ,
\ee
and the dispersion equation (\ref{dis-eq-A}) is written in the form
\be
\label{dis-eq-77}
\gamma^2 + \frac{\lambda}{k} |\gamma | - k_{\rm A}^2 + k^2 = 0,
\ee
where $ k_{\rm A}$ is defined by the formula (\ref{k-crit-A}) and 
\be
\label{wqwq}
\lambda \equiv \frac{\pi}{4}\Big[1 -\frac{\xi}{2}\Big(\frac{1}{3} - 3\cos^2\theta \Big)\Big] m^2 .
\ee
Eq. (\ref{dis-eq-77}) has no roots for an isotropic or prolate system, since $k_A^2 \le 0$ when $\xi\le 0$. For oblate systems, $\xi$ and $k_A^2$ are positive and there are two solutions which read
\be
\label{solution-77}
\gamma ({\bf k}) = \pm \frac{1}{2} \Big(\sqrt{\frac{\lambda^2}{k^2} + 4 (k_{\rm A}^2 -k^2)} - \frac{\lambda}{k} \Big) .
\ee
Equations (\ref{k-crit-A}) and (\ref{wqwq}) show that in the limit of weak anisotropy $\lambda \gg k_{\rm A}^2$, and therefore the expression (\ref{solution-77}) can be approximated as
\be
\label{solution-78}
\gamma ({\bf k}) \approx \pm \frac{1}{\lambda} \, k(k_{\rm A}^2 -k^2).
\ee
The solutions  (\ref{solution-77}) or  (\ref{solution-78}) represent the unstable and overdamped transverse modes which exist only for oblate plasmas ($\xi >0$) provided the condition (\ref{k-crit-A}) is satisfied.

\begin{figure}[t]
\begin{minipage}{8.5cm}
\center
\includegraphics[width=1.04\textwidth]{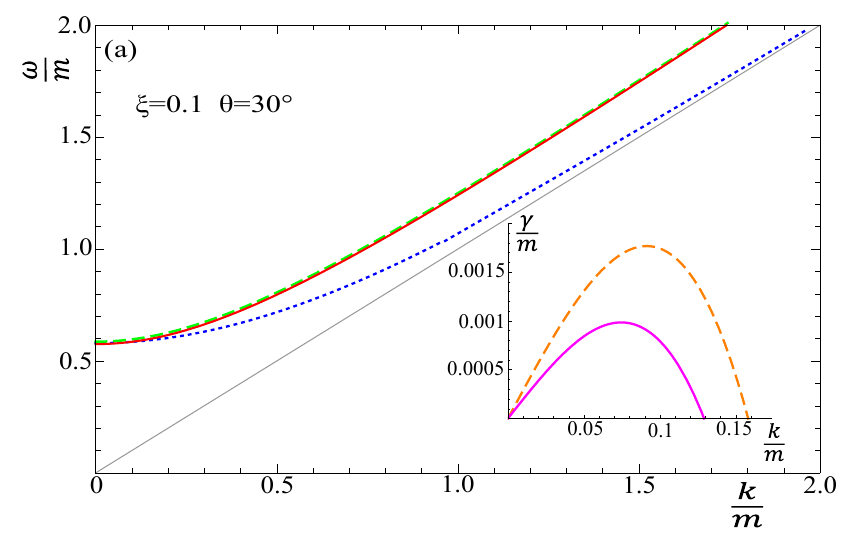}
\end{minipage}
\hspace{2mm}
\begin{minipage}{8.5cm}
\center
\includegraphics[width=1.03\textwidth]{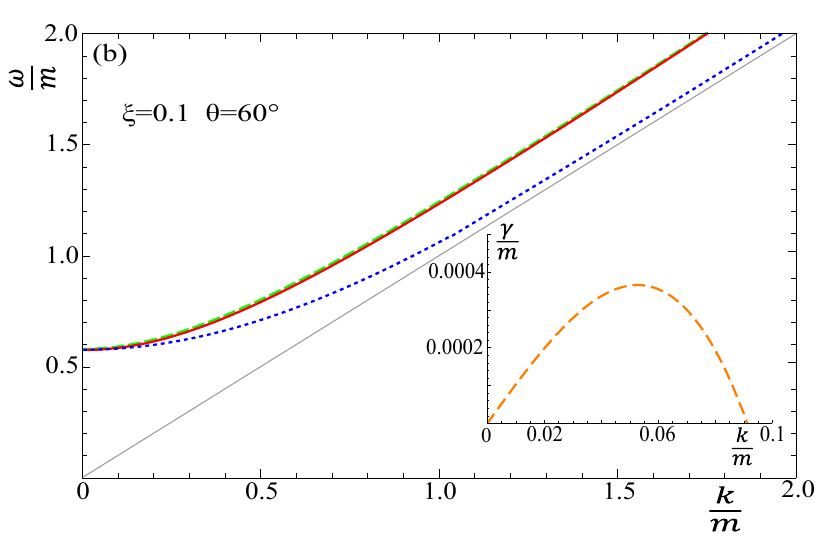}
\end{minipage}
\caption{(Color online) Dispersion curves of plasmons in a weakly oblate plasma with $\xi=0.1$ for $\theta = 30^\circ$ (a) and $\theta = 60^\circ$ (b). The line code is explained in the text. Figure from \cite{Carrington:2014bla}.}
\label{fig-weak-oble-30-60}
\end{figure}

\subthreesection{$B$-modes}

The $B$-mode dispersion equation (\ref{dis-eq-B}), which describes longitudinal modes, always has two solutions. In the limit $\omega^2 \gg k^2$ one can find these solutions analytically. The coefficient $b(\omega,{\bf k})$ is approximated by Eq.~(\ref{beta-small-k}) and the dispersion equation (\ref{dis-eq-B}) is solved by
\be
\label{sol-weak-aniso-L}
\omega^2({\bf k}) = \frac{m^2}{3} \Big[1 + \frac{\xi}{5}\Big(-\frac{1}{3} + \cos^2\theta \Big) \Big]  +
\frac{3}{5} \Big[ 1 + \frac{4\xi}{35}  \big(1 - 3 \cos^2\theta \big) \Big] k^2 
+  {\cal O}\Big(\frac{k^4}{m^2} \Big) ,
\ee
which reduces to the well-known result for the longitudinal plasmon (\ref{omega-long}) when $\xi=0$. The first term on the right side gives the plasmon mass which depends on the anisotropy parameter $\xi$ and the orientation of wave vector ${\bf k}$.  Analogously to the formula (\ref{omega-long-large-k}), the longitudinal mode approaches the light cone as $k \rightarrow \infty$.

\subthreesection{$C$-modes}

The $C$-mode dispersion equation  (\ref{dis-eq-C}) has the richest structure. There are four solutions \cite{Carrington:2014bla} when 
\be
\label{cond-insta-alpha-gamma}
k^2 + \xi \, \frac{m^2}{3} \,\big( 1 - 2 \cos^2\theta \big) < 0
\ee
and two solutions otherwise. The condition (\ref{cond-insta-alpha-gamma}) can be fulfilled for an oblate plasma ($\xi >0$) when $1/2 < \cos^2\theta$ and for a prolate plasma ($\xi <0$) when $1/2> \cos^2\theta$. In both cases the wave vector must satisfy
\be
\label{k-crit-C}
k < k_{\rm C} \equiv m \, \Re\sqrt{\frac{\xi}{3}\,\big(2 \cos^2\theta - 1 \big)} .
\ee
When the argument of the square root is negative, the real part of the root is zero and the critical wave vector $k_{\rm C}$ vanishes. 

In the long wavelength limit ($\omega^2 \gg k^2$), when the coefficients $a(\omega,{\bf k})$ and $c(\omega,{\bf k})$ are approximated by the formulas (\ref{alpha-small-k}) and (\ref{gamma-small-k}), the dispersion equation (\ref{dis-eq-C}) is solved by 
\be
\label{sol-weak-aniso-T2}
\omega^2({\bf k}) = \frac{m^2}{3} \Big[ 1 + \frac{\xi}{5} \Big( \frac{2}{3}  - \cos^2\theta  \Big) \Big]
 +
\frac{6}{5} \Big[ 1 - \frac{\xi }{5} \Big(\frac{23}{42} - \cos^2\theta \Big) \Big] k^2 
+  {\cal O}\Big(\frac{k^4}{m^2} \Big) ,
\ee
which reduces to the well-known transverse plasmon (\ref{omega-trans}) when $\xi=0$. The plasmon mass, which is given by the first term on the right side, depends on the anisotropy parameter $\xi$ and on the orientation of ${\bf k}$. 

One  also finds purely imaginary solutions by substituting $\omega = i \gamma$ with $\gamma \in \mathbb{R}$ and assuming $ \gamma^2 \ll k^2$. The dispersion equation and its solutions have the same form as in the previous section, see Eqs.~(\ref{dis-eq-77}), (\ref{solution-77}), and (\ref{solution-78}), but the coefficient $\lambda$ is now defined as
\be
\lambda \equiv \frac{\pi}{4}\Big[1 - \frac{\xi}{2} \Big( \frac{7}{3} -  5 \cos^2\theta \Big) \Big] m^2 ,
\ee
and $k_{\rm A}$ is replaced by $k_{\rm C}$ given in Eq.~(\ref{k-crit-C}).

\begin{figure}[t]
\begin{minipage}{8.5cm}
\center
\includegraphics[width=1.01\textwidth]{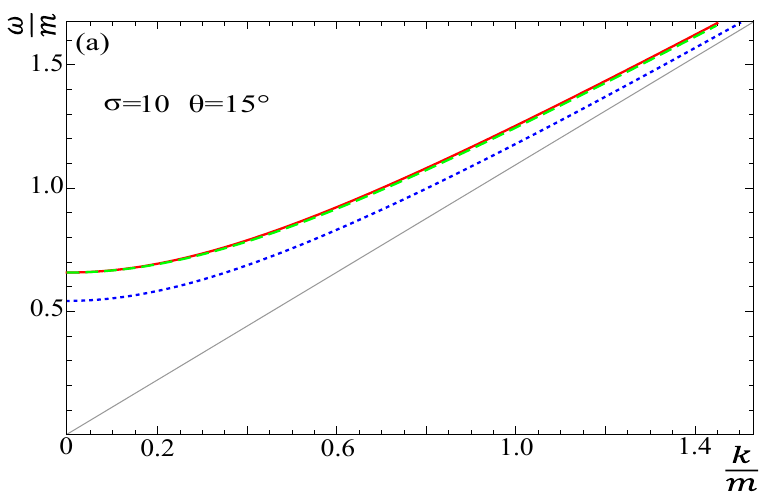}
\end{minipage}
\hspace{5mm}
\begin{minipage}{8.5cm}
\includegraphics[width=1.01\textwidth]{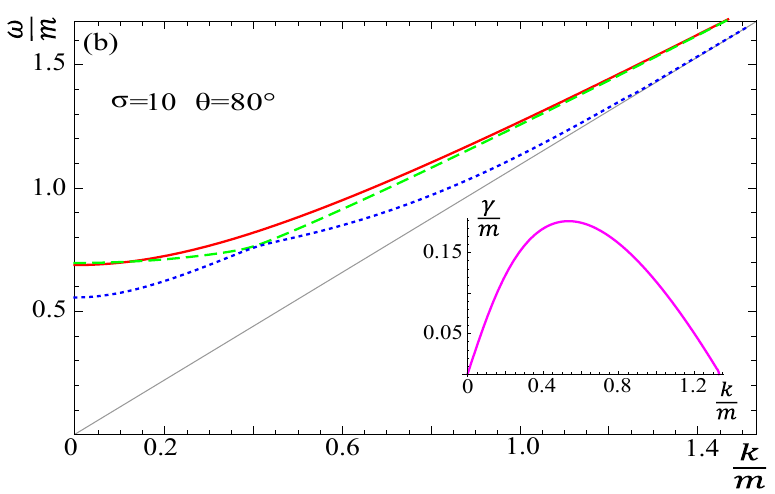}
\end{minipage}
\caption{(Color online) Dispersion curves of plasmons in a prolate plasma with $\sigma = 10$ for $\theta = 15^\circ$ (a) and $\theta = 80^\circ$ (b). The line code is explained in the text. Figure from \cite{Carrington:2014bla}.}
\label{fig-prolate-80-89}
\end{figure}

\subthreesection{Discussion}

A complete spectrum of plasmons in a weakly anisotropic QGP can be found numerically solving the dispersion equations  (\ref{dis-eq-A}), (\ref{dis-eq-B}), and (\ref{dis-eq-C}). The numerical solutions obviously agree with the approximated analytical ones (\ref{sol-weak-aniso-L}), (\ref{sol-weak-aniso-T1}), (\ref{solution-77}), and (\ref{sol-weak-aniso-T2}) in the domains of their applicability. Fig.~\ref{fig-weak-pro-60} shows the spectrum for a weakly prolate plasma ($\xi = - 0.1$) at $\theta=60^\circ$  and  Fig.~\ref{fig-weak-oble-30-60} presents the spectra for a weakly oblate plasma ($\xi = 0.1$) at $\theta=30^\circ$ and $\theta=60^\circ$. The main part of each figure shows the dispersion curves of the positive real modes and the insets present the positive imaginary solutions. 

For weakly prolate and oblate systems, real $A$-, $B$- and $C$-modes exist for all wave vectors and depend only weakly on the angle. The real $A$- and $C$-modes look very much like the real isotropic transverse mode. In Figs.~\ref{fig-weak-pro-60} and \ref{fig-weak-oble-30-60} these modes are represented by the red (solid) and green (dashed) curves which almost overlay each other. The real $B$-mode looks like the real isotropic longitudinal mode and is represented by the blue (dotted) line. 

In addition to the real modes, for a weakly prolate plasma there is an imaginary $C$-mode, seen in Fig.~\ref{fig-weak-pro-60}, which exists for $k < k_C$. The critical wave vector $k_C$ is maximal for $\theta=90^\circ$. When $\theta$ decreases, $k_C$ also decreases until it reaches zero at $\theta=45^\circ$ and the imaginary $C$-mode disappears. In a weakly oblate system there are two imaginary modes seen in Fig.~\ref{fig-weak-oble-30-60}a when $k < k_C < k_A$. Both $k_A$ and $k_C$ are maximal when $\theta=0^\circ$. As $\theta$ increases from $0^\circ$, $k_A$ and $k_C$ decrease. At $\theta=45^\circ$, $k_C$ goes to zero and the imaginary $C$-mode disappears. The regime of the imaginary $A$-mode shrinks to zero at $\theta=90^\circ$. 

\begin{figure}[t]
\begin{minipage}{8.5cm}
\center
\includegraphics[width=1.02\textwidth]{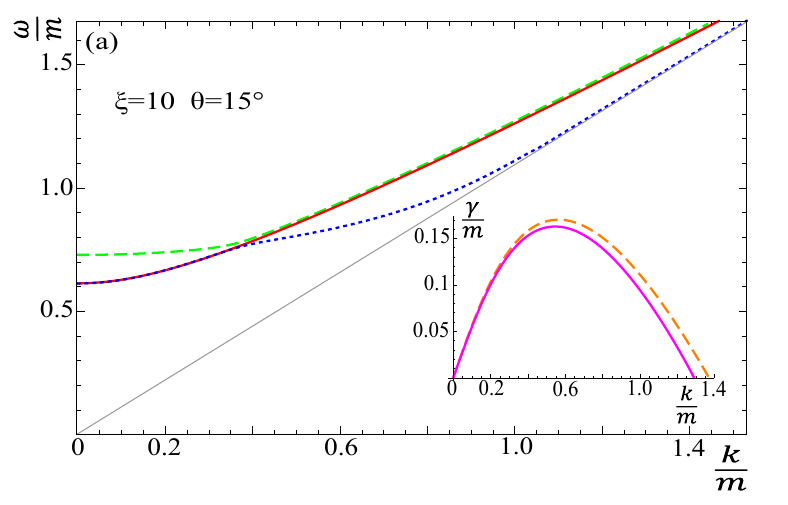}
\end{minipage}
\hspace{5mm}
\begin{minipage}{8.5cm}
\includegraphics[width=1.02\textwidth]{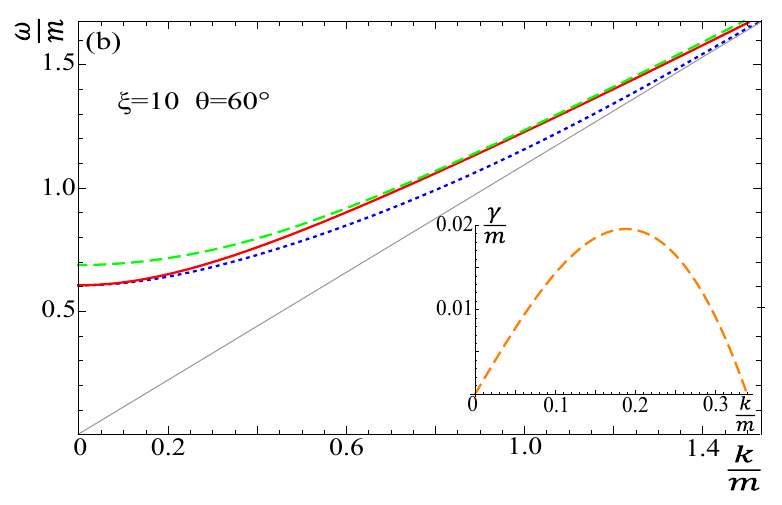}
\end{minipage}
\caption{(Color online) Dispersion curves of plasmons in an oblate plasma with $\xi = 10$ for $\theta = 15^\circ$ (a) and $\theta = 60^\circ$ (b). The line code is explained in the text. Figure from \cite{Carrington:2014bla}.}
\label{fig-oblate-15-60}
\end{figure}

In comparison with the spectra of an isotropic system, the weakly anisotropic plasma exhibits the following important differences.

\begin{itemize}

\item The transverse real mode, which is doubled in the isotropic case, is  now split into two slightly different modes, the $A$-mode and $C$-mode, which are given in Eqs. (\ref{sol-weak-aniso-T1}, \ref{sol-weak-aniso-T2}). In Figs. \ref{fig-weak-pro-60} and \ref{fig-weak-oble-30-60} the curves that correspond to these modes are represented by the red (solid) and green (dashed) curves which lie almost on top of each other. 

\item In isotropic plasmas longitudinal and transverse plasmons have the same plasma frequency $\omega_p = m/\!\sqrt{3}$, but in anisotropic plasmas there are three different minimal frequencies for the three real modes. 

\item In isotropic plasmas there are no imaginary solutions. In anisotropic plasmas the number of imaginary solutions depends on the magnitude and orientation of the wave vector ${\bf k}$. In prolate plasmas the number of imaginary solutions is zero or two (one pair) and in oblate plasmas there are zero, two (one pair) or four (two pairs) imaginary modes.

\end{itemize}

Using Eqs.~(\ref{k-crit-A}) and (\ref{k-crit-C}), the number of modes can be written as
\ba
\label{weak-fin-A}
&&A-{\rm modes}\!:~ \left\{ \begin{array}{lll} 
2+ 2\Theta(k_{\rm A}-k) 
&& \textrm {for oblate plasma},
\\[2mm]
2
&& \textrm{for prolate plasma}, 
\end{array} \right.
\\
\label{weak-fin-B}
&&B-{\rm modes}\!:~~~~~ 2,
\\[1mm]
\label{weak-fin-C}
&&C-{\rm modes}\!:~~~~~ 2+ 2\Theta(k_{\rm C}-k),
\ea
which show that there is a maximum of 8 solutions for a prolate plasma and 10 for an oblate plasma. 

The analysis in this section could equally well have been done using the $\sigma$-distribution (\ref{alter-ansatz}) in the limit $|\sigma| \ll 1$. This would reproduce the results expressed by Eqs.~(\ref{cond-insta-alpha}) and (\ref{cond-insta-alpha-gamma}) with $\xi\to-\sigma$. Since the weakly prolate and weakly oblate systems correspond to  $\sigma>0$ and $\sigma<0$, respectively, the number of modes in Eqs.~(\ref{weak-fin-A}), (\ref{weak-fin-B}), and (\ref{weak-fin-C}) is obviously reproduced. 

There is no anisotropy threshold for the existence of unstable modes, and even an infinitesimal anisotropy produces an instability. However, when $\xi \rightarrow 0$ (or $\sigma \rightarrow 0$) the growth rate of instability ($\gamma$) decreases and the domain of unstable modes shrinks. In this sense, the system becomes less and less unstable as it tends to isotropy. When the effect of inter-parton collisions is taken into account \cite{Schenke:2006xu}, the growth rates of unstable modes are reduced and systems of small anisotropy are effectively stabilized.

\subsubsection{Finite Anisotropy}
\label{sec-finite-aniso}

When the anisotropy parameter is not small, the coefficients $a, \, b, \, c,$ and $d$ of the decomposition (\ref{Sigma-A-B-C-D}) and the solutions of the dispersion equations must be computed numerically. However, the spectrum of plasmons has the same structure as in the case of the weakly anisotropic plasma discussed in the previous section - the number of modes is the same and the behavior of the dispersion curves is very similar. 

One can consider both the $\xi$-distribution (\ref{R-S-ansatz}) and the $\sigma$-distribution (\ref{alter-ansatz}), which together describe deformations of an isotropic distribution with arbitrary prolateness and oblateness. If the anisotropy parameter is not assumed small, the coefficient $\delta$ cannot be neglected, which means that the dispersion equation for the $G$-modes (\ref{dis-eq-G}) does not factorize into equations (\ref{dis-eq-B}) and (\ref{dis-eq-C}). However it can be factorized as \cite{Romatschke:2003ms}
\be
\label{Delta-G-factor}
 \Delta^{-1}_G(\omega, {\bf k}) = \big(\omega^2 -\Omega_+^2(\omega, {\bf k})\big)\big(\omega^2 -\Omega_-^2(\omega, {\bf k})\big) = 0 ,
\ee
where
\be
\label{Omega-pm-def}
\Omega_{\pm}^2(\omega, {\bf k})  \equiv \frac{1}{2} \Big( 2 \omega^2 - a - b 
\pm \sqrt{(b- a -c)^2 + 4 {\bf k}^2 {\bf n}_T^2 d^2}\;\Big) .
\ee
The square root in Eq.~(\ref{Omega-pm-def}) is undefined if its argument is purely real and negative. When all coefficients $a,~b,~c,$ and $d$ are real, the argument of the root is positive definite. When these coefficients are complex, the root argument is also complex. Therefore, there is no case for which the argument of the root is real and negative, which means that one can find the dispersion relations by solving the equations $\omega=\pm \Omega_+(\omega, {\bf k})$ and  $\omega=\pm \Omega_-(\omega, {\bf k})$ self-consistently. 

\begin{figure}[t]
\includegraphics[width=0.48\textwidth]{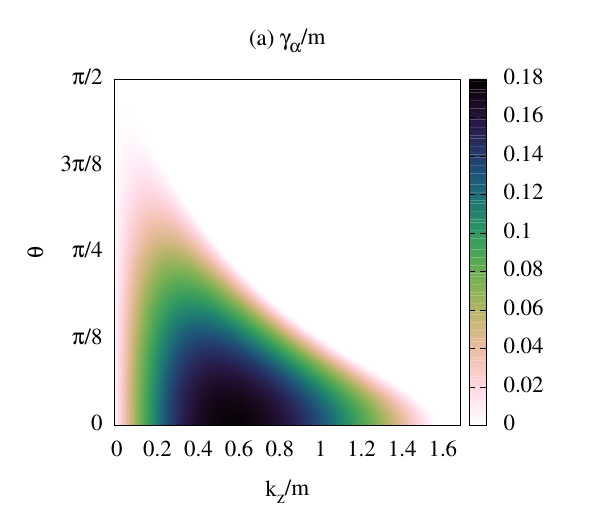}
\includegraphics[width=0.48\textwidth]{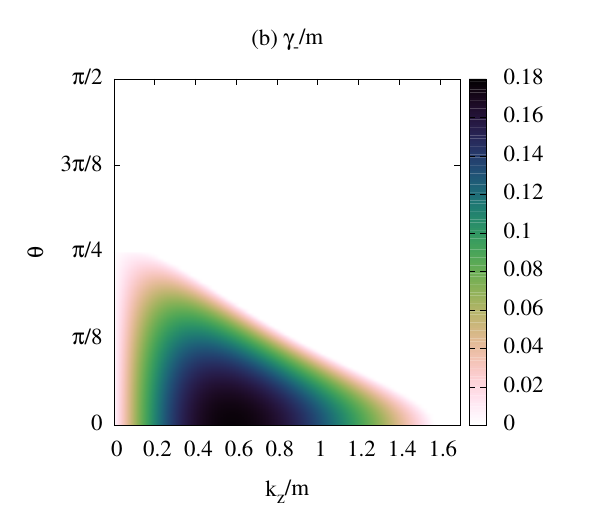}
\vspace{- 3mm}
\caption{(Color online)
The growth rates of unstable $A-$mode (a) and $G-$mode (b) for $\xi=10$ as a function of  $k_z/m$ and $\theta = \arctan(k_T/k_z)$. }
\label{fig-unstablemodes2d}
\end{figure}

Characteristic examples of the complete spectra of plasmons in prolate and oblate plasmas, computed with the $\xi$- and $\sigma$-distribution, respectively,  are shown in Figs.~\ref{fig-prolate-80-89} and \ref{fig-oblate-15-60} for fixed values of $\theta$. For both prolate and oblate cases, there are six (three pairs) of real modes for all ${\bf k}$ which change slowly with $\theta$. For prolate plasmas, there is at most one pair of imaginary modes. For small angles these modes are absent. As the angle increases, the imaginary modes appear at small $k$, and extend to larger and larger $k$ as the angle increases. In oblate systems, there are at most two pairs of imaginary modes. They are both absent at $\theta=90^\circ$. When the angle decreases, the $A$-mode shows up first, and both pairs extend to larger and larger $k$ as $\theta$ continues to decrease. These features, which are the same as for the weakly anisotropic plasma discussed in Sec.~\ref{sec-weakly-aniso}, are nicely seen in Fig.~\ref{fig-unstablemodes2d} where we show the magnitude of the growth rate $\gamma$ of unstable $A-$ and $G-$modes  as a function of $k_z$ and $\theta$ 
for $\xi = 10$. 

\begin{figure}
\includegraphics[width=0.5\textwidth]{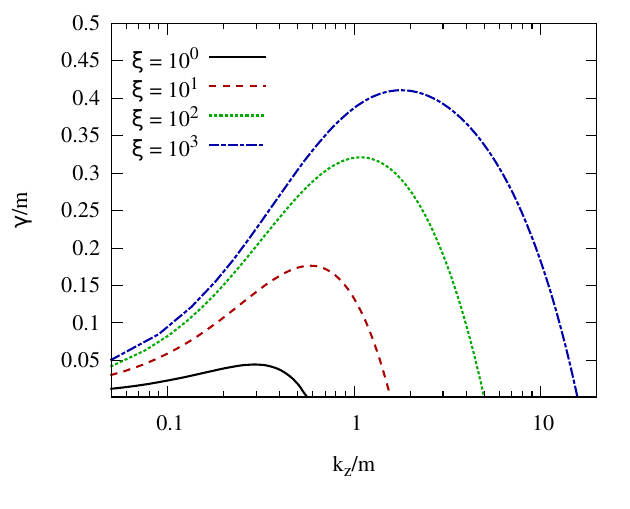}
\vspace{-5mm}
\caption{(Color online) The growth rate of the fastest unstable mode in an oblate system for fixed $\xi$ as a function of $k_z/m$. }
\label{fig-largexiGamma}
\end{figure}

One asks how the growth rate of unstable modes depends on plasma anisotropy. To address the question we consider the mode of the largest growth rate in an oblate system when ${\bf k} \parallel {\bf n}$. Then, the $A-$ and $G-$modes coincide and we deal with the double unstable mode.  In Fig.~\ref{fig-largexiGamma} we plot the growth rate of the unstable modes with ${\bf k} \parallel {\bf n}$ for $\xi \in \{10^0,10^1,10^2,10^3\}$.  The anisotropy vector is chosen along the $z-$axis and thus ${\bf k} =(0,0,k_z)$. From this figure we can see that there is a band of unstable modes for $k_z \in (0,k_z^{\rm max})$ and there is a well-defined maximum growth rate $\gamma^{\rm max}$ at each value of $\xi$.  As $\xi$ increases, both $\gamma^{\rm max}$ and $k_z^{\rm max}$ increase monotonically.  

\subsubsection{Extremely prolate plasma}
\label{sec-ex-prolate}

\begin{figure}[t]
\begin{minipage}{8.5cm}
\center
\includegraphics[width=1.04\textwidth]{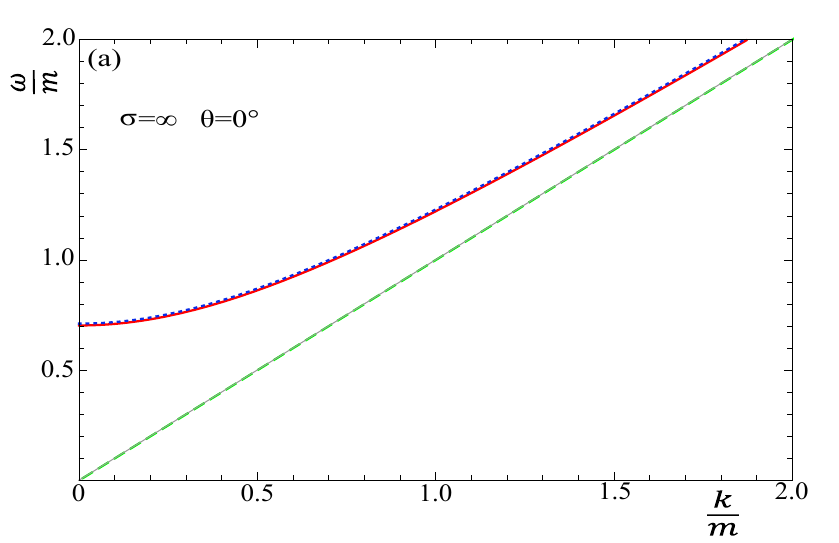}
\end{minipage}
\hspace{2mm}
\begin{minipage}{8.5cm}
\center
\includegraphics[width=1.02\textwidth]{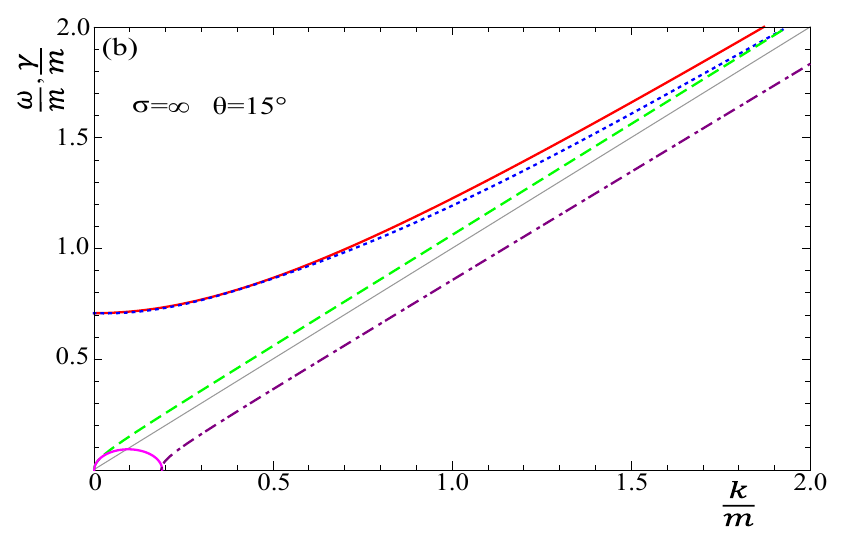}
\end{minipage}
\vspace{2mm}
\begin{minipage}{8.5cm}
\center
\includegraphics[width=1.02\textwidth]{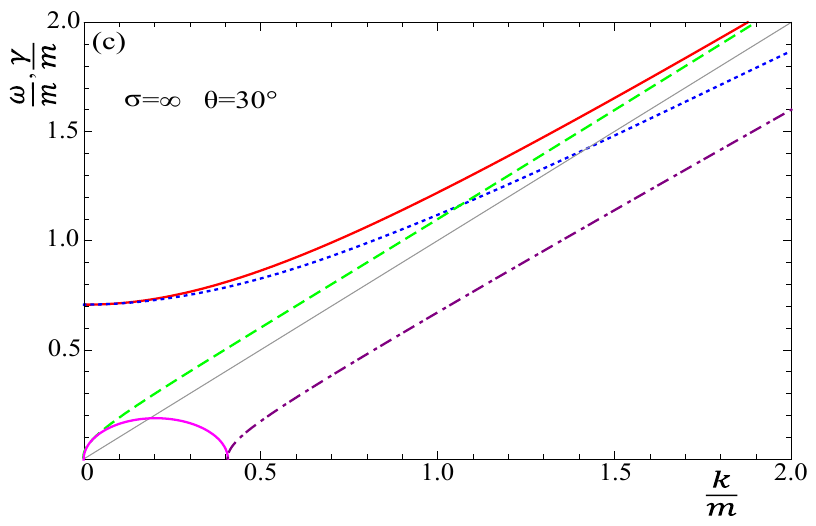}
\end{minipage}
\hspace{2mm}
\begin{minipage}{8.5cm}
\center
\includegraphics[width=1.02\textwidth]{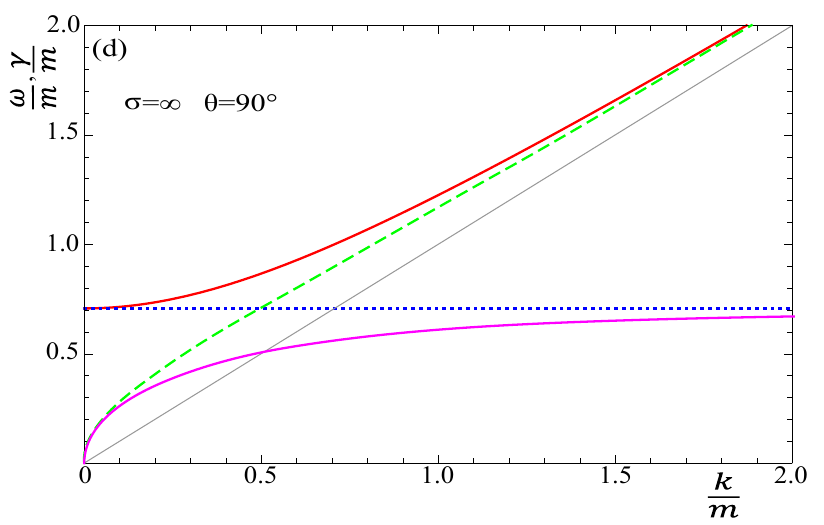}
\end{minipage}
\caption{(Color online) Dispersion curves of plasmons in an extremely prolate plasma  for $\theta = 0^\circ$ (a), $\theta = 15^\circ$ (b), $\theta = 30^\circ$ (c) and $\theta = 90^\circ$ (d). The line code is explained in the text. Figure from \cite{Carrington:2014bla}.}
\label{fig-ex-prolate-0-15-30-90}
\end{figure}

The extremely prolate system with the momentum distribution (\ref{extreme-prolate}) is the only one which can be treated analytically to the very end \cite{Carrington:2014bla}. It was solved in \cite{Arnold:2003rq} using a different method. The coefficients $a, \, b, \, c, \, d$ and the inverse propagator $\Xi$ defined by Eq.~(\ref{Xi-def}) can be computed analytically without even specifying the coordinate system.  Since the velocity ${\bf v}$ of a massless parton is ${\bf v}={\bf n}$ for ${\bf p} \cdot {\bf n} > 0$ and ${\bf v} = - {\bf n}$ for ${\bf p} \cdot {\bf n} < 0$, the matrix $\Xi$, which is purely real, is found to be
\ba
\label{Sigma-final-ex-pro}
\Xi^{ij}(\omega,{\bf k}) = 
(\omega^2 - \frac{m^2}{2} -{\bf k}^2) \delta^{ij} +k^ik^j
-\frac{m^2 {\bf k}\cdot {\bf n}}
{2\big(\omega^2 - ({\bf k}\cdot {\bf n})^2\big)}
(k^i n^j + n^i k^j)
-\frac{m^2 \big(\omega^2 + ({\bf k}\cdot {\bf n})^2\big)
({\bf k}^2 - \omega^2)}
{2\big(\omega^2 - ({\bf k}\cdot {\bf n})^2\big)^2}
n^i n^j ,
\ea
and the coefficients $a, \, b, \, c, \, d$ are
\ba
\label{a-ex-pro}
a(\omega,{\bf k}) &=& \omega^2 - {\bf k}^2 - \frac{m^2}{2}  ,
\\[2mm]
\label{b-ex-pro}
b(\omega,{\bf k}) &=& \omega^2 - \frac{m^2}{2} - \frac{m^2  ({\bf k}\cdot {\bf n})^2}
{\omega^2 - ({\bf k}\cdot {\bf n})^2}
- \frac{m^2 \big(\omega^2 + ({\bf k}\cdot {\bf n})^2\big)
({\bf k}^2 - \omega^2)}
{2\big(\omega^2 - ({\bf k}\cdot {\bf n})^2\big)^2}
\frac{({\bf k}\cdot {\bf n})^2}{{\bf k}^2} ,
\\[2mm]
\label{c-ex-pro}
c(\omega,{\bf k}) &=& 
- \frac{m^2(\omega^2+(\mathbf{k}\cdot\mathbf{n})^2)({\bf k}^2-\omega^2)}
{2(\omega^2-(\mathbf{k}\cdot\mathbf{n})^2)^2}
\left(1 - \frac{(\mathbf{k}\cdot\mathbf{n})^2}{{\bf k}^2} \right) ,
\\[2mm]
\label{d-ex-pro}
d(\omega,{\bf k}) &=& 
- \frac{m^2  ({\bf k}\cdot {\bf n})}
{2\big(\omega^2 - ({\bf k}\cdot {\bf n})^2\big)}
- \frac{m^2 \big(\omega^2 + ({\bf k}\cdot {\bf n})^2\big)
({\bf k}^2 - \omega^2)}
{2\big(\omega^2 - ({\bf k}\cdot {\bf n})^2\big)^2}
\frac{({\bf k}\cdot {\bf n})}{{\bf k}^2} .
\ea

The dispersion equation for the $A$-modes (\ref{dis-eq-A}) has the simple solution 
\be
\label{prolate-Amode}
\omega_\alpha^2({\bf k}) = \frac{m^2}{2} + {\bf k}^2 .
\ee
Although the dispersion equation for the $G$-modes (\ref{dis-eq-G}) is rather complicated, it also has three relatively simple solutions \cite{Carrington:2014bla}
\ba
\label{general-solution-1}
\omega_2^2({\bf k}) &=&  \frac{m^2}{2} + ({\bf k}\cdot {\bf n})^2 ,
\\ [2mm]
\label{general-solution-2}
\omega_{\pm}^2({\bf k}) &=& 
\frac{1}{2}\Big({\bf k}^2 + ({\bf k}\cdot {\bf n})^2
\pm
\sqrt{{\bf k}^4 + ({\bf k}\cdot {\bf n})^4 
+ 2m^2 {\bf k}^2 - 2m^2 ({\bf k}\cdot {\bf n})^2
-2 {\bf k}^2({\bf k}\cdot {\bf n})^2} \; \Big) .
\ea
The modes $\omega_\alpha$, $\omega_2$, and  $\omega_+$ are real and exist for any ${\bf k}$. The solutions $\omega_\alpha$ and $\omega_+$ always lie above the light cone. The mode $\omega_2$ lies above the light cone for $k<\frac{m}{\sqrt{2}\sin\theta}$ and below for $k>\frac{m}{\sqrt{2}\sin\theta}$. The modes $\omega_+$ and $\omega_2$ cross each other at $k=\frac{m}{2\sin\theta}$. The solution $\omega_-$ can be either purely real or purely imaginary. It is imaginary for 
\be
\label{k-crit-pro}
k < k_{\rm pG} \equiv \frac{m}{\sqrt{2}} |\tan\theta| ,
\ee
and real for $k > k_{\rm pG}$. The solution  $i\gamma$, where $\gamma \equiv |\omega_-|$,  is the Weibel unstable mode, and $-i\gamma$ is its  overdamped partner. When ${\bf k} \perp {\bf n}$ or $\theta = 90^\circ$, the unstable mode exists for all values of $k$, as $k_{\rm pG}$ given by Eq.~(\ref{k-crit-pro}) goes to infinity. When ${\bf k} || {\bf n}$ or $\theta = 0^\circ$ the configuration is cylindrically symmetric and  there is no instability, since $k_{\rm pG} \to 0$. The real modes are  $\omega_\alpha^2({\bf k}) = \omega_2^2({\bf k})=m^2/2 + k^2$ and $\omega_+^2({\bf k}) = \omega_-^2({\bf k})=k^2$ in this limit.  

Some spectra of plasmons in an extremely prolate plasma are shown in Fig.~\ref{fig-ex-prolate-0-15-30-90} for different orientations of the wave vector ${\bf k}$. We use the following color scheme: red (solid) is $\omega_\alpha$, green (dashed) is $\omega_+$, blue (dotted) is $\omega_2$, pink (solid) is $\Im\omega_-$ and purple (dotted-dashed) is $\Re \omega_-$. The imaginary mode emerges at finite $\theta$ and it extends to infinite $k$ at $\theta =90^\circ$. The mode $\omega_\alpha({\bf k})$ is independent of $\theta$, and $\omega_2({\bf k})$ changes qualitatively when $\theta$ grows from $0^\circ$ to $90^\circ$. The mode $\omega_+({\bf k})$ is massless, that is $\omega_+(0) =0$, and its dispersion curve is everywhere concave, in contrast to other real dispersion curves which are usually convex. 

There is a qualitative difference between the plasmon spectra of the extremely prolate system, which is discussed here, and that of a system with prolateness characterized by the parameter $\sigma\gg 1$. In an extremely prolate plasma, the mode $\omega_-$ given by the formula (\ref{general-solution-2}) exists for any wave vector ${\bf k}$: it is real for $k>k_{\rm pG}$ and imaginary for $k<k_{\rm pG}$. For a very large but finite $\sigma$, only the imaginary piece at $k<k_{\rm pG}$ is found. One could suspect that a solution has been missed in the numerical calculation, but the Nyquist analysis proves that this is not the case. The key point is that when $\sigma \to \infty$ there is a change in the analytic properties of the left-hand-side of the $G$-mode dispersion equation (\ref{dis-eq-G}) as a function of $\omega$. The cut singularity at $\omega \in [-k,k]$ is replaced by double poles at  $\omega = \pm {\bf k} \cdot {\bf n}$ and the number of modes in extremely prolate plasmas equals 8 for any ${\bf k}$. It also appears that the limit of extreme prolateness is approached very slowly as $\sigma \to \infty$ \cite{Carrington:2014bla}.

\subsubsection{Extremely oblate plasma}
\label{sec-ex-oblate}

In this section we consider the second limiting case - the extremely oblate plasma with the momentum distribution given by  Eq.~(\ref{extreme-oblate}). The coefficients $a, \, b, \, c,$ and $d$, which have a much more complicated structure than for an extremely prolate plasma, equal 
\ba
\label{alpha-extremely-oblate}
a(\omega,{\bf k}) &=& \omega^2 - {\bf k}^2 - \frac{m^2}{2(1-x^2)}
\Big[\hat\omega ^2-x^2+\frac{\hat\omega(1 -\hat\omega ^2)}{r_+ r_-}\Big],
\\[2mm]
\label{beta-extremely-oblate}
b(\omega,{\bf k})  &=&\omega^2 +\frac{m^2\hat\omega^2}{2}
\Big[1 - \frac{\hat\omega (2 x^2+\hat\omega ^2-1)}{r_+^3 r_-^3}\Big],
\\[2mm]
\label{gamma-extremely-oblate}
c(\omega,{\bf k})  &=& - \frac{m^2 (\hat\omega ^2-1)}{2(1-x^2)} 
\Big[\frac{\hat\omega  \big(2 x^4+(x^2+1) \hat\omega ^2-x^2-1\big)}{r_+^3r_-^3}-x^2-1\Big],
\\[2mm]
\label{delta-extremely-oblate}
k\,d(\omega,{\bf k})  &=&-\frac{m^2\hat\omega \, x }{2(1-x^2)}
\Big[\frac{-2 (x^2-1) \hat\omega ^2+x^2-\hat\omega ^4-1}{r_+^3 r_-^3} + \hat\omega\Big] ,
\ea
where $\hat\omega \equiv \omega/k$, $x \equiv \cos\theta$ and
\be
 r_+ r_- \equiv \Big(\hat\omega  + \sqrt{1-x^2} +i0^+ \Big)^{1/2} \Big(\hat\omega - \sqrt{1-x^2} +i0^+\Big)^{1/2} .
\ee
The results analogous to the formulas (\ref{alpha-extremely-oblate}) - (\ref{delta-extremely-oblate}) were first derived in \cite{Romatschke:2004jh}.

The dispersion equations  (\ref{dis-eq-A}) and (\ref{dis-eq-G}) with the coefficients (\ref{alpha-extremely-oblate}) - (\ref{delta-extremely-oblate})  cannot be solved analytically. Using a Nyquist analysis, one shows \cite{Carrington:2014bla} that the $A$-mode dispersion equation (\ref{dis-eq-A}) has a pair of real solutions for all ${\bf k}$ and a pair of imaginary solutions if the wave vector obeys
\be
\label{k-crit-obl-A}
k < k_{\rm oA}  \equiv \frac{m}{\sqrt{2}} |\cot \theta| .
\ee
The $G$-mode dispersion equation (\ref{dis-eq-G}) has two pairs of real solutions for all ${\bf k}$ and a pair of imaginary solutions when the wave vector satisfies the condition
\be
\label{k-crit-obl-G}
k < k_{\rm oG}  \equiv \frac{m}{2}\Re \sqrt{\frac{|\cos\theta|\sqrt{\cos^2\theta + 4}+\cos^2\theta-2}{\sin^2\theta}} .
\ee
When $\cos^2\theta < 1/2$ (that is $90^\circ > \theta > 45^\circ$), the argument of the square root is negative, the real part of the root is zero, and the critical wave vector $k_{\rm oG}$ vanishes. One observes that $k_{\rm oA}$ is obtained from $k_{\rm pG}$ by changing the tangent function into a cotangent. As explained in \cite{Carrington:2014bla}, the critical values (\ref{k-crit-obl-A}) and (\ref{k-crit-obl-G}) are the values of $k$ for which the inverse propagators $\Delta_A^{-1}$ and $\Delta_G^{-1}$, given by Eqs.~(\ref{dis-eq-A}) and (\ref{dis-eq-G}), vanish at $\omega=0$.  

The total number of modes is 6, 8, or 10 exactly as in the weakly oblate case (\ref{weak-fin-A}), (\ref{weak-fin-B}), and (\ref{weak-fin-C}). The numbers  can be written in a compact form as
\ba
\label{ob-fin-A}
A-{\rm modes}:~~ 2+ 2\Theta(k_{\text{oA}}-k),
\\[2mm]
\label{ob-fin-G}
G-{\rm modes}:~~ 4+ 2\Theta(k_{\text{oG}}-k).
\ea

\begin{figure}[t]
\begin{minipage}{8.5cm}
\center
\includegraphics[width=1.05\textwidth]{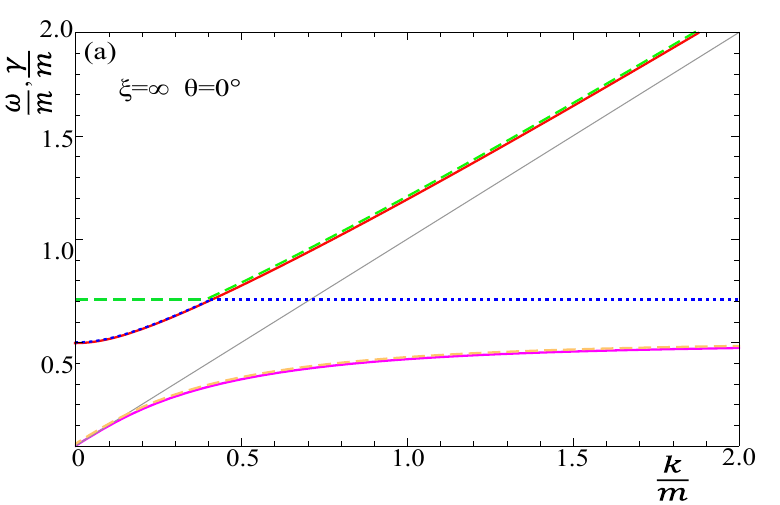}
\end{minipage}
\hspace{2mm}
\begin{minipage}{8.5cm}
\center
\includegraphics[width=1.02\textwidth]{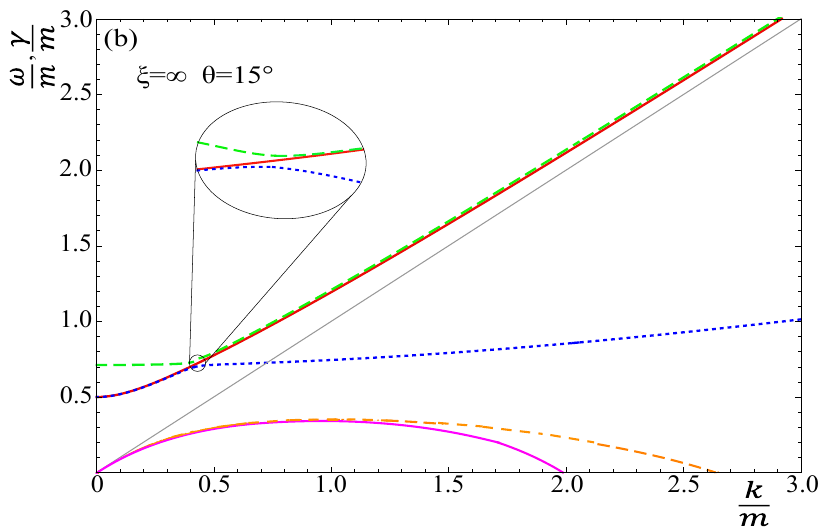} 
\end{minipage}
\vspace{2mm}
\begin{minipage}{8.5cm}
\center
\includegraphics[width=1.02\textwidth]{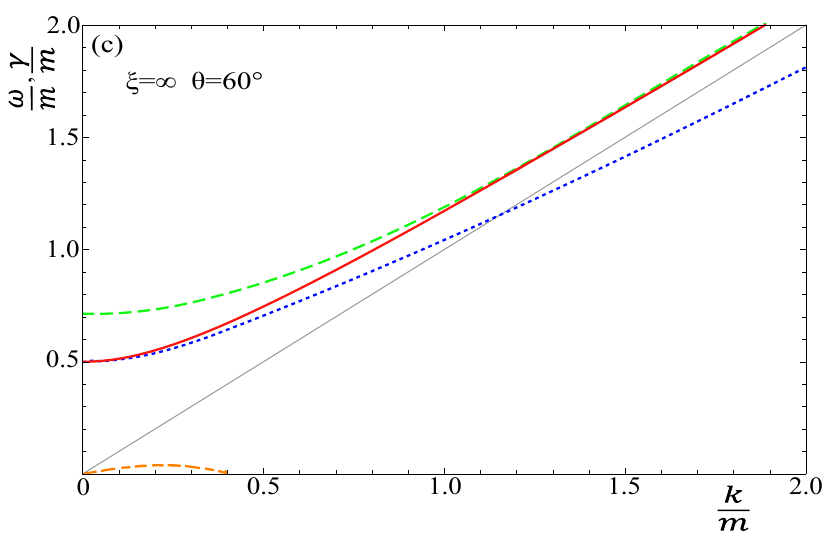}
\end{minipage}
\hspace{2mm}
\begin{minipage}{8.5cm}
\center
\includegraphics[width=1.02\textwidth]{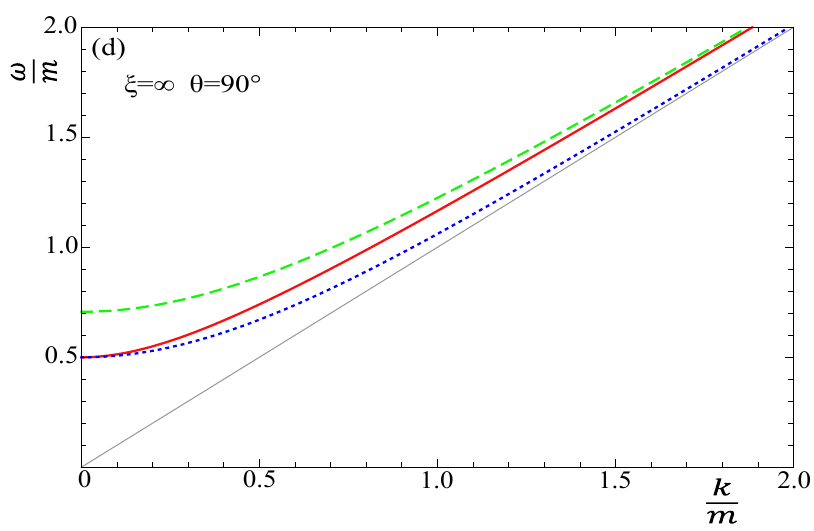}
\end{minipage}
\caption{(Color online) Dispersion curves of plasmons in an extremely oblate plasma  for $\theta = 0^\circ$ (a), $\theta = 15^\circ$ (b), $\theta = 60^\circ$ (c) and $\theta = 90^\circ$ (d). The line code is explained in the text. Figure from \cite{Carrington:2014bla}.}
\label{fig-extreme-oblate}
\end{figure}

In Fig.~\ref{fig-extreme-oblate} we show the dispersion curves obtained numerically from Eqs.~(\ref{dis-eq-A}) and (\ref{dis-eq-G}) for the angle $\theta$ equal $0^\circ,\; 15^\circ,\; 60^\circ$, and $90^\circ$.  When $\theta=0^\circ$ the real solutions $\omega_-$ and $\omega_+$ exhibit sharp kinks at the same value of $k$. The $\omega_\alpha$ solution lies on top of the $\omega_-$ solution at small $k$ and on top of $\omega_+$ at large $k$. The two imaginary solutions extend through all values of $k$ and lie on top of each other, which is consistent with the observation that $k_{\rm oA}$ and $k_{\rm oG}$ both go to infinity at $\theta=0$. At $\theta=15^\circ$ we see that increasing the angle softens the kinks in the real modes and causes the imaginary modes to retreat. The inset shows a blow-up of the region where the real modes approach each other. When $\theta$ has increased to $60^\circ$, the imaginary $G$-mode has dropped out, and at $90^\circ$ both imaginary modes are gone. 

The structure of the plasmon spectrum in Fig.~\ref{fig-extreme-oblate} is rather complicated. To understand it better, we consider following \cite{Carrington:2014bla} three special limits which can be treated analytically. We start with ${\bf k}\parallel{\bf n}$ ($\theta=0^\circ$), then we discuss the situation when ${\bf k}$ is almost parallel to ${\bf n}$ ($|\sin\theta|\ll 1$), and finally we analyze the limit ${\bf k}\perp{\bf n}$ ($\theta=90^\circ$).

\subthreesection{Special case: ${\bf k} || {\bf n}$ }
\label{xequalspm1}

We first consider ${\bf k}=(0,0,k)$ parallel to ${\bf n}=(0,0,1)$. In this case, the vector ${\bf n}_T$, which is defined by Eq.~(\ref{nT-def}), vanishes. The decomposition using the basis $A,B,C,D$, which is introduced in Sec.~\ref{sec-dis-eqs-deformed}, is therefore singular.  However, since there is only one independent vector in this case, one can decompose the inverse propagator or dielectric tensor using the same basis as in the isotropic case. One easily finds the coefficients $a$ and $b$, which equal 
\ba
\label{sf-parallel}
a(\omega,{\bf k}) &=&  \omega^2 - {\bf k}^2 - \frac{m^2}{2} + \frac{m^2( \omega^2 - k^2) }{4 \omega^2},
\\[2mm]
\label{sf-parallel-beta}
b(\omega,{\bf k}) &=& \omega^2 - \frac{m^2}{2} ,
\ea
and the matrix $\Xi (\omega,{\bf k})$ is
\ba
\label{sigma-k-||-n}
\Xi(\omega, {\bf k}) = \left[
\begin{array}{ccc}
a (\omega, {\bf k})  & 0 &0
\\[2mm] 
0 & 
a (\omega, {\bf k}) & 0
\\[2mm]
0 & 0 & b(\omega,{\bf k}) 
\end{array}
\right] .
\ea
As seen, there are two double solutions of the dispersion equation $a(\omega,{\bf k})=0$ of transverse modes and the double solution of the equation $b(\omega,{\bf k}) =0$ of longitudinal ones. Using Eqs.~(\ref{sf-parallel}) and (\ref{sf-parallel-beta}), one easily finds the dispersion relations which are  
\ba
\label{omega-pm-ex-oblate}
\omega^2_{\alpha} (k) &=& \frac{1}{2} \bigg( \frac{1}{4} \, m^2 + k^2 + \sqrt{ \Big( \frac{1}{4}\,m^2 + k^2 \Big)^2 + m^2 k^2} \; \bigg) 
\approx 
\left\{ \begin{array}{ccc} 
\frac{1}{4}\, m^2 +  2 k^2 \;\;\;\;\;
 & {\rm for} &  \;\;\;\;\; m^2 \gg k^2 ,
\\ [2mm]
k^2 
 & {\rm for} & \;\;\;\;\;  m^2 \ll k^2 ,
\end{array} \right.
\\[2mm]
\label{omega-pm-ex-oblate2}
\omega^2_{\alpha i} (k) &=& \frac{1}{2} \bigg( \frac{1}{4} \, m^2 + k^2 - \sqrt{ \Big( \frac{1}{4}\,m^2 + k^2 \Big)^2 + m^2 k^2} \; \bigg) 
\approx 
\left\{ \begin{array}{ccc} 
-k^2 \;\;\;\;\;
 & {\rm for} &  \;\;\;\;\; m^2 \gg k^2 ,
\\ [2mm]
- \frac{1}{4}\, m^2 
 & {\rm for} & \;\;\;\;\;  m^2 \ll k^2 ,
\end{array} \right.
\\[2mm]
\label{omega-pm-ex-oblate3}
\omega^2_\beta (k) &=& \frac{1}{2}\,m^2 .
\ea
Both $\omega_\alpha$ and $\omega_\beta$ are real solutions which exist for all $k$, and $\omega_{\alpha i} = i\gamma $ is an imaginary solution which also exists for all $k$. The maximum of the imaginary frequency is $\gamma_{\rm max} =  m/2$. 

The solutions $\omega_\alpha^2$ and $\omega_\beta^2$ cross each other  at 
\be
\label{kc-ex-oblate}
k^2 = k_c^2 = \frac{m^2}{6}. 
\ee
Let us define two combinations of the real solutions:
\ba
\label{omega-2-zero-eps}
\omega_-^2 (k)  
=
\left\{ \begin{array}{ccc} 
\omega_\alpha^2 (k)   \;\;\;\;\; & {\rm for} &  \;\;\;\;\; k < k_c  ,
\\ [2mm]
\omega_\beta^2 (k)   \;\;\;\;\; & {\rm for} &  \;\;\;\;\; k > k_c  ,
\end{array} \right.
\ea
\ba
\label{omega-3-zero-eps}
\omega_+^2 (k)  
=
\left\{ \begin{array}{ccc} 
\omega_\beta^2 (k)   \;\;\;\;\; & {\rm for} &  \;\;\;\;\; k < k_c  ,
\\ [2mm]
\omega_\alpha^2 (k)   \;\;\;\;\; & {\rm for} &  \;\;\;\;\; k > k_c  .
\end{array} \right.
\ea
The dispersion curves are shown in Fig.~\ref{fig-extreme-oblate}. The modes denoted $\omega_-$ and $\omega_+$ are represented, respectively, by the blue (dotted) and green (dashed) lines. As will be explained in the next subsection, the modes $\omega_+$ and $\omega_-$ are physical in the sense that one can obtain them by taking the limit $\theta \to 0^\circ$ of the solutions with the same names which were found at  $\theta > 0^\circ$.

\subthreesection{Special case: ${\bf k}$ almost  parallel to $ {\bf n}$ }

When the wave vector is not exactly along the $z$-axis but is slightly tilted, the spectrum of collective modes is changed qualitatively. To discuss this case we assume that the wave vector has a small $x$ component $k_x = k \sin \theta \approx k \theta$. The matrix $\Xi$, which for $\theta = 0^\circ$ is given by Eq.~(\ref{sigma-k-||-n}), now contains small off-diagonal components $\sim k^2 \theta$ and is given by
\ba
\label{sigma-k-alomost-||-n}
\Xi (\omega, {\bf k}) = \left[
\begin{array}{ccc}
-k^2 + \omega^2 - \frac{m^2}{2} + \frac{m^2( \omega^2 - k^2) }{4 \omega^2} & 0 & k^2 \theta 
\\[2mm]
0 & -k^2 + \omega^2 - \frac{m^2}{2} + \frac{m^2( \omega^2 - k^2) }{4 \omega^2} & 0
\\[2mm] 
 k^2 \theta & 0 & 
\omega^2 - \frac{m^2}{2}
\end{array}
\right] .
\ea

Computing the determinant of $\Xi$, one finds two dispersion equations. The first reproduces the $\alpha$ modes in Eq.~(\ref{omega-pm-ex-oblate}), and the solutions are doubled as was the case for ${\bf k}$ parallel to ${\bf n}$. The second dispersion equation can be written as 
\be
\label{dis-eq-kT-small-3}
\frac{1}{\omega^2} \big(\omega^2 - \omega^2_\alpha (k) \big)\big(\omega^2 - \omega^2_{\alpha i} (k) \big)
\big(\omega^2 - \omega^2_\beta (k) \big) = k^4 \theta^2.
\ee
When $\theta=0^\circ$ we clearly recover the solutions of the previous section. Since the mode $\omega_{\alpha i}^2$ crosses neither $\omega_\alpha^2$ nor $\omega_\beta^2$, we express it as $\omega_{\alpha i}^2 = - \gamma^2$ and rewrite Eq.~(\ref{dis-eq-kT-small-3}) in the form
\be
\label{dis-eq-kT-small-4}
\big(\omega^2 - \omega^2_{\alpha} (k) \big)\big(\omega^2 - \omega^2_\beta (k) \big) = \epsilon ,
\ee
where $\epsilon \equiv \frac{\omega^2 k^4 \theta^2}{\omega^2 + \gamma^2}$. We want to look at the modes $\omega_\alpha$ and $\omega_\beta$ in the vicinity of the point where they cross. To lowest order in deviations from the solutions with $\theta=0^\circ$, we take $\epsilon$ as constant and solve the quadratic equation to obtain
\ba
\label{omega-2}
\omega^2_-  &=& \frac{1}{2} \Big(\omega^2_\alpha + \omega^2_\beta 
- \sqrt{\big(\omega^2_\alpha - \omega^2_\beta\big)^2 + 4\epsilon } \; \Big) ,
\\[2mm]
\label{omega-3}
\omega^2_+  &=& \frac{1}{2} \Big(\omega^2_\alpha + \omega^2_\beta
+ \sqrt{\big(\omega^2_\alpha - \omega^2_\beta\big)^2 + 4\epsilon } \; \Big) .
\ea
From these expressions, it is clear that the small parameter $\epsilon$ plays a role only in the vicinity of the crossing point where $\omega_\alpha = \omega_\beta$. Since both $\omega_\alpha^2$ and $\omega_\beta^2$ are positive, we have $\epsilon \ge 0$. Assuming that $(\omega^2_\alpha - \omega^2_\beta\big)^2 \gg \epsilon$, we expand the square roots in the formulas (\ref{omega-2}) and (\ref{omega-3}) to obtain
\ba
\omega_-^2 (k)  
=
\left\{ \begin{array}{ccc} 
\omega_\alpha^2 (k) - \frac{\epsilon}{| \omega^2_\alpha - \omega^2_\beta |}  \;\;\;\;\; & {\rm for} &  \;\;\;\;\; k < k_c  ,
\\ [2mm]
\omega_\beta^2 (k)  - \frac{\epsilon}{| \omega^2_\alpha - \omega^2_\beta |}  \;\;\;\;\; & {\rm for} &  \;\;\;\;\; k > k_c  ,
\end{array} \right.
\ea
\ba
\omega_+^2 (k)  
=
\left\{ \begin{array}{ccc} 
\omega_\beta^2 (k) + \frac{\epsilon}{| \omega^2_\alpha - \omega^2_\beta |}  \;\;\;\;\; & {\rm for} &  \;\;\;\;\; k < k_c  ,
\\ [2mm]
\omega_\alpha^2 (k)  + \frac{\epsilon}{| \omega^2_\alpha - \omega^2_\beta |}  \;\;\;\;\; & {\rm for} &  \;\;\;\;\; k > k_c  .
\end{array} \right. 
\ea
This result shows that the modes $\omega_-^2$ and $\omega_+^2$ approach each other at $k=k_c$ but do not cross. This is referred to as {\it mode coupling}, which is a general phenomenon that is explained in \S 64  of  \cite{Landau-Lifshitz-1981}. One can also show that the double imaginary mode $\omega_{\alpha i}$ splits into two different modes when $\theta$ is finite. 

The complete spectrum is presented in Fig.~\ref{fig-extreme-oblate}b for $\theta=15^\circ$.  As shown in the inset, the $\omega_+$ and $\omega_-$ modes  approach each other at $k = k_c$ but do not cross. The number of modes is the same as for the extremely oblate distribution with arbitrary values of $\theta$. 

\subthreesection{Special case: ${\bf k} \perp {\bf n}$ }

When ${\bf k} \perp {\bf n}$, the system can be treated as effectively two-dimensional and {\em isotropic} in the $x\!-\!y$ plane. From Eqs. (\ref{k-crit-obl-A}) and (\ref{k-crit-obl-G}) we see that both of the critical wave vectors $k_{\text{oA}}$ and $k_{\text{oG}}$ go to zero in the limit $\theta \to 90^\circ$ and therefore the two imaginary modes disappear, as expected for an isotropic system. There should be two real solutions (one pair) from the $A$-mode dispersion equation (\ref{dis-eq-A}) and four real solutions (two pairs) from the $G$-mode equation (\ref{dis-eq-G}).

For ${\bf k} \perp {\bf n}$, the coefficients (\ref{alpha-extremely-oblate}) - (\ref{delta-extremely-oblate}) simplify to
\ba
\label{a-kL0}
a(\omega, {\bf k}) &=& \omega^2 - {\bf k}^2 
- \frac{m^2}{2} \frac{\omega^2}{k^2} \bigg( 1 -  \frac{\sqrt{\omega^2 - k^2}}{\omega} \bigg),
\\ [4mm]
\label{b-kL0}
b(\omega, {\bf k}) &=& \omega^2 
- \frac{m^2}{2} \frac{\omega^2}{k^2} \bigg(\frac{\omega}{\sqrt{\omega^2 - k^2}} - 1 \bigg) ,
\\ [4mm]
\label{c-kL0}
c(\omega, {\bf k}) &=&
 - \frac{m^2}{2} \frac{\omega^2 - k^2}{ k^2}
\bigg(\frac{\omega}{\sqrt{\omega^2 - k^2}} -1 \bigg) ,
\\ [4mm]
\label{d-kL0}
d(\omega, {\bf k}) &=& 0 ,
\ea
where $\omega \in \mathbb{R}$ and $\omega^2 > k^2$. 

Since $d(\omega, {\bf k}) = 0$, the dispersion equation (\ref{dis-eq-G}) factors into two equations, as in the case of the weakly anisotropic plasma, and we solve the dispersion equations for $A$-modes, $B$-modes, and $C$-modes (\ref{dis-eq-A}), (\ref{dis-eq-B}), and (\ref{dis-eq-C}). The $A$-mode dispersion equation (\ref{dis-eq-A}) has the form
\be
(\omega^2 - k^2)k^2 +  \frac{m^2}{2} \Big(  \omega  \sqrt{\omega^2 - k^2}  - \omega^2 \Big) = 0 ,
\ee
which is quadratic in $\omega^2$ and can be solved analytically. The solution is 
\ba
\omega_\alpha^2 (k) = \frac{m^4 + 4 m^2k^2 -8 k^4 + m^3 \sqrt{ m^2 + 8 k^2}}{8(m^2-k^2)}  
\approx 
\left\{ \begin{array}{ccc} 
\frac{1}{4}\, m^2 +  \frac{5}{4}\, k^2 \;\;\;\;\;
 & {\rm for} &  \;\;\;\;\; m^2 \gg k^2 ,
\\ [2mm]
k^2 
 & {\rm for} & \;\;\;\;\;  m^2 \ll k^2 .
\end{array} \right.
\ea
The $B$-mode dispersion equation (\ref{dis-eq-B}) simplifies to
\be
k^2 + \frac{m^2}{2} \bigg(1 -  \frac{\omega}{\sqrt{\omega^2 - k^2}} \bigg) = 0 ,
\ee
and the solution gives the longitudinal mode
\ba
\omega_\beta^2 (k) = \frac{\big(\frac{m^2}{2} + k^2 \big)^2}{m^2 + k^2} \approx 
\left\{ \begin{array}{ccc} 
\frac{1}{4}\, m^2 +  \frac{3}{4}\, k^2 \;\;\;\;\;
 & {\rm for} &  \;\;\;\;\; m^2 \gg k^2 ,
\\ [2mm]
k^2 
 & {\rm for} & \;\;\;\;\;  m^2 \ll k^2 .
\end{array} \right.
\ea
Finally, the $C$-mode dispersion equation (\ref{dis-eq-C}) becomes 
\be
\omega^2 - k^2 - \frac{m^2}{2} =  0 ,
\ee
which produces the solution
\be
\omega_{\alpha\gamma}^2 (k) = \frac{1}{2}\, m^2 + k^2 .
\ee
The $B$-mode and $C$-mode solutions are the limits $\theta \to 90^\circ$ of those found for arbitrary angles by numerically solving the $G$-mode dispersion equation (\ref{dis-eq-G}). The solution $\omega_{\alpha\gamma}$ is the larger of the two real $G$-modes (which we call $\omega_+$) and $\omega_\beta$ is the smaller $G$-mode (called $\omega_-$) which stays above the light cone for all $k$ when $\theta=90^\circ$. The dispersion curves for ${\bf k} \perp {\bf n}$ are shown in Fig.~\ref{fig-extreme-oblate}d. 

We finally mention that, in contrast to the case of the $\sigma-$distribution, the spectrum of the extremely oblate system coincides with that of large but finite $\xi$. 
\begin{table}[b]
\caption{\label{table-modes} Number of modes }
\begin{ruledtabular}
\label{tab-modes}
\begin{tabular}{ccccc}
 Momentum &  Number  & Number   & Total number & Maximal number
\\
distribution &  of real modes & of imaginary modes & of modes & of modes 
\\ \hline \hline 
extremely prolate & $6 + 2\Theta(k-k_{\rm p})$ & $2\Theta(k_{\rm p}-k)$  &  8 &  8
\\[2mm]
weakly prolate   & 6 & $2\Theta(k_{\rm C}-k)$ & $6 + 2\Theta(k_{\rm C}-k)$ & 8
\\[2mm]
isotropic & 6 & 0 & 6 & 6
\\[2mm]
weakly oblate    & 6 & $2\Theta(k_{\rm A}-k) + 2\Theta(k_{\rm C}-k)$  & $6 + 2\Theta(k_{\rm A}-k) + 2\Theta(k_{\rm C}-k)$ & 10
\\[2mm]
extremely oblate  & 6 & $2\Theta(k_{\rm oA}-k) + 2\Theta(k_{\rm oG}-k)$  & $6 + 2\Theta(k_{\rm oA}-k) + 2\Theta(k_{\rm oG}-k)$ & 10
\end{tabular}
\end{ruledtabular}
\end{table}

\begin{figure}[t]
\center
\includegraphics[width=13cm]{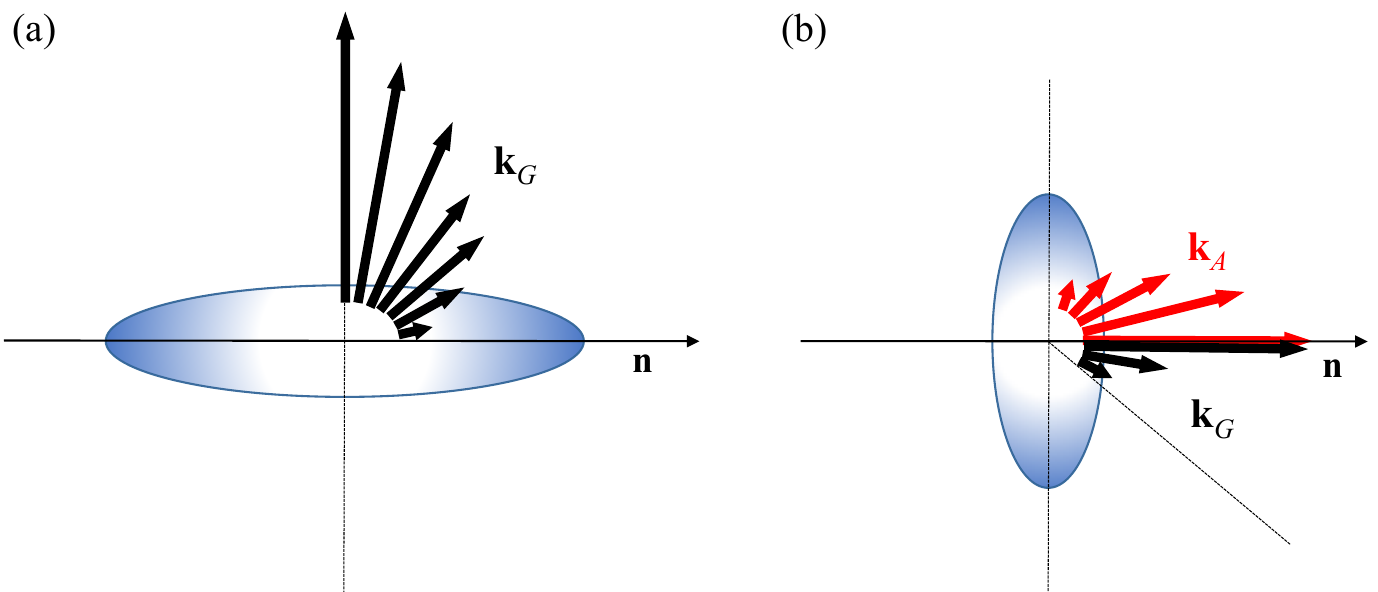}
\caption{(Color online) The largest possible wave vectors of unstable modes at different orientations. In the prolate plasma (a) there is one unstable $G$-mode, which exists for $0^\circ <\theta \le 90^\circ$, and in oblate plasma (b) there are two unstable $A$- and $G$-modes which exist for $0^\circ \le \theta < 90^\circ$ and $0^\circ \le \theta < 45^\circ$, respectively. Figure from \cite{Carrington:2014bla}.}
\label{fig-configurations}
\end{figure}

\subsubsection{Discussion}

We have considered in this section distributions with all degrees of deformation along the beam axis, from the extremely prolate distribution, which is infinitely elongated along the beam, to the extremely oblate distribution, which is infinitely squeezed along the beam axis. Solving the dispersion equations analytically or numerically, we have found the full spectrum of plasmons. All modes are either purely real or purely imaginary, and they always appear as pairs of partners with opposite sign. In all systems under consideration, except the isotropic plasma, there are unstable modes (positive purely imaginary solutions). Imaginary solutions exist only for certain wave vectors. The number of modes for each system, which was determined by means of a Nyquist analysis, is  summarized in Table~\ref{tab-modes}.

The orientation of the largest possible wave vectors of unstable modes is schematically shown in Fig.~\ref{fig-configurations}. Since the $A$-modes are transverse, the chromoelectric field of the maximally unstable $A$-mode in oblate plasmas is perpendicular to ${\bf n}$. The $G$-modes are, in general, neither transverse nor longitudinal. However, the maximally unstable $G$-modes (with ${\bf k} || {\bf n}$ in oblate plasma and ${\bf k} \perp {\bf n}$ in prolate plasma) are transverse. Therefore, for $G$-modes in prolate systems, and both $A$-modes and $G$-modes in oblate plasmas, the chromoelectric field of the maximally unstable modes is aligned with the direction where the momentum of plasma particles is maximal. This observation is important because the early-time dynamics of unstable systems are dominated by the fastest growing unstable modes.


\section{Numerical simulations}
\label{sec-simulate}

A drop of quark-gluon plasma is produced in high-energy nuclear collisions. The system is initially out of equilibrium but it equilibrates very fast, as suggested by the tremendous success of the hydrodynamical description of relativistic heavy-ion collisions, see the reviews \cite{Voloshin:2008dg,Heinz:2013th,Gale:2013da}. A natural question is how the process of equilibration proceeds. What are the microscopic mechanisms responsible for it? The problem of the evolution of a system of relativistic quantum fields governed by non-Abelian interactions is certainly very complex. To study such non-linear dynamics, numerical simulations are not only necessary but they seem to be the only theoretical tool to attack the problem in its whole complexity.  

Various approaches with varying approximations can be found in the literature, and we will discuss in detail the main frameworks. These are the hard-loop approximation, in which the hard degrees of freedom are assumed to be unaffected by the soft field modes and the two classical approaches. We will discuss hard loop simulations in the more academic scenario of a static box in various spatial dimensions (Sec.~\ref{subsec-hardloop}) and in the scenario more closely related to heavy ion collisions, where the system is expanding in one dimension (Sec.~\ref{subseq-expanding-HL}).

We begin the discussion of hard loop simulation results by presenting 1+1 dimensional calculations (one time and one spatial dimension), because those were the first ones to be done. We then continue with the discussion of fully 3+1 dimensional simulations, which show qualitative differences to the 1+1 dimensional case in the late stage behavior of instability growth, where field amplitudes are large enough for non-linear effects to be important. In order to understand these differences we review the intermediate case of 2+1 dimensions.

The hard-loop approach relies on a large separation of momentum scales between the soft and hard modes. This requires a sufficiently small gauge coupling $g$ and thus limits applicability of the results to the weakly coupled regime. To go beyond this regime, nonperturbative studies can be performed in the framework of fully classical simulations. In the first classical approach we discuss, the system's dynamics is fully represented by the classical Yang-Mills equations (Sec.~\ref{subsec-purefield-class}), and in the second one uses the Yang-Mills-Wong framework (Sec.~\ref{Wong-Yang-Mills}) where soft modes are still described as classical fields, but hard degrees of freedom are included as particles, propagating in the background of the fields.

Our discussion of the purely classical simulations begins with a review of the classical statistical framework and the study of a simple scenario with anisotropic initial conditions. We then continue with the introduction of the more realistic color glass condensate framework, which provides initial conditions for the classical background fields and can be augmented by superimposed rapidity fluctuations that will trigger instabilities. We close that section with a review of recent improvements in particular in the treatment of initial fluctuations.

Finally, we discuss the Yang-Mills-Wong picture in which hard momentum modes are treated as particles propagating in the background of soft field modes. The advantage of this formulation is the possibility to study directly the interaction of color currents and fields to {\it e.g.} analyze the momentum diffusion of hard particles in an unstable plasma.



\subsection{Hard-Loop simulations}
\label{subsec-hardloop}

One possibility to follow the dynamical evolution of a non-Abelian system with an anisotropic momentum distribution is the hard-loop approach defined by the effective action discussed in detail in Sec.~\ref{subsec-actions}. The approach was employed for both non-expanding \cite{Rebhan:2004ur,Arnold:2005vb,Rebhan:2005re,Arnold:2005ef} and expanding \cite{Romatschke:2006wg,Rebhan:2008uj,Attems:2012js} systems. We explain in this section how numerical hard-loop simulations are performed for systems of fixed volume and we present the most important results obtained by the different groups. For the presentation of the concept of discretization of the hard-loop dynamics, we follow \cite{Rebhan:2004ur,Rebhan:2005re} and briefly explain the differences to the work \cite{Arnold:2005vb} along the way. The simulations of expanding systems are discussed in Sec.~\ref{subseq-expanding-HL}.

The quark-gluon plasma under consideration is initially spatially homogeneous and colorless, but the momentum distribution of plasma constituents is, in principle, arbitrary. Homogeneity and color neutrality can be broken by randomly fluctuating charges, currents and chromodynamic fields. When the plasma is in equilibrium, the fluctuations, which remain small, constitute a stationary noise. One asks what happens when the momentum distribution of plasma constituents is anisotropic and a spontaneous generation of exponentially growing modes becomes possible. This is the question to be addressed by hard-loop simulations.

The quark, antiquark, and gluon distribution functions are assumed to be of the form 
\begin{subequations}
\begin{align}
\label{f-dis-q-aq-g}
f^{q}(X,\mathbf{p}) = f^{q}(\mathbf{p}) +  \delta f^{q}(X,\mathbf{p}),
\\[1mm]
f^{\bar{q}}(X,\mathbf{p}) = f^{\bar{q}}(\mathbf{p}) +  \delta f^{\bar{q}}(X,\mathbf{p}),
\\[1mm]
f^{g}(X,\mathbf{p}) = f^{g}(\mathbf{p}) +  \delta f^{g}(X,\mathbf{p}),
\end{align}
\end{subequations}
where the deviations $\delta f^{q}(X,\mathbf{p}), \; \delta f^{\bar{q}}(X,\mathbf{p}), \; \delta f^{g}(X,\mathbf{p})$ from the initial homogeneous state defined by the functions $f^{q}(\mathbf{p}), \; f^{\bar{q}}(\mathbf{p}),$ and $f^{g}(\mathbf{p})$ are small throughout the whole system's evolution. The capital $X$ denotes here the four position, that is $X=(t,x,y,z)$. The deviations are responsible for non-vanishing color charge densities and currents which in turn generate  chromodynamic fields. The hard-loop framework describes the regime where the backreaction of soft collective fields on the hard modes - particles - is still weak but where the self-interaction of the former may already be strongly nonlinear. 

Since constituents of the QGP are approximately massless,  and consequently move with the speed of light, the momentum dependence of the distribution functions can be factored out into the dependence on momentum magnitude  $|\textbf{p}|$ and the dependence on orientation of the velocity  $\textbf{v} \equiv \textbf{p}/|\textbf{p}|$. The $|\textbf{p}|$-independent part of the quark, antiquark, and gluon distribution functions is described by a single auxiliary field $W^{\mu}(X,\mathbf{v})$, defined by the relations
\begin{subequations}
\begin{align}\label{Wdefs}
  \delta f^{q}(X,\mathbf{p})=g \frac{\partial f^{q}(\mathbf{p})}
  {\partial p^\mu}W^\mu(X,\mathbf{v}),
\\
  \delta f^{\bar{q}}(X,\mathbf{p})=-g \frac{\partial f^{\bar{q}}(\mathbf{p})}
  {\partial p^\mu}W^\mu(X,\mathbf{v}),
\\
  \delta f^{g}(X,\mathbf{p})=g \frac{\partial f^{g}_a(\mathbf{p})}
  {\partial p^\mu} T^a \, \text{Tr} \big[ t^a W^\mu(X,\mathbf{v})\big].
\end{align}
\end{subequations}
Note that the derivatives with respect to $p^0$ vanish identically, because the distribution functions are independent of $p^0$. The $W$'s satisfy
\begin{equation}
    p_\mu D^\mu W^\nu(X,\mathbf{v})=p_\rho F^{\nu\rho}(X),
\label{Weq}
\end{equation}
and the Yang-Mills equation reads
\begin{equation}
    D_\mu F^{\mu\nu}(X)=J^\nu(X)=-g^2\int \frac{d^3p}{(2\pi)^3}\frac{p^\nu}{2 |\mathbf{p}|}
    \frac{\partial f(\mathbf{p})}{\partial p^\rho}
    W^\rho(X,\mathbf{v}) .
\label{WfieldYM}
\end{equation}
The hard-loop action expressed through the auxiliary fields $W$ is given by Eq.~(\ref{Saniso-glue}). 

In the following, we consider the special case of cylindrically symmetric anisotropies in momentum space, so that there is only one direction of anisotropy, which in heavy-ion collisions would be given by the collision axis. We can then parameterize $f(\mathbf p)= \tilde f(|\mathbf p|,p^z)$ and write
\begin{eqnarray}
{\partial f(\mathbf p) \over \partial p^\rho}
&=&
{\partial \tilde f(|\mathbf p|,p^z) \over \partial |\mathbf p|}
{p^i \delta_{i\rho}\over|\mathbf p|}
+{\6 \tilde f(|\mathbf p|,p^z) \over \6 p^z}\delta_{z\rho}\nonumber\\
&\equiv&
-\tilde f_1(|\mathbf p|,p^z)
{p^i \delta_{i\rho}\over|\mathbf p|}
-\tilde f_2(|\mathbf p|,p^z)\delta_{z\rho}.
\end{eqnarray}

From Eq.~(\ref{Weq}) and $p_\rho p_\nu F^{\rho\nu} =0$ one finds that $p^\mu W_\mu=0$, which leads to
\begin{equation}
\label{Jind2}
J^\mu(X)={1\over2}g^2
\int {d^3p\over(2\pi)^3} 
\frac{p^\mu}{|\mathbf{p}|}\, \left(\tilde f_1(|\mathbf p|,p^z) W^0(X,\mathbf{v})+\tilde f_2(|\mathbf p|,p^z)W^z(X,\mathbf{v})\right).
\end{equation}
In the isotropic case, $\tilde{f_2}=0$, and only $W^0$ appears, whose equation of motion (\ref{Weq}) involves only electric fields. In the anisotropic case, however, $W^z$ enters, whose equation of motion contains the $z$ component of the Lorentz force.

The structure of Eqs.~(\ref{Weq}) and (\ref{WfieldYM}) is such that only $W^0$ and $W^z$ participate nontrivially in the dynamical evolution. A closed set of gauge-covariant equations is thus obtained when the hard-loop integral in Eq.~(\ref{Jind2}) is discretized with respect to the directions $\mathbf v$
\begin{eqnarray}
\label{Jindd}
J^\mu(X)&=&g^2
\int {p^2 dp\over(2\pi)^2} {1\over\mathcal N_{W}} \sum_{\bf v} 
\frac{p^\mu}{|\mathbf{p}|}\, \left(\tilde f_1(|\mathbf p|,p^z) W^0_{\mathbf v}(X)+
\tilde f_2(|\mathbf p|,p^z) W_{\mathbf v}^z(X)\right)
\nonumber\\
&\equiv& {1\over \mathcal N_{W}} \sum_{\bf v} v^\mu \left(a_{\mathbf v} W^0_{\mathbf v}(X)+
b_{\mathbf v} W^z_{\mathbf v}(X)\right),
\end{eqnarray}
where the set of unit vectors $\mathbf v$ define a partition of the unit sphere in $\mathcal N_{W}$ patches of equal area and  $v^\mu \equiv p^\mu /|\mathbf{p}| = (1, {\bf v})$. Although the quantity $v^\mu$ is written as a four-vector it does not transform under Lorentz transformations as a four-vector. It has been introduced here to stress the dependence of the color current (\ref{Jindd}) on the velocity ${\bf v}$. The coefficients $a_{\mathbf v}$, $b_{\mathbf v}$ are constants defined by
\begin{eqnarray}
\label{abv}
a_{\mathbf v}&=&-g^2\int_0^\infty {p^2 dp\over(2\pi)^2} 
f_1(|\mathbf p|,|\mathbf p|v^z), 
\\
b_{\mathbf v}&=&-g^2\int_0^\infty {p^2 dp\over(2\pi)^2} 
f_2(|\mathbf p|,|\mathbf p|v^z). \end{eqnarray}
Isotropic distribution functions $f$ are characterized by $a_{\mathbf v}$'s which are independent of $\mathbf v$ and vanishing $b_{\mathbf v}$'s.

The special choice for an anisotropic distribution function given by Eq.~(\ref{R-S-ansatz}), 
\begin{equation}
\label{faniso-hl}
\tilde f(|\mathbf p|,p^z)=C_\xi f_{\rm iso}(\sqrt{\mathbf p^2+\xi p_z^2}),
\end{equation}
provides
\begin{equation}
\label{avbvxi}
a_{\mathbf v}={C_\xi m_D^2\over (1+\xi v_z^2)^2},\qquad
b_{\mathbf v}=\xi v_z a_{\mathbf v},
\end{equation}
where $m_D$ is the Debye mass of the isotropic distribution from Eq.~(\ref{faniso-hl}), that is
\be
\label{Debye-mass-HL}
m_D^2 \equiv g^2 \int \frac{d^3p}{(2\pi)^3} \frac{f_{\rm iso}(|\mathbf{p}|)}{|\mathbf{p}|} .
\ee
In the following we shall often absorb the normalization constant into the Debye mass and use the abbreviation 
\begin{equation}
\label{m2norm}
m^2\equiv C_\xi m_D^2 ,
\end{equation}
which equals the mass parameter defined by Eq.~(\ref{mass2}), if the normalization constant $C_\xi$ is chosen according to Eq.~(\ref{norm-constant-xi}).

Discretization of the directions $\mathbf v$ can spoil the reflection properties of $f_{1,2}$ and thus the covariant conservation of the current (\ref{Jindd}). The latter can still be achieved when $a_{\mathbf v}$ and $b_{\mathbf v}$ satisfy
\begin{equation}
\label{avsum}
\sum_{\mathbf v} a_{\mathbf v} \mathbf v = 0,\qquad
\sum_{\mathbf v} b_{\mathbf v}  = 0,\qquad
\sum_{\mathbf v} b_{\mathbf v} \mathbf v_\perp  = 0,
\end{equation}
where $\mathbf v_\perp = \mathbf v - v^z \mathbf e_z$.

A suitable discretization of the sphere that fulfills the conditions (\ref{avsum}) is given by a set of unit vectors $\mathbf v$ following from regular spacing in $z$ (the $z$-component of $\mathbf{v}$) and $\varphi$ according to
\begin{equation}
\label{disco}
z_i=-1+(2i-1)/N_z,\quad i=1\ldots N_z,\qquad
\varphi_j=2\pi j/N_\varphi,\quad j=1\ldots N_\varphi.
\end{equation}
The resulting `disco balls' are such that they are covered with $\mathcal N_{W}=N_z\times N_\varphi$ curved tiles of equal area.

Given a set of $\mathbf v$, $a_{\mathbf v}$, and $b_{\mathbf v}$ satisfying the conditions (\ref{avsum}), one needs to solve Eq.~(\ref{Jindd}) together with the Yang-Mills field equations and
\begin{equation}
\label{Weqd}
v_\mu D^\mu\,W_{\mathbf v}^\rho(X)
= v_\nu\,F^{\rho\nu}(X), \qquad \rho=0,3.
\end{equation}

In temporal axial gauge $A^0=0$, which is used in the simulations discussed here, the dynamical variables are $A_i$, $E_i=-\dot A_i$, $W_{\mathbf v}^0$, and $W_{\mathbf v}^z$, and Eq.~(\ref{Weqd}) becomes
\begin{equation}
\6_t W_{\mathbf v}^\rho(X)=-\mathbf v\cdot \mathbf D\; W_{\mathbf v}^\rho(X)
+v_\nu F^{\rho\nu}(X) .
\end{equation}
The gauge choice $A^0=0$ is merely a convenience for the numerical simulations. When considering only gauge invariant observables this gauge choice is not necessary, because all equations are gauge covariant and the current is covariantly conserved.

Through Eq.~(\ref{Jindd}) $W^0_{\mathbf v}$ and $W^z_{\mathbf v}$ enter the dynamical evolution only in a linear combination. So in addition to the gauge fields, only the $\mathcal N_{W}$ auxiliary fields
\begin{equation}
\mathcal W_{\mathbf v}(X)=a_{\mathbf v} W^0_{\mathbf v}(X)+
b_{\mathbf v} W^z_{\mathbf v}(X) ,
\end{equation}
need to be considered. The full hard-loop dynamics is then approximated by the following set of matrix-valued equations
\begin{eqnarray}
\label{DHLW}
&&v_\mu D^\mu\,\mathcal W_{\mathbf v}=(a_{\mathbf v} F^{0\mu}+b_{\mathbf v}
F^{z\mu})v_\mu ,
\\[2mm]
\label{DHLF}
&&D_\mu F^{\mu\nu}=J^\nu={1\over \mathcal N_{W}}\sum\limits_{\mathbf v} v^\nu 
\mathcal W_{\mathbf v},
\end{eqnarray}
which can be systematically improved by increasing $\mathcal N_{W}$. The Gauss law constraint, which reads
\begin{equation}
D_i F^{i0}=-D_i \dot A^i=J^0=
{1\over \mathcal N_{W}} \sum_{\bf v} \mathcal W_{\mathbf v},
\end{equation}
is solely a redundant equation of motion in the temporal axial gauge. It can be used to monitor the quality of numerical solutions in the course of simulated temporal evolution.

A similar discretization of the hard-loop dynamics, which uses a discrete set of vectors $\mathbf v$, has been considered before in \cite{Rajantie:1999mp} in the case of an isotropic plasma. A different possibility for discretization is a decomposition of the slightly different auxiliary fields
\begin{equation}
\label{AMY-W}
 \tilde{W}(X,\mathbf{v}) = g\,\int_0^\infty \frac{4\pi p^2 dp}{(2 \pi)^3} \delta f(X,\mathbf{p}) 
\end{equation}
into spherical harmonics \cite{Bodeker:1999gx,Rajantie:1999mp,Arnold:2005vb} and truncating at a maximal angular momentum $l_{\rm max}$.

Using the discretized auxiliary-field formalism as described above, a full hard-loop simulation for an SU(2) plasma was carried out in \cite{Rebhan:2004ur}. The initial field configurations considered here evolve 1+1-dimensionally and contain the most unstable modes for a given oblate momentum distribution (see Fig.~\ref{fig-largexiGamma}).  This study was extended to the gauge group SU(3) and to 3+1 dimensions for SU(2) in \cite{Rebhan:2005re} and then to SU(4) and SU(5) in \cite{Ipp:2010uy}. The 3+1-dimensional SU(2) simulations were also performed in \cite{Arnold:2005vb}, using the discretization method based on spherical harmonics. Results of these studies are discussed in the two subsequent subsections. 

\subsubsection{1+1-dimensional simulations}
\label{sec-1+1}

By considering initial fields and currents, which are constant in the plane transverse to the beam or anisotropy direction along the axis $z$, the complete dynamics of the unstable modes that appear for oblate momentum space distributions at wave-vector $k_\perp=0$ were studied in a 1+1-dimensional setting in \cite{Rebhan:2004ur,Rebhan:2005re}. These modes discussed in Sec.~\ref{subsec-deformed} dominate the system's dynamics at late times. The coordinate space is effectively dimensionally reduced, $(X)\to(t,z)$, and only $A_z(t,z)$ plays the role of a gauge field, while $A_{x,y}(t,z)$ behave as adjoint matter.

Using temporal axial gauge, $A^0=0$, the equations of motion for the dynamical fields can then be simplified to\footnote{Note that we use the conventions of Refs.~\cite{Rebhan:2004ur,Rebhan:2005re} here. The sign in ${\mathbf D}={\nabla}+i g [{\mathbf A},\dots]$ is different from our earlier definition in Eq.~(\ref{covariant-derivative}). Changing the sign in the field strength tensor as well leaves the physics unaffected.}
\begin{subequations}
\label{1+1eoms}
\begin{align}
\partial_{t}^{2} \, A_x &= D_z^2 A_x-g^2 [A_y,[A_y,A_x]]+j_x ,
\\
\partial_{t}^{2} \, A_y &= D_z^2 A_y-g^2 [A_x,[A_x,A_y]]+j_y ,
\\
\partial_t^2 \, A_z &=-i g [A_x,D_z A_x]-i g [A_y,D_z A_y]+j_z ,
\\
(\partial_t +  {\mathbf v}\cdot{\mathbf D} )
\mathcal W_{\mathbf v}
&= a_{\mathbf v} ( -{\mathbf v}\cdot{\partial_t {\mathbf A}} )
+b_{\mathbf v} [
-\partial_t A_z+ D_z(v_x A_x+v_y A_y) ],
\end{align}
\end{subequations}
where ${\mathbf D}={\nabla}+i g [{\mathbf A},\dots]$, $\nabla=(0,0,\partial_z)$, and 
${\mathbf j} ={1\over \mathcal N_{W}}\sum\limits_{\mathbf v} \mathbf v \mathcal W_{\mathbf v}$.

For oblate momentum distributions, $\xi>0$, there are, as discussed in Sec.~\ref{subsec-deformed}, unstable modes for the wave vectors $|k_z|<k_{\rm max}$ and sufficiently small $k_\perp$. In the linear regime, where field amplitudes $A$ are much smaller than $k/g$, the time evolution is determined by the spectrum of plasmons (see Sec.~\ref{subsec-deformed}) and all modes evolve independently. Unstable modes grow exponentially with growth rate $\gamma(\mathbf k)$, which for a given value of $k_z$ is largest when $k_\perp=0$. We denote the scale of maximal growth by $k_*$ and the corresponding maximal growth rate by $\gamma_*$.

In the following we will review results from \cite{Rebhan:2004ur,Rebhan:2005re} for the evolution of collective soft fields starting from very small random fluctuations. If the initial field amplitudes are sufficiently small to allow for growth by a large number of e-folds in the linear regime, the emerging field configurations are dominated by the modes of largest growth rate. For $\xi>0$ this will favor modes with $|k_z|\approx k_*$ and $k_\perp \approx 0$. The special initial conditions with strictly $k_\perp=0$, {\it i.e.} modes which are constant in the {\it x-y} plane, should therefore be an idealization of particular interest. All fields are then functions of only one spatial coordinate $z$, and the equations of motion are simplified according to Eqs.~(\ref{1+1eoms}).

Non-Abelian dynamics with such initial conditions was studied numerically in \cite{Arnold:2004ih}, but a  drastically simplified model for the induced current was used, namely
\begin{equation}
\label{jAL}
j_\nu^{\rm AL}=\mu^2 A_\nu,\quad \nu=x,y,
\end{equation}
where the mass parameter $\mu^2$ approximates the gluon polarization tensor. As shown in \cite{Arnold:2004ih}, this correctly reflects the static limit of the anisotropic hard-loop effective action for fields that vary only in the anisotropy direction, but it neglects its general frequency dependence and dynamical nonlinearity. Already at the linearized level the former complication means that modes with vanishing wave vector, $\mathbf k=0$, are stable, see Fig.~\ref{fig-largexiGamma}, whereas the toy model (\ref{jAL}) implies a growth rate $\gamma=\sqrt{\mu^2-\mathbf k^2}$, which is maximal at $\mathbf k=0$, where it equals $\mu$.  As follows from the results presented in Sec.~\ref{subsec-deformed}, the anisotropic distribution function (\ref{faniso-hl}) with $\xi>0$ leads to $\gamma_*<k_*<\mu$ instead.

In simulations of the toy model defined by Eq.~(\ref{jAL}), the authors of \cite{Arnold:2004ih} have found that unstable non-Abelian modes, which are constant in the {\it x-y} plane, might behave very similarly to Abelian Weibel instabilities also in the nonlinear regime. In the latter, they observed rapid Abelianization\footnote{A discussion of the process of Abelianization is given in Appendix~\ref{app-abelianization}.}, both locally and globally. 

Use of the $\mathbf v$-discretized equations of motion, (\ref{DHLW}) and (\ref{DHLF}), allows to extend the lattice simulations of \cite{Arnold:2004ih} to the full hard-loop effective theory \cite{Rebhan:2004ur,Rebhan:2005re}. We discuss results for the `disco-ball' discretization with $\mathcal N_{W} = N_z\times N_\varphi=100\times 20 = 2000$, and mention results for coarser grids. For a detailed description of the lattice implementation please refer to Appendix B of \cite{Rebhan:2005re}.

When studying plasmas with massless constituents, one needs only a single dimensionful parameter - the Debye mass (\ref{Debye-mass-HL}) for instance - to define a system of units where all dimensionful quantities of interest are expressed by the appropriate powers of the parameter. The authors of \cite{Rebhan:2004ur,Rebhan:2005re} expressed dimensionful quantities in terms of the asymptotic mass  of transverse plasmons $m_\infty$, which in isotropic plasmas equals $m_D/\sqrt{2}$, see the dispersion relation (\ref{T-plasmon-iso}). With the momentum distribution of plasma constituents given by Eq.~(\ref{faniso-hl}), the asymptotic mass is
\be
\label{asym-mass}
m_\infty^2 = \frac{C_\xi}{2} \frac{ {\rm Arctan}\sqrt{\xi}}{\sqrt{\xi} } m_D^2 ,
\ee
and we note that the authors of \cite{Rebhan:2004ur,Rebhan:2005re} used $C_\xi = \sqrt{1 +\xi}$.

\begin{figure}[t]
\hspace{-7mm}
\begin{minipage}{8cm}
\center
\includegraphics[width=0.95\textwidth]{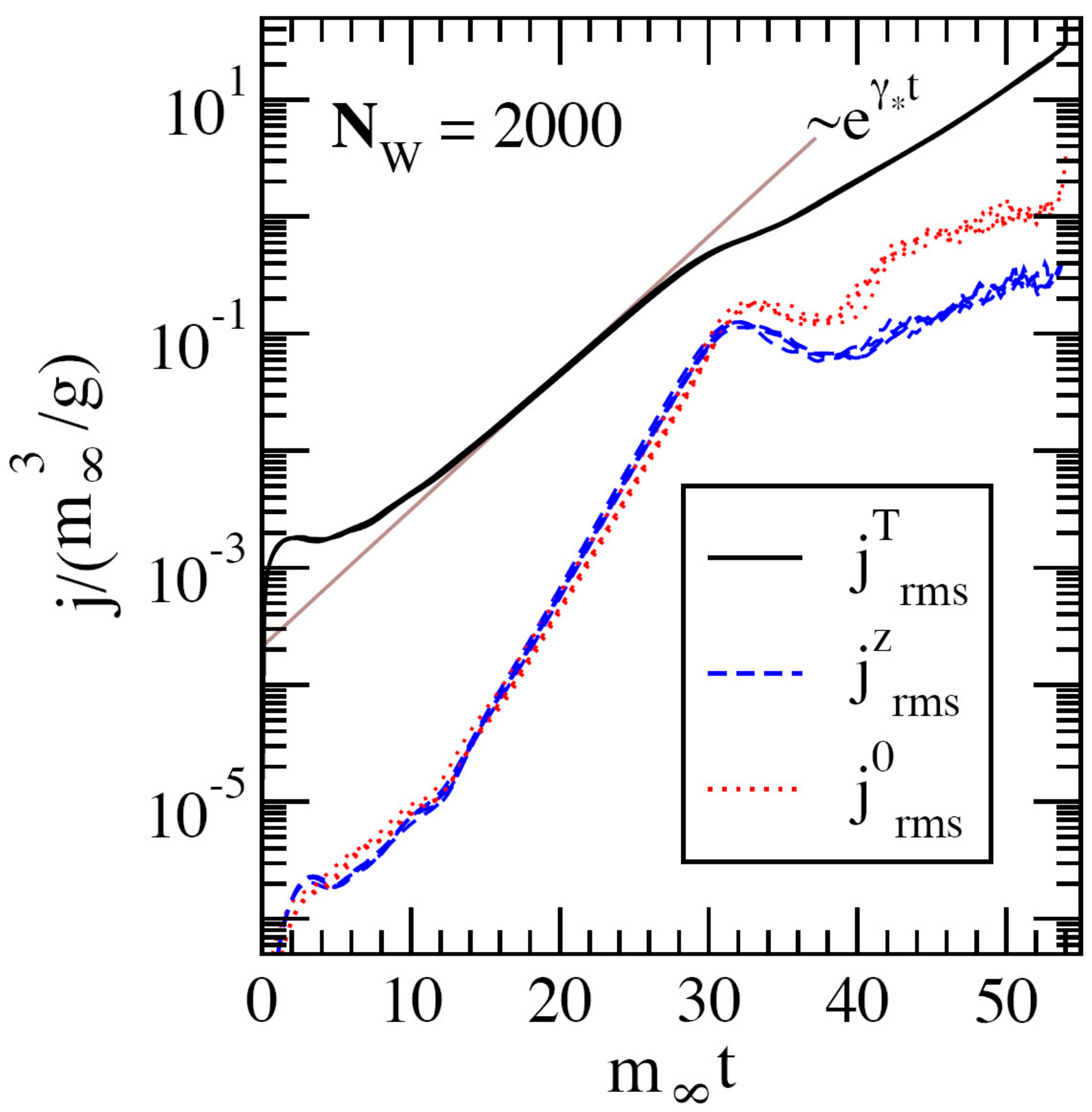}
\vspace{-2mm}
\caption{The average current $j_{rms}^\perp$ (black), $j_{rms}^z$ (blue) and rms charge density $j^0_{rms}$ (red) for $\mathcal N_{W}=2000$ using a collection of 4 runs with different random initial conditions. Figure from \cite{Rebhan:2005re}.}
\label{fig-jrms}
\end{minipage}
\hspace{3mm}
\begin{minipage}{9cm}
\center
\includegraphics[width=1.0\textwidth]{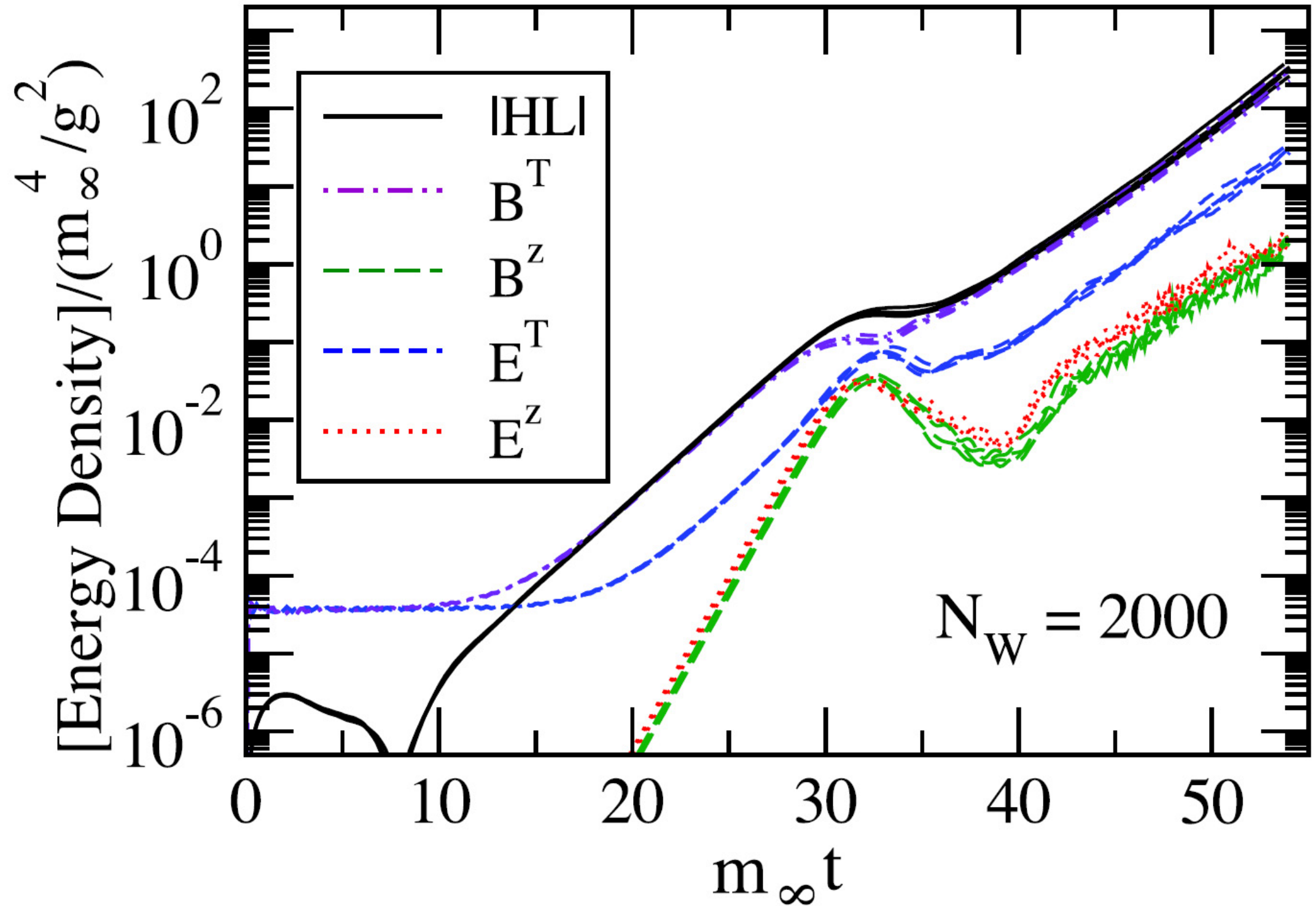}
\vspace{-3mm}
\caption{Average energy densities in transverse/longitudinal chromomagnetic/electric fields and the total energy density contributed by hard particles. Each plot shows four different runs using a different random seed. Figure from \cite{Rebhan:2005re}.}
\label{fig-energy-denisties}
\end{minipage}
\end{figure}

In the simulations  \cite{Rebhan:2004ur,Rebhan:2005re} a one-dimensional spatial lattice with periodic boundary conditions and 5,000 sites and lattice spacing $a=0.0707 m_\infty^{-1}$ was used, so that the physical size is $L\approx 350\,m_\infty^{-1}$. 

\vspace{2mm}
\underline{\it Currents and charge density}
\vspace{1mm}

We start with a discussion of the growth of the currents and charge density caused by the unstable modes. Fig.~\ref{fig-jrms} shows the evolution of the average root-mean-square transverse and longitudinal (with respect to the $z$-direction) currents and charge density defined by
\be
j_{\rm rms}^{\perp,z,0}=
\bigg[ \int_0^L {dz\over L} 2 \tr \big[(\mathbf j^{\perp,z,0})^2\big] \bigg]^{1/2}.
\ee
The transverse currents are seen to grow exponentially with a growth rate that is most of the time only slightly below $\gamma_*$, except for a transitory reduction at the beginning of the nonlinear regime, when $j_{\rm rms}\sim m_\infty^3/g$. This happens because the unstable modes with the wave vector along the axis $z$ are responsible in the oblate plasma for an exponential growth of $A_x$ and $A_y$ which are coupled to $j_x$ and $j_y$, see Eqs.~(\ref{1+1eoms}). The leading behavior of $j_x$ and $j_y$ is as in Abelian plasmas that is it is determined by the terms in Eqs.~(\ref{1+1eoms}a) and (\ref{1+1eoms}b) which are linear in $A$. Growth of the currents along the $z$ direction and the charge densities arises solely due to non-Abelian interactions. Because only electric fields $E_x$ and $E_y$ were initialized at $t=0$, the magnitudes of $j^z$ and $j^0$ are smaller than $j^\perp$ but their growth rates are double the one of $j^\perp$, as the growth is driven by the gauge potentials $A_x$ and $A_y$  squared, see Eq.~(\ref{1+1eoms}c). The charge density $j^0$  behaves as $j^z$ because they are coupled through the continuity equation. 

Calculations using smaller $\mathcal N_{W}$ lead to only slightly different behavior in the nonlinear regime, where higher $\mathcal N_{W}$ seems to be required to capture the precise variations of the subdominant components $j^z$ and $j^0$. 

\vspace{2mm}
\underline{\it Energy density}
\vspace{1mm}

Fig.~\ref{fig-energy-denisties} shows how the exponentially growing energy that is transferred from hard modes (particles) to soft modes (fields) gets distributed among chromomagnetic and chromoelectric fields. As the mechanism of Weibel instability, which is explained in Sec.~\ref {subsec-mechanism-trans}, suggests, the dominant contribution is in $B_x$ and $B_y$, and it grows roughly with the maximum rate $\gamma_*$, both in the linear as well as in the highly nonlinear regime, with a transitory slowdown in between. Transverse electric fields behave similarly, and are suppressed by a factor of the order of $(\gamma_*/k_*)^2$. Note that in the model discussed in \cite{Arnold:2004ih} the situation is different: Because $k_*$ is zero there, the dominant energy component is from transverse electric fields, whereas the relative importance of magnetic fields drops with time. In the 1+1-dimensional evolution, the transverse magnetic field component is the dominant one, both in the linear and in the nonlinear regime.

The appearance of longitudinal contributions is again a purely non-Abelian effect. Their growth rate is double the one in the transverse sector, and they begin to catch up with the latter when the nonlinear regime is reached. At this point the general growth stalls for a time of order $\gamma_*^{-1}$. When the transverse magnetic field resumes its growth (with the transverse electric field following with some delay), the respective longitudinal components drop for some time before also starting to grow again. 

\vspace{2mm}
\underline{\it Color correlations and Abelianization}
\vspace{1mm}

For initial conditions where all fields point in the same color direction all over the spatial lattice, only strictly exponential growth would occur once the stable modes have become of negligible importance. The behavior would then be exactly the same as with Abelian Weibel instabilities, which grow exponentially until they come into conflict with the assumptions of the hard-loop approximation (namely that the fields have only small effects on the trajectories of the hard particles), whereupon isotropization is supposed to set in.

\begin{figure}[t]
\hspace{-10mm}
\begin{minipage}{6.5cm}
\center
\includegraphics[width=1.05\textwidth]{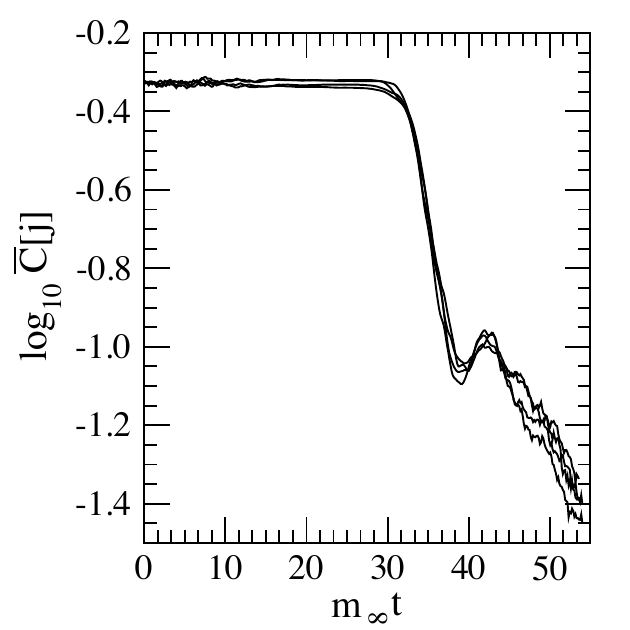}
\vspace{-2mm}
\caption{The measure of `non-Abelianness' $\bar C$ defined by Eq.~(\ref{cbareq}) as a function of time for $\mathcal N_{W}=2000$. Figure from \cite{Rebhan:2005re}.
\label{figCbars}}
\end{minipage}
\hspace{4mm}
\begin{minipage}{10cm}
\center
\includegraphics[width=1.0\textwidth]{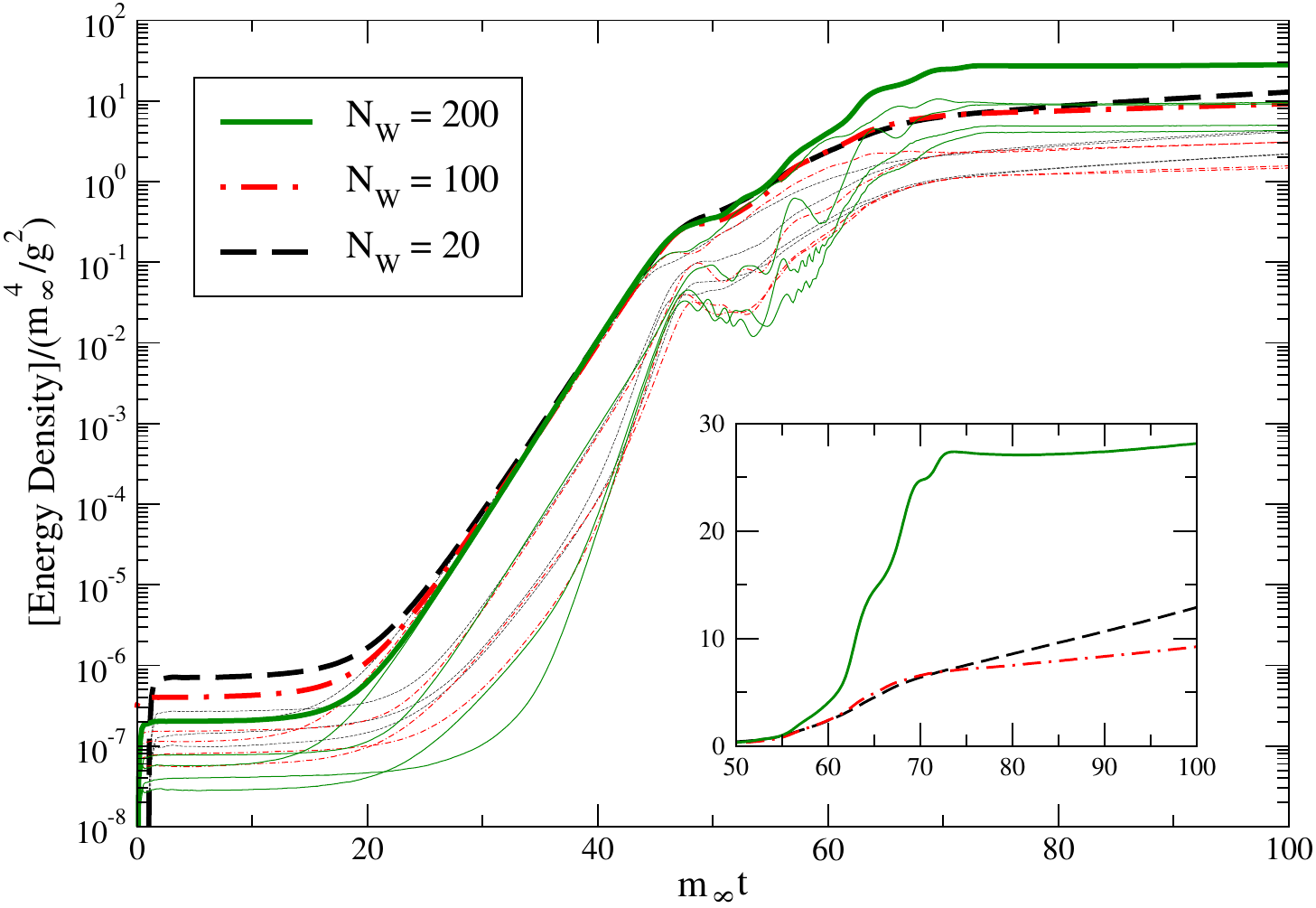}
\vspace{-3mm}
\caption{Comparison of average energy densities for 3+1-dimensional simulations with $\mathcal N_{W}=20,100,200$ on $96^3, 88^3, 69^3$ lattices. The thick lines represent the total field energy density and the thin lines the components of the field energy density. The inset shows late-time behavior of the hard-loop energy density on a linear scale. Figure from \cite{Rebhan:2005re}.
\label{figNenlog32}}
\end{minipage}
\end{figure}

Apart from a brief transitory slowdown, the 1+1-dimensional simulations thus evolve similarly to Abelian instabilities both in the linear and the strongly nonlinear domain (at least as far as the dominant transverse components are concerned). In \cite{Rebhan:2005re} a measure of local `non-Abelianness' was defined by 
\be
\label{cbareq}
\bar C[j] = \int_0^L {dz\over L} { \left\{ \tr \left( (i[j_x,j_y])^2 \right) 
\right\}^{1/2}\over \tr (j_x^2+j_y^2) },
\ee
which coincides with the definition in \cite{Arnold:2004ih} in the simplifying model (\ref{jAL}), but is gauge invariant in the full theory also when the restriction to 1+1-dimensional configurations is relaxed. If the field configurations are Abelian (aligned in one color direction), $\bar C[j]$ vanishes because of the commutator in the numerator. When the nonlinear regime is entered,  $\bar C[j]$ is found to decay roughly exponentially, as seen in Fig.~\ref{figCbars}, but not as fast as in the toy model of \cite{Arnold:2004ih}, which implies strong Abelianization in a 1+1-dimensional system. Studies of the global Abelianization can be found in \cite{Arnold:2004ih} and \cite{Rebhan:2005re} but we omit their detailed discussion here because they are not very relevant for the more physical 3+1-dimensional simulations.

\subsubsection{3+1-dimensional simulations}
\label{sec-3+1}

We now turn to the more realistic 3+1-dimensional dynamics which turns out to be qualitatively different from the 1+1-dimensional case. 3+1-dimensional systems were studied both in \cite{Rebhan:2005re}, using the discretization described above, as well as in \cite{Arnold:2005vb}, where an expansion in spherical harmonics of Eq.~(\ref{AMY-W}) was used. In the following we will present results from both groups.

\vspace{2mm}
\underline{\it Energy density}
\vspace{1mm}

The energy densities obtained using ${\cal N}_W = 20, 100,$ and $200$, with respective lattice sizes $96^3, \; 88^3,$ and $69^3$ are compared in Fig.~\ref{figNenlog32}. The initial conditions were taken to be vanishing color electric and magnetic fields and small uncorrelated fluctuations in the $\mathcal W$ fields. The thick lines are the field energy densities provided by the particles (hard modes) and the other various components of the field energy are shown for comparison but not labeled explicitly. The fluctuations in the overall amplitude are due to the different random initial conditions used for each run.

In the linear regime, {\it i.e.}, for amplitudes of gauge potentials much smaller than $m_\infty/g$, the differences are only due to the different amount of initial energy densities in longitudinal and transverse components. In both the 3+1- and 1+1-dimensional simulations, the dominant component is from transverse magnetic fields, followed by transverse electric fields (suppressed by a factor of $(\gamma_*/k_*)^2<1$). Longitudinal components are initially absent in the 1+1-dimensional preparation of the system, but not in the full 3+1-dimensional case because of the different initial conditions. Nevertheless, at the later stages of the linear evolution, they behave rather similarly. The regime where specifically non-Abelian effects first come into play is also very similar in the two cases. However, whereas the unstable modes in the 1+1-dimensional system eventually recover their initial growth rates of the linear regime, in the 3+1-dimensional case the transversely non-constant modes seem to be the determining factor deeper into the nonlinear regime. In Fig.~\ref{figNenlog32} one sees subexponential behavior at late times, which in fact seems to tend to a linear growth. 

\begin{figure}[t]
\hspace{-10mm}
\begin{minipage}{9.5cm}
\center
\includegraphics[width=1.05\textwidth]{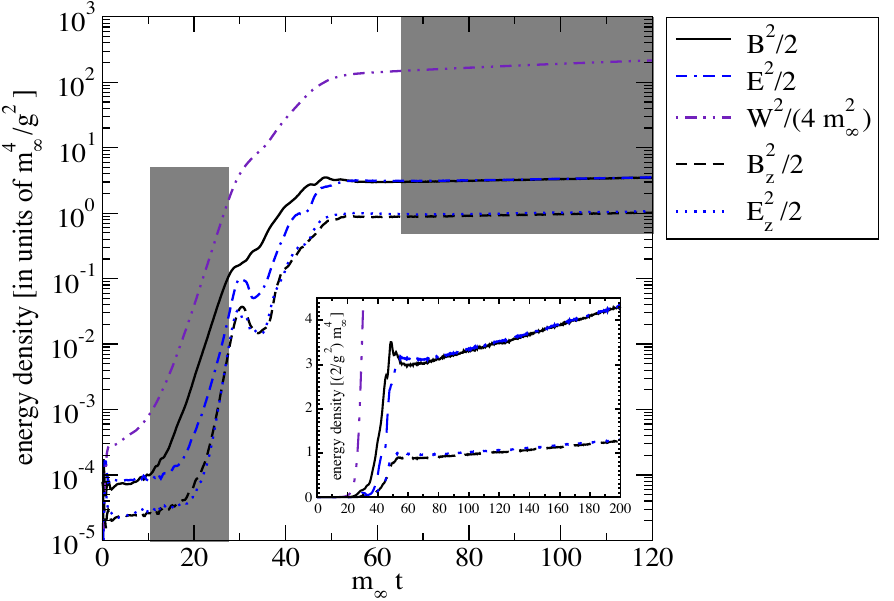}
\vspace{-2mm}
\caption{Various components of the energy density as a function of time. The inset shows the same plot on a linear scale. Figure from  \cite{Arnold:2005vb}. \label{fig:energies}}
\end{minipage}
\hspace{4mm}
\begin{minipage}{7cm}
\center
\includegraphics[width=1.05\textwidth]{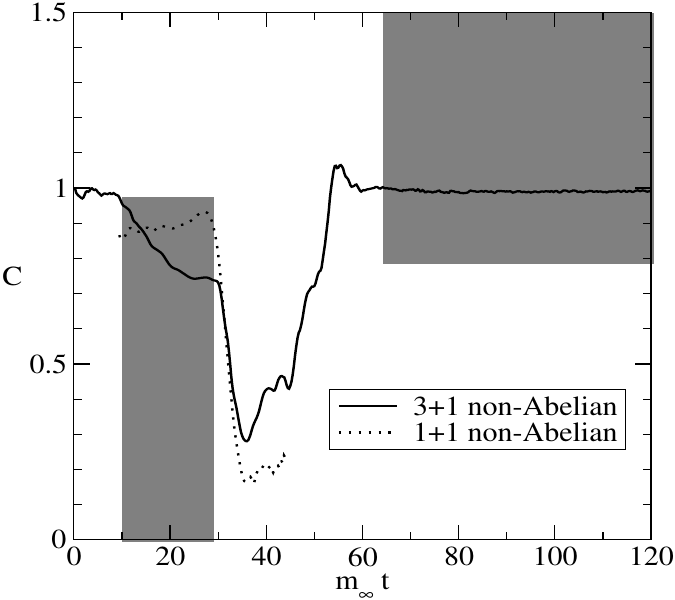}
\vspace{-3mm}
\caption{\label{fig:Jcomm} 
The measure of `non-Abelianness' $C$ defined by Eq.~(\ref{eq:C}) as a function of time. Figure from \cite{Arnold:2005vb}.}
\end{minipage}
\end{figure}

Fig.~\ref{fig:energies} shows the results for the same quantities determined in SU(2) simulations by \cite{Arnold:2005vb}. Here the initial conditions chosen were Gaussian noise for the gauge fields $\textbf{A}$ with an exponential fall-off in $\textbf{k}$. $W$ represents the energy density in the $\tilde{W}$ fields defined by Eq.~(\ref{AMY-W}). Again, there is an initial period of non-perturbative growth that looks quite similar to the 1+1-dimensional case and to early conjectures about Abelianization, but eventually the instabilities in 3+1-dimensions settle into a period characterized by linear rather than exponential growth of the field energy density.

\vspace{2mm}
\underline{\it Color correlation and Abelianization}
\vspace{1mm}

To study local Abelianization, the observable $C$, the three dimensional extension to the quantity (\ref{cbareq}), was computed in \cite{Arnold:2005vb}. It is defined by
\begin {equation}
C \equiv \frac3{\sqrt2} \>\frac{ \left[\int \frac{d^3x}{V} \>
      \Bigl( ([j_x,j_y])^2 + ([j_y,j_z])^2 + ([j_z,j_x])^2 \Bigr)
    \right]^{1/2} }{ \int \frac{d^3x}{V} \> |\mathbf{j}|^2  },
\label {eq:C}
\end {equation}
where
$[j_x,j_y]^2 \equiv \epsilon^{abc} j^b_x j^c_y \, \epsilon^{amn} j^m_x j^n_y$, {\it etc}. The normalization of $C$ has been chosen so that $C$ would be unity if the components of $\mathbf{j}$ were independent random numbers with the same distribution.  For an Abelian configuration, $C$ would be zero.  Fig.~\ref{fig:Jcomm} shows the time evolution of $C$ in the SU(2) simulations of \cite{Arnold:2005vb}.  $C$ drops suddenly when non-linear (non-Abelian) interactions first become important, in agreement with the Abelianization conjecture. However, in the 3+1-dimensional simulations, $C$ later rises again all the way to unity, showing no local Abelianization in the linear growth regime.

\begin{figure}[t]
\hspace{-10mm}
\begin{minipage}{7cm}
\center
\includegraphics[width=1.05\textwidth]{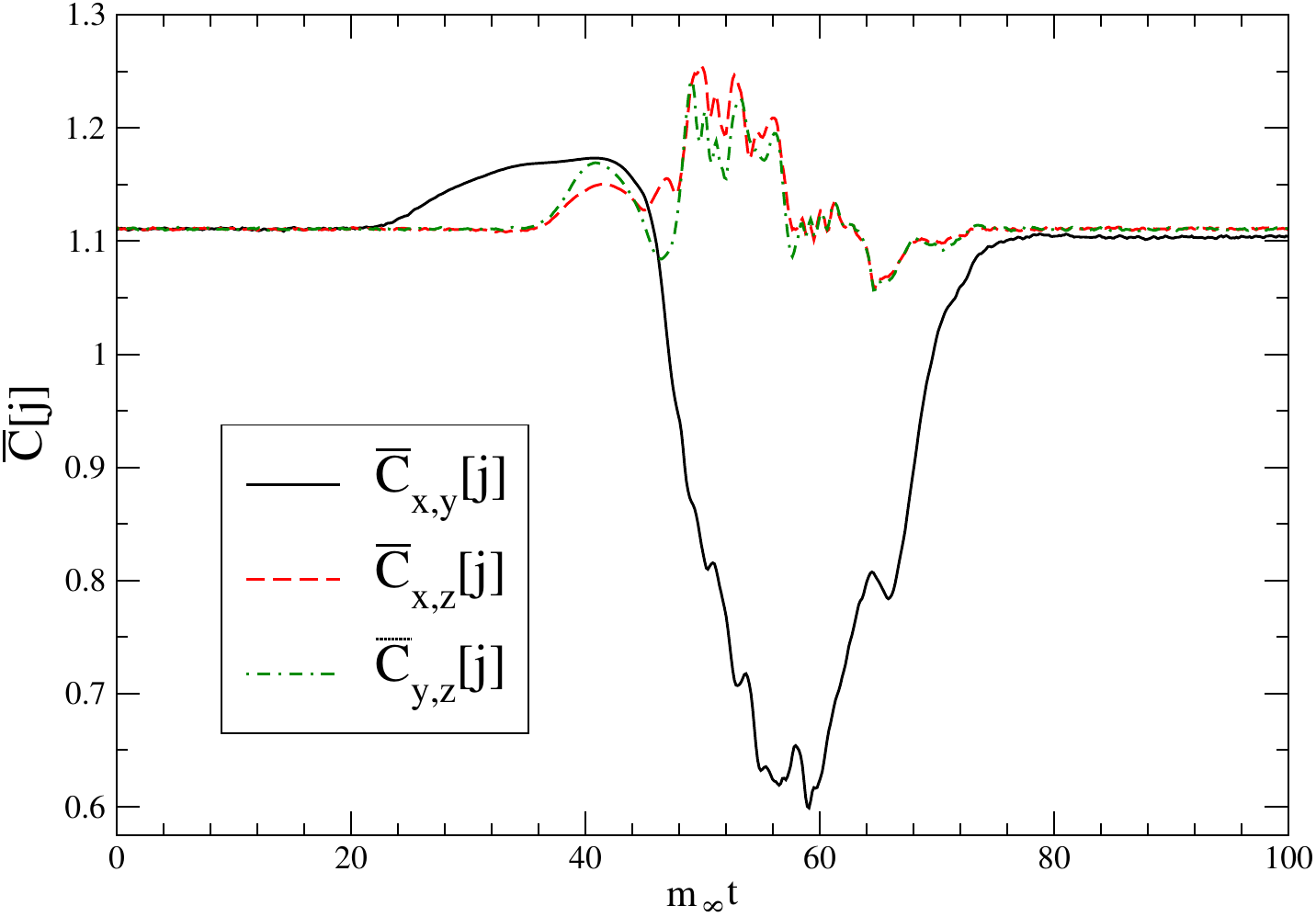}
\vspace{-4mm}
\caption{The relative rms size of commutators ${\bar C}^2_{i,j}={\rm tr} ([j_i,j_j])^2/({\rm tr} j_i^2\,{\rm tr} j_j^2)$. Figure from \cite{Rebhan:2005re}.
\label{figcbarj31}}
\end{minipage}
\hspace{4mm}
\begin{minipage}{9.5cm}
\center
\includegraphics[width=0.95\textwidth]{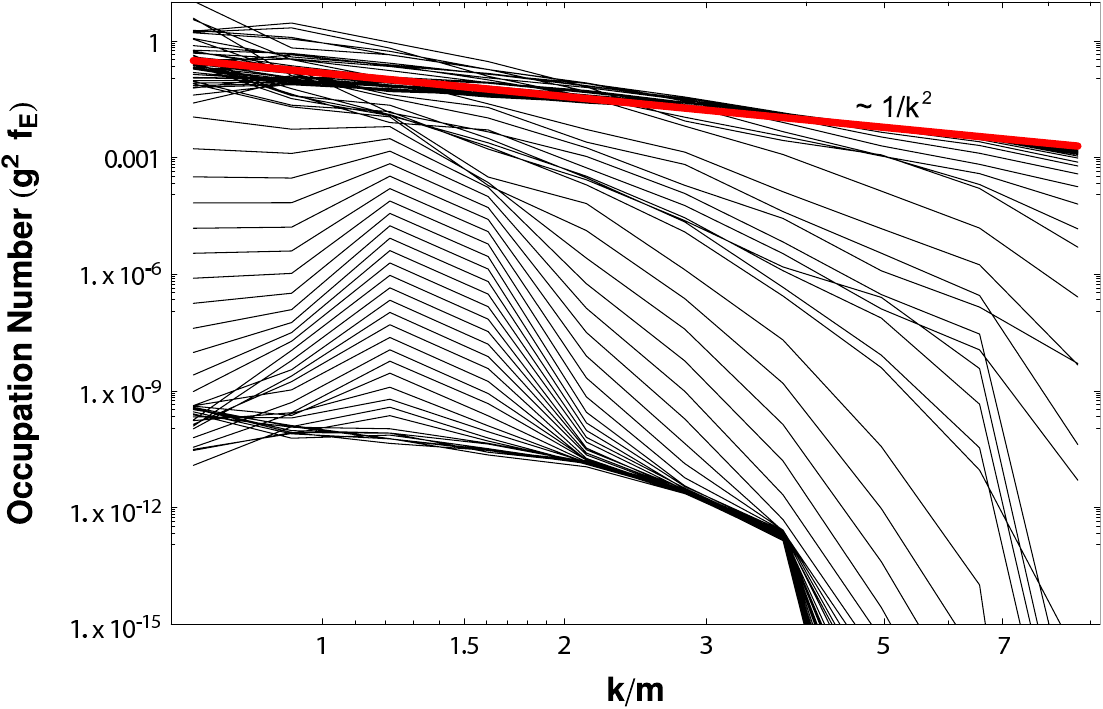}
\vspace{-3mm}
\caption{Field mode spectrum for the SU(2) simulation showing saturation of soft field growth at $f\sim 1/g^2$ and an associated cascade of energy to the UV as the simulation time increases. Figure from \cite{Strickland:2007fm}.
\label{fig:spectrumMike}}
\end{minipage}
\end{figure}

Furthermore, in Fig.~\ref{figcbarj31} we display results from \cite{Rebhan:2005re} for the normalized size of the commutators ${\bar C}^2_{i,j}={\rm tr} ([j_i,j_j])^2/({\rm tr} j_i^2\,{\rm tr} j_j^2)$ which give a measure of local Abelianization with respect to the three possible spatial directions.  As can be seen  from this figure, there is initially Abelianization in the $xy$ commutator starting at $m_\infty t \sim 40$, while the other two commutators show no evidence for Abelianization.  At later times ($m_\infty t \gtrsim 60$) the $xy$ commutator becomes less and less Abelian in agreement with Fig.~\ref{fig:Jcomm} from \cite{Arnold:2005vb} and at late times, in the linear-growth phase, the soft-current commutators are nearly isotropic. We will discuss the reason for differences between the 1+1- and 3+1-dimensional simulations in Sec.~\ref{sec:2p1}.
 
\vspace{2mm}
\underline{\it Kolmogorov cascade}
\vspace{1mm}

The regime of linear growth in non-expanding plasmas has been studied in more detail in \cite{Arnold:2005ef} as well as in \cite{Strickland:2007fm} in the hard-loop framework. It was found that when the exponential growth of the energy of magnetic fields ends, the long-wavelength modes associated with the instability stop growing, but the energy \emph{cascades} toward the ultraviolet (UV) in the form of plasmon excitations. This way a quasi-stationary state with a power-law distribution $k^{-2}$ of the plasmon mode population is created. This phenomenon was argued to be very similar to Kolmogorov wave turbulence in hydrodynamics, where long-wavelength modes transfer their energy to shorter ones without dissipation.

\begin{figure}[t]
\begin{minipage}{\textwidth}
  \includegraphics[width=3.2in]{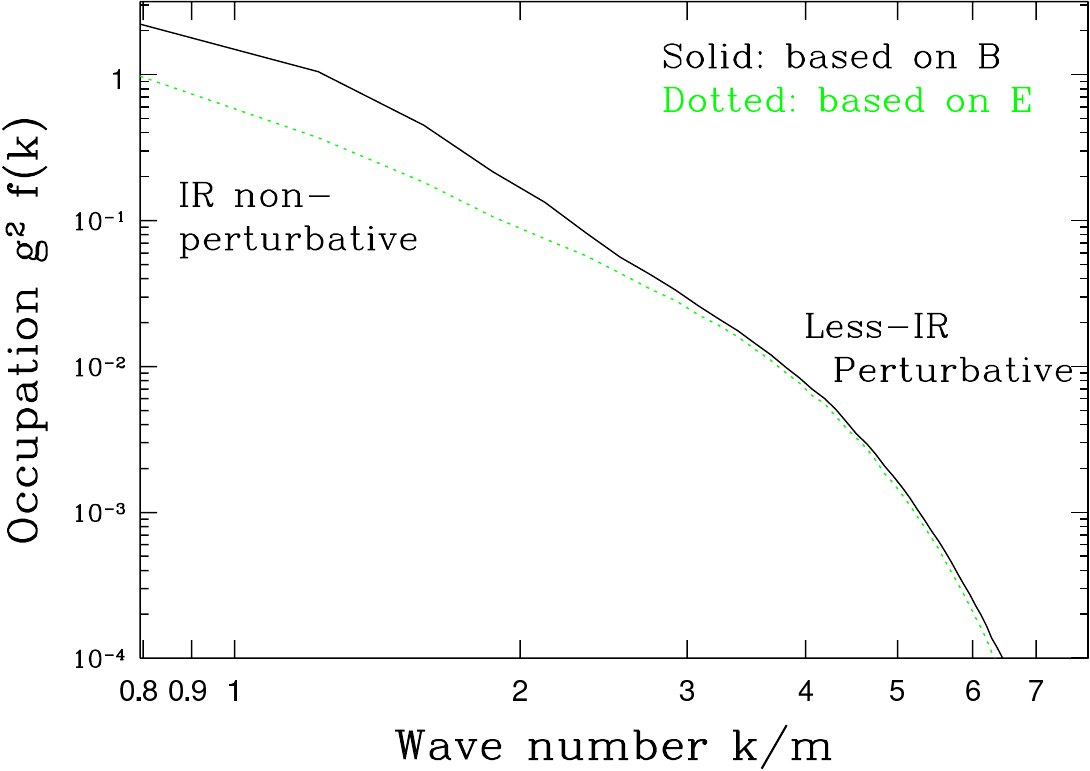}
  \hspace{0.5cm}
  \includegraphics[width=3.2in]{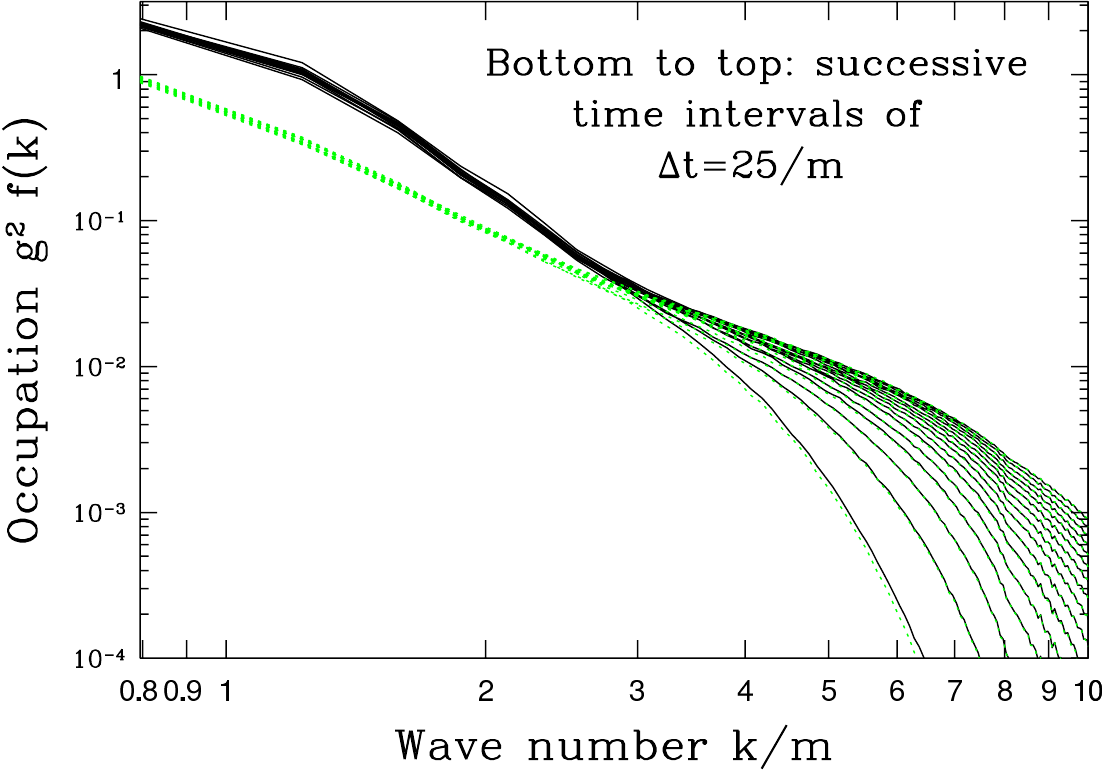}
\end{minipage}
\caption{Soft gauge field power spectra: initial (left panel) and as a function of time (right panel).  The IR fields are in a quasi-steady state, and the energy cascades toward more UV modes.  Figure from \cite{Arnold:2005ef}. \label{fig:coulomb1}}
\end{figure}

The mode occupancy after initial transients, as found in \cite{Arnold:2005ef}, is displayed on the left in Fig.~\ref{fig:coulomb1}, showing that, in the infrared (IR), there are nonperturbatively large fields, while fields become perturbative at larger wave number.  On the right of Fig.~\ref{fig:coulomb1} we show the time evolution.  Each curve is time-averaged over a certain time interval, and the central times of consecutive curves are spaced apart equally. Fig.~\ref{fig:spectrumMike} presents the mode spectrum obtained in \cite{Strickland:2007fm}. The IR occupancies remain nonperturbative but with stable amplitude, while the UV occupancy increases.  At any $k$, the occupancy rises and eventually saturates. The saturation point progresses to larger $k$.  

Both hard-loop frameworks provide a power-law fall-off spectrum $f(k) \propto k^{-\nu}$ with spectral index $\nu \simeq 2$. This same power law behavior was found for SU(2), SU(3), SU(4), and SU(5) \cite{Ipp:2010uy}.  For comparison, a thermal spectrum is $f(k)\propto k^{-1}$ and cascades in scalar field theories during `preheating' after cosmological inflation, for example, typically display a power spectrum with various power laws at different stages, such as $f(k) \propto k^{-3/2}$ and $k^{-5/3}$ \cite{Son:1996uv,Micha:2004bv}.

\vspace{2mm}
\underline{\it Anisotropy dependence}
\vspace{1mm}

We have described above how, in the 3+1-dimensional case, for small initial field fluctuations one observes the crossover from perturbative, exponential growth of instabilities to the limiting late-time behavior. In fact, for tiny initial conditions there is a significant spurt of continued exponential-like growth even after the field strength reaches non-perturbatively large values. It is only later, at much higher energy, that linear growth finally sets in. In the simulations presented in \cite{Bodeker:2007fw}, which extended those by \cite{Arnold:2005vb} to larger anisotropies, this spurt of post-non-perturbative exponential growth for tiny initial conditions was found to become much more significant for extreme anisotropy. Here, even at late times only exponential-like growth is found and late-time linear behavior is absent. Given that the final spectrum behaves like $\sim k^{-1}$ and that saturation depends on the lattice spacing, it is possible that the late-time behavior is ultimately linear but sets in at such large field energy that the simulations cannot reproduce it because the compactness bound of the lattice is reached.

To determine how the late time behavior of the chromodynamic Weibel instabilities scales with the anisotropy of the system, a detailed analysis was performed in \cite{Arnold:2007cg}. It was found that the limiting magnetic field strength of unstable modes scales with the anisotropy parameter $\xi$ as
\begin{equation}
\label{eq:limitingB}
B_*\sim \frac{m_\infty^2}{g}\,\xi^{1/2},
\end{equation}
showing that the chromomagnetic fields can be very large in strongly anisotropic systems. 

Let us explain how the result (\ref{eq:limitingB}) was obtained. The exponential growth of the magnetic field due to unstable modes ends when non-linear non-Abelian effects are no longer negligible - that is when the magnitude of the gauge potential becomes as large as $k/g$, where $k$ is absolute value of the characteristic wave vector of unstable soft modes. Consequently, the order of magnitude estimate of the limiting magnetic field $B_*$ is 
\be
\label{lim-B-small-aniso}
B_*\sim \frac{k^2}{g} \sim \frac{m_\infty^2}{g}.
\ee
This estimate holds in case of weak anisotropy, when the characteristic momentum of plasma constituents, which is denoted as $p$, is of the same order in every direction. Then, the characteristic wave-vector of unstable modes $k$ is $\sim gp$. The situation changes when the plasma is strongly anisotropic ($\xi \gg 1$), as the unstable modes can extend to large wave-vectors. In case of extremely oblate systems discussed in Sec.~\ref{sec-ex-oblate}, the Weibel mode exists up to infinite $k_z$, see Eq.~(\ref{omega-pm-ex-oblate2}). Therefore, in case of strong anisotropy, the $z$-component of the magnetic field is properly estimated by Eq.~(\ref{lim-B-small-aniso}) but the $x$- and $y$-components, where the wave-vector $k_z$ enters, can be much bigger. Therefore, the estimate (\ref{lim-B-small-aniso}) is modified as 
\be
\label{lim-B-big-aniso}
B_*\sim \frac{m_\infty^2}{g} \,\xi^\alpha ,
\ee
where the exponent $\alpha$ is to be determined by the slope of the late-time linear growth of the magnetic field energy density and its dependence on the anisotropy. The argument goes as follows.

The increasing energy of magnetic fields comes from the unstable modes, which take energy from the hard particles and dump it into the cascade of plasmons. Imagine that a large fraction of the magnetic energy density ${\cal E}^B_*$ stored in the unstable modes were abruptly transferred to the cascade of plasmons. One asks how long would it take the unstable modes to grow back to their original, limiting size?  The time should be of order the inverse instability growth rate $t \sim \gamma^{-1} \sim m_\infty^{-1}$. So, the rate at which the total magnetic energy changes due to the cascade is expected to be of order $\gamma {\cal E}^B_*$:
\begin {equation}
   \frac{d {\cal E}^B_{\rm tot} }{dt} \equiv
   \frac{d}{dt} \Big( \frac{1}{2} B^2 \Big)_{\rm tot}
   \sim \gamma {\cal E}^B_*
   \sim \gamma B_*^2
   \sim \frac{m_\infty^5}{g^2}\,\xi^{2\alpha},
\label {eq:dEdtn}
\end {equation}
where the estimate (\ref{lim-B-big-aniso}) is used. The simulations presented in \cite{Arnold:2007cg} allowed one to measure how the time derivative (\ref{eq:dEdtn}) scales with $\xi$ , providing the desired exponent $\alpha=1/2$ which enters Eq.~(\ref{lim-B-big-aniso}) and gives the final formula (\ref{eq:limitingB}).

\subsubsection{2+1-dimensional simulations} 
\label{sec:2p1}

Why does one find strong qualitative differences in the late time behavior of the non-Abelian instability development between one- and three-dimensional simulations? Three-dimensional simulations include variations of the fields and distributions in all spatial directions, while in one-dimensional simulations one assumes that fields and distributions vary only in one spatial direction $z$, as $A_\mu(t,z)$ and $\delta f(t,z,\mathbf{p})$. However, particle momenta $\mathbf{p}$ are still treated three-dimensionally and all three spatial components of the gauge field $(A_x,A_y,A_z)$ are included (the gauge condition is $A_0=0$). It is necessary to include the $x$- and $y$-components in order to describe magnetic physics at all. Otherwise Weibel instabilities could not be studied. The origin of the differences between one- and three-dimensional simulations can be found by investigating the intermediate case of two spatial dimensions \cite{Arnold:2007tr}. 

\begin{figure}[t]
\hspace{-10mm}
\center
\includegraphics[width=0.45\textwidth]{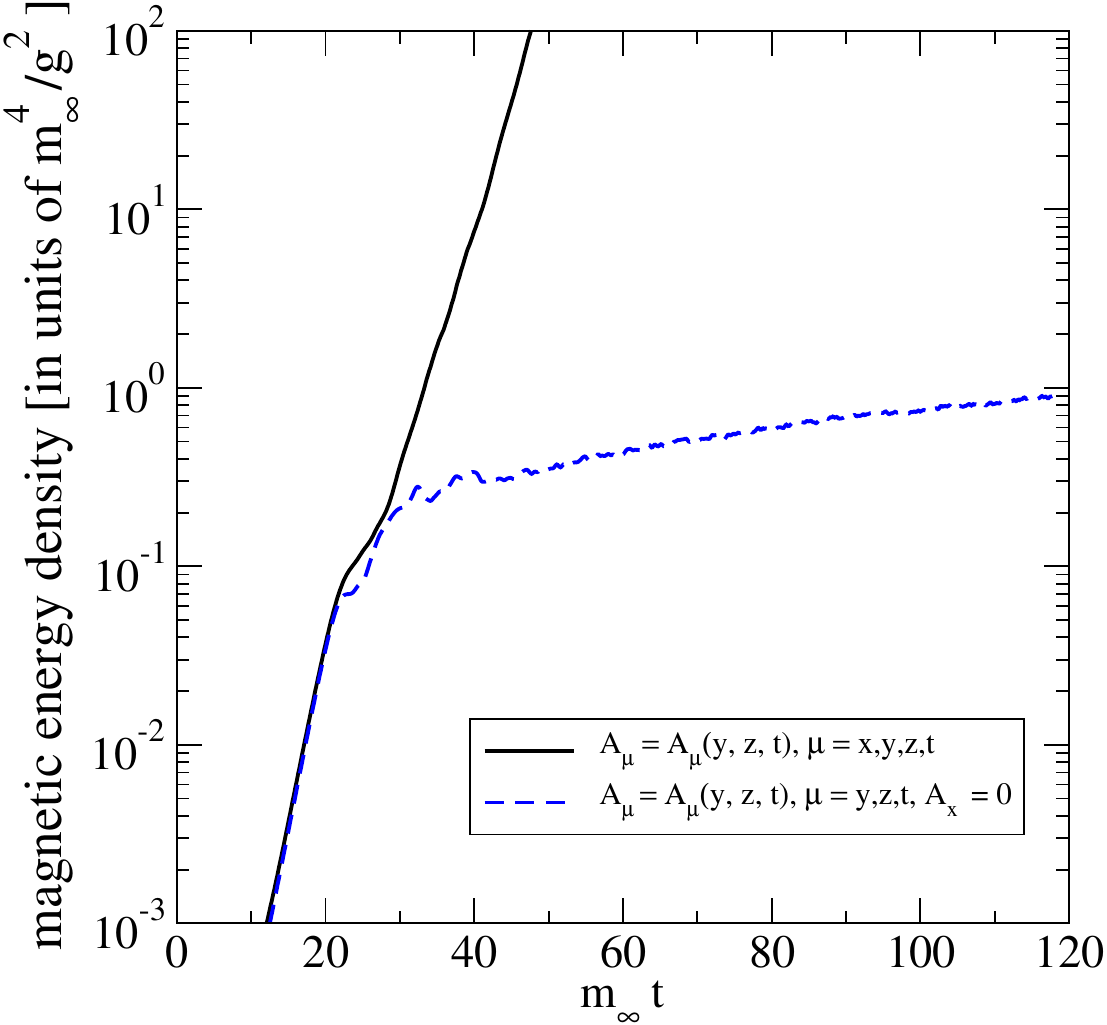}
\vspace{-2mm}
\caption{\label{fig:results2d} Magnetic field energy density as a function of time for the 2D+3V SU(2) simulation. The solid line is for the straightforward dimensional reduction of (\ref{eq:2d}).  The dashed line is a simulation that does not include the out-of-plane component $A_x$. Figure from \cite{Arnold:2007tr}.}
\end{figure}

Let us adopt the classification of dimensionalities from traditional plasma physics $n$D+$m$V, where $n$ indicates the spatial dimensionality and $m$ that of the velocity (momentum). For the 2D+3V dimensional simulations one has two possibilities for dimensional reduction. First, one can proceed as in the 1D+3V case and retain all components of the gauge potential, but have the fields and distributions only depend on two spatial dimensions:
\begin{subequations}
\begin{align}\label{eq:2d}
A_\mu&=A_\mu(t,y,z)\,,~~~~~\mu=t,x,y,z,
\\
\delta f &=\delta f(t,y,z,\mathbf{p}) .
\end{align}
\end{subequations}
For this case, it was found in the SU(2) simulations by \cite{Arnold:2007tr} that the system evolves similarly to the 1D+3V case: instability growth continues exponentially even after non-Abelian interactions become relevant (black line in Fig. \ref{fig:results2d}). On the other hand, if one eliminates the component $A_x$, which lies outside of the {\it y-z}-plane, by setting $A_x=0$, one obtains behavior similar to the 3D+3V simulation, where energy growth at late times is linear rather than exponential (dashed blue line in Fig. \ref{fig:results2d}).

For the moderately anisotropic particle distributions simulated in \cite{Arnold:2007tr}, the authors explain the results the following way. Non-abelian effects will prevent late-time exponential growth if and only if no gauge-field components are included outside of the subset of spatial dimensions simulated. In one dimension one has no choice: unless such components are included, one cannot simulate magnetic fields and so cannot study the Weibel instability at all. In three dimensions, one automatically includes all components.  For two dimensions one has a choice leading to the two qualitatively different results described above.

Formally, the difference between gauge field components in and out of the subset of spatial dimensions is that the ones within the subset (like $A_y$ and $A_z$ in two dimensions) are gauge fields in the dimensionally reduced theory, while those outside (like $A_x$) represent  adjoint-charge scalars. There is a significant difference between gauge fields and scalars. Gauge fields can always be transformed locally to $A=0$, but scalars cannot. In the two-dimensional theory, the squared amplitude $A_x^a A_x^a$ is locally gauge invariant and can have physical consequences. The authors of \cite{Arnold:2007tr} argue that it suppresses {\it via} the Higgs mechanism all color fields except those in color directions commuting with $A_x^a$. If the unsuppressed fields commute with each other, non-Abelian effects will not prevent instability growth, equivalent to the `abelianization' mechanism discussed above and in Appendix~\ref{app-abelianization}.

\subsection{Expanding Hard-Loop simulations}
\label{subseq-expanding-HL}

The simulations discussed thus far were all performed in a static box with fixed momentum-space anisotropy.
The quark-gluon plasma produced in relativistic heavy-ion collisions is not kept in a fixed box, but instead rapidly expands in vacuum. In particular, at early times the system primarily expands in the longitudinal direction which drives the emergence of large momentum-space anisotropies in the first place.  To take into account the effect of expansion, the hard-loop formalism for anisotropic plasmas in a stationary geometry was extended to the case of a boost-invariant longitudinally expanding distribution of plasma particles in \cite{Romatschke:2006wg}. In this way, the hard-expanding-loop (HEL) effective theory was formulated. The essentially Abelian weak-field regime was derived semi-analytically in  \cite{Romatschke:2006wg} and further worked out in \cite{Rebhan:2009ku}. The main finding is that the counterplay of increasing anisotropy and decreasing plasma density lets Weibel instabilities grow exponentially in the square root of proper time, with more and more modes becoming unstable as time goes on, but each one experiencing a certain delay before growth kicks in. However, the onset of growth appears to be much earlier when the unstable modes are initiated with fluctuations of chromodynamic currents than of chromodynamic fields.  

The first numerical study of the evolution of genuinely non-Abelian plasma instabilities in a longitudinally expanding plasma was performed in \cite{Rebhan:2008uj}, using a lattice discretization of the HEL effective theory and doing 1D+3V simulations. The setting was extended to the 3D+3V case in \cite{Attems:2012js} where a rich sample of simulation results was presented. Here we review the HEL effective theory and present the most important numerical results.

In a stationary but possibly anisotropic geometry, the background distribution $f_0$ of plasma constituents only depends on momenta. We now consider the generalization to a plasma expanding longitudinally, which should be a good approximation for the initial stage of a parton system produced in a heavy ion collision as long as the transverse dimension of the system is sufficiently large. Assuming, furthermore, boost invariance in rapidity \cite{Bjorken:1983qr} and isotropy and homogeneity in the transverse directions, the unperturbed distribution function $f_0$, being a Lorentz scalar, has the form \cite{Baym:1984np,Mueller:1999pi}
\be
\label{eq:f0}
f_0(t,\mathbf{x},\mathbf p)=f_0(t,z,p_\perp,p_z)=f_0(\tau, p_\perp,p'_z),
\ee
where the transformed longitudinal momentum is
\be
p'_z=\gamma(p_z-\beta p_0), 
\quad
\beta=z/t,
\quad 
\gamma=t/\tau,
\quad 
\tau=\sqrt{t^2-z^2},
\ee
with $p_0=\sqrt{p_\perp^2+p_z^2}$ for ultrarelativistic (massless) particles. The distribution function (\ref{eq:f0}) satisfies the free transport equation
\be
\label{vdf0}
p \cdot \6\, f_0(t,\mathbf{x},\mathbf{p})=0 .
\ee

In the following, we use comoving coordinates with the proper time $\tau$ and space-time rapidity $\eta$ defined as
\be
t=\tau \cosh\eta, ~~~~~~~~~~ z=\tau \sinh\eta .
\ee
We follow the notation from \cite{Rebhan:2008uj} where $\tilde x^\alpha=(x^\tau,x^{\9{i}},x^\eta)= (\tau,x^1,x^2,\eta)$ with indices from the beginning of the Greek alphabet for these new coordinates. The indices $i,j,\ldots = 1,2$ are restricted to the two transverse spatial coordinates. The space-time interval $ds^2=d\tau^2-d\mathbf x_\perp^2-\tau^2 d\eta^2$ defines the diagonal metric tensor $g^{\alpha \beta}$ as $g^{\tau \tau}{\rm =1}$, $g^{ij} =-\delta^{ij}$, $g^{\eta\eta}=-\tau^{-2}$. 

The Yang-Mills or non-Abelian Maxwell equations can be written compactly as
\be
\label{eq:maxwellcomoving}
{1\0\tau}\tilde D_\alpha(\tau \tilde F^{\alpha\beta}) \equiv
{1\0\tau}\tilde D_\alpha\left[\tau g^{\alpha\gamma}(\tau) g^{\beta\delta}(\tau)
\tilde F_{\gamma\delta}\right]=\tilde j^\beta.
\ee
The metric tensor is explicitly written in Eq.~(\ref{eq:maxwellcomoving}) to emphasize that it is important now to keep track of indices being `up' or `down'.

The momentum-space rapidity $y$ for the massless particles is defined as
\be
p^\mu=p_\perp(\cosh y,\cos\phi,\sin\phi,\sinh y) ,
\ee
and in comoving coordinates, we then have
\bea
\tilde p^\tau&=&\sqrt{p_\perp^2+\tau^2(\tilde p^\eta)^2}=\cosh\eta\, p^0-\sinh\eta\, p^z
=p_\perp \cosh(y-\eta),
\\
\tilde p^\eta&=&-\tilde p_\eta/\tau^2=(\cosh\eta\, p^z-\sinh\eta\, p^0)/\tau=p'^z/\tau
=p_\perp \sinh(y-\eta)/\tau.
\eea

Instead of $v^\mu=p^\mu/p^0$ containing a unit 3-vector ${\bf v}$ that enters Eqs.~(\ref{DHLW}) and (\ref{DHLF}), we will use the new quantity 
\be
\tilde V^\alpha = {\tilde p^\alpha \0 p_\perp} =
\Big(\cosh (y-\eta),\,\cos\phi,\,\sin\phi,\,{1\0\tau}\sinh (y-\eta)\Big),
\label{velocityDef1}
\ee
which is normalized so that it has a unit 2-vector in the transverse plane and $\tilde V^\alpha \tilde V_\alpha = 0$.

The transport equation (\ref{vdf0}), involving space-time derivatives at fixed $\mathbf p_\perp$ and $p_z$, can be rewritten as
\begin{equation}
(\tilde{p} \cdot \tilde{\partial}) f_0 \Big|_{y,\mathbf{p}_\perp} = 0 .
\end{equation}
Because
\bea
\tilde p^\tau \6_\tau \tilde p_\eta(\tilde x)\Big|_{y,\1p_\perp}
&=&-p_\perp^2 \sinh(y-\eta)\cosh(y-\eta)=-\tilde p^\eta\6_\eta \tilde p_\eta(\tilde x)\Big|_{y,\1p_\perp}
\label{petaconstancy}
\eea
the equation is solved by $f_0(t, \1x, \1p)=f_0(\1p_\perp,\tilde p_\eta(x))
=f_0(\1p_\perp,-p'^z(x)\tau(x))$.

Following \cite{Romatschke:2006wg} and \cite{Rebhan:2008uj}, we introduce
\be\label{faniso}
f_0(t, \1x,\mathbf p)=f_{\rm iso}\bigg(\sqrt{p_\perp^2+\Big({p'_z\tau\0\tau_{\rm iso}}\Big)^2}\,\bigg)
=f_{\rm iso}\left(\sqrt{p_\perp^2+\tilde p_\eta^2/\tau_{\rm iso}^2}\,\right) ,
\ee
which corresponds to local isotropy on the hypersurface $\tau=\tau_{\rm iso}$, and increasingly oblate momentum space anisotropy at $\tau>\tau_{\rm iso}$ (but prolate anisotropy for $\tau<\tau_{\rm iso}$). Since a plasma description does not make sense at arbitrarily small times and so time evolution will have to start at a nonzero proper time $\tau_0$, the time $\tau_{\rm iso}$ may be entirely fictitious in the sense of pertaining to the pre-plasma (``glasma'')  phase \cite{Lappi:2006fp,Romatschke:2006nk}. This is the case in the numerical simulations reviewed, where the starting point was an already oblate anisotropy, using $\tau_{\rm iso} < \tau_0$.

The distribution function (\ref{faniso}) has the same form as the ansatz (\ref{R-S-ansatz}) but the anisotropy parameter $\xi$ is now space-time dependent according to 
\be
\label{xi}
\xi(\tau)=(\tau/\tau_{\rm iso})^2-1 .
\ee
The behavior $\xi\sim\tau^2$ at large $\tau$ is a consequence of having a free-streaming background distribution. In a more realistic collisional plasma, $\xi$ will grow more slowly than this. In the first stage of the original bottom-up scenario \cite{Mueller:2005un}, where plasma instabilities are ignored, one would have had $\xi\sim\tau^{2/3}$. In \cite{Bodeker:2005nv} it was advocated that plasma instabilities reduce the exponent to $\xi\sim\tau^{1/2}$, whereas \cite{Arnold:2007cg} argued in favor of $\xi\sim\tau^{1/4}$. All these scenarios have $\xi\gg1$, so \cite{Rebhan:2008uj} concentrated on the case $\tau_{\rm iso}<\tau_0$ and thus high anisotropy for all $\tau>\tau_0$, but in the idealized case of a collisionless free-streaming expansion.

Transforming the gauge-covariant Vlasov equation to comoving coordinates one can write 
\be
\label{VDf}
\tilde V\cdot \tilde D\, \delta f^a(\tilde x, \1p_\perp,\tilde p_\eta) \big|_p = g \tilde V^\alpha 
\tilde F_{\alpha\beta}^a \tilde \6_{(p)}^\beta f_0(\1p_\perp,\tilde p_\eta),
\ee
where the derivative on the left-hand side has to be taken at fixed $p^\mu$ as opposed to fixed $\tilde p^\alpha$. On the right-hand side the derivative with respect to momenta is at fixed $x$, but the transformation from $x$ to $\tilde x$ does not depend on momenta anyway. However, in the following it will be important to write the right-hand side in terms of $\tilde \6_{(p)}^\beta f_0(\1p_\perp,\tilde p_\eta)$ with index up so that this factor depends only on $\1p_\perp$ and $\tilde p_\eta$ and not on $\tau$. This means, in particular, that $p\cdot\6 \,(\tilde \6_{(p)}^\beta f_0)|_p=
\tilde p\cdot\tilde\6 \,(\tilde \6_{(p)}^\beta f_0)|_p=0$.

As done above in the static case ({\it cf.} Eqs.~(\ref{Wdefs}) and (\ref{Weq})), one can solve Eq.~(\ref{VDf}) by introducing auxiliary fields $\tilde W_\beta(\tilde x,\phi,y)$ defined by 
\be
\label{Wdef}
\delta f (\tilde x, \1p_\perp,\tilde p_\eta) =-g \tilde W_\beta (\tilde x,\phi,y) 
\tilde\6_{(p)}^\beta f_0(\1p_\perp,\tilde p_\eta),
\ee
which satisfy 
\be
\label{VDW}
\tilde V\cdot \tilde D\, \tilde W_\beta (\tilde x,\phi,y) \big|_{\phi,y}=\tilde V^\alpha \tilde F_{\beta\alpha} (\tilde x) .
\ee
Note that the equation of motion (\ref{VDW}) is formally the same as in the static geometry (\ref{Weq}) only when written for $\tilde W_\beta(\tilde x,\phi,y)$ with lower index. 

Expressed in terms of the auxiliary field $\tilde W$, the induced current in comoving coordinates reads
\bea
\label{tjind}
\tilde j^\alpha (\tilde x)  &=& - \frac{g^2 t_R}{2}
\int {d^3p\over(2\pi)^3} 
{1\over p^0} \,\tilde p^\alpha\, 
{\partial f_0 (\1p_\perp,\tilde p_\eta) \over \partial \tilde p_\beta}
\tilde W_\beta (\tilde x,\phi,y) \nn
\\
&=&-g^2t_R
\int_0^\infty {p_\perp dp_\perp \over 8\pi^2}
\int_0^{2\pi} {d\phi \0 2\pi}
\int_{-\infty}^\infty dy \,
\tilde p^\alpha\, {\partial f_0 (p_\perp,\tilde p_\eta) \over \partial \tilde p_\beta}
\tilde W_\beta (\tilde x,\phi,y) ,
\eea
where for each $(\phi,y)$, {\it i.e.} fixed $\1v$, the transverse momentum $p_\perp \equiv |\1p_\perp|$ (related to energy by $p^0=p_\perp\cosh y$) can be integrated out. Here $t_R$ is a suitably normalized group factor. 

\begin{figure}[t]
\hspace{-10mm}
\begin{minipage}{8.8cm}
\center
\includegraphics[width=1.0\textwidth]{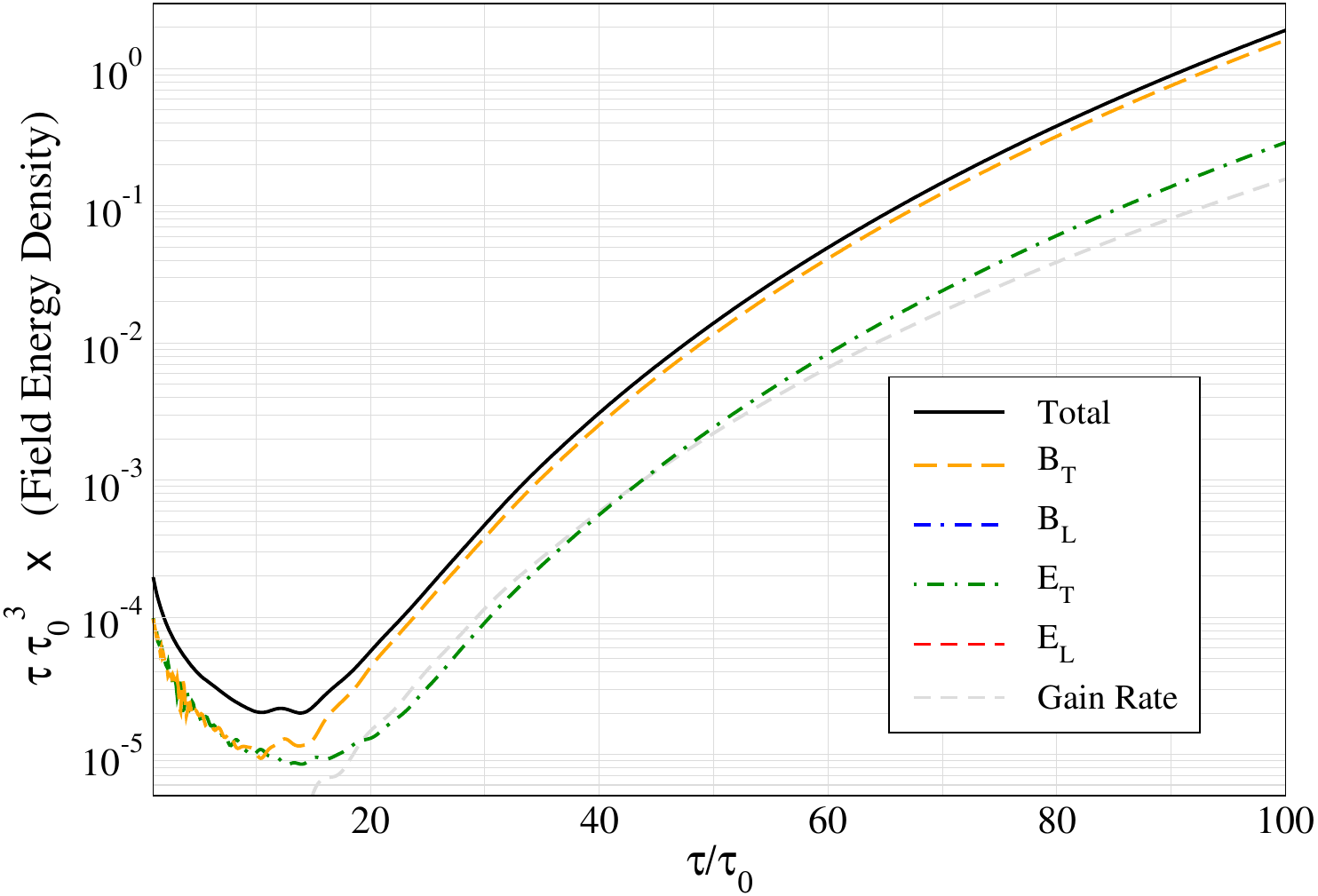}
\vspace{-3mm}
\end{minipage}
\begin{minipage}{8.8cm}
\center
\includegraphics[width=1.02\textwidth]{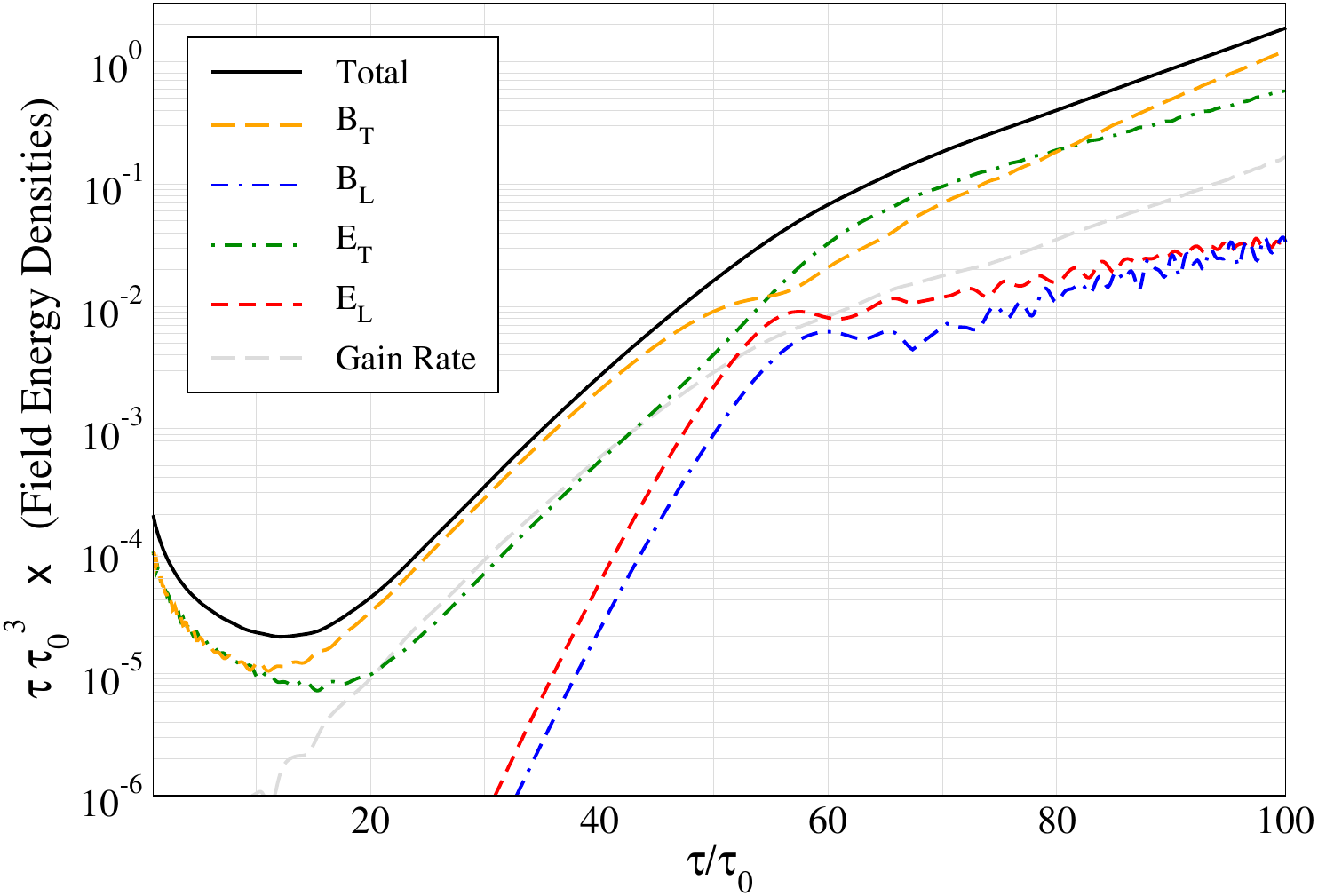}
\vspace{-5mm}
\end{minipage}
\caption{\label{fig6}
Temporal evolution of the field energy density and energy gain rate (\ref{REG}) multiplied by an extra factor of $\tau_0$. The results come from Abelian (left panel) and non-Abelian (right panel) 1D+3V simulations initialized with the FGM initial conditions. Figure from \cite{Rebhan:2008uj}.}
\end{figure}

Now the Yang-Mills equation (\ref{eq:maxwellcomoving}) with the current (\ref{tjind}) and the Vlasov equation (\ref{VDW}) are solved numerically. For this purpose the comoving temporal axial gauge $A^\tau=0$ is chosen and one introduces canonical conjugate field momenta defined as
\be
\label{conjugate-mom}
\Pi^i (\tilde x) \equiv \tau \partial_\tau A_i (\tilde x), 
~~~~~~~~~~~~~~
\Pi^\eta (\tilde x) \equiv \frac{1}{\tau} \partial_\tau A_\eta (\tilde x) .
\ee
We note that the comoving electric fields in the {\it x-y-}plane equals $E^i = \tau^{-1}\Pi^i = \partial_\tau A_i$. For more details on the particular numerical implementation for the 1+1- and 3+1-dimensional simulations see \cite{Rebhan:2008uj} and \cite{Attems:2012js}, respectively.

Below we review the 1D+3V-simulations \cite{Rebhan:2008uj}.  Results on energy density and pressure as well as on energy spectra obtained in more complete 3D+3V simulations \cite{Attems:2012js} are qualitatively very similar. Color correlators have not been studied in the expanding 3D+3V setting. 

\vspace{2mm}
\underline{\it Energy density and pressure}
\vspace{1mm}

\begin{figure}[t]
\hspace{-10mm}
\begin{minipage}{8.8cm}
\center
\includegraphics[width=1.0\textwidth]{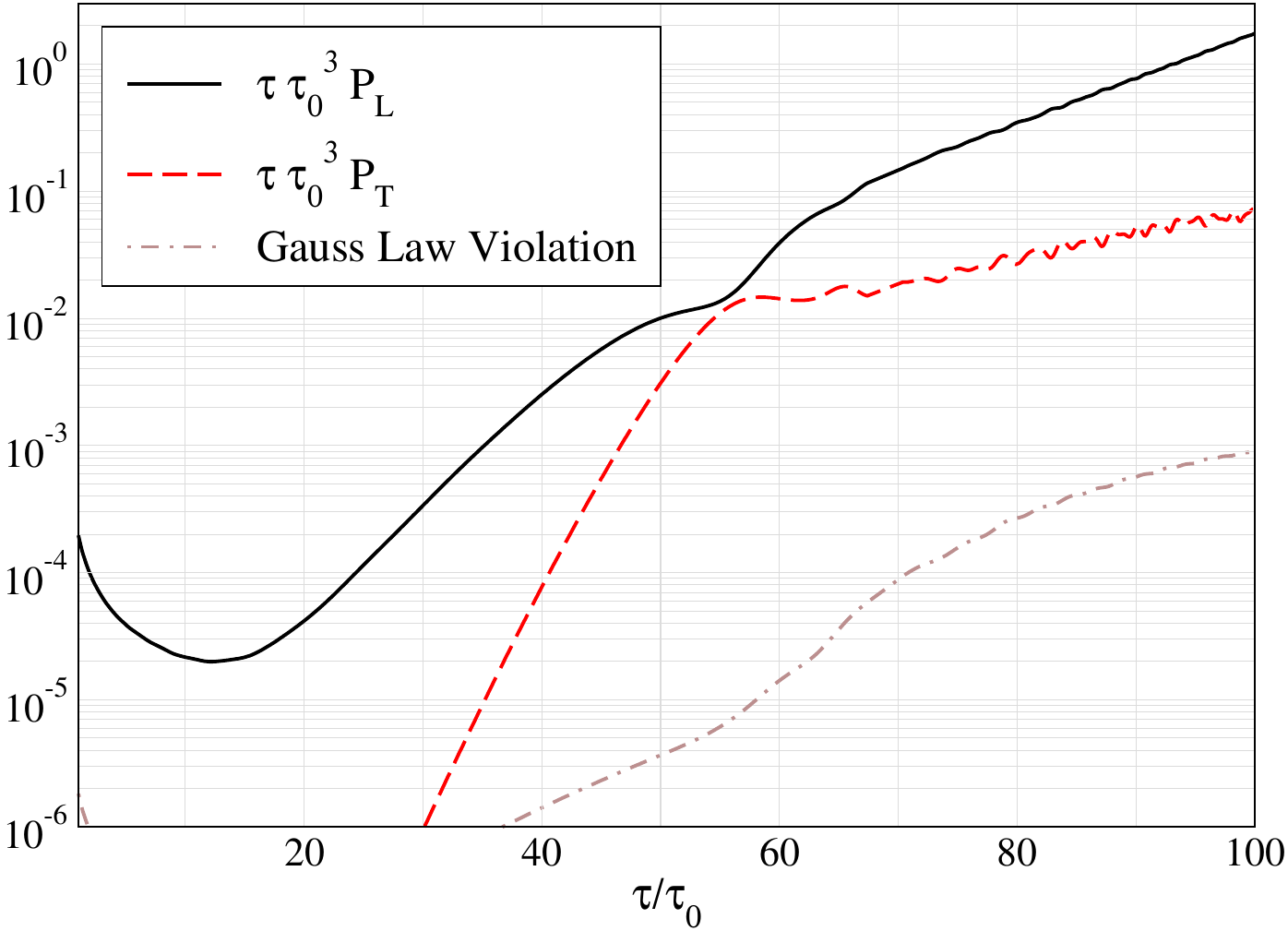}
\vspace{-5mm}
\end{minipage}
\begin{minipage}{8.6cm}
\center
\includegraphics[width=1.01\textwidth]{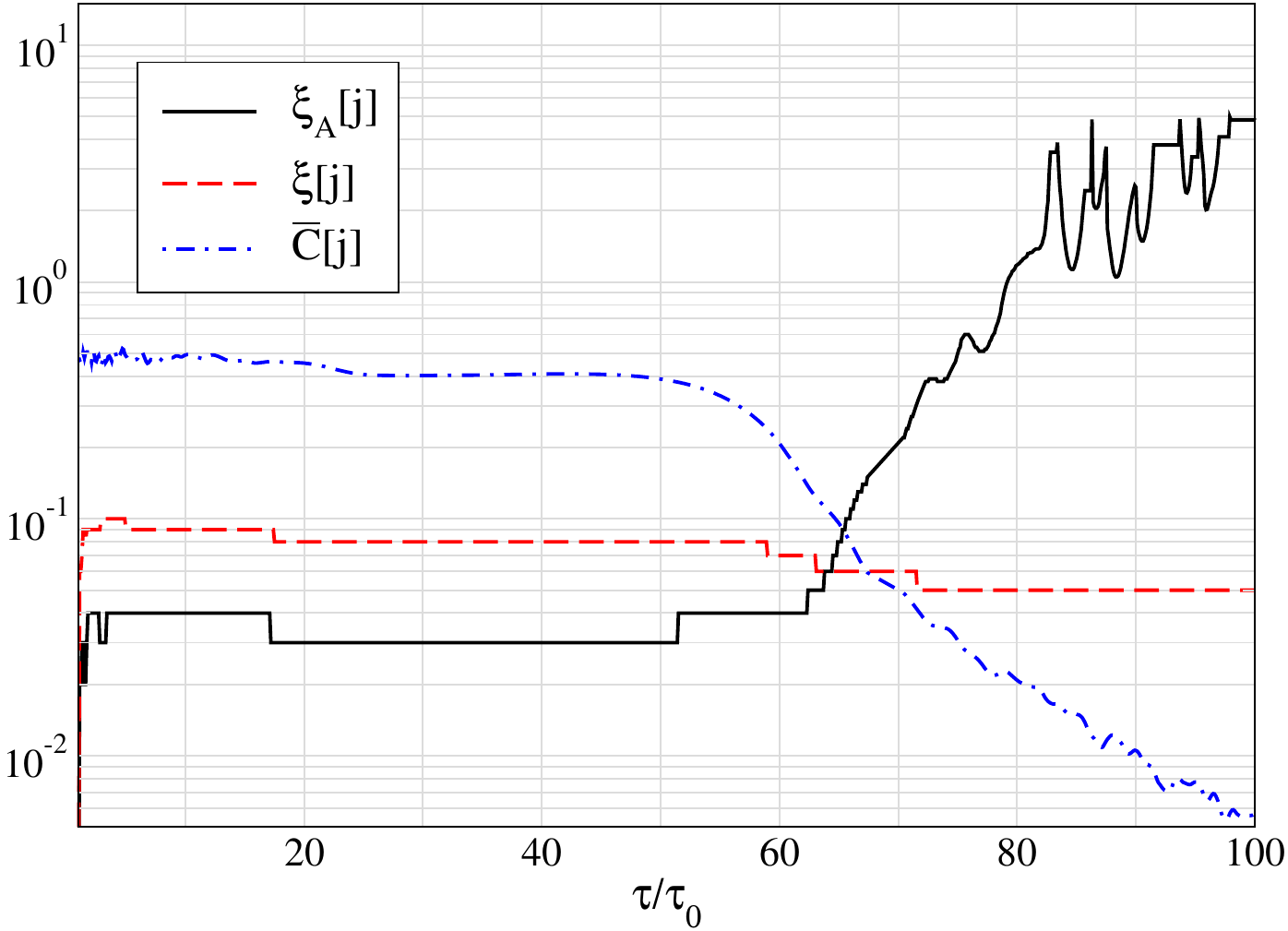}
\vspace{-3mm}
\end{minipage}
\caption{\label{fig7}
Results from non-Abelian 1D+3V simulations initialized with the FGM initial conditions. The left panel shows the longitudinal and transverse pressures along with the numerical Gauss law violation. The right panel presents the correlations $\xi_A[j]$, $\xi[j]$, and ${\bar C}[j]$. Figure from \cite{Rebhan:2008uj}.}
\end{figure}

Figs.~\ref{fig6} and \ref{fig7} show results obtained using initial seed fields which reflect the spectral properties derived by Fukushima, Gelis, and McLerran (FGM) within the Color-Glass-Condensate (CGC) framework \cite{Fukushima:2006ax}, with a Debye mass corresponding to the estimates of the `gluon liberation factor' $c$ found in \cite{Kovchegov:2000hz,Lappi:2007ku} (see Appendix in \cite{Rebhan:2008uj} for details).

Various components of energy density and pressure, which are obtained in the simulations, are defined through the energy-momentum tensors of fields and of massless particles 
\ba
\label{energy-mom-field}
T^{\mu \nu}_{\rm field} &\equiv& {\rm tr}\big[F^\mu_{\;\;\rho} F^{\rho \nu}
+ \frac{1}{4}g^{\mu \nu} F_{\rho \sigma} \, F^{\rho \sigma}\big],
\\[2mm]
\label{energy-mom-part}
T^{\mu \nu}_{\rm particle} &\equiv& \int \frac{d^3p}{(2\pi )^3} \frac{p^\mu p^\nu}{E_{\bf p}} f({\bf p}) ,
\ea
which are equated to the energy-momentum tensor of an anisotropic fluid in a local rest frame
\ba
T^{\mu \nu}_{\rm fluid} &\equiv& 
\left[
\begin{array}{cccc}
\varepsilon & 0 & 0 & 0
\\
0 & P_T & 0 & 0
\\
0 & 0 & P_T & 0
\\
0 & 0 & 0& P_L 
\end{array}
\right] .
\ea
The three energy-momentum tensors are assumed to be traceless $T^\mu_{\;\;\mu}=0$. Since the fields are split into longitudinal and transverse components with respect to the anisotropy axis $z$, the field energy density is decomposed as $\varepsilon = \varepsilon_L + \varepsilon_T$. Computing the diagonal components of $T^{\mu \nu}_{\rm fluid}$ one finds that $P_T = \varepsilon_L$ and $P_L = \varepsilon_T - \varepsilon_L$.

The energy densities presented in the left panel of Fig.~\ref{fig6} were obtained from an Abelian run of the 1D+3V simulation, in which all fields were constrained to initially point in the same direction in color space. The $x$- and $y$-components of electric and magnetic fields (transverse fields) vanish and the Gauss law is satisfied within machine precision. The field pressures are $P_T= - P_L = \varepsilon = \varepsilon_L$. The right panel of Fig.~\ref{fig6} shows analogous results of non-Abelian SU(2) simulations. As can be seen, at early times there is equal partitioning between chromoelectric and chromomagnetic fields which both initially decrease and then begin to grow exponentially with transverse chromomagnetic fields dominating for nearly the entire run. Longitudinal field energies, which vanish initially, grow exponentially with a rate about twice of that of the transverse fields, but almost saturate when the nonlinear regime is reached. This translates into the generation of exponentially large longitudinal pressure as shown in the left panel of Fig.~\ref{fig7}. In contrast to the non-expanding system, we do not observe a linear growth regime of energy density in the non-Abelian regime. Instead, there is only a small reduction of the exponential growth which persists to much longer times. The reason of this behavior is presumably the same as in the case of extreme anisotropy in static geometry discussed in Sec.~\ref{sec-3+1}, because the free streaming expansion leads to the strong increase of anisotropy. 

In the right panel of Fig.~\ref{fig6}, there is also shown the gain rate of the energy density. We note that because of the expansion, the total energy is not conserved, even when the induced current is identically zero. The energy gain shown in the figure is defined as the gain in energy density of the field modes minus the change in energy density at  vanishing current, caused by the expansion \cite{Romatschke:2006nk,Rebhan:2008uj}. The rate equals
\be
\label{REG}
R_{\,\rm energy\;gain} \; \equiv \; 
  \frac{d{\cal E}}{d\tau} + \frac{2}{\tau}{\cal E}_T,
\ee
where ${\cal E}_T$ is the transverse component of the field energy density. 

\vspace{2mm}
\underline{\it Color correlations and Abelianization}
\vspace{1mm}

The right panel of Fig.~\ref{fig7} shows various measures of the color correlations of the chromodynamic fields.  The measure of `non-Abelianness' $\bar C[j]$ is defined as in the static case Eq.~(\ref{cbareq}), but $dz$ is replaced by $d\eta$ and $L$ by $L_\eta$. In order to further study the color correlations of the chromo-fields in spatial rapidity, $\eta$, the quantity
\be
\chi_A(\xi)={N_c^2-1\02N_c}
\int_0^{L_\eta} {d\eta\0L_\eta} { \tr \left\{
(i[j_i(\eta+\xi),\mathcal U(\eta+\xi,\eta)j_j(\eta)])^2 \right\}
\0 \tr\{ j_k^2(\eta+\xi) \} \tr \{ j_l^2(\eta) \} } ,
\ee
was also defined. $\mathcal U(\eta',\eta)$ is the adjoint-representation parallel transport from $\eta$ to $\eta'$. When colors are completely uncorrelated over a distance $\xi$, this quantity equals unity; if they point in the same direction, this quantity vanishes. Then the Abelianization correlation length $\xi_A$ is defined as the smallest distance where $\chi_A$ is larger than 1/2,
\be
\xi_A[j]=\min_{\chi_A(\xi)\ge 1/2}  (\xi) ,
\ee
and compared with a general correlation length, which does not focus on color, defined through the gauge invariant function 
\be
\chi(\xi)={ \int_0^{L_\eta} {d\eta} \tr\{ j_i(\eta+\xi)
\mathcal U(\eta+\xi,\eta)j_i(\eta)\} \0
\int_0^{L_\eta} {d\eta} \tr\{ j_l(\eta) j_l(\eta)\} }  .
\ee
This function vanishes when fields are uncorrelated over a distance $\xi$, and it is normalized such that $\chi(0)=1$. Thus, the general correlation length is defined by
\be
\xi[j]=\min_{\chi(\xi)\le 1/2}  (\xi)  .
\ee
The right panel of Fig.~\ref{fig7} shows that the system becomes Abelianized with large color correlation length, $\xi_A[j]$, when the fields have grown such that nonlinear self-interactions would become important. 

\begin{figure}
\centerline
{\includegraphics[width=0.48\linewidth]{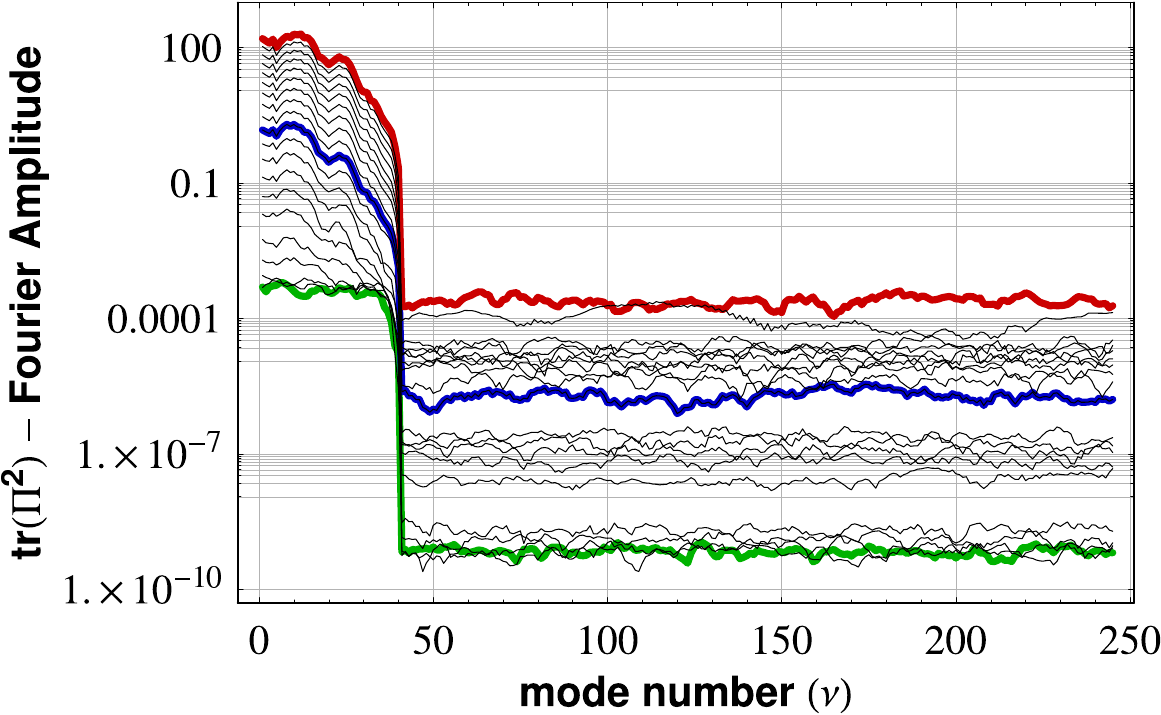}
\hspace{0.02\linewidth}
\includegraphics[width=0.48\linewidth]{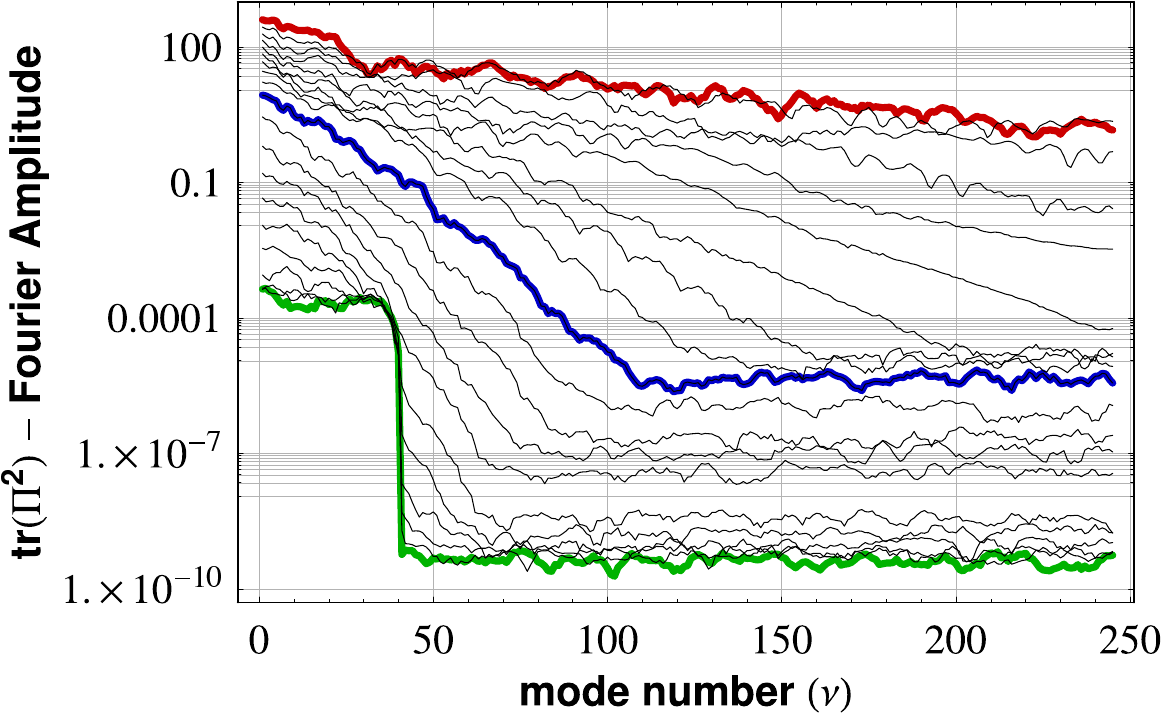}}
\caption{\label{fig9}
Fourier spectrum of the color-traced conjugate field momenta, obtained from (left) Abelian and (right) non-Abelian runs with FGM initial conditions.  The lowest (bold green) line indicates the starting spectrum and the uppermost (bold red) line 
indicates the final spectrum.  In the right panel the bold blue line indicates the `non-Abelian point' at $\tau/\tau_0 \sim$  55 when all field components become approximately the same order of magnitude. The Abelian and non-Abelian spectra were obtained by analyzing the currents produced during the runs shown in Figs.~\ref{fig6} and \ref{fig7}, respectively. Figure from \cite{Rebhan:2008uj}.}
\end{figure}

\vspace{2mm}
\underline{\it Spectral analysis}
\vspace{1mm}

In order to gain more understanding of the momentum-space dynamics of the fields, the authors of
\cite{Rebhan:2008uj} considered the quantity 
\be
\label{trPi2}
{\rm tr}\big[\mathbf{\Pi}^2 (\tau,\nu) \big] = {\rm tr}\big[\Pi^i(\tau,\nu)\, \Pi^i(\tau,\nu) 
+ \tau^2 \Pi^\eta(\tau,\nu) \, \Pi^\eta(\tau,\nu) \big],
\ee
where $\Pi^\alpha(\tau,\nu)$ is the Fourier transformed conjugate field momentum (\ref{conjugate-mom}), that is
\be
\Pi^\alpha (\tau,\nu) = \int d\eta \, e^{i\nu \eta}\, \Pi^\alpha (\tau,\eta) ,
\ee
and the trace is taken over colors. Since $\Pi^\alpha (\tau,\eta)$ is real and even as a function of $\eta$, $\Pi^\alpha (\tau,\nu)$ is also real. The quantity (\ref{trPi2}) corresponds to the energy of mode $\nu$ of the electric field in a comoving frame. 

The left panel of Fig.~\ref{fig9} shows the $\nu$-spectrum of ${\rm tr}[\mathbf{\Pi}^2]$ resulting from analysis of the induced current from the Abelian run shown in the left panel of Fig.~\ref{fig6}. In the right panel of Fig.~\ref{fig9}, the spectrum resulting from analysis of the induced current from the non-Abelian run shown in the right panel of Fig.~\ref{fig6} is given. The lowest (bold green) line indicates the starting spectrum, the bold blue line indicates the `non-Abelian point' at which all field components become approximately equal in magnitude, and the uppermost (bold red) line shows the final spectrum obtained. As can be seen from this figure, there is a strong qualitative difference between the Abelian and non-Abelian spectra with the former maintaining the spectral cutoff imposed on the initial condition and the latter `cascading' energy to higher and higher momentum modes starting already at very early times. This is similar to earlier results for the spectra induced by instability growth in a static geometry \cite{Arnold:2005qs}. 

Surprisingly, in the right panel of Fig.~\ref{fig9} one sees that, at the `non-Abelian point' indicated by the bold blue line, the low frequency modes have generated a quasi-thermal (Boltzmann) distribution up to $\nu \sim 80$. In the static case the distribution also becomes quasi thermal at intermediate times but then turns into a power-law at late times. In the expanding case the evolution is delayed due to the dynamic weakening of the fields such that the power law spectrum is expected to be achieved much later. Simulations of the classical Yang-Mills dynamics, to be discussed in the following section, indeed show that the spectrum develops a power law tail at later times, even in the expanding case. The late time regime can however not be simulated within the hard loop approximation because the assumption of negligible back reaction on the hard modes becomes invalid once the fields are sufficiently strong.



\subsection{Pure field classical simulations}
\label{subsec-purefield-class}

The hard-loop approach discussed in Secs.~\ref{subsec-hardloop} and \ref{subseq-expanding-HL} relies on a large separation of momentum scales between the soft modes corresponding to fields and the hard modes represented by quasiparticles. This requires a sufficiently small gauge coupling $g$ and thus limits applicability of the results to the weakly coupled regime. One wonders what happens when the scale separation between the hard and soft modes is no longer so large or when the initial fields have sizable amplitudes. Nonperturbative studies addressing this question have been performed in the framework of fully classical simulations. Their applicability to quark-gluon plasma from relativistic heavy-ion collisions is justified, at least partially, by an expectation of large occupation numbers of soft plasma modes which drive the system's dynamics. In the subsequent two sections we briefly review two classes of such approaches: the pure field simulations, where particles are completely absent but their role is taken by high momentum field modes, and the Wong-Yang-Mills simulations where classical color particles interact with classical chromodynamic fields.

\subsubsection{Classical-statistical lattice gauge theory}
\label{sub-sec-class-stat}

Our discussion of the pure field classical approaches starts with the classical-statistical lattice gauge theory, as formulated by Berges {\it et al.} \cite{Berges:2007re,Berges:2008zt} who studied anisotropic non-Abelian plasmas. The work begins from the Wilsonian lattice action for an ${\rm SU}(N_c)$ gauge theory in a discretized Minkowski space-time \cite{Wilson:1974sk}. Since the work's objective is to simulate the temporal evolution of the plasma, the action is written in the form appropriate for the Hamiltonian formulation by  \cite{Kogut:1974ag} where the lattice spacing in the time direction $\Delta t$ differs from that in the spatial directions $a$. The action reads
\begin{equation}
\label{eq:latticeaction}
S[U]=-\beta_0\sum_x\sum_{i=1}^3
\left[\frac{1}{2{\rm Tr}\mathds{1}}\left({\rm Tr}\,U_{0,i}(x)+ {\rm Tr}\,U^\dag_{0,i}(x)\right)-1\right]
+\beta_s\sum_x \sum_{\genfrac{}{}{0pt}{1}{i,j=1}{i<j}}^3 \bigg[\frac{1}{2{\rm Tr}\mathds{1}}\left({\rm Tr}\,U_{i,j}(x)+{\rm Tr}\,U^\dag_{i,j}(x)\right)-1\bigg] ,
\end{equation}
where the plaquette $U_{\mu,\nu}(x)$ is defined as
\begin{equation}
\label{eq:plaquette}
U_{\mu,\nu}(x)=U_\mu(x)U_\nu(x+\hat{\mu})U_\mu^\dag(x+\hat{\nu})U_\nu^\dag(x),
\end{equation}
with the link or parallel transporter given by
\begin{equation}\label{linkdef}
U_\mu(x)=e^{ig a^\mu A_\mu(x)} .
\end{equation}
The index $\mu$ on the link $U_\mu$ is not a Lorentz index, but merely indicates the link's direction. The shift $\hat{\mu}$ is one lattice spacing $a^\mu$ in length with $a^0 = \Delta t$ and $a^i =a, \;\; i=1,2,3$. The gauge potential is in the fundamental representation, see Sec.~\ref{subsec-QCD-class}, and so are the links (\ref{eq:plaquette}). Consequently, ${\rm Tr}\mathds{1}= N_c$. The sum in Eq.~(\ref{eq:latticeaction}) is performed over all elementary squares - plaquettes - in the discretized Minkowski space-time. The parameters $\beta_0$ and $\beta_s$  are chosen as
\begin{equation}
\beta_0 \equiv \frac{2\,a\,{\rm Tr}\mathds{1}}{g^2\,\Delta t} \, ,
~~~~~~~~~~
\beta_s \equiv \frac{2\,\Delta t\,{\rm Tr}\mathds{1}}{g^2\,a} \, ,
\label{eq:ganisoM}
\end{equation}
to obtain a proper continuum limit. Indeed, when $\Delta t \rightarrow 0$ and $a \rightarrow 0$, the action (\ref{eq:latticeaction}) with the constants (\ref{eq:ganisoM}) reduces to the classical action of gauge fields corresponding to the first term of Eq.~(\ref{lagrangian-QCD}). As is well known, the lattice action (\ref{eq:latticeaction}) violates translational and rotational symmetries of space-time but the gauge symmetry is exactly preserved. 

To proceed toward the Hamiltonian formulation, one goes to the temporal axial gauge, $A_0=0$. In the continuum limit, the canonical variables and conjugate momenta are then the gauge potentials and electric fields. In the lattice formulation, the gauge $A_0=0$ fixes all temporal links to identity, $U_0(x)=\mathds{1}$, and one uses as canonical variables and conjugate momenta  the spatial links $U_j(x)$ and temporal plaquettes $U_{0,j}(x)$ which are related to the gauge potentials and electric fields in the  adjoint representation as
\ba
\label{gauge-field-FCS}
A_j^b(x) &=& \frac{2 i}{a g} \, \tr \big(\tau^b U_j(x) \big)  ,
\\
\label{electric-field-FCS}
E^b_j(x)  &=& \frac{2 i}{a \Delta t \, g} \, \tr \big(\tau^b U_{0,j}(x) \big) .
\ea
The relations (\ref{gauge-field-FCS}) and (\ref{electric-field-FCS}) hold, strictly speaking, when $\Delta t \rightarrow 0$ and $a \rightarrow 0$. We note that the lower and upper spatial indices $i,j$ are not distinguished in this subsection.

Varying the action (\ref{eq:latticeaction}) with respect to $U_j(x)$ yields a leapfrog-type update rule for the electric field
\begin{eqnarray}
\nn
E^b_j(t + \Delta t, {\bf x})  =  E^b_j(t, {\bf x}) 
&+& \frac{i \Delta t}{g a^3} \sum_{k}  \Big[ \tr \big[\tau^b U_j(x) U_k (x +\hat{j}) U^{\dagger}_j (x + \hat{k}) U^{\dagger}_k(x) \big] 
\\ \label{eq:bergesEOM}
&&~~~~~~~~~ 
+  \tr \big[\tau^b U_j(x) U^{\dagger}_{k}(x+\hat{j}-\hat{k}) U^\dagger_{j}(x-\hat{k}) U_{k}( x -\hat{k}) \big] \Big]   .
\end{eqnarray}
Therefore, $E^b_j(t + \Delta t, {\bf x})$ can be computed from the electric fields and the spatial links at time $t$. Then, one computes the temporal plaquette $U_{0,j}(x)$ satisfying Eq.~(\ref{electric-field-FCS}) which is further used to evolve the links as
\begin{equation}
 U_j(t+\Delta t,\mathbf{x})=U_{0,j}(x) \,U_j(x).
\end{equation}
Repeating the three steps $n$-times, one gets the electric fields and links at the time $t+n\Delta t$. It should be remembered that, throughout the whole temporal evolution, the fields and links must obey the Gauss law constraint
\begin{equation}
\label{Gauss-law-lattice}
  \sum_{j=1}^3\left(E_j^b(x)-U^\dag_{j}(x-\hat{j})E_j^b(x-\hat{j})U_j(x-\hat{j})\right)=0,
\end{equation}
which is obtained by varying the action (\ref{eq:latticeaction}) with respect to temporal links $U_0(x)$ before the gauge condition $A^0=0$ is imposed. The coupling $g$ can be scaled out of the equations of motion and one typically puts
$g = 1$ for the simulations. While $g$ is irrelevant for the classical dynamics, it reappears when calculating physical observables.

From the action (\ref{eq:latticeaction}) one can derive the energy density (see {\it e.g.} \cite{Montvay:1997}) expressed through the spatial links and temporal plaquette.  In the case of the $\text{SU}(2)$ gauge group, the energy density is
\begin{equation}
\label{DefEnergyDensitySU2}
\varepsilon (x) = \frac{1}{4g^2 a^4}\bigg[ \frac{a^2}{\Delta t^2}\sum_{j=1}^3 \Big( 2- \tr U_{0,j}(x) \Big)
  + \sum_{\genfrac{}{}{0pt}{1}{j,k=1}{j<k}}^3 \Big( 2 - \tr U_{j,k}(x)\Big)\bigg] ,
\end{equation}
and for $\text{SU}(3)$ it reads
\begin{equation}
\label{DefEnergyDensitySU3}
\varepsilon (x) 
=\frac{1}{g^2 a^4}\bigg[ \frac{a^2}{\Delta t^2}
\sum_{j=1}^3\Big(6 - \tr U_{0,j}(x) - \tr U_{j,0}(x) \Big)
+ \sum_{\genfrac{}{}{0pt}{1}{j,k=1}{j<k}}^3 \Big(6 - \tr U_{j,k}(x) - \tr U_{k,j}(x) \Big)\bigg] .
\end{equation}

\begin{figure}[t]
\vspace{-3mm}
\begin{minipage}{8.5cm}
\center
\includegraphics[width=1.0\textwidth]{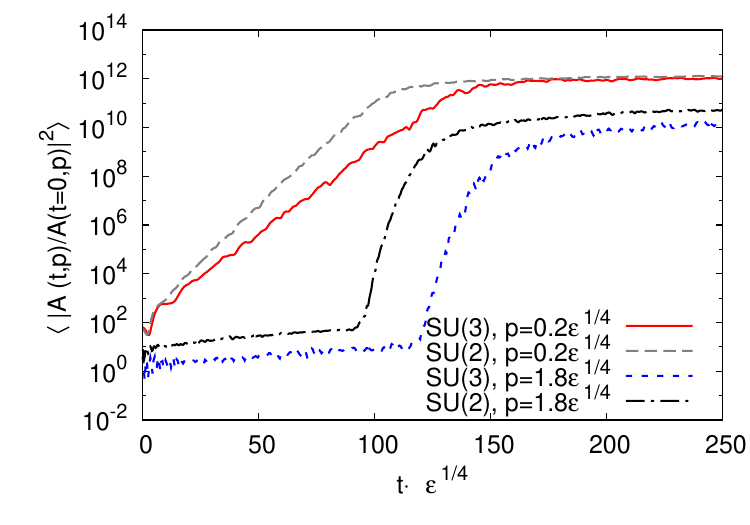}
\vspace{-5mm}
\caption{\label{fig-A-vs-t}
Fourier coefficients of the squared modulus of the gauge field versus time for two different momenta. Compared is the time evolution of $\text{SU}(2)$ and $\text{SU}(3)$ gauge fields for the same energy density $\varepsilon$. Figure from \cite{Berges:2008zt}.}
\end{minipage}
\hspace{3mm}
\begin{minipage}{8.5cm}
\vspace{-7mm}
\center
\includegraphics[width=1.05\textwidth]{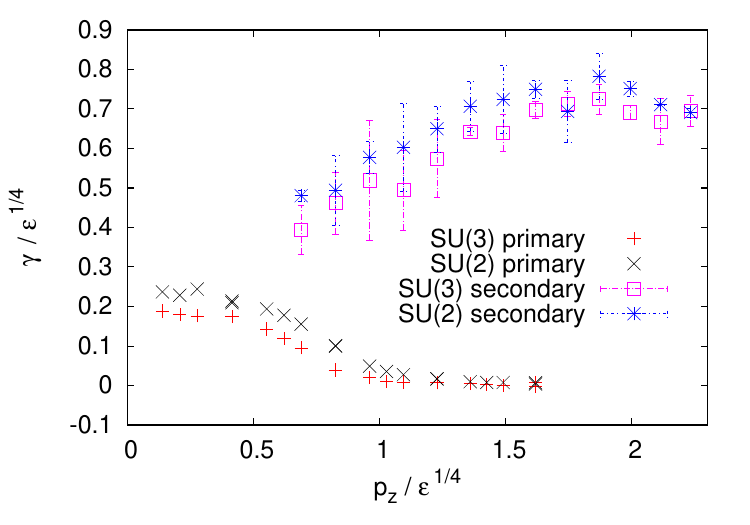}
\vspace{-8mm}
\caption{\label{fig-rates}
The primary and secondary growth rates for the $\text{SU}(2)$ and $\text{SU}(3)$ gauge theory for the same energy density. Figure from \cite{Berges:2008zt}.}
\end{minipage}
\end{figure}

To generate an unstable system, the gauge potentials, which are Fourier transformed with respect to the spatial variable ${\bf x}$, are initialized in \cite{Berges:2007re,Berges:2008zt} with an anisotropic distribution 
\begin{equation}
\label{initial-dis-A}
\langle |A_j^b(t=0,\mathbf{p})|^2\rangle=\frac{\tilde{A}^2}{(2\pi)^{3/2}\Delta^2\Delta_z}
\exp\left(-\frac{p_x^2+p_y^2}{2\Delta^2}-\frac{p_z^2}{2\Delta_z^2}\right) ,
\end{equation}
with $\Delta_z\ll\Delta$, which corresponds to the strongly oblate momentum distribution. The parameter $\Delta$ controls the transverse momentum of gluons and can be associated with the saturation scale $Q_s$. The equations of motion are solved numerically for a set of initial configurations sampled according to Eq.~(\ref{initial-dis-A}) and expectation values are computed as averages over the results from the individual runs.

Fig.~\ref{fig-A-vs-t} shows the temporal evolution of the color-averaged squared modulus of the Fourier transform of the gauge field $A(t, {\bf p})$ in three spatial dimensions for two different momenta ${\bf p}$ parallel to the $z$-axis. They are displayed as a function of time, normalized by the corresponding initial values. For comparison, the $\text{SU}(3)$ results are shown together with the corresponding $\text{SU}(2)$ results. The plotted low-momentum modes clearly show exponential growth starting at the very beginning of the simulation. Initially, only the band of unstable modes (see {\it e.g.} Fig.~\ref{fig-largexiGamma}) grows but at later times the cascading toward higher modes begins. This manifests itself in two stages of growth for the intermediate modes, which have been dubbed `primary' and `secondary' instabilities by \cite{Berges:2007re,Berges:2008zt}. For both gauge groups the behavior is qualitatively very similar. 

Fig.~\ref{fig-rates} displays the momentum dependence of the growth rates for $\langle |A(t, \mathbf{p})|^2 \rangle$ obtained from an exponential fit, which is done separately for the primary and secondary growth rates. One observes that while the primary rates are approximately 25\% bigger for $\text{SU}(2)$, the secondary rates agree within the given errors. The primary growth rates in $\text{SU}(2)$ and $\text{SU}(3)$ differ because, when initializing at the same energy density, the initial field amplitudes in the two theories are different. This leads to a different characteristic self-energy contribution to the primary growth rate, explaining the observed difference between the gauge theories.

In order to obtain an estimate of the growth rate in physical units for the $\text{SU}(3)$ case, the simulations were performed in \cite{Berges:2007re,Berges:2008zt} for initial energy densities from the interval $5\div 100$ GeV/fm$^3$ which covers the expected range of RHIC and LHC experiments. The inverse of the maximum primary growth rate for $| A(t, \mathbf{p}) |^2$ is 
\be
\label{inverse-primary-gr}
\gamma_{\text{max}}^{-1}  \simeq  1.3 \div 2.4 \;\; {\rm fm}/c .
\ee
The result is rather insensitive to the initial energy density because it scales with the fourth root of $\varepsilon$. 

Because the secondary growth rates are larger, a certain range of higher momentum modes can `catch up' with initially faster growing infrared modes before the exponential growth stops, as seen in Fig.~\ref{fig-A-vs-t}. This leads to a relatively fast effective isotropization of a finite momentum range, while higher momentum modes do not isotropize on a time scale characterized by plasma instabilities.
 
\begin{figure}[t]
\begin{center}
\vspace{-3mm}
\includegraphics[width=0.5\linewidth]{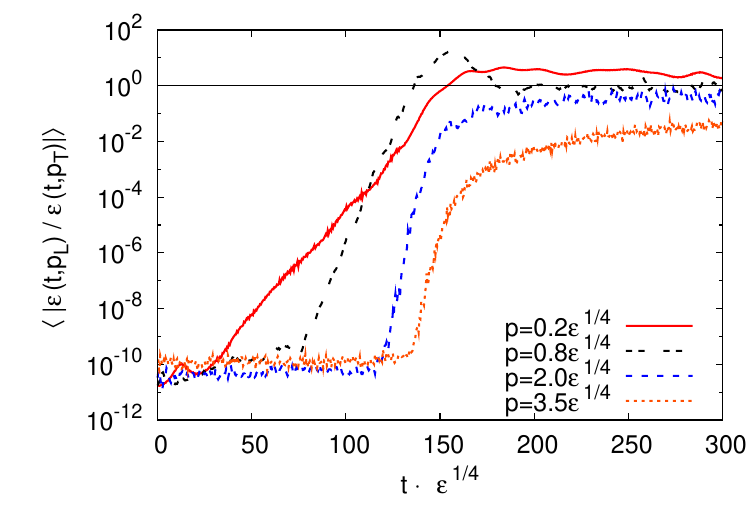}
\vspace{-5mm}
\caption{The ratio of the longitudinal and transverse modes of the Fourier transformed energy density for several momenta. Figure from \cite{Berges:2008zt}. }
\label{fig-pressure}
\end{center}
\end{figure}

To be more quantitative, let us review the results on isotropization from \cite{Berges:2008zt}. In terms of the local energy density~\eqref{DefEnergyDensitySU3}, the isotropization can be quantified by the ratio of spatially Fourier transformed energy densities 
\begin{equation}
\label{eq-edens-ratios}
\frac{\varepsilon (t, \mathbf{p}_L ) }{\varepsilon (t, \mathbf{p}_T)} 
~~ \textrm{with } ~~
\mathbf{p}_L \parallel \hat{\mathbf{z}} 
~~~ \& ~~~
 \mathbf{p}_T \perp \hat{\mathbf{z}} 
~~~ \& ~~~
|\mathbf{p}_L | = |\mathbf{p}_T|.
\end{equation}
If, at some time $t$, the system reaches a truly isotropic state, the mean absolute value of the ratio~\eqref{eq-edens-ratios} is one for all momenta. For the $\text{SU}(3)$ case the time evolution of the ratio is depicted in Fig.~\ref{fig-pressure} for several momenta. One observes that the low-momentum modes of the energy density isotropize, while the high-momentum modes remain anisotropic after the instability growth ends. The Fourier components of $\varepsilon$ become isotropic up to approximately $|\mathbf{p}|/\varepsilon^{1/4} \simeq 1 \div 2$ at the time of saturation of $t \simeq 150 \; \varepsilon^{-1/4}$. Using the same values for the physical energy density $\varepsilon$ as for the maximal growth rate \eqref{inverse-primary-gr}, one finds effective isotropization up to a characteristic momentum of about
\begin{eqnarray}
|{\bf p}|  \lesssim  1 \;\; {\rm GeV} ,
\end{eqnarray}
which is approximately the same as in the case of SU(2) gauge theory \cite{Berges:2007re}. Although complete isotropization does not occur on short timescales, the effectively isotropic pressure in the low-momentum range is a crucial ingredient for the validity of the hydrodynamic description of the plasma. Complete isotropization may not be necessary to account for its successful applications to experimental data from RHIC and LHC. 

Recent developments in the classical-statistical lattice simulations are briefly reviewed in Sec.~\ref{subsec-recent-prog}.

\subsubsection{Color glass condensate and glasma}
\label{sub-sec-CGC}

We are now going to discuss not just a method to simulate unstable quark-gluon plasma but an effective theory of the early stage of relativistic heavy-ion collisions which is known as the Color Glass Condensate (CGC). For a broad presentation, see the review articles \cite{Iancu:2003xm,Gelis:2010nm,Gelis:2012ri}. 

The term CGC refers to the state of matter produced in high-energy collisions: `color' because gluons are colored, `glass' in analogy to the disordered systems like actual glasses, and `condensate' due to very high occupation numbers of soft gluons resembling a Bose condensate. Since the density of small $x$ or `wee' partons per transverse area of colliding nuclei is expected to be large, the corresponding momentum scale $Q_s$ is also large, when compared to the QCD scale parameter $\Lambda_{\rm QCD}$. Consequently, the coupling constant $\alpha_s$ is presumably sufficiently small for an applicability of perturbative methods. However, the system is not weakly but rather strongly interacting because of the high occupation numbers of color charges.

Our attention here will be focused on the `glasma' - the phase created by two colliding CGCs that precedes the formation of an equilibrated quark-gluon plasma. It is rather a system of strongly interacting classical chromodynamic fields than a dilute gas of partons. Similarly to the systems discussed in Sec.~\ref{subseq-expanding-HL}, the glasma is strongly anisotropic and experiences rapid longitudinal expansion. There is no surprise that plasma instabilities, which were studied in \cite{Romatschke:2005pm,Romatschke:2006nk,Fukushima:2007ja}, play an important role in the glasma dynamics. We review the work of \cite{Romatschke:2005pm,Romatschke:2006nk} where 3+1-dimensional numerical simulations of the ${\rm SU}(2)$ Yang-Mills equations with CGC initial conditions were performed. Remarkably, arbitrarily weak violations of boost invariance were found to trigger the unstable Weibel modes which grow in the same way, as seen in the case of the hard-loop simulations of an expanding QGP discussed in Sec.~\ref{subseq-expanding-HL}.

We begin this section by briefly introducing the framework of color glass condensate which is used further on. In nuclear collisions at very high energies, the large-$x$ valence partons in each of the nuclei, where $x$ is the fraction of the nucleon's momentum carried by a parton, act as highly Lorentz contracted color charge sources for small $x$ soft gluon modes. These modes, which are in the first approximation treated within the Weizs\"acker-Williams approach, are coherent across the longitudinal extent of the nucleus. With increasing collision energy, the scale separating soft and hard modes shifts toward smaller values of $x$. The modification of the sources with this change is quantified by a Wilsonian renormalization group procedure.

Let us consider more quantitatively a collision of two identical nuclei, which move toward one another along the axis $z$, in the center of mass frame. Both nuclei are so energetic that they can be treated as infinitesimally thin sheets moving with the speed of light. They are large enough that these sheets can be taken to be of infinite extent in the transverse direction. The process is naturally described using the light-cone variables. Then, the four-position $x^\mu =(t, {\bf x}_\perp, z)$, as any other four-vector, is written as  
\be 
x^\mu = (x^+, x^-,{\bf x}_\perp) ~~~~~ {\rm with} ~~~~~ x^\pm \equiv \frac{t\pm z}{\sqrt{2}}. 
\ee
Since the four-position with the lower index is 
\be 
x_\mu = (x_+, x_-,- {\bf x}_\perp) = (x^-, x^+,- {\bf x}_\perp) ,
\ee
the scalar product of two four-vectors $x$ and $y$ equals
\be
x \cdot y = x^+y_+ + x^- y_- - {\bf x}_\perp \cdot {\bf y}_\perp = x^+y^- + x^- y^+ - {\bf x}_\perp \cdot {\bf y}_\perp .
\ee 
The d'Alembertian is expressed as
\be
\label{partial-LC}
\square = 2  \partial^+  \partial^- - \nabla_\perp^2 
~~~~~ {\rm with} ~~~~~
\partial^\pm \equiv \frac{\partial}{\partial x_\pm} =\frac{\partial}{\partial x^\mp} .
\ee
For a highly boosted particle in the positive direction of the $z$-axis, $x^+$ plays the role of time.

The large-$x$ partons of incoming nuclei - mostly valence quarks - are described by the current  
\beq
j^{\mu,a}(x)=\delta^{\mu +} \rho_{1}^a({\bf x}_\perp) \,\delta(x^-)
+ \delta^{\mu -} \rho_{2}^a({\bf x}_\perp) \, \delta(x^+) ,
\label{current-CGC}
\eeq
where the color charge densities $\rho_{1,2}^a$ of the two nuclei act as independent sources of gluons. The sources are static in a sense that they do not depend on the light cone time $x^+$ or $x^-$. The functions $\delta(x^\pm)$ appear because Lorentz contraction squeezes the nuclei to infinitesimally thin sheets. Under this assumption, the gauge fields generated by the current (\ref{current-CGC}), which are found by solving the classical Yang-Mills equations (\ref{YM-eq}), are boost invariant, that is they are independent of the space-time rapidity 
\be
\label{space-time-eta}
\eta \equiv \frac{1}{2}\ln \frac{x^+}{x^-}.
\ee

Gluon distributions are obtained from the Fourier transform of the solution of the Yang-Mills equations $A_\mu({\bf k}_\perp)$ using $\langle A_\mu({\bf k}_\perp) A_\mu^*({\bf k}_\perp)\rangle_\rho$, where the averaging over classical charge distributions is defined by
\be
\langle O\rangle_\rho = \int {\cal D}\rho_{1}\, {\cal D}\rho_{2}\, O[\rho_1,\rho_2] \,
\exp\bigg( -\int d^2 x_\perp {{\rm Tr}\left[\rho_1^2({\bf x}_\perp)+\rho_2^2({\bf x}_\perp) \right] \over {2\mu^2}}\bigg) .
\ee
${\cal D}\rho$ denotes here functional integration. The averaging is performed independently for each nucleus with equal Gaussian weight $\mu^2$ which is the average color charge squared per unit area in each of the nuclei. Since the average color charge vanishes and the average color charge squared results from random fluctuations,  $\mu^2$ scales with the atomic number $A$ as $\mu^2 \sim A^{1/3}$. 

For very large nuclei, $\mu$ can be much larger than the QCD scale parameter, $\mu^2 \gg \Lambda_{\rm QCD}^2$, which allows for the application of weak coupling techniques to the problem. Such a Gaussian weight is justified \cite{McLerran:1993ni,Kovchegov:1996ty,Jeon:2004rk,Jeon:2005cf} in the limit of $A \gg 1$ and $\alpha_s Y \ll 1$, where $Y$ is the momentum-space rapidity, providing a window $\ln(A^{1/3}) < Y < A^{1/6}$ in $Y$, which exists for large $A$. This window of applicability can be extended to larger values of $Y$ by using the Balitsky-Kovchegov \cite{Balitsky:1995ub,Kovchegov:1999ua} or JIMWLK \cite{Balitsky:1995ub,Jalilian-Marian:1997xn,Jalilian-Marian:1997jx,JalilianMarian:1998cb,Iancu:2000hn,Ferreiro:2001qy} equation to evolve the sources $\rho_{1,2}$ to higher rapidities. 

Before discussing a solution of the Yang-Mills equations with the current (\ref{current-CGC}), we consider, following \cite{McLerran:1993ka}, a simplified Abelian problem of a single nucleus moving along the axis $z$ with the speed of light. In other words, we are going to solve the Maxwell equations $\partial_\mu F^{\mu \nu}=j^\nu$ with the current given by the first term of  Eq.~(\ref{current-CGC}).  The solution is not only instructive but it played an important role in formulating the CGC approach. Choosing the gauge condition $A^+(x)=0$, it follows that $A^-(x) =0$, and the three Maxwell equations corresponding to $\nu = \pm, \perp$ are
\ba
\label{Max-LC-1}
\partial^+ \nabla_\perp \!\cdot\!  {\bf A}_\perp(x) &=& -\rho ({\bf x}_\perp) \,\delta(x^-),
\\
\label{Max-LC-2}
\partial^- \nabla_\perp \!\cdot\! {\bf A}_\perp(x) &=& 0 ,
\\
\label{Max-LC-3}
\big(2  \partial^+  \partial^- - \nabla_\perp^2 \big) {\bf A}_\perp(x)  
+ \nabla_\perp \big( \nabla_\perp \!\cdot\! {\bf A}_\perp(x) \big)&=& 0 .
\ea
Keeping in mind the formulas (\ref{partial-LC}) and the relation $\partial^+ \theta(x^-) = \delta (x^-)$, one easily checks that the equations (\ref{Max-LC-1}), (\ref{Max-LC-2}), and (\ref{Max-LC-3}) are solved by
\be
\label{solution-MV}
{\bf A}_\perp(x) = \frac{1}{g} \,\theta(x^-) \, \nabla_\perp \Lambda ({\bf x}_\perp) ,
\ee
if the function $\Lambda ({\bf x}_\perp)$ obeys the two-dimensional Poisson equation 
\be
\nabla_\perp^2 \Lambda ({\bf x}_\perp) = - g \rho ({\bf x}_\perp). 
\ee

Let us discuss the solution (\ref{solution-MV}). Due to causality, it vanishes for $x^- < 0$ that is the field ${\bf A}_\perp(x)$ is nonzero only `behind' the nucleus for $z < t$ which in physical units would read $z < ct$. In the transverse plane the potential (\ref{solution-MV}) is a pure gauge, that is $F^{ij}=0$. The electric and magnetic fields corresponding to the potential (\ref{solution-MV}) are purely transverse and they read
\be
\label{E-B-MV}
{\bf E}_\perp(x) = -\frac{\delta(x^-) }{\sqrt{2}} \, \nabla_\perp \Lambda ({\bf x}_\perp) ,
~~~~~~~~
{\bf B}_\perp(x) = \hat{\bf z} \times {\bf E}_\perp(x) ,
\ee
where $\hat{\bf z}$ is the unit vector along the axis $z$. The fields ${\bf E}_\perp(x)$ and ${\bf B}_\perp(x)$ are confined to the sheet $z=t$ of the infinitely contracted nucleus moving with the speed of light. We also note that since the solution (\ref{solution-MV}) linearly depends on the charge density $\rho$, it vanishes everywhere when averaged over an ensemble of charge densities because $\langle \rho \rangle =0$. 

We are now ready to consider the Yang-Mills equations (\ref{YM-eq}) with the current (\ref{current-CGC}). Before the collision, at $t<0$, and in the domains which are causally disconnected with the collision point $t=z=0$, that is, everywhere except the forward light cone, where $x^+ > 0$ and $x^- > 0$, the field should be a superposition of fields generated by each incoming nucleus separately.  Indeed, the solution in this region can be written as \cite{Kovner:1995ts,Kovner:1995ja} 
\be
A^{\pm}(x) =0 , ~~~~
A^i(x) = \theta_\epsilon(x^-)\theta_\epsilon(-x^+)\alpha_1^i({\bf x}_\perp)
+ \theta_\epsilon(x^+)\theta_\epsilon(-x^-) \alpha_2^i({\bf x}_\perp) ,
\label{befsoln}
\ee
where the spatial coordinates transverse to the $z$-axis are labeled by $i$. The $\epsilon$ subscripts on the $\theta$-functions denote that they are smeared by an amount $\epsilon$ in the respective $x^\pm$ light cone directions. The functions $\alpha_{m}^i({\bf x}_\perp)$, where the index $m=1,2$ labels the colliding nuclei, obey $F^{ij}=0$, and they satisfy the equation
\beq
D_i \alpha_m^i({\bf x}_\perp) = - \rho_m({\bf x}_\perp) ,
\eeq
where $D_i = \partial_i - i g A_i = \partial_i -i \partial_i\Lambda$, which has an analytical solution given by 
\beq
\alpha^i_m({\bf x}_\perp) = \frac{-i}{g} e^{i \Lambda_m({\bf x}_\perp)} \partial^i e^{- i \Lambda_m({\bf x}_\perp)}, 
~~~~~~
\nabla^2_\perp \Lambda_m({\bf x}_\perp) = -g\, \rho_m ({\bf x}_\perp).
\eeq
Note that formally $\Lambda_m$ depends on $x^+$ or $x^-$, depending on the nucleus, with support over the range $\epsilon$. 
Thus, path ordering in $x^\pm$ for nuclei 1 and 2 is assumed and the limit $\epsilon\rightarrow 0$ is taken in the end.

To solve the Yang-Mills equations in the forward light cone, where the actual interaction of colliding nuclei proceeds, the initial conditions for the evolution of the gauge field have to be chosen. They are formulated on the proper time surface $\tau =0$, with 
\be
\label{proper-time-LC}
\tau \equiv \sqrt{t^2-z^2}=\sqrt{2 x^+x^-}. 
\ee
The initial fields are obtained by generalizing the ansatz (\ref{befsoln}) to \cite{Kovner:1995ts,Kovner:1995ja} 
\bqa
A^{\pm}(x) &=&\pm x^\pm \theta_\epsilon(x^-)\theta_\epsilon(x^+) \beta(x),
\label{genansatz-1}
\\[1mm]
A^i(x) &=& \theta_\epsilon(x^-)\theta_\epsilon(-x^+) \alpha_1^i({\bf x}_\perp)
+\theta_\epsilon(-x^-) \theta_\epsilon(x^+) \alpha_2^i({\bf x}_\perp) 
+\theta_\epsilon(x^-)\theta_\epsilon(x^+) \alpha^i_3(x) 
\label{genansatz-2}
\eqa
with the Fock-Schwinger gauge condition $A^\tau \equiv x^+ A^- + x^- A^+ =0$. To obtain the unknown gauge fields $\alpha_3(x)$, $\beta(x)$ in the forward light cone from the known gauge fields $\alpha_{1,2}$ of the respective nuclei before the collision, one has to invoke a physical `matching condition' which requires that the Yang-Mills equations (\ref{YM-eq}) be regular at $\tau=0$. The $\delta$-functions of the current in the Yang-Mills equations have to cancel identical terms in spatial derivatives of the field strengths. This leads to the unique solution \cite{Kovner:1995ts,Kovner:1995ja} 
\beq
\alpha_3^i(x)=\alpha^i_1({\bf x}_\perp)+\alpha^i_2({\bf x}_\perp),
~~~~~~~~ 
\beta(x)=-\frac{i}{2}\, g
\big[\alpha_{i,1}({\bf x}_\perp),\alpha_2^i( {\bf x}_\perp) \big].
\label{matcond}
\ee
In addition, the conditions on the derivatives of the fields that lead to regular solutions are $\partial_\tau \beta|_{\tau=0} = \partial_\tau \alpha_m^i |_{\tau=0} =0$. 

The Yang-Mills equations (\ref{YM-eq}) with the current (\ref{current-CGC}) can be solved perturbatively \cite{Kovner:1995ts,Kovner:1995ja,Kovchegov:1997ke,Gyulassy:1997vt} in the limit $\alpha_s \mu \ll k_\perp$. However, the general classical solutions for $\tau>0$ have to be determined numerically. To do this, one can solve Hamilton's equations of motion on the lattice. Therefore, one first has to construct a lattice Hamiltonian.

The gluonic part of the QCD action in general coordinates takes the form
\beq
S =\int d\tau \, d\eta \, d^2x_\perp \, {\mathcal L}
=- \frac{1}{2}\int d\tau \, d\eta \, d^2x_\perp \, \sqrt{-{\rm det} g_{\mu \nu}} \;
{\rm Tr} \left[F_{\mu \nu} g^{\mu \alpha} g^{\nu \beta} F_{\alpha \beta}  + 
j_{\mu} g^{\mu \nu} A_{\nu} \right] ,
\label{action}
\eeq
where the trace is taken over color indices. Working as in Sec.~\ref{subseq-expanding-HL} in comoving coordinates, where $x = (\tau, {\bf x}_\perp, \eta)$ with the proper time (\ref{proper-time-LC}) and space-time rapidity (\ref{space-time-eta}), and using the gauge condition $x^+ A^-+x^- A^+=0$, which is equivalent to $A^\tau=0$, the Lagrangian density reads
\beq
\label{Lagran-CGC}
{\mathcal L} = \tau \, {\rm Tr} \Big[
\frac{1}{\tau^2}\, F_{\tau \eta}F^{\tau \eta} + F_{\tau i} F^{\tau i} 
-\frac{1}{\tau^2} \, F_{\eta i} F^{\eta i} -\frac{1}{2}\,F_{i j}F^{i j} 
+\frac{1}{\tau^2} \, j_\eta A_\eta \Big] .
\eeq
We note that because the integration measure $d\tau \,d\eta \, d^2x_\perp$ has the dimension $m^{-3}$, where $m$ is any mass parameter, the Lagrangian density (\ref{Lagran-CGC}) is not of the dimension $m^4$ but $m^3$, and so is the Hamiltonian density (\ref{contHamil}). The contribution of the hard valence current to ${\mathcal L}$ drops out in the boost-invariant case and can be neglected as long as one only considers a small region around central space-time rapidities $\eta=0$. 

Choosing the potentials $A_i$ and $A_\eta$ as canonical variables, the conjugate momenta are
\beq
E_i=\frac{\partial {\mathcal L}}{\partial (\partial_\tau A_i)}=\tau
\partial_\tau A_i\,,\qquad E_\eta=\frac{\partial {\mathcal L}}{\partial
(\partial_\tau A_\eta)}=\frac{1}{\tau}
\partial_\tau A_\eta\,,
\eeq
and the Hamiltonian density equals
\beq
{\mathcal H}= {\rm Tr} \left[ E_i(\partial_\tau A_i) 
+E_\eta(\partial_\tau A_\eta) - {\mathcal L} \right]
={\rm Tr} \bigg[ \frac{E_i^2}{\tau} + \frac{F_{\eta i}^2}{\tau} 
+ \tau E_\eta^2 + \tau F_{ij}^2 \bigg].
\label{contHamil}
\eeq
The canonical equations of motion 
\beq
\partial_\tau A_\mu  = \frac{\partial {\mathcal H}}{\partial E_\mu} ,
~~~~~~
\partial_\tau E_\mu = - \frac{\partial {\mathcal H}}{\partial A_\mu} ,
\eeq
where $\mu = i$ or $\eta$, read explicitly
\ba
\partial_\tau A_i &=& \frac{E_i}{\tau},  
~~~~~~~~~~~~~~~~~~~~~~~~~~~
\partial_\tau A_\eta = \tau E_\eta ,
\\[2mm]
\partial_\tau E_i &=& \tau D_j F_{ji} + \frac{1}{\tau} D_\eta F_{\eta i} ,
~~~~~~~
\partial_\tau E_\eta =\frac{1}{\tau} D_j F_{j\eta}.
\ea
The Gauss law constraint is simply
\beq
D_i E_i + D_\eta E_\eta =0 \, .
\eeq

As already mentioned, the system is boost invariant, if the color current is delta-like in $x^+$ and $x^-$, as the current (\ref{current-CGC}). Then, the gauge fields are independent of the space-time rapidity $\eta$. The boost-invariant initial conditions for the fields, which are derived from Eqs.~(\ref{genansatz-1}) - (\ref{matcond}) and the requirement that the fields are regular at $\tau=0$, are
\ba
A^i(\tau=0, {\bf x}_\perp) &=& \alpha_1^i({\bf x}_\perp) + \alpha^i_2({\bf x}_\perp), 
~~~~~~~~~~
A_\eta(\tau=0, {\bf x}_\perp)=0, 
\label{initcond-1}
\\[1mm]
E^i(\tau=0, {\bf x}_\perp) &=& 0, 
~~~~~~~
E_\eta(\tau=0, {\bf x}_\perp) = i g\, [\alpha_1^i({\bf x}_\perp),\alpha_2^i({\bf x}_\perp)] \,.
\label{initcond-2}
\ea
The components of the magnetic field corresponding to these initial conditions are $B_\eta \neq 0$ and $B_i =0$. Therefore, the initial transverse electric and magnetic fields vanish while $E_\eta, B_\eta$ are nonzero. This is in sharp contrast to the electric and magnetic fields before the collision (\ref{E-B-MV}) which are purely transverse.

The boost invariance cannot be exact in heavy-ion collisions, because a heavy nucleus will never be Lorentz contracted to an infinitely thin sheet and quantum fluctuations, which cannot be infinitely well localized, are important as well. Hence, one should take into account the finite width of the nuclei in $x^\pm$. Relaxing of the boost-invariance requirement is also crucial to investigate the role of plasma instabilities in heavy-ion collisions. The point is that the wave vectors of unstable modes in a plasma system with oblate momentum distribution have a component parallel to the beam direction, see Fig.~\ref{fig-configurations}. Therefore, the assumption of boost invariance which eliminates any inhomogeneity along the beam also eliminates the unstable modes. For this reason the assumption was relaxed in \cite{Romatschke:2005pm,Romatschke:2006nk}. 

Since nuclei of finite size are still highly localized on the light cone, the initial conditions (\ref{initcond-1}) and (\ref{initcond-2}) are expected to be approximately valid. The qualitative difference to the idealized boost-invariant case is the appearance of rapidity fluctuations. These were included in \cite{Romatschke:2005pm} by modifying the boost-invariant initial conditions to 
\be
E_i(\tau=0, {\bf x}_\perp,\eta) = \delta E_i({\bf x}_\perp,\eta), 
~~~~~~~
E_\eta(\tau=0, {\bf x}_\perp,\eta) = i g\, [\alpha_1^i({\bf x}_\perp),\alpha_2^i({\bf x}_\perp)] 
+\delta E_\eta ({\bf x}_\perp,\eta) .
\label{initcond-NBI}
\ee
while keeping $A_i$ and $A_\eta$ unchanged. The perturbations $\delta E_i$ and $\delta E_\eta$ were assumed to satisfy the Gauss' law
\beq
\label{Gauss-RV}
D_i \delta E_i+D_\eta E_\eta = 0  .
\eeq

\begin{figure}[t]
\begin{minipage}{8.5cm}
\vspace{-10mm}
\begin{center}
\includegraphics[width=1.07\textwidth]{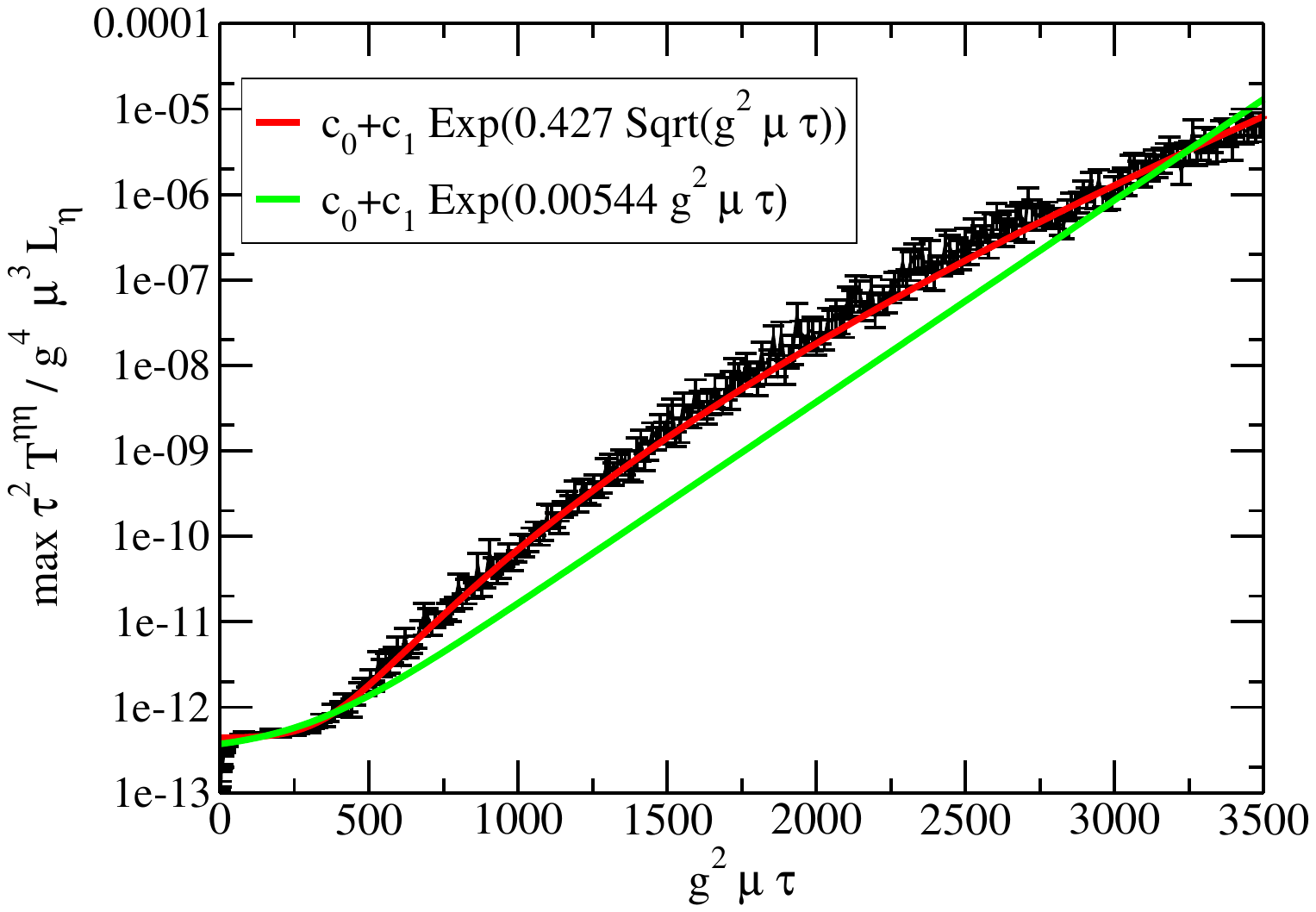}
\end{center}
\vspace{-9mm}
\caption{The maximal Fourier amplitude of $P_L = \tau^2 T^{\eta\eta}$ as a function of proper time $\tau$. The best fits with $e^{\tau}$ and $e^{\sqrt{\tau}}$ behavior are also shown. Figure from \cite{Romatschke:2005pm}.}
\label{fig:maxFM}
\end{minipage}
\hspace{2mm}
\begin{minipage}{8.5cm}
\vspace{0mm}
\begin{center}
\includegraphics[width=1.0\textwidth]{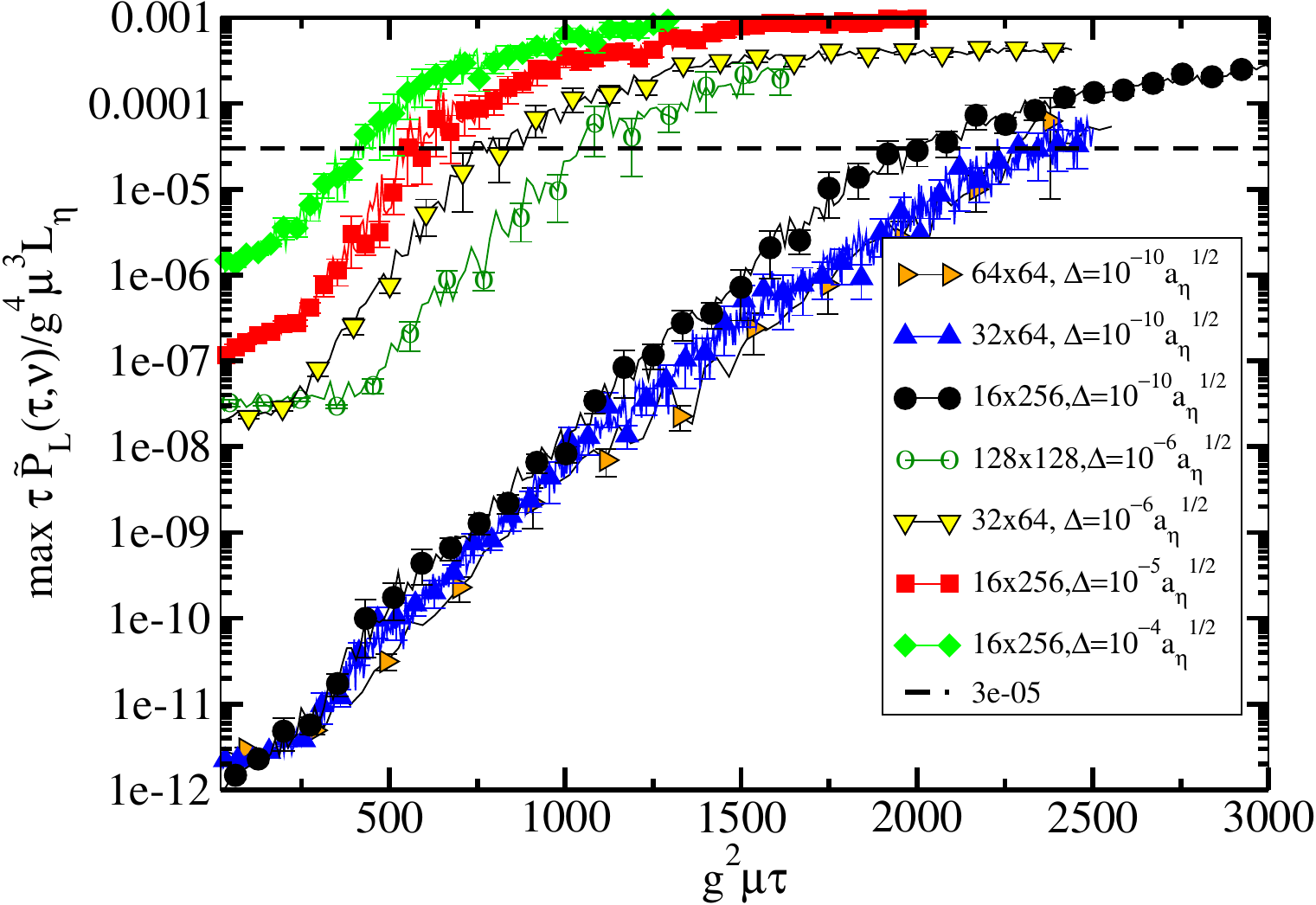}
\end{center}
\vspace{-6mm}
\caption{Temporal evolution of the maximum amplitude $\tau \tilde P_L(\tau,\nu)$ for various size of initial rapidity fluctuations controlled by $\Delta$. Figure from \cite{Romatschke:2006nk}.}
\label{fig:maxamp}
\end{minipage}
\end{figure}

The perturbations were constructed in the following three steps:
\begin{itemize}
\item
generate a random field $\delta \bar{E}_i( {\bf x}_\perp)$ from the Gaussian distribution with 
$$\langle \delta\bar{E}^i( {\bf x}_\perp) \, \delta\bar{E}^j( {\bf x}_\perp') \rangle =\delta^{ij} \delta^{(2)}({\bf x}_\perp - {\bf x}_\perp');$$
\item
generate a random function $F(\eta)$ from the Gaussian distribution with 
$$\langle F(\eta) \, F(\eta^\prime) \rangle =\Delta^2 \delta(\eta-\eta^\prime),$$ 
where $\Delta$ is the parameter which controls the size of rapidity fluctuations;
\item
compute the fluctuating part of the electric field as
\beq
\delta E^i({\bf x}_\perp,\eta) = \partial_\eta F(\eta) \, \delta \bar{E}^i({\bf x}_\perp), 
~~~~~~~
E_\eta({\bf x}_\perp,\eta) = - F(\eta) \, D_i \delta \bar{E}^i({\bf x}_\perp) ,
\eeq
which explicitly satisfies the Gauss' law (\ref{Gauss-RV}).
\end{itemize}

For small initial rapidity fluctuations, the system under study is strongly anisotropic in momentum space and 
can consequently develop a Weibel instability. To observe the development of the instability the evolution of longitudinal and transverse pressures $P_T$ and $P_L$ were investigated in \cite{Romatschke:2006nk}. The quantities are defined as
\beq
P_T = \frac{1}{2} \left(T^{xx}+T^{yy}\right), 
~~~~~~~~
P_L=\tau^2 T^{\eta \eta} ,
\eeq
where $T^{\mu \nu}$ is the energy-momentum tensor of the chromodynamic field given by Eq.~(\ref{energy-mom-field}). The observable
\beq
\tilde{P}_L(\tau,\nu,{\bf k}_\perp=0)=\int d \eta \, e^{i \eta \nu} \langle P_L(\tau,{\bf x}_\perp,\eta)\rangle_\perp\,,
\label{FTdef}
\eeq
which is a Fourier transform of $P_L(\tau,{\bf x}_\perp,\eta)$ with respect to rapidity averaged over transverse coordinates $(x,y)$, was also studied. The quantity (\ref {FTdef}) would be strictly zero in the boost-invariant ($\Delta =0$) case, except at $\nu = 0$.

\begin{figure}[t]
\begin{minipage}{8.5cm}
\vspace{0mm}
\begin{center}
\includegraphics[width=1.0\textwidth]{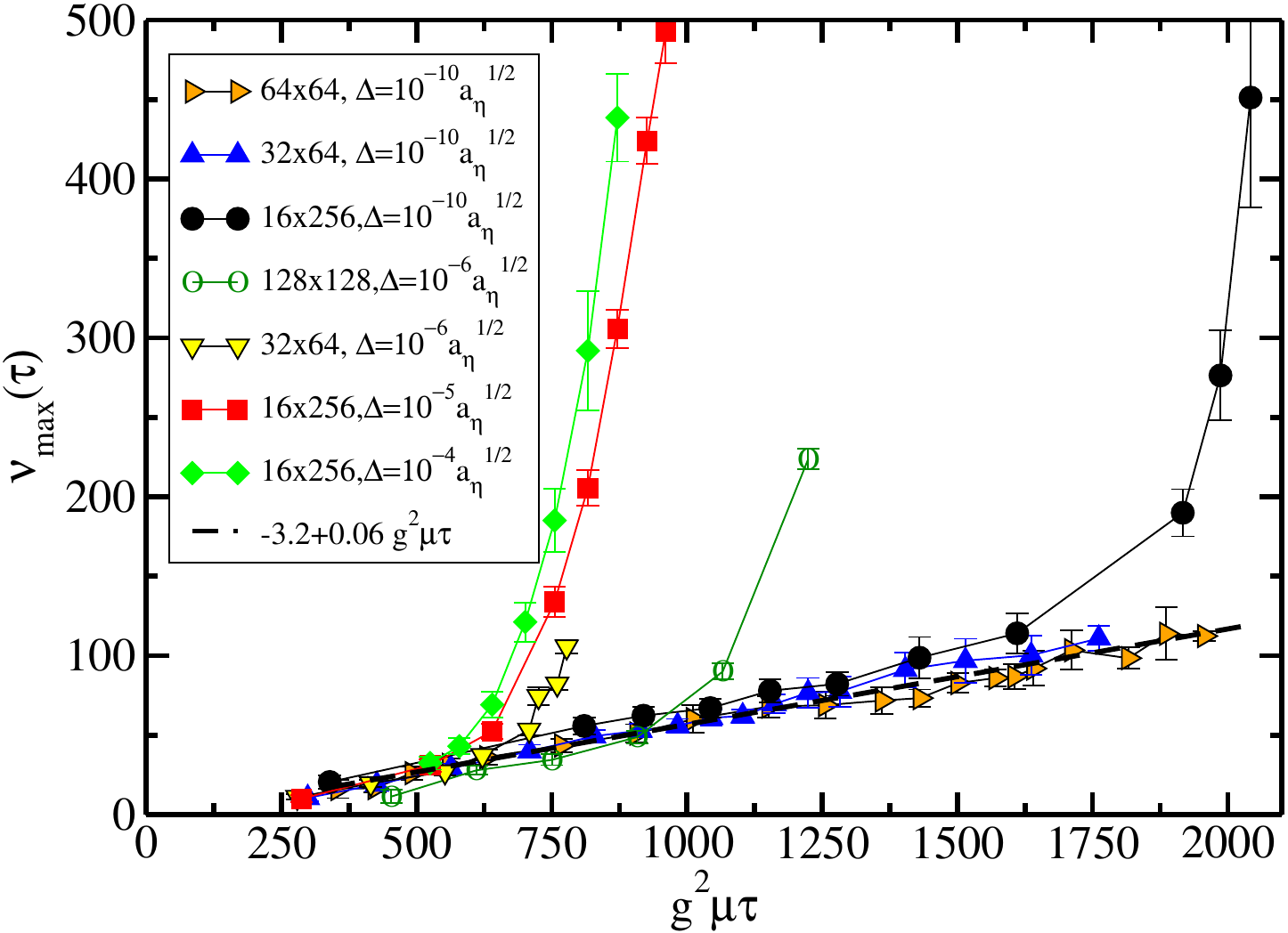}
\end{center}
\vspace{-5mm}
\caption{Temporal evolution of $\nu_{\rm max}$ for various size of initial rapidity fluctuations controlled by $\Delta$. The dashed line represents the linear scaling behavior. Figure from \cite{Romatschke:2006nk}.}
\label{fig:numax}
\end{minipage}
\hspace{2mm}
\begin{minipage}{8.5cm}
\vspace{-4mm}
\begin{center}
\includegraphics[width=1.02\textwidth]{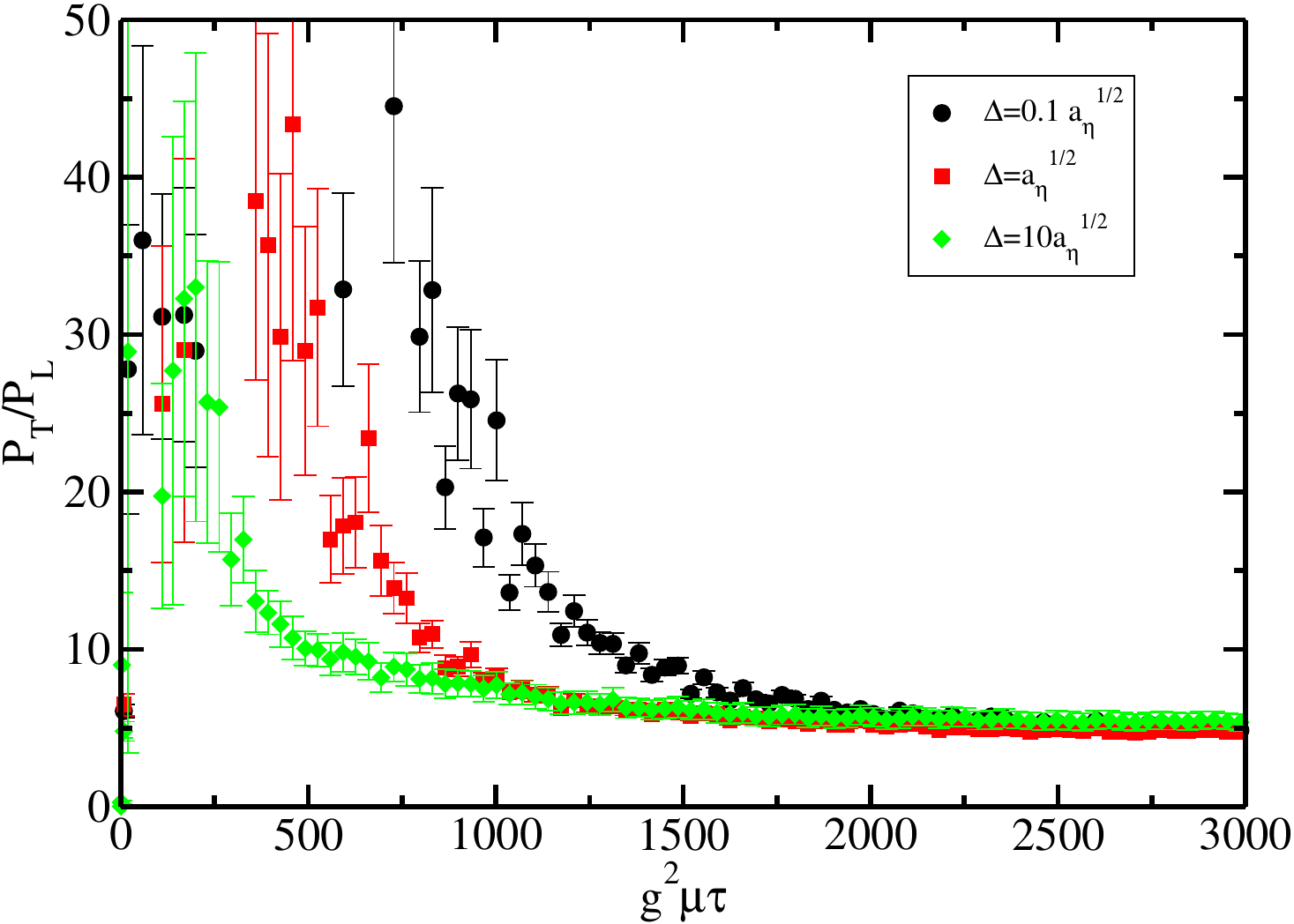}
\end{center}
\vspace{-5mm}
\caption{The mean system anisotropy as measured by $P_T/P_L$ for various size of initial rapidity fluctuations controlled by $\Delta$.  Figure from \cite{Romatschke:2006nk}.}
\label{fig:glasmaaniso}
\end{minipage}
\end{figure}

We now concentrate on reviewing the results of \cite{Romatschke:2005pm,Romatschke:2006nk} and for details on the lattice discretization refer the reader to Appendix A in \cite{Romatschke:2006nk}. We only note that $L$ and $L_\eta$ are the total lattice lengths in the transverse and longitudinal directions. 

The main finding of \cite{Romatschke:2005pm} is that a Weibel instability is present in the system, and appears as a rapidly growing fluctuation of $P_L$. Fig.~\ref{fig:maxFM} shows the maximal value of $\tilde{P}_L$ (\ref{FTdef}) at each proper time step, as a function of $g^2\mu \tau$. The maximal value remains nearly constant until $g^2\mu \tau \approx 250$, beyond which it grows rapidly. A best fit to the functional form $c_0 + c_1 \exp({c_2 \tau^{c_3}})$ gives $c_2 =0.427\pm 0.01$ for $c_3 = 0.5$; the coefficients  $c_0$, $c_1$ are small numbers when compared to the initial seed. It is evident from Fig.~\ref{fig:maxFM} that the form $e^{\sqrt{\tau}}$ not $e^\tau$ is preferred, clearly showing an effect of the system's expansion already observed in the hard-loop simulations discussed in Sec.~\ref{subseq-expanding-HL}. 

In Fig.~\ref{fig:maxamp}, which is taken from \cite{Romatschke:2006nk}, one sees that the exponential growth of the maximum Fourier amplitude $\tau \tilde P_L(\tau,\nu)$ saturates at a certain critical size which is denoted in the figure by the dashed horizontal line. This effect is analogous to the phenomenon of non-Abelian saturation of unstable mode growth observed in the hard-loop simulations \ref{subsec-hardloop}. The larger initial rapidity fluctuations controlled by $\Delta$, the earlier the saturation. Tracking the time evolution of the hardest unstable mode, $\nu_{\rm  max}$, it was deduced that when the maximum Fourier amplitude $\tau \tilde P_L(\tau,\nu)$ reaches the critical size, $\nu_{\rm max}$ starts to increase dramatically, filling up the mode spectrum. As shown in Fig.~\ref{fig:numax}, the rapid growth of $\nu_{\rm max}$ appears earlier for larger initial rapidity fluctuations. Because the growth of $\nu_{\rm max}$ is accompanied by a drop in the transverse pressure, see Fig.~\ref{fig:glasmaaniso}, this result can be interpreted as the effect of the Lorentz force exerted by soft modes of the transverse magnetic fields on the hard transverse gauge modes. 

The magnetic fields effectively bend the hard transverse gauge modes into the longitudinal direction. The total amplitude of the transverse magnetic field does not increase significantly, but the `bending' can be accomplished since the instability has populated predominantly modes with ${\bf k}_\perp=0$ and the corresponding transverse magnetic fields thus can act over long transverse distances. The simulations on lattices with large longitudinal UV cutoff suggest that significant longitudinal pressure is built up in this process, leading to isotropization of the system, which is also depicted in Fig.~\ref{fig:glasmaaniso}. Isotropization within a weak coupling framework is thus possible but its time scale presumably cannot account for full isotropization at times as short as $0.5 \div 1.0 \; {\rm fm}/c$, as suggested by hydrodynamic models of heavy-ion collisions. However, the results from \cite{Romatschke:2005pm,Romatschke:2006nk} indicate that isotropization time scales are much shorter if longitudinal fluctuations are already larger initially than if they have to be built up by the instability.

Recent developments in the CGC simulations are briefly reviewed in the subsequent section \ref{subsec-recent-prog}.

\subsubsection{Recent progress}
\label{subsec-recent-prog}

Instabilities and their role for the thermalization and isotropization of the non-equilibrium system of classical chromodynamic fields have been recently under active investigations in both classical-statistical lattice and CGC frameworks introduced in Secs.~\ref{sub-sec-class-stat} and \ref{sub-sec-CGC}, respectively. Here we briefly review the recent developments. 

Classical statistical simulations of the non-equilibrium dynamics have been significantly extended and analyzed in more detail in  \cite{Berges:2011sb,Berges:2012iw}. Fermion production in a system with unstable bosonic modes via Yukawa-type interactions was studied in \cite{Berges:2009bx} and in \cite{Berges:2012iw} the parametric resonance in scalar $N$-component quantum field theories with boost-invariant initial conditions was investigated and many aspects analyzed analytically. This work was extended to SU(2) dynamics in \cite{Berges:2012cj}. In \cite{Berges:2013eia,Berges:2013fga} the classical-statistical lattice simulations were used to demonstrate the existence of a nonthermal fixed point. The authors find that after an initial transient regime dominated by plasma instabilities and free streaming, the ensuing overpopulated non-Abelian plasma exhibits the universal self-similar dynamics characteristic of wave turbulence. In contrast to the static case, the simulations including longitudinal expansion \cite{Berges:2013eia,Berges:2013fga} show that interactions are not strong enough to compensate for the longitudinal expansion, such that the anisotropy keeps increasing at late times. Surprisingly, it was found that, despite the increasing anisotropy of the system, instabilities do not play a significant role for the late time dynamics although they were assumed to have already the highly populated longitudinal modes in their initial condition.  In \cite{Kurkela:2012tq} a new algorithm for solving the Yang-Mills equations in an expanding box was developed, allowing for the evolution to proceed to late times in boxes of large transverse extent without encountering the problem that the longitudinal lattice spacing becomes too coarse. This is achieved by periodically cropping the box in the longitudinal direction and refining the mesh. 
 
As in the hard-loop and classical statistical simulations, CGC lattice simulations presented in \cite{Fukushima:2011nq,Fukushima:2013dma} show an energy flow from low to high wave number modes, resulting in a spectrum consistent with Kolmogorov's power law form reminiscent of turbulent behavior. Turbulent phenomena were also studied in a static box by \cite{Kurkela:2012hp} and \cite{Schlichting:2012es,Berges:2012ev} and a cascade of energy toward higher momenta and the existence of a scaling solution were observed. The work by \cite{Kurkela:2012hp} also studied the formation of condensates. Electric condensates were not found to carry occupancies larger than $1/g^2$ and they were damped out quickly. Magnetic condensates are found to be unstable due to the Nielsen-Olesen instability. 

Some work has been focused on the computation of the initial quantum fluctuations on top of the classical background field that trigger the growth of the instabilities. This includes analysis of a scalar $\phi^4$ theory, where all exponentially growing modes have been resummed \cite{Dusling:2010rm,Epelbaum:2011pc}. This study was extended to include the longitudinal expansion of the system in \cite{Dusling:2012ig} where it was found that the microscopic processes that drive the system toward equilibrium are able to keep up with the expansion of the system. The pressure tensor was found to become isotropic despite the anisotropic expansion.

A detailed derivation of the initial quantum fluctuations in a gauge theory was performed in \cite{Epelbaum:2013waa}. These initial conditions were used in simulations performed in \cite{Gelis:2013rba} where it was found that the system can isotropize on short time scales.  While there seems to be a disagreement between the work in \cite{Berges:2013eia,Berges:2013fga} and \cite{Gelis:2013rba} in the fact that the former finds that the expansion will dominate and prevent instability driven isotropization and the latter finds a rather fast isotropization, it is important to note that the approaches differ in various regards. In \cite{Berges:2013eia,Berges:2013fga} calculations of the late time behavior are performed in the extremely weak coupling limit and then extrapolated to more realistic couplings. In \cite{Gelis:2013rba} calculations were directly performed at larger values of the coupling, starting at the earliest times after the collision. The resolution of the disagreement is the topic of active ongoing research.

The later stages of longitudinally expanding plasmas after instabilities have stopped growing was studied in \cite{Berges:2013eia,Berges:2013fga,Berges:2014bba,Berges:2013lsa} and a non-thermal fixed point was found. The authors concluded that the study of how and on what time scales the system isotropizes and approaches thermal equilibrium is beyond the reach of classical-statistical simulations since this seems to occur later in the quantum regime. Based on the observation that plasma instabilities do not play a significant role at late times, they argue that a kinetic theory approach would be more appropriate to answer the question of isotropization and thermalization.  

Such studies have been performed in \cite{Kurkela:2014tea,Kurkela:2015qoa}, which solves 2+1D effective kinetic theory of weakly coupled QCD \cite{Arnold:2002zm} subject to longitudinal expansion \cite{Kurkela:2015qoa}. Agreement with classical Yang-Mills simulations (and hydrodynamics) was found in the regime where they should be applicable. Furthermore, the effective kinetic theory provides a link between the early-time classical Yang-Mills regime and hydrodynamics at later times, by correctly describing systems where the typical occupancies of gluons are not nonperturbative and momenta are significantly larger than the in-medium screening scale. Although no instabilities are present in the system (because isotropic screening is assumed), fast isotropization and `hydrodynamization' seems to be achieved with large enough coupling.




\subsection{Wong-Yang-Mills simulations}
\label{Wong-Yang-Mills}

In the simulations presented in Sec.~\ref{subsec-purefield-class}, not only the highly populated soft modes are described as classical fields but the hard modes are treated in the same manner. While the description of soft modes is fully reliable, the treatment of hard modes is rather uncertain. The approach can be improved by representing the hard modes as classical colored particles propagating in the background of classical gauge fields. Motion of particles is then governed by the Wong equations \cite{Wong:1970fu} with the chromodynamic field generated self-consistently by the particles' current {\it via} the Yang-Mills equations. Such a framework, which was initiated in \cite{Hu:1996sf,Moore:1997sn,Bass:1998nz}, is equivalent in the weak coupling limit to QCD in the hard-loop approximation as shown in \cite{Kelly:1994dh,Kelly:1994ig}, see also the review \cite{Litim:2001db}. Moving from the solid ground of the hard-loop regime, the Wong-Yang-Mills approach can be naturally extended to the less certain domain of strong coupling. 

Here we first introduce the Wong-Yang-Mills approach and then briefly describe the update algorithm used to solve the coupled system of equations of motion.  Finally, we present results on instability growth and isotropization in this framework.  Several groups have performed Wong-Yang-Mills simulations \cite{Hu:1996sf,Moore:1997sn,Bass:1998nz,Krasnitz:1998ns,Krasnitz:1999wc,Krasnitz:2001qu,Krasnitz:2002mn,Krasnitz:2003jw,Krasnitz:2002ng,Dumitru:2005gp,Dumitru:2006pz}. 
In our presentation we mostly follow the works \cite{Moore:1997sn,Dumitru:2006pz} where simulations with the SU(2) gauge group were preformed. 

The goal of these simulations is to study the behavior of $N$ massless partons moving in a chromodynamic field along the trajectories ${\bf x}_i(t)$ with momenta ${\bf p}_i(t)$ and color charges $q^a_i(t)$ where $i = 1,\;2, \dots N$. The particles' variables ${\bf x}_i(t),\;{\bf p}_i(t),$, and $q^a_i(t)$ obey the three Wong equations (\ref{wong1}), (\ref{wong2}), and (\ref{wong3}) with the chromodynamic fields described by the Yang-Mills equations (\ref{YM-eq}) and the current of the form (\ref{j1y}), (\ref{j2y}) summed over particles. The lattice realization of the approach is far from trivial. The time evolution of the chromodynamic fields is determined by the Hamiltonian method \cite{Ambjorn:1990pu,Hu:1996sf,Moore:1997sn} which has been already encountered in Sec.~\ref{sub-sec-class-stat} where the classical-statistical approach is presented.  The method takes advantage of the temporal axial gauge $A^0=0$, which is particularly useful because of a simple choice of the gauge potential $A^i$ as the field canonical variable and the electric field $E^i = - \dot{A}^i$ as the conjugate momentum. However, in the lattice formulation this choice needs to be appropriately modified.

\subsubsection{Lattice formulation}

The simulations are realized on three-dimensional spatial lattices. There are the lattice {\it sites},  {\it links}, which connect nearest sites, and {\it cells} surrounded by links. The dynamical variables which describe the fields are parallel transporters $U_i$ defined by Eq.~(\ref{linkdef}) and electric fields $E_i$ given by Eq.~(\ref{electric-field-FCS}) which take values on the lattice's links. Particles' positions $x_i$ are specified within the cells and momenta are real discrete numbers. In the original numerical formulation, particles move freely except when crossing a cell boundary. The charge is then transported and a current is induced. The fields and particle variables obey periodic boundary conditions. For every physical particle there are $N_{\text{test}}$ test particles and one uses dimensionless lattice variables
\begin{align}
\label{latticevariables}
\mathbf{E}^a_L \equiv \frac{ga^2}{2}\mathbf{E}^a , 
&& \mathbf{B}^a_L \equiv \frac{ga^2}{2}\mathbf{B}^a , 
&& \mathbf{p}_L \equiv \frac{a}{4}\mathbf{p}, 
&& Q_L^a \equiv \frac{1}{2}q^a\,, 
&& N_{\text{test}\,L} \equiv \frac{1}{g^2}N_{\text{test}},
\end{align}
where $a$ is the lattice spacing. To convert lattice variables to physical units one fixes the lattice length $L$ in fm, which will then determine the physical scale for $a$. All other dimensionful quantities can then be determined by using Eqs.~(\ref{latticevariables}).  Following the work \cite{Dumitru:2006pz}, we chose here a different normalization from the previous sections and for SU(2) gauge group the generators $\tau_a$ equal the Pauli matrices $\sigma_a$, without the usual factor of $1/2$, {\it i.e.}, the commutation relation reads $[\tau^a,\tau^b]=2\delta^{ab}$. Another factor of $1/2$ is absorbed into the potential $\mathbf{A}^a$, which has to be taken into account when calculating the electric and magnetic fields.

The lattice Hamiltonian of the Kogut-Susskind form \cite{Kogut:1974ag} is
\begin{equation}
\label{KSH}
H_L \equiv \frac{g^2 a}{4} H = \frac{1}{2}\sum_i \mathbf{E}_{L\,i}^{a\,2}
+\frac{1}{2}\sum_\square\left(N_c-\text{Re}\text{Tr}U_\square\right)
+\frac{1}{N_{\text{test}\,L}}\sum_j|\mathbf{p}_{L\,j}| ,
\end{equation}
where $\sum_\square$ is the summation over all plaquettes which are defined as in Eq.~(\ref{eq:plaquette}). The last term in the Hamiltonian (\ref{KSH}) represents the kinetic energy of particles and there is no explicit term of particle-field interaction, as the particles are free except when they cross cell boundaries. The lattice equations of motion are constructed in such a way that the energy corresponding to the Hamiltonian (\ref{KSH}) is conserved and the Gauss law constraint (\ref{Gauss-law-lattice}) is obeyed when the dynamical variables satisfy the equations of motion.

The equations of motion of the chromodynamic fields are treated similarly to the case of classical statistical simulations discussed in Sec.~\ref{sub-sec-class-stat}. However, the equations include the color currents of the form (\ref{j1y}), (\ref{j2y}) due to particles carrying color charges. The currents are smeared as it is common for particle-in-cell (PIC) simulations in Abelian plasma physics \cite{Hockney:1981,Birdsall:1985} in order to avoid numerical noise. Throughout the temporal evolution of the system, one keeps track of every particle's position $\mathbf{x}_i(t)$, momentum $\mathbf{p}_i(t)$ and charge $q^a_i(t)$ with $i=1,\;2,\dots N$. Knowing the $i$-th particle's momentum, its coordinate is updated according to Eq. (\ref{wong1}) for a time step of length $\Delta t$. Specifically, 
\begin{equation}
\label{coordupdate}
\mathbf{x}_i(t+\Delta t)=\mathbf{x}_i(t)+\mathbf{v}_i(t+\Delta t /2) \Delta t ,
\end{equation}
where $\mathbf{v}_i=\mathbf{p}_i/E_i$, and $E_i =|\mathbf{p}_i|$. Since the velocity is defined at time $t+\Delta t/2$, the momentum update has to provide $\mathbf{p}_i$ at time $t+\Delta t/2$.

The update of the particles' momenta is done according to the so-called Buneman-Boris method \cite{Hockney:1981,Birdsall:1985} that satisfies the requirement of time reversibility. The procedure is aimed to solve the finite difference form of the Wong equation (\ref{wong2}) which is written as
\be
\label{Wong2-diff}
\frac{\mathbf{p}(t + \Delta t/2) - \mathbf{p}(t - \Delta t/2)}{\Delta t} = \mathbf{E}(t) +
\frac{\mathbf{p}(t + \Delta t/2) + \mathbf{p}(t - \Delta t/2)}{2 E(t )} \times \mathbf{B}(t) ,
\ee
where the chromomagnetic and chromoelectric fields enter through $\mathbf{E} \equiv - gq^a \mathbf{E}^a$ and $\mathbf{B} \equiv - gq^a \mathbf{B}^a$. The solution of Eq.~(\ref{Wong2-diff}) with respect to $\mathbf{p}(t + \Delta t/2)$ and $\mathbf{p}(t)$ is found in four steps. One first calculates
\begin{equation}
\label{p(t)}
\mathbf{p} (t)=\mathbf{p}\Big(t-\frac{\Delta t}{2}\Big)+\frac{\Delta t}{2} \mathbf{E}(t) ,
\end{equation}
taking into account only the effect of the electric field. In the second step, the momentum is rotated by the magnetic field to obtain
\begin{equation}
\mathbf{p}' (t)=\mathbf{p} (t) + \frac{\Delta t}{2} \frac{\mathbf{p} (t)}{E(t)} \times \mathbf{B}(t) .
\end{equation}
In the third step one defines 
\begin{equation}
\mathbf{p}''(t) =\mathbf{p} (t) 
+ \frac{\Delta t}{E(t) } \mathbf{p}' (t) \times \mathbf{B}(t)  ,
\end{equation}
and finally we obtain
\begin{equation}
\label{p(t+dt/2)}
\mathbf{p}\Big(t + \frac{\Delta t}{2}\Big) = \mathbf{p}'' (t)+\frac{\Delta t}{2} \mathbf{E}(t) .
\end{equation}
One can verify that the momenta (\ref{p(t)}) and (\ref{p(t+dt/2)}) solve Eq.~(\ref{Wong2-diff}) to second order in $\Delta t$, that is when the terms of order ${\cal O}\big((\Delta t)^3\big)$ and higher are neglected. 

\begin{figure}[t]
\begin{minipage}{8.5cm}
\center
\vspace{-2mm}
\includegraphics[width=1.0\textwidth]{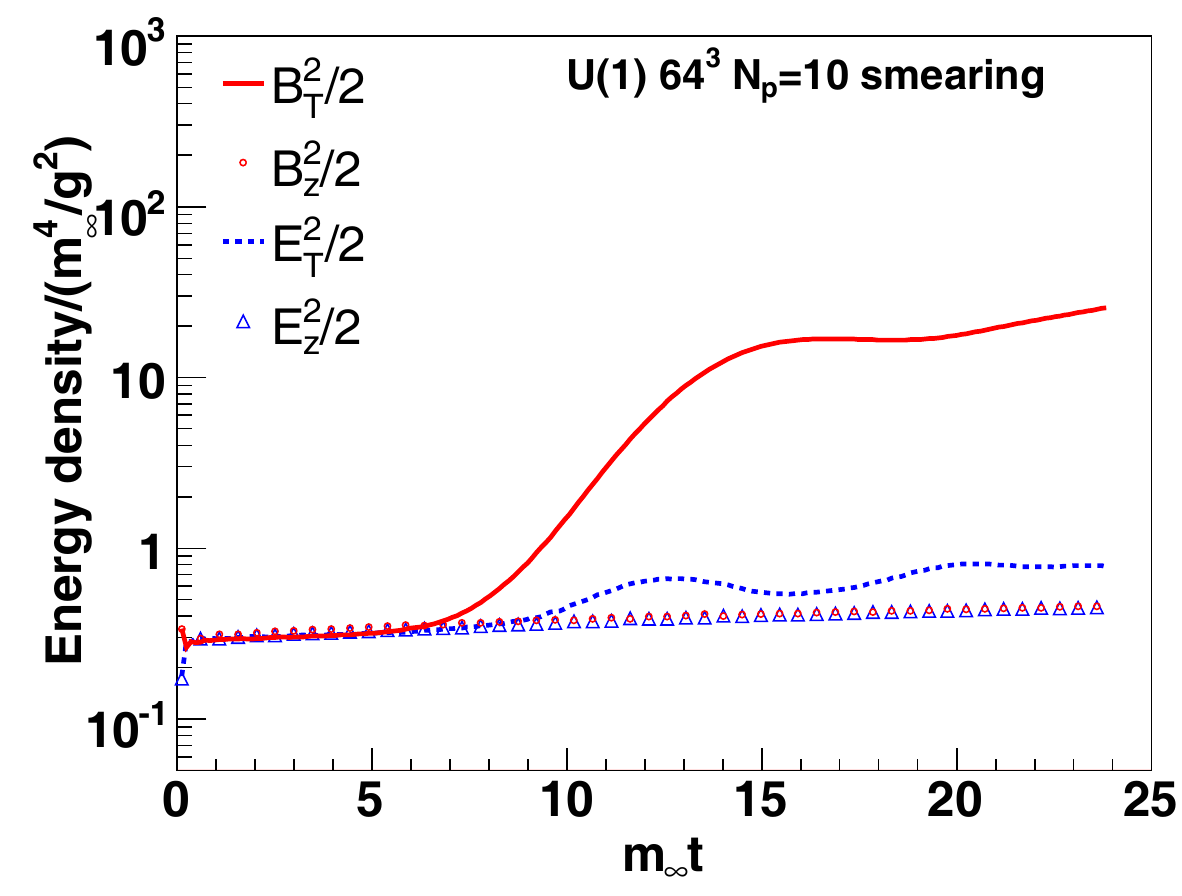}
\vspace{-6mm}
\caption{Time evolution of the field energy densities in an Abelian anisotropic plasma. Figure from \cite{Dumitru:2006pz}. 
\label{fig:fieldw64u1_n10}}
\end{minipage}
\hspace{5mm}
\begin{minipage}{8.5cm}
\center
\vspace{-2.5mm}
\includegraphics[width=1.01\textwidth]{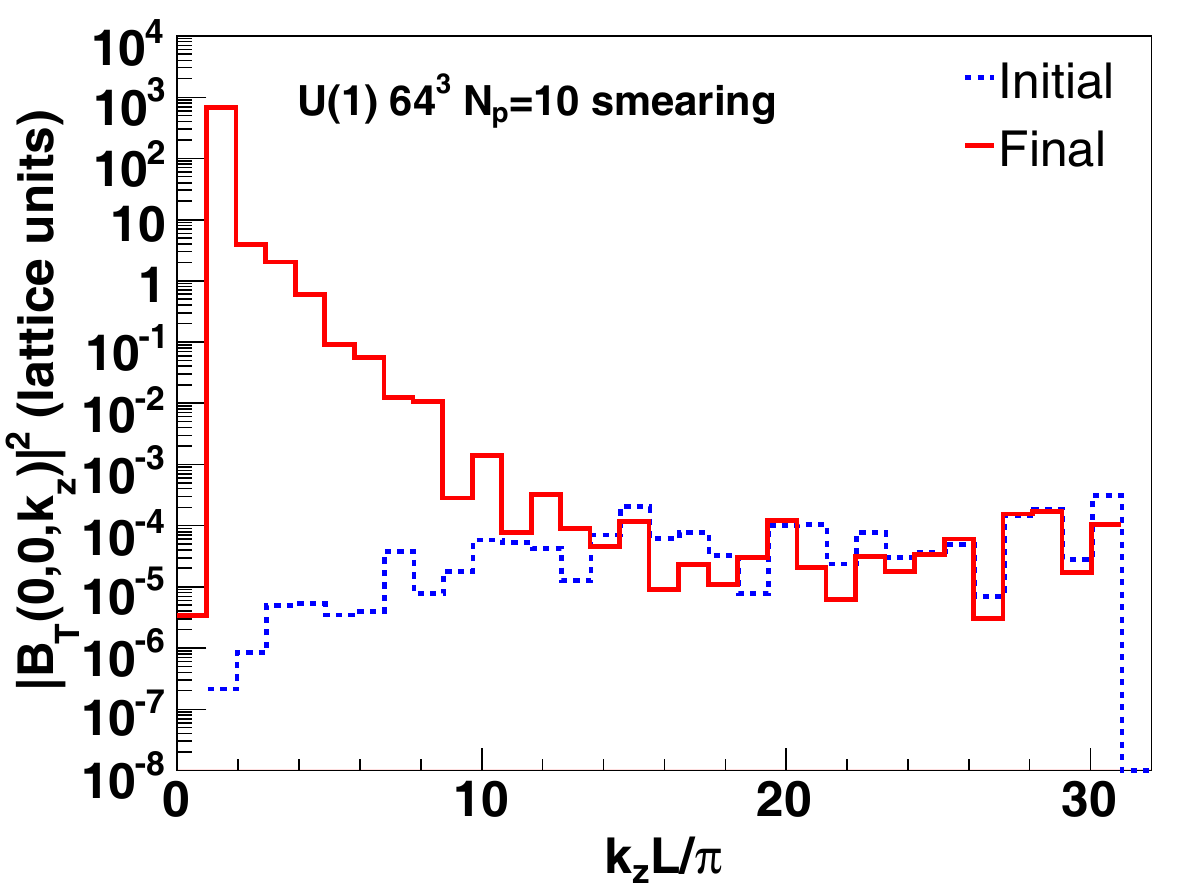}
\vspace{-6.5mm}
\caption{Fourier spectra of the initial and final transverse magnetic field in an Abelian anisotropic plasma. Figure from \cite{Dumitru:2006pz}. 
\label{fig:FT_Bt_U1}}
\end{minipage}
\end{figure}

\subsubsection{Initial conditions}

At the initial time $t=0$ the momentum distribution for the hard plasma gluons is taken to be
\be
\label{anisof}
f(\mathbf{p}) = n_g \left(\frac{2\pi}{p_{\text{h}}}\right)^2 \delta(p_z)  \exp(-p_\perp/p_{\text{h}}),
\ee
where $p_\perp \equiv \sqrt{p_x^2+p_y^2}$ and the parameter  $p_{\text{h}}$ controls the momentum scale of hard gluons. The formula (\ref{anisof}) represents a quasi-thermal distribution in two dimensions with average momentum equal $2p_{\text{h}}$. The mass parameter (\ref{mass2}) computed with the distribution (\ref{anisof}) becomes 
\begin{equation}
\label{estimate-aniso-m}
m^2 = g^2 \, \frac{n_g}{p_h} . 
\end{equation}
As we know from Sec.~\ref{sec-collective}, the parameter sets the scale for the growth rate of unstable field modes. In \cite{Dumitru:2005hj} and \cite{Dumitru:2006pz} the mass parameter  (\ref{mass2}) is defined with an additional factor of $N_c$ and it is denoted as $m^2_{\infty}$. This is rather unfortunate, as $m_{\infty}$ usually denotes the asymptotic gluon mass which equals $m/\sqrt{2}$. Nevertheless, we will use the  symbol  $m_{\infty}$ here. 

The initial field amplitudes are sampled from a Gaussian distribution with a two-point function
\begin{equation}
\langle A_i^a(x)A_j^b(y)\rangle=\frac{4\mu^2}{g^2}\delta_{ij}\delta^{ab}\delta(\mathbf{x}-\mathbf{y}) ,
\end{equation}
which is tuned to the initial energy density through the parameter $\mu$. At the initial time $t=0$ one also randomly samples $N_p = N_{\rm test}\, n_g\, a^3$ particles from the distribution (\ref{anisof}) in each lattice cell. When $N_p$ is not very large it is useful to ensure explicitly that the sum of particle momenta in each cell vanishes by adjusting, for example, the momentum of the last particle accordingly.

\subsubsection{Results}

Below we summarize the most important results form \cite{Dumitru:2005hj} and \cite{Dumitru:2006pz} and we extend the study, showing explicitly the generation of filaments discussed in Sec.~\ref{subsec-mechanism-trans}.

\begin{figure}[t]
\begin{minipage}{8.5cm}
\center
\vspace{-2mm}
\includegraphics[width=1.0\textwidth]{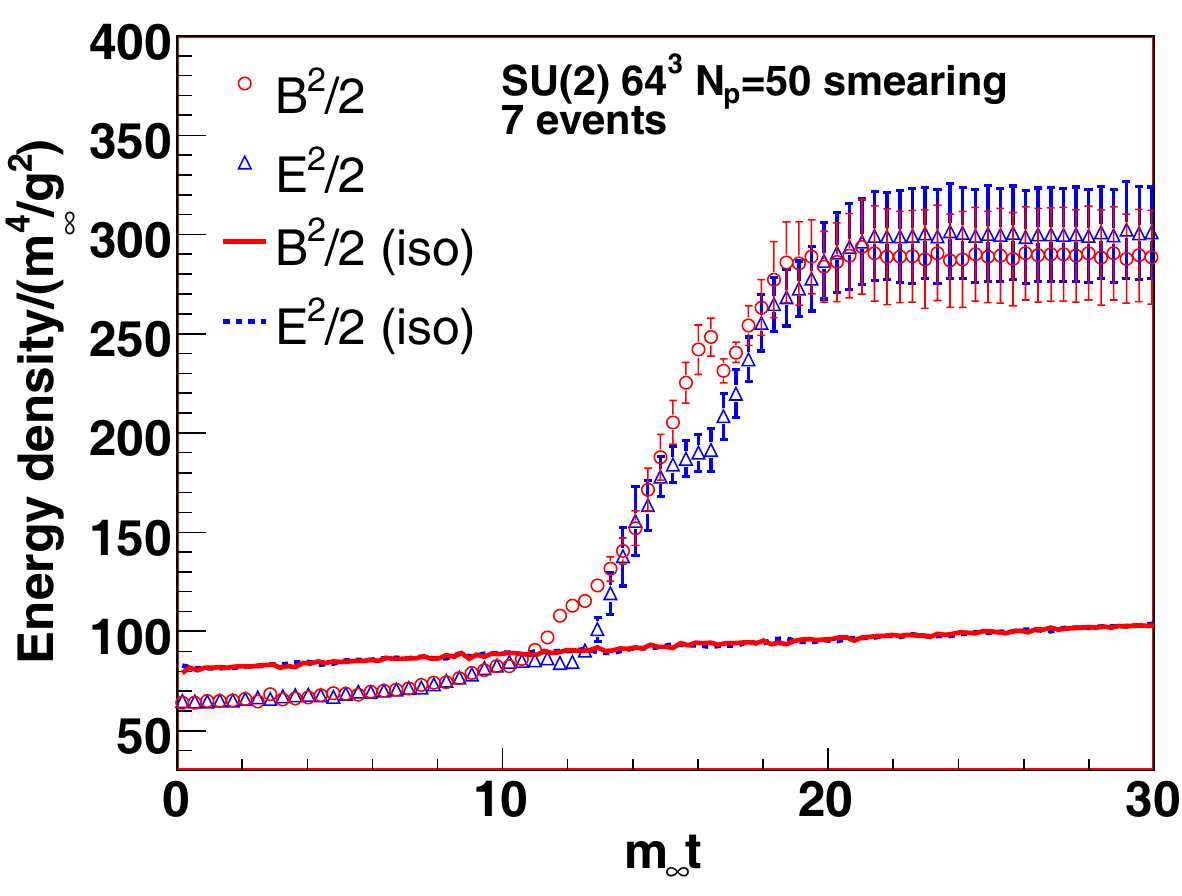}
\vspace{-6mm}
\end{minipage}
\hspace{2mm}
\begin{minipage}{8.5cm}
\center
\vspace{-2.5mm}
\includegraphics[width=1.01\textwidth]{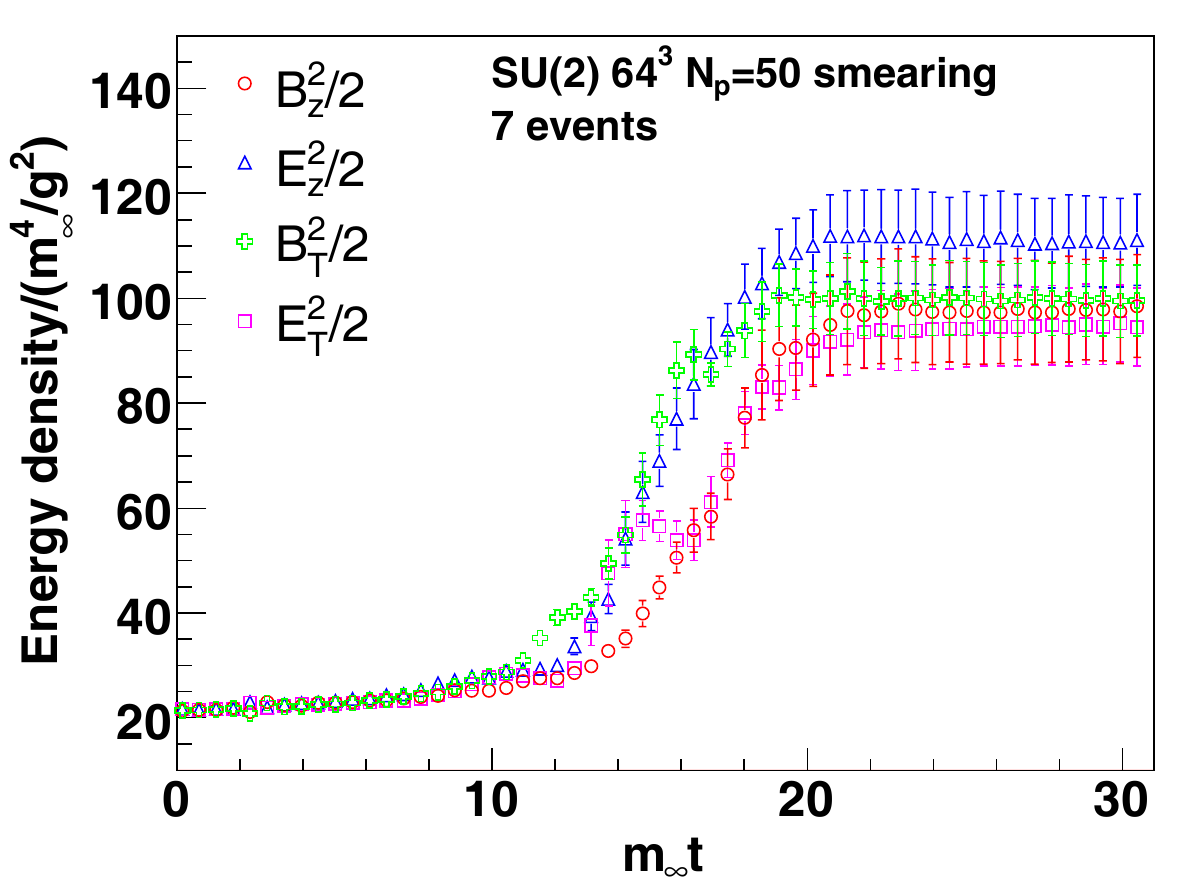}
\vspace{-6.5mm}
\end{minipage}
\caption{Time evolution of the energy densities of various field components in anisotropic SU(2) plasma.  Runs with an isotropic particle distribution are also shown in the left panel. Figure from \cite{Dumitru:2006pz}. 
\label{fig:field-SU2}}
\end{figure}

\vspace{2mm}
\underline{\it Growth of instabilities}
\vspace{1mm}

First, we discuss the temporal evolution of anisotropic Abelian plasmas. For weak initial fields of energy density on the order of $0.1\, m_\infty^4/g^2$ (with the simulation parameters: $L=5$ fm, $p_h=16$ GeV, $g^2\,n_g=20$ fm$^{-3}$, and $m_\infty=0.1$ GeV), the energy densities related to transverse magnetic and electric fields, which are shown in Fig.~\ref{fig:fieldw64u1_n10}, exponentially grow for some time and then saturate. Both the magnitude and growth rate of the energy density of the electric field are significantly smaller than the magnetic ones. The mode spectra of the transverse magnetic field before the unstable modes start to grow (initial spectrum) and after the growth is saturated (final spectrum) are presented in Fig.~\ref{fig:FT_Bt_U1}. The modes above $k_z\simeq 10\pi/L$ are stable as opposed to the result in the linear approximation, where for extreme anisotropy the spectrum of unstable modes extends to infinite $k_z$, as discussed in Sec.~\ref{sec-ex-oblate}. 

In the non-Abelian SU(2) case, it is difficult to address weak fields with the PIC method, and consequently the authors of \cite{Dumitru:2006pz} focused on initial field energy densities of order $10\, m_\infty^4/g^2$ and above, which are beyond the regime of the hard-loop approximation discussed in Sec.~\ref{subsec-hardloop}. The initial condition was taken to be Gaussian random chromoelectric fields which were `low-pass' filtered such that only the lower half of available lattice modes were populated. 

Fig.~\ref{fig:field-SU2} presents the temporal evolution of energy densities related to various field components in an anisotropic SU(2) plasma. The simulation parameters are $L=5$~fm, $p_{\rm hard}=16$~GeV, $g^2\,n_g=10/$fm$^3$, $m_\infty=0.1$~GeV. Runs with an isotropic particle distribution, where no unstable modes are expected, are also shown in the left panel of Fig.~\ref{fig:field-SU2} as an indication of numerical accuracy.  In the isotropic runs, the field energy densities are indeed nearly constant over the time interval $t\le 30 \,m^{-1}_\infty$. In the anisotropic plasma, there is a period of rapid growth of energy densities of both electric and magnetic fields which later on settle to essentially constant values. Therefore, also beyond the hard-loop regime, the time evolution of non-Abelian fields differs significantly from that in the Abelian weak-field limit. In the former case, a sustained exponential growth is absent even during the stage where the back-reaction of fields on the particles is weak.

\begin{figure}[t]
\vspace{-5mm}
\includegraphics[width=14cm]{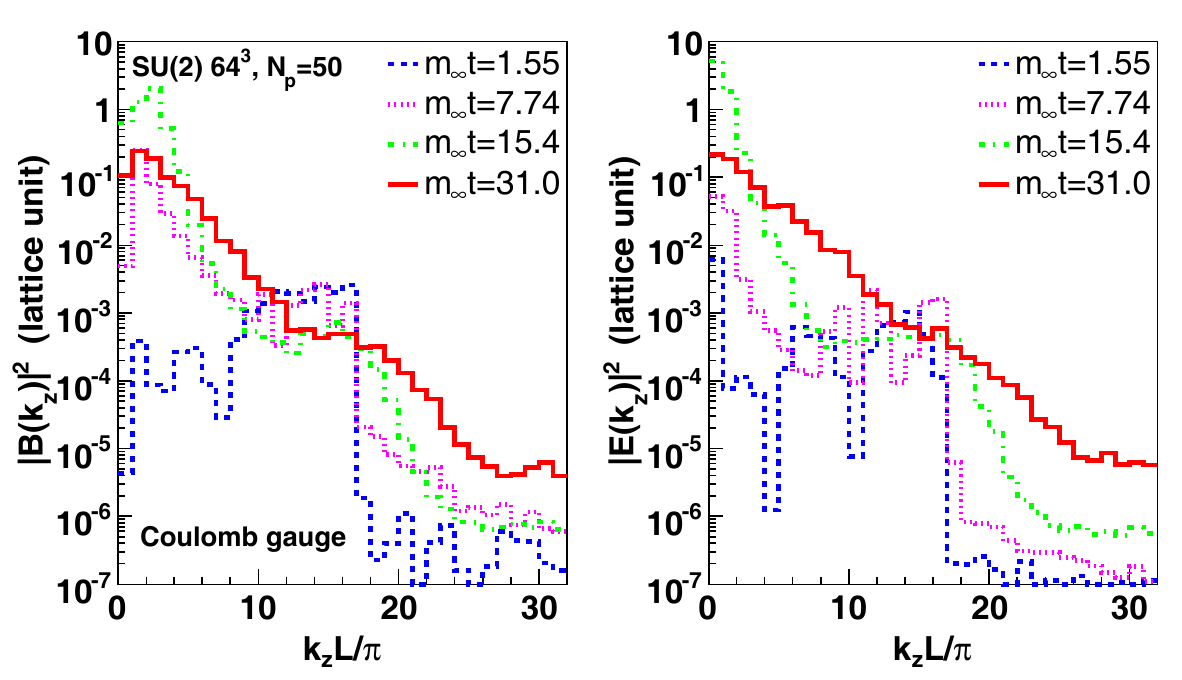}
\vspace{-3mm}
\caption{Fourier transformed electric and magnetic fields in an anisotropic SU(2) plasma at four different times. Figure from \cite{Dumitru:2006pz}.}
\label{fig:fieldWsu2_weak_FFT}
\end{figure}

\begin{figure}[b]
\vspace{-7mm}
\begin{center}
\includegraphics[width=8cm]{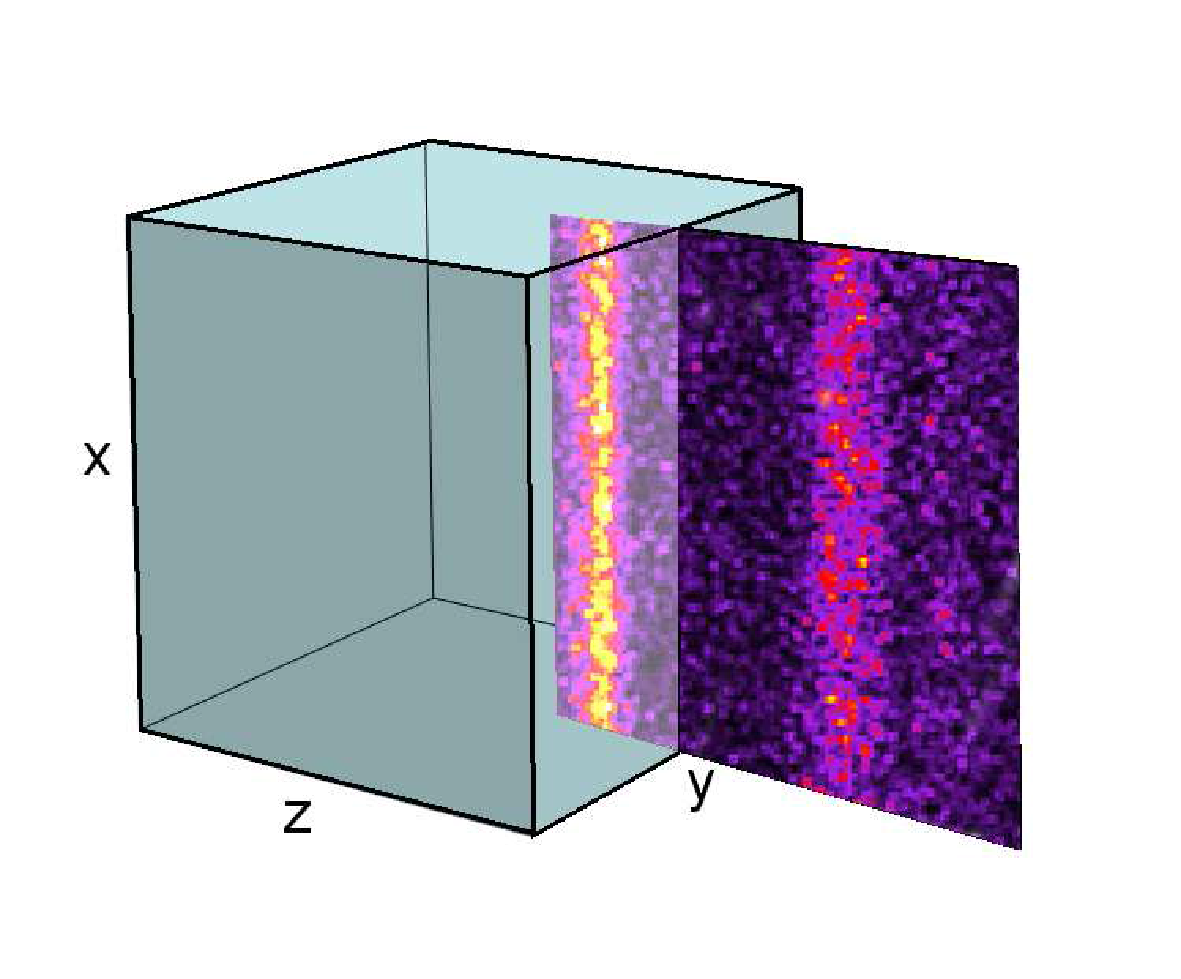}
\vspace{-10mm}
\caption{The cubic lattice with the slice we will visualize pulled out.}
\label{fig:lsfig}
\end{center}
\end{figure}

For very strong anisotropies a linear analysis predicts, as discussed in Sec.~\ref{sec-3+1}, that the exponential field growth in the weak-field limit can perhaps continue until the magnetic field reaches the value given by Eq.~(\ref{eq:limitingB}). Although the simulations shown in Fig.~\ref{fig:field-SU2} are performed with the initial momentum distribution (\ref{anisof}), which is extremely oblate, that is $\xi =\infty$, such strong fields are not seen. This is presumably related to the above-mentioned back-reaction effects which prevent instability of modes of large $k_z$.

The results by \cite{Dumitru:2006pz} indicate a sensitivity to hard field modes at the ultraviolet end of the Brioullin zone, when $k={\cal O}(a^{-1})$, in contrast to the U(1) and earlier 1D+3V SU(2) simulations \cite{Dumitru:2005gp,Dumitru:2005hj,Nara:2005fr}. The energy density contained in the fields at late times increases by a factor of 1.5 when going from a $16^3$ to $32^3$ to $64^3$ lattice with the same physical size $L$. Hence, the dynamics of SU(2) instabilities is not dominated entirely by a band of unstable modes in the infrared but clearly involves a cascade of energy from those modes to a harder scale $\Lambda$ \cite{Arnold:2005qs}. However, the simulations show that $\Lambda$ grows to ${\cal O}(a^{-1})$ during the period of rapid growth of the field energy density; otherwise, the final field energy density would not depend on the lattice spacing. 

In Fig.~\ref{fig:fieldWsu2_weak_FFT} we show the Fourier transformed chromoelectric and chromomagnetic fields at four different times obtained by \cite{Dumitru:2006pz}.  In this figure, the authors plotted the quantities ${\bf E}^a({\bf k}) \cdot {\bf E}^a(-{\bf k})$ and ${\bf  B}^a({\bf k})\cdot{\bf B}^a(-{\bf k})$ as functions of $k_z$ at $k_x=k_y=0$. The field configurations at the corresponding time were gauge transformed to satisfy the Coulomb gauge condition. The field spectra from Fig.~\ref{fig:fieldWsu2_weak_FFT} clearly show the avalanching to the UV: after the unstable soft modes have grown, the energy is rapidly transferred to the higher modes.

\begin{figure}[t]
\includegraphics[width=17cm]{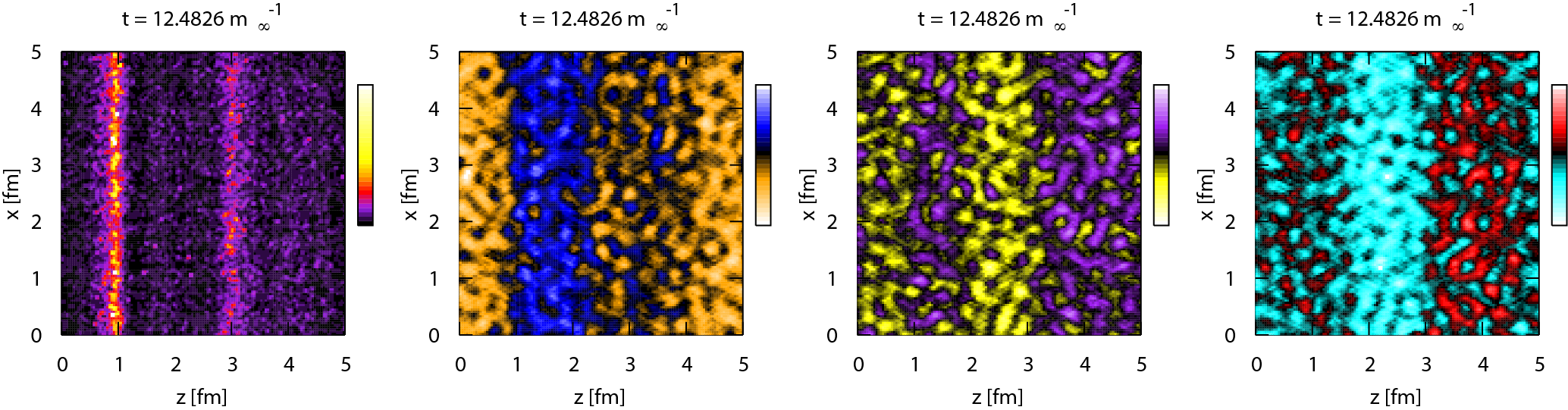}
\includegraphics[width=17cm]{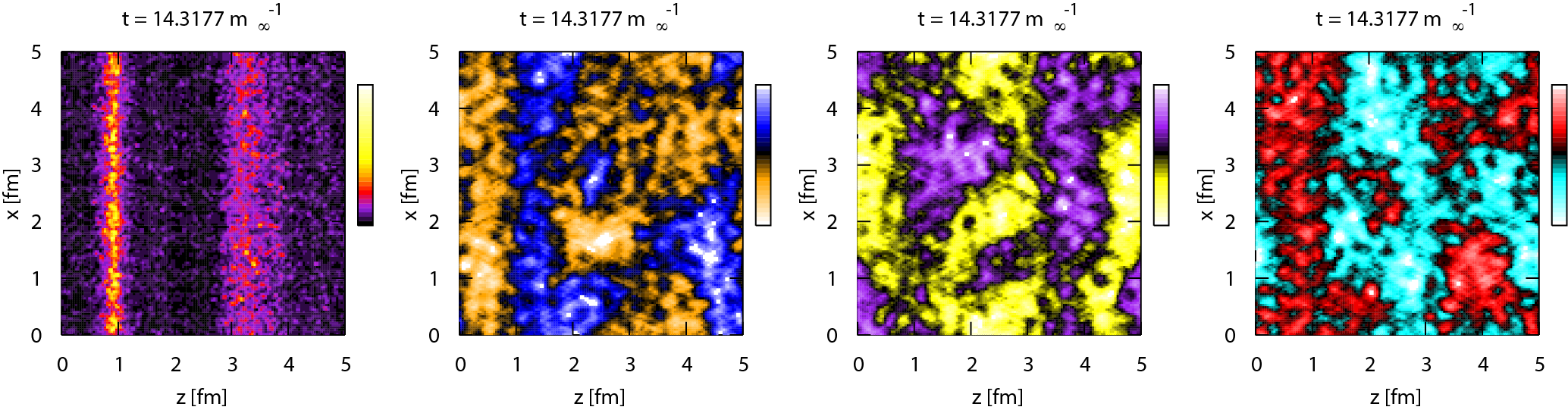}
\includegraphics[width=17cm]{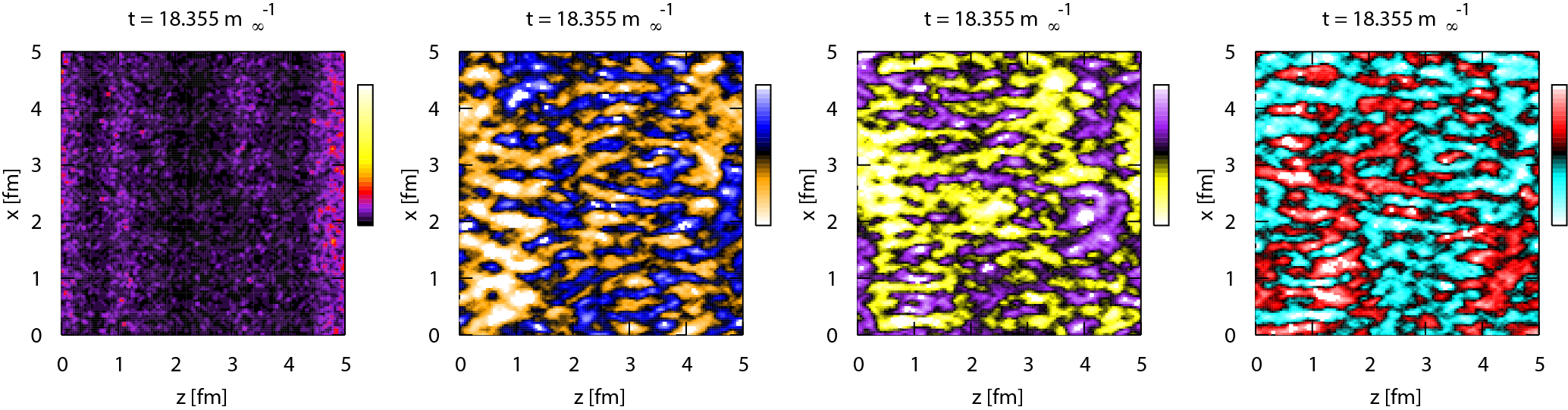}
\vspace{-1mm}
\caption{Slices in the $x$-$z$-plane at fixed $y$ of (from left to right) the current $j_x$ and three color components of the chromomagnetic field $B_y$ for three times equal (from top to bottom) $12.5, \, 14.4$, and $18.5 \, m_\infty^{-1}$. The scales are in lattice units and reach from 0 to $5 \cdot 10^{-8}$ for the current and from $-4 \cdot 10^{-3}$ to $4 \cdot 10^{-3}$ for the chromomagnetic fields.}
\label{fig:slices}
\end{figure}

Although energy conservation will eventually stop the growth of the soft fields as the lattice spacing decreases toward the continuum, it does not solve the  following problem. When $\Lambda \sim 1/a$, the hard field modes have reached the momentum scale of the particles, $p_h={\cal O}(a^{-1})$, and so the clean separation of scales of particles and fields, which is crucial for the Wong-Yang-Mills approach, is lost. In fact, since the occupation number (or phase space density) at that scale is of order of unity or less by construction, it is inappropriate to describe modes at that scale as classical fields. Those perturbative modes should be converted dynamically into particles at a lower scale, so that the field energy density and the entire coupled field-particle evolution is independent of the artificial lattice spacing. This problem reveals the principle limitation of any classical approach. It also shows how difficult it is to simulate the quark-gluon plasma which is a system of quantum fields of non-Abelian dynamics and that the useful concepts of classical fields and classical particles are of limited validity.

\vspace{2mm}
\underline{\it Filamentation}
\vspace{1mm}

It has been explained in Sec.~\ref{subsec-mechanism} that the Weibel instability occurs in anisotropic plasmas due to current filaments which are amplified by the self-consistently generated magnetic field. The mechanism of instability is illustrated in Fig.~\ref{fig-mechanism}. If the unstable modes, which are clearly seen in the simulations presented above, are indeed of the Weibel type, one expects to observe the pattern seen in Fig.~\ref{fig-mechanism} with the current filaments and magnetic fields varying on a length-scale of the order $m_\infty^{-1}$. The length is roughly 0.5 fm with the used simulation parameters which, as previously, are $L=5$~fm, $p_{\rm h}=16$~GeV, $g^2\,n_g=10/$fm$^3$, $m_\infty=0.1$~GeV. 

To see whether the pattern is present in the simulations we have studied the spatial distribution of fields on a cut through the lattice at fixed $y$ after the instability growth started, presented in Fig.~\ref{fig:lsfig}. Fig.~\ref{fig:slices} shows  the current $j_x$ in the left column, as well as the three color components of the chromomagnetic field $B_y$ in the remaining three columns. The uppermost row corresponds to the earliest time $t \approx 12.5 \, m_\infty^{-1}$, the middle one to $t \approx 14.3 \, m_\infty^{-1}$, and the bottom row to the time $t \approx 18.4 \, m_\infty^{-1}$. The filaments are nicely seen in the early stage of the instability growth, displayed in Fig.~\ref{fig:field-SU2}. However, already during the period of growth, which lasts until about $t\approx 18 \, m_\infty^{-1}$, the filaments break and the domains of strong aligned fields become smaller. Finally, at the time when the saturation, seen in Fig.~\ref{fig:field-SU2}, sets in, the fields become turbulent and  the populated high momentum field modes become visible. 

Since we are looking at quantities which are not gauge invariant, the plots of the current and fields have to be taken with care. Also the observed pattern depends to some extent on the lattice size. Nevertheless the qualitative structure of the current and fields is very illuminating and it clearly shows that the instability observed in the simulations is of the Weibel kind.

\begin{figure}[t]
\includegraphics[width=3in]{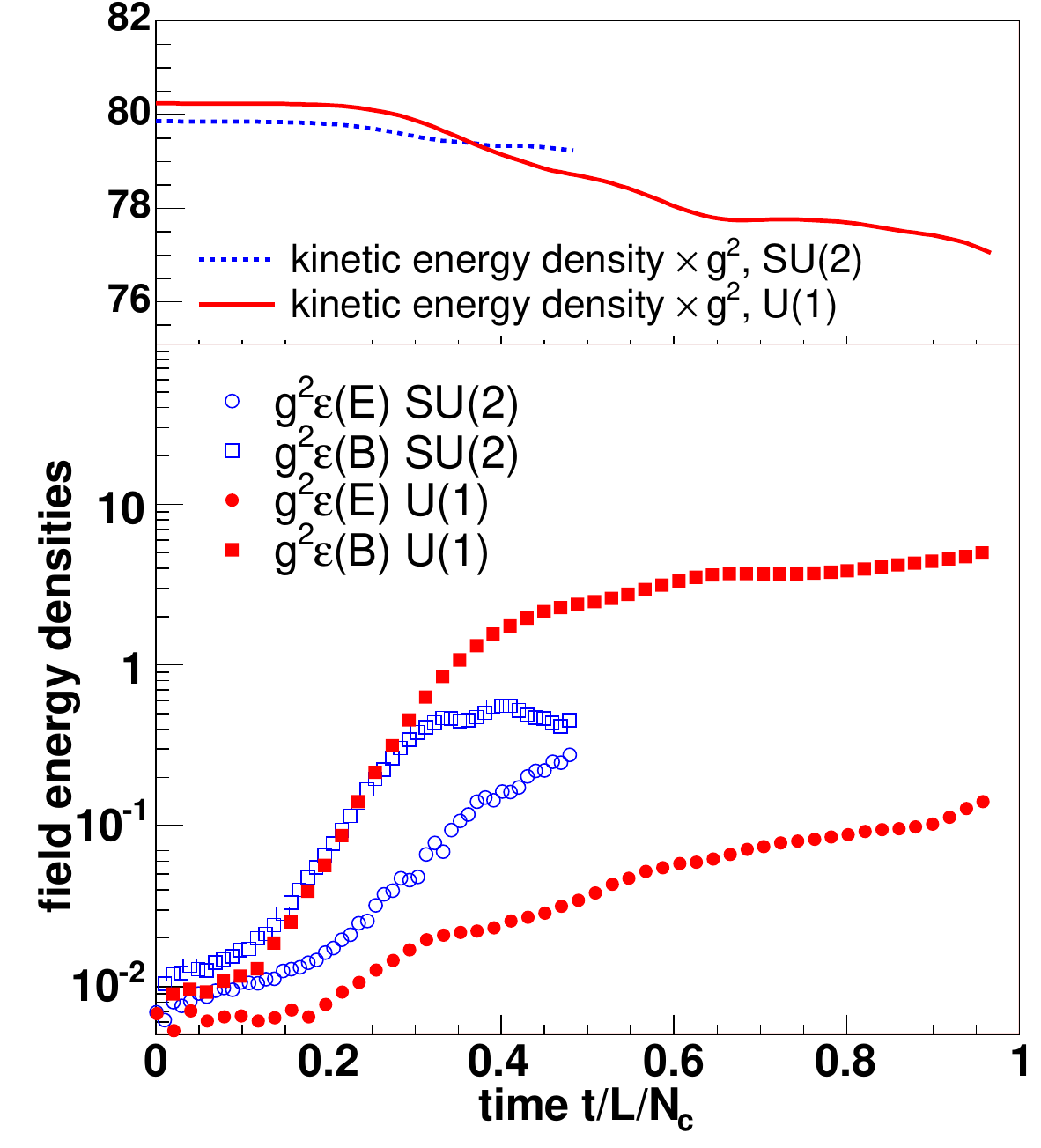}
\includegraphics[width=3.4in]{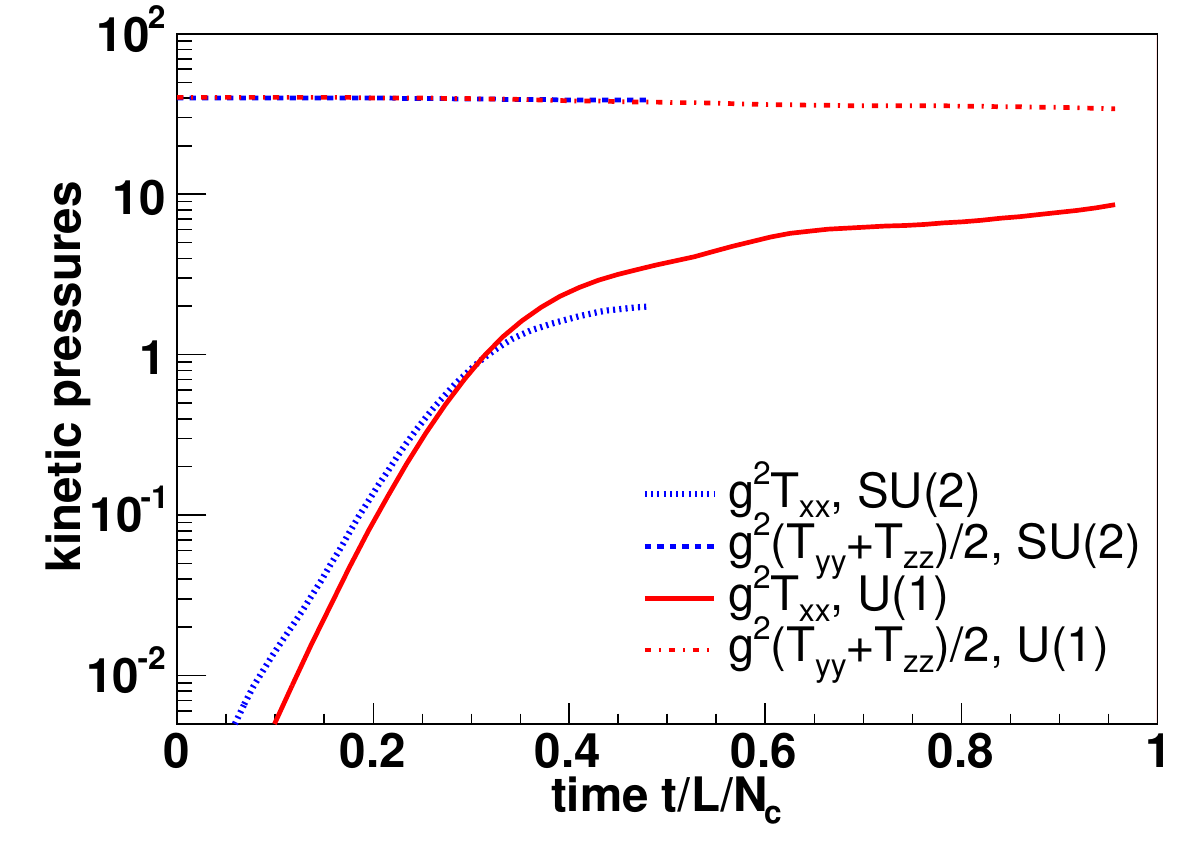}
\caption{Time evolution of the energy densities related to particles and fields (magnetic and electric) expressed in GeV$/$fm$^3$ for U(1) and SU(2) gauge groups (left panel). Transverse and longitudinal components of the energy-momentum tensor of the particles for a simulation with weak initial fields (right panel). Figure from \cite{Dumitru:2005hj}.}
\label{fig:kineticW}
\end{figure}

\vspace{2mm}
\underline{\it Isotropization}
\vspace{1mm}

Plasma instabilities are expected to speed up the process of the system's isotropization because the Lorentz force (due to growing magnetic fields) transfers the momentum of plasma particles from the direction of momentum surplus to the direction of momentum deficit. This effect of interplay of fields and particles can be studied within the Wong-Yang-Mills simulations. We note that the simulations presented in the preceding sections do not offer such a possibility because plasma particles are either absent or they remain unaffected by the fields.

Fig.~\ref{fig:kineticW}, which is taken from \cite{Dumitru:2005hj}, shows the time evolution of energy density and pressure of particles (kinetic energy density and kinetic pressure) together with the field energy density of U(1) and SU(2) plasmas which appear to be qualitatively rather similar. The fields only depend on time and on one spatial coordinate, $x$, which reduces the Yang-Mills equations to $1+1$-dimensions. The classical particles are allowed to propagate in three spatial dimensions and the system does not expand. In the left panel of Fig.~\ref{fig:kineticW}, we present the energy densities of particles and fields for Abelian and non-Abelian plasmas. As expected, the growth of the field energy density is associated with the decrease of the particle energy density because, as explained in Sec.~\ref{subsec-mechanism}, the instability is driven by the energy transferred from particles to fields. In the right panel of Fig.~\ref{fig:kineticW}, we show the longitudinal and transverse components of the particle energy-momentum tensor (\ref{energy-mom-part}) which give the kinetic pressure. The process of isotropization of the kinetic pressure is clearly seen and it is correlated with the stage of exponential growth of the soft fields. However, the component $T_{xx}$ remains significantly smaller than the sum of transverse components for times smaller than $L$, meaning that full isotropization is not achieved.\footnote{Complete isotropization of the QGP is not achieved in any known framework.  In viscous hydrodynamics, for example, there are large momentum-space anisotropies in the initial stage which persist throughout the evolution.  The level of isotropization implied by these simulation results is in rough agreement with that seen in viscous hydrodynamics itself \cite{Strickland:2013uga}.}

\begin{figure}[t]
\begin{minipage}{8.5cm}
\center
\vspace{-2mm}
\includegraphics[width=1.0\textwidth]{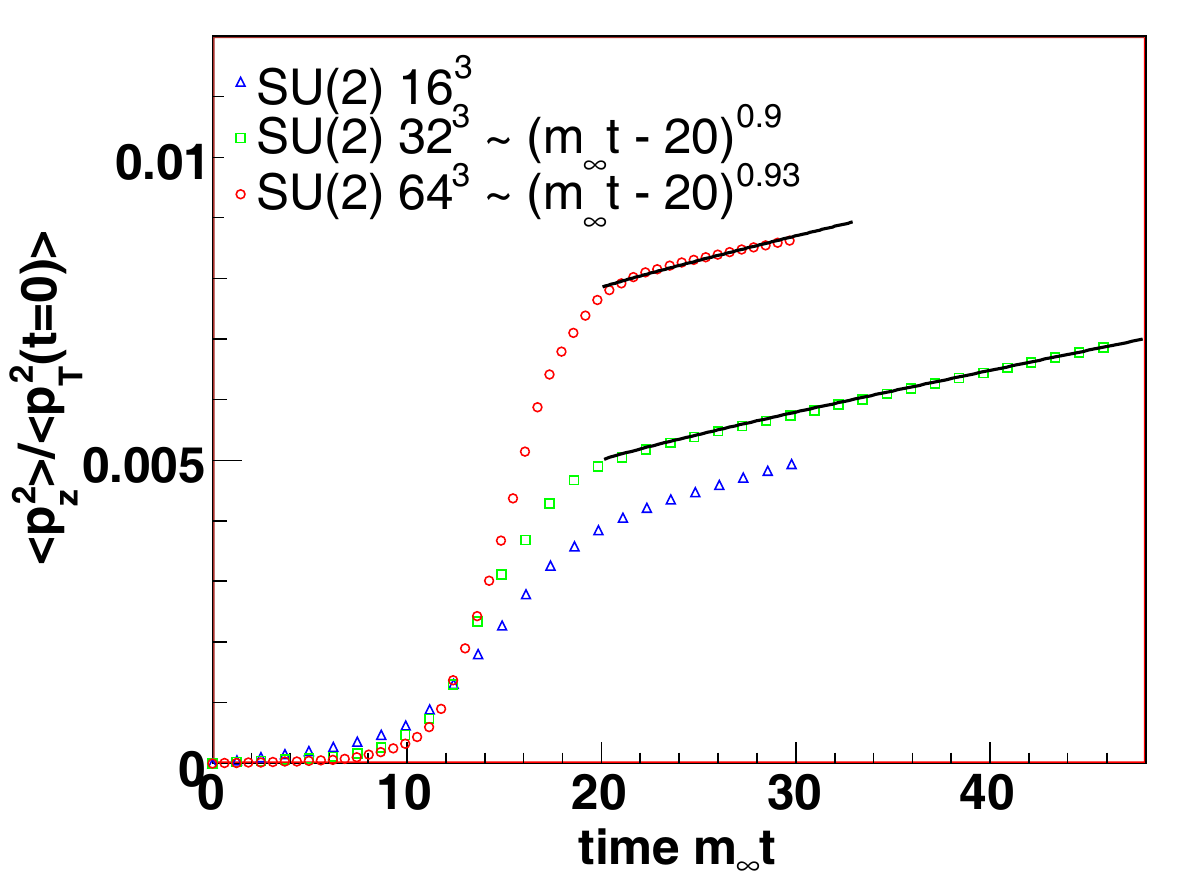}
\vspace{-6mm}
\caption{Broadening of the longitudinal particle momentum distribution for the simulations from Fig.~\ref{fig:field-SU2}. The solid lines represent the power-law fits. Figure from \cite{Dumitru:2006pz}.}
\label{fig:thetaSU2_weak}
\end{minipage}
\hspace{2mm}
\begin{minipage}{8.5cm}
\center
\vspace{-9.5mm}
\includegraphics[width=1.04\textwidth]{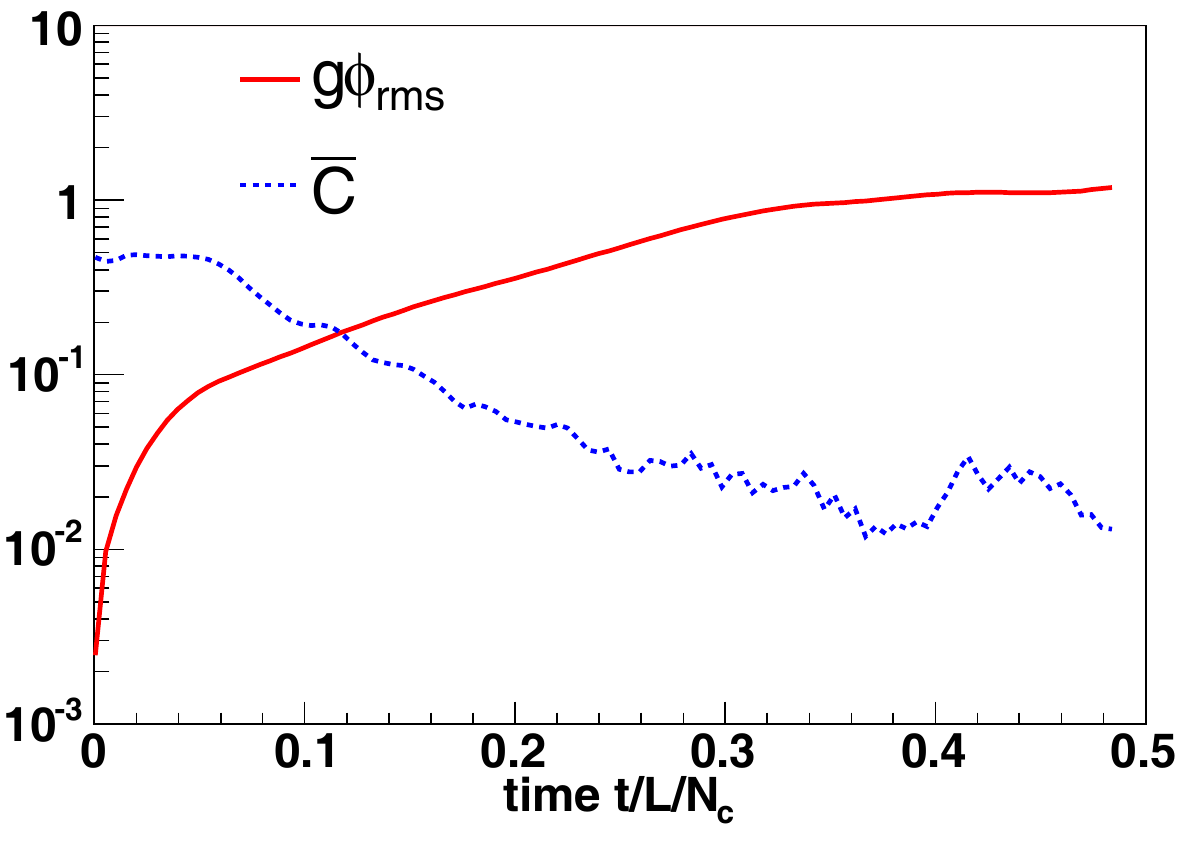}
\vspace{-6mm}
\caption{Time evolution of the dimensionless functional $\bar{C}$ and the functional $g\phi_\mathrm{rms}$ expressed in GeV. Figure from \cite{Dumitru:2005gp}. 
\label{fig:abel-WYM}}
\end{minipage}
\end{figure}

The average longitudinal momentum of plasma particles was further investigated in \cite{Dumitru:2006pz}. Fig.~\ref{fig:thetaSU2_weak} depicts its time evolution for lattices of different sizes. The broadening of the longitudinal distribution during the initial transient times $t \lesssim 10 \, m_\infty^{-1}$ is rather independent of the lattice spacing but during the stage of steep field growth, the mean $p_z$ increases more rapidly the finer the lattice. A power-law fit $t^\alpha$ to the late-time evolution of $\langle p_z^2 \rangle$ yields $\alpha \approx 1$, corresponding to a random walk in momentum space with a constant collision rate and fixed momentum transfer. 

\vspace{2mm}
\underline{\it Abelianization}
\vspace{1mm}

The effect of Abelianization, which has been already discussed in Sec.~\ref{subsec-hardloop} in the context of hard-loop simulations (see also Appendix \ref{app-abelianization}), was also studied in the Wong-Yang-Mills simulations \cite{Dumitru:2005gp}. To check whether Abelianization occurs, the time evolution of the average field
\begin{equation}
\label{eq:phirms}
\phi_\mathrm{rms}= \bigg[ \int_0^L \frac{dx}{L} \big(A^a_yA^a_y + A^a_zA^a_z \big)\bigg]^{1/2} ,
\end{equation}
and the average of the relative size of the field commutator defined analogously to the quantity (\ref{cbareq}), that is 
\begin{equation}
\label{eq:cbar}
\bar{C} = \int_0^L \frac{dx}{L} \frac{\sqrt{\mathrm{tr}\big( (i[A_y,A_z])^2 \big)}}{\mathrm{tr} (A_y^2 + A_z^2)} ,
\end{equation}
were investigated in  \cite{Dumitru:2005gp}. The fields only depended on time and on one spatial coordinate, $x$, which reduced the Yang-Mills equations to $1+1$-dimensions. The classical particles were allowed to propagate in three spatial dimensions. The temporal evolution of the gauge invariant functionals (\ref{eq:phirms}) and (\ref{eq:cbar}) is shown in Fig.~\ref{fig:abel-WYM}. Initially, $\bar{C}$ is constant but then starts dropping exponentially when the magnetic instability sets in, even though the field magnitude as indicated by $ \phi_\mathrm{rms}$ grows. This proves the partial Abelianization already observed in the hard-loop simulations \cite{Arnold:2004ih,Rebhan:2004ur}. The Abelianization, however, appears to stop after $\bar{C}$ dropped by about one order of magnitude, at about the same time when the exponential growth of the fields and of $\phi_\mathrm{rms}$ saturates. The effect of Abelianization is anyway not very physical because, as we know from the hard-loop simulations presented in Sec.~\ref{subsec-hardloop}, the effect hardly survives in the 
$3+1$-dimensional simulations, see Figs.~\ref{fig:Jcomm} and \ref{figcbarj31}. 



\section{Concluding remarks and outlook}
\label{sec-conclusions}

The primary goal of this review was to (i) present the theoretical tools which are used to describe unstable quark-gluon plasma in a systematic manner and (ii) collect known results related to the dynamics of an unstable QGP plasma which are scattered in numerous original papers. We also intended to rederive or explain some results using simple physical arguments. The theoretical approaches to study an unstable QGP, which are introduced in Secs.~\ref{sec-field-theory}-\ref{sec-chromo-fluid}, include: field theory Keldysh-Schwinger formalism, kinetic theory, and fluid techniques. The methods have been further applied to derive the spectra of collective excitations of anisotropic QGP, which were discussed in great detail in Sec.~\ref{sec-collective} with a particular emphasis on unstable modes. In Sec.~\ref{sec-simulate} we presented various numerical approaches to simulate the temporal evolution of an unstable QGP with a plethora of results which clearly show an important dynamical role of instabilities. Our attention has been focused on the process of plasma equilibration which is very complex and strongly depends on initial conditions. Obviously not all efforts - analytical or numerical - to understand the dynamics of unstable QGP have been presented in our review. In this concluding section we intend to partially compensate the deficiency. 

One should mention here a whole series of papers \cite{Kurkela:2011ti,Kurkela:2011ub,York:2014wja,Kurkela:2014tea} where parametric estimates of the thermalization time have been derived for various initial anisotropies and densities. A quasilinear transport approach to equilibration of quark-gluon plasmas, which allows for an analytic treatment of the problem, has been developed in \cite{Mrowczynski:2009gf}. Since the process of equilibration is very complex, the problem is far from being completely understood and more studies in this direction are expected. 

Historically, it was presumed that the process of equilibration and isotropization of the QGP is very fast, and as a result, that the QGP spends most of its lifetime in a state of local equilibrium. In this case, the characteristics of an equilibrium QGP, such as equation of state or viscosity, become of crucial importance for models of relativistic heavy-ion collisions. However, even in this case there would be a short period of time when the plasma is anisotropic, and hence potentially unstable and populated with strong chromodynamic fields which might have a significant effect on various observables measured in relativistic heavy-ion collisions. An analysis of such effects is difficult because the unstable QGP evolves quickly and the problem has to be treated as an initial value problem.  That being said, our current understanding of the dynamics of the QGP is more nuanced and it is now understood that the QGP is momentum-space anisotropic over its entire lifetime.  This opens the possibility that unstable dynamics could play a role over a longer period of time than one naively assumes.

The phenomenon of jet broadening has been studied in \cite{Dumitru:2007rp} for an unstable QGP simulated in the framework of the Wong-Yang-Mills approach introduced in Sec.~\ref{Wong-Yang-Mills}. The exponentially growing width of the transverse momentum distribution of jet partons clearly shows a strong effect of unstable modes. An analytic formulation of this problem is given in \cite{Majumder:2009cf} and more results are expected soon.
A spectrum of fluctuations of chromodynamic fields in an unstable QGP has been found as a solution of the initial value problem in  \cite{Mrowczynski:2008ae}. The unstable modes grow exponentially and after some time they strongly dominate the spectrum. 

The collisional energy loss of high-energy parton traversing an unstable QGP has been studied in detail in \cite{Carrington:2015xca}. The test parton has been shown to lose or gain energy depending on the initial conditions. The energy transfer grows exponentially in time and its magnitude can greatly exceed the absolute value of energy loss in an equilibrium plasma. The energy transfer is also strongly directionally dependent.  A computation of radiative energy loss in an unstable QGP is still missing and a link with phenomenology of jet quenching in relativistic heavy-ion collisions needs to be established. 

The dynamics of unstable systems of non-Abelian quantum fields is a rich domain of research. Despite the progress presented in this article, there is a long list of unanswered questions which will require the development of new theoretical methods and numerical techniques. Let us hope that this review will be useful in these efforts. 

\section*{Acknowledgments}

Fruitful discussions with numerous colleagues are gratefully acknowledged. Special thanks are addressed to Margaret Carrington, Edmond Iancu, Anton Rebhan, and S\"oren Schlichting. B. Schenke and M. Strickland were supported by the U.S. DOE under Contracts Nos. DE-SC0012704 and  No. DE-SC0013470, respectively.  B. Schenke also acknowledges a U.S. DOE Office of Science Early Career Award.


\appendix

\section{Dielectric vs. polarization tensor}
\label{app-eps-vs-pi}
We derive here Eq.~(\ref{epsilon-Pi}) which relates the dielectric tensor to the polarization tensor.  For simplicity, the derivation is  performed in terms of electrodynamics. It is thus relevant for QCD in the hard-loop approximation when the fields are sufficiently weak. 
 
The electromagnetic action in a medium can be written as
\be
\label{action-1}
S  = \frac{1}{2} \int d^4x \Big({\bf D}(x) \cdot {\bf E}(x) - {\bf B}^2(x) \Big),
\ee 
where ${\bf E}$, ${\bf D}$, and ${\bf B}$ are, respectively, the electric field, electric induction, and the magnetic induction. The Fourier  transformed electric induction vector is expressed through the electric field by means of the electric permeability $\varepsilon^{ij} (\omega,{\bf k})$ as
\be
\label{induction-def}
D^i (\omega,{\bf k}) = \varepsilon^{ij} (\omega,{\bf k}) \: E^j (\omega,{\bf k}) .
\ee
Because the material relation (\ref{induction-def}) is nonlocal, the action (\ref{action-1}), similarly as the hard-loop effective action (\ref {g-action}), is nonlocal. The condition of a retarded response is implemented not at the level of the action, but at the level of field equations. 

One often introduces also a magnetic field ${\bf H}$ which is related to ${\bf B}$ via the magnetic permeability $\mu$ as $B^i (\omega,{\bf k}) = \mu^{ij} (\omega,{\bf k}) \: H^j (\omega,{\bf k})$. However, the magnetic field ${\bf H}$ is not really needed, as the electric permeability $\varepsilon^{ij} (\omega,{\bf k})$ is sufficient to encompass both electric and magnetic properties of a medium. This is most easily seen when the medium is isotropic. If one uses the three fields ${\bf E}$, ${\bf D}$, and ${\bf B}$, the medium is fully described by longitudinal and transverse components of the dielectric tensor defined by Eq.~(\ref{e-L-T}). If one uses the four fields ${\bf E}$, ${\bf D}$, ${\bf B}$, and ${\bf H}$, electromagnetic properties of the medium are encoded in the electric and magnetic permeabilities  $\varepsilon (\omega, {\bf k})$ and  $\mu (\omega,{\bf k})$ which in an isotropic medium can be defined through the relations
 ${\bf D} (\omega,{\bf k}) = \varepsilon (\omega,{\bf k}) \: {\bf E} (\omega,{\bf k})$ and ${\bf B} (\omega,{\bf k}) = \mu (\omega,{\bf k}) \: {\bf H} (\omega,{\bf k})$. Comparing to each other the two versions of Maxwell equations in a medium, one finds 
\be
\varepsilon (\omega, {\bf k}) = \varepsilon_L(\omega, {\bf k}) , ~~~~~~~~~~~~ 
1 - \frac{1}{ \mu (\omega,{\bf k})} = \frac{\omega^2}{{\bf k}^2}\Big(\varepsilon_T(\omega, {\bf k}) - \varepsilon_L(\omega, {\bf k})\Big),
\ee
which shows that the two descriptions of a medium are related to each other and actually they are equivalent. However, one should remember that the displacement vector ${\bf D}$ is defined differently in the two versions of electrodynamics. For more details, see the review article \cite{Rukhadze-Silin-1961}.

Since the polarization tensor $\Pi^{\mu \nu}(x,y)$ can be defined as a second functional derivative of the action with respect to $A^\mu (x)$, the electromagnetic action can be written as
\be
\label{action-2}
S = - \frac{1}{4} \int d^4x \,F^{\mu \nu}(x) F_{\mu \nu}(x) + \frac{1}{2} \int d^4x \, d^4y \; A_\mu (x) \, \Pi^{\mu \nu}(x,y) \, A_\nu (y) .
\ee
Requiring the equality of the actions (\ref{action-1}) and (\ref{action-2}) and remembering that the electromagnetic fields are real, one finds 
\be
\label{action-1-2}
\varepsilon^{ij} (\omega,{\bf k}) \, E^i (\omega,{\bf k}) \, E^{*j} (\omega,{\bf k}) = E^i (\omega,{\bf k}) \, E^{*i} (\omega,{\bf k})  + A_\mu (\omega,{\bf k}) \, \Pi^{\mu \nu}(\omega,{\bf k}) \, A_\nu^* (\omega,{\bf k}) ,
\ee
where the Fourier transformation is performed and Eq.~(\ref{induction-def}) is used. 

The consequences of Eq.~(\ref{action-1-2}) are easily derived in the temporal axial gauge where $A^0(x) = 0$. Then, ${\bf E}(t, {\bf r}) = - \dot{\bf A}(t, {\bf r})$ and 
\be
\label{E-Coulomb}
{\bf E}(\omega, {\bf k}) =  i \omega \; {\bf A}(\omega, {\bf k}) .
\ee
Substituting the chromoelectric field in the form (\ref{E-Coulomb}) into Eq.~(\ref{action-1-2}), one finds the desired relation (\ref{epsilon-Pi}).

\section{Abelianization}
\label{app-abelianization}

In this appendix we explain the phenomenon of `Abelianization' that we referred to several times in the main text. We follow the derivation given in \cite{Arnold:2004ih} but our discussion is much simplified. 

Self interactions of the soft fields are expected to stop the instability growth once they become strong. However, in certain situations the system is inclined to approach an Abelian configuration, which allows for continued growth. An important assumption in this argument is that the dimensionality of the system is reduced. We refer the reader to Sec.~\ref{sec:2p1} and \cite{Arnold:2007tr} for a discussion on the dimensional reduction.

Let us analyze an effective potential of the form
\be
\label{eff-pot}
V(x) = - \frac{1}{4} F^{\mu \nu}_a (x) \, F_{a \mu \nu} (x) 
- \frac{1}{2}\int d^4y \, A_{a \mu} (x) \, \Pi^{\mu \nu}_{ab} (x-y)\, A_{a \nu} (y) ,
\ee
where in the first term one recognizes the fundamental Lagrangian density of gluodynamics, see Eq.~(\ref{lagrangian-QCD}), while the second term, which is known from Eq.~(\ref{g-2-action1}), provides after some manipulations the Hard Loop Action (\ref{g-action}). We consider the potential (\ref{eff-pot}) in the temporal axial gauge $A^0(x)=0$,  assuming a strongly oblate momentum distribution of plasma constituents with the characteristic momentum $p_z$ which is much smaller that $p_x,p_y$. Then, the spectrum of unstable modes has typical wave vectors obeying $k_z\gg k_\perp$, see Sec.~\ref{sec-ex-oblate}. There, the fields vary more rapidly with $z$ than with $x$ or $y$ and thus we can consider a gauge field that only depends on $z$ that is
\begin{equation}
\label{A-z}
\mathbf{A}^a =\mathbf{A}^a (z),
\end{equation}
which is the crucial step to allow for Abelianization. Because of the assumption (\ref{A-z}), the wave vectors of unstable modes are along the axis $z$ and the only non-vanishing components of $\mathbf{A}^a$ are $A^a_x$ and $A^a_y$. The polarization tensor, which enters Eq.~(\ref{eff-pot}), can be then approximated as
\begin {equation}
\label{Pi-mu2}
\Pi = \begin{pmatrix}
 \mu^2 & 0 &  0 \\
  0 & \mu^2 & 0 \\
  0 & 0 & 0
\end{pmatrix} ,
\end {equation}
where $ \mu$ is a positive real parameter of the dimension of mass. With the polarization tensor of the form (\ref{Pi-mu2}), the dispersion equation (\ref{dispersion-pi}) gives the two transverse modes $\omega^2(k) = k^2 - \mu^2$ which are unstable for $k^2 \le \mu^2$. 

\begin{figure}[t]
\includegraphics[width=8cm]{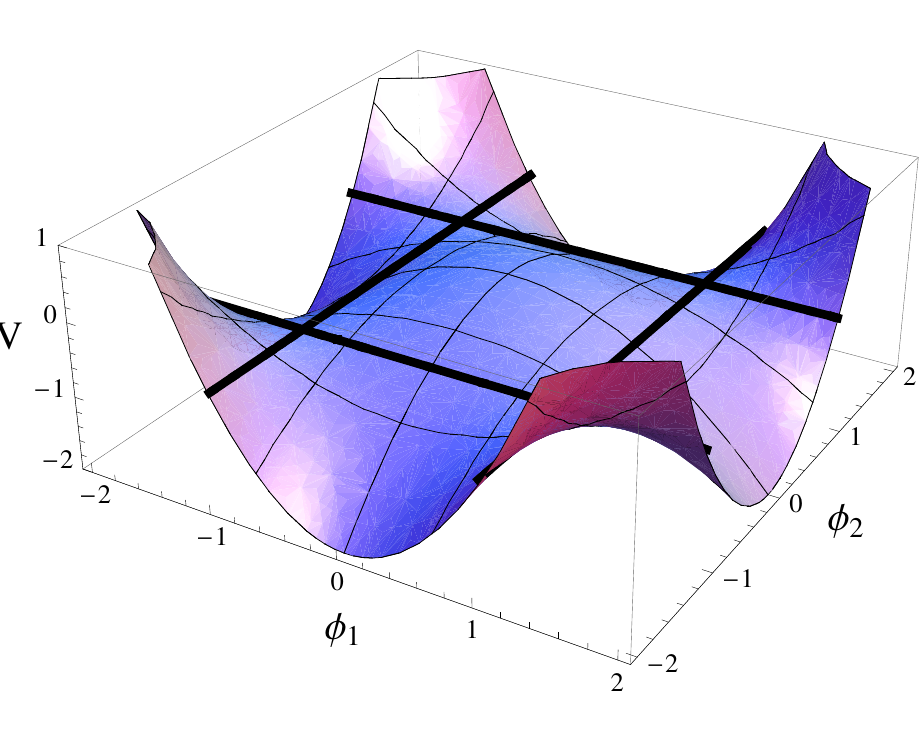}
\vspace{-5mm}
\caption{The effective potential (\ref{eq:Vphi}) as a function of $\phi_1$ and $\phi_2$. The fields $\phi_1$ and $\phi_2$ are expressed in units of $\mu/g$, and the values of $V$ are in units of $\mu^4/g^2$. The lines $\phi_1=0$ and $\phi_2=0$, which are of the steepest decent, correspond to Abelian configurations. Figure from \cite{Arnold:2004ih}.}
\label{fig:potential}
\end{figure}

Using Eq.~(\ref{Pi-mu2}), the effective potential (\ref{eff-pot}) in the static and long wavelength limit, where the terms with space-time derivatives are neglected, is 
\begin {equation}
V =  \frac{1}{4} \, g^2 f^{abc} f^{ade} A^b_i A^c_j A^d_i A^e_j
       - \frac{1}{2} \, \mu^2 A^a_i A^a_i 
=  \frac{1}{4} \, g^2 \left[(A_i^b A_i^b)^2 - A^b_i A^c_j A^c_i A^b_j\right]
       - \frac{1}{2} \, \mu^2 A^a_i A^a_i ,
\label {eq:pota}
\end {equation}
where the second equality holds for the SU(2) group which is chosen for simplicity. The potential (\ref{eq:pota}), which can be written as
\begin{equation}
V = \frac{1}{4} g^2 \left\{\left[\text{tr}\left({\cal A}^\top{\cal A}\right)\right]^2
-\text{tr}\left[\left({\cal A}^\top{\cal A}\right)^2\right]\right\}
-\frac{1}{2} \mu^2\, \text{tr}\left({\cal A}^\top P^{(xy)}{\cal A}\right) ,
\end{equation}
with
\begin {equation}
\label{cal-A}
   {\cal A} = \begin{pmatrix}
                 A_x^1 & A_x^2 & A_x^3 \\
                 A_y^1 & A_y^2 & A_y^3 \\
                 A_z^1 & A_z^2 & A_z^3
               \end{pmatrix} \,, ~~~~   P^{(xy)} = \begin{pmatrix} 1 & & \\ & 1 & \\ & & 0 \end{pmatrix},
\end {equation}
is invariant under spatial rotations in the $x$-$y$-plane, due to the symmetry of momentum distribution of plasma constituents, and color rotations, because of gauge freedom. Therefore, we can assume without loss of generality that ${\cal A}$ is a symmetric real matrix, since this can always be achieved by a color rotation. And because only non-vanishing components of $\mathbf{A}^a (z)$ are $A^a_x(z)$ and $A^a_y(z)$, the matrix (\ref{cal-A}) can be chosen in the diagonal form
\begin {equation}
\label{cal-A-diag}
   {\cal A} = \begin{pmatrix}
                 A_x^1 & 0 & 0 \\
                 0 & A_y^2 & 0 \\
                 0 & 0 & 0
               \end{pmatrix} .
\end {equation}
Then, the effective potential is written as
\begin{equation}
\label{eq:Vphi}
 	V =\frac{1}{2}g^2 \phi_1^2\phi_2^2-\frac{1}{2}\mu^2\left(\phi_1^2+\phi_2^2\right),
\end{equation}
where the fields $\phi_1$ and $\phi_2$ are defined through the equation 
\begin{equation}	
 	A_i^a=\phi_1\delta_{ix}\delta_{a1}+\phi_2\delta_{iy}\delta_{a2} .
\end{equation}
The potential (\ref{eq:Vphi}) is depicted in Fig.~\ref{fig:potential}. The Abelian configurations correspond to the $\phi_1$ axis with $\phi_2=0$ and the $\phi_2$ axis with $\phi_1 = 0$. The static non-Abelian solutions are indicated by the intersection points of the four straight lines in the figure. At these points, the amplitude of the gauge field is $A \sim \mu/g$. Recalling that unstable modes typically have $k \sim \mu$, this corresponds to the non-Abelian scale $A \sim k/g$. However, these solutions are unstable to rolling down and subsequently growing in amplitude along one of the axes. The picture suggests that, if we start from $A$ near zero, the system might possibly at first roll toward one of these configurations with $A \sim k/g$, but its trajectory would eventually roll away, growing along either the $\pm \phi_1$ or $\pm \phi_2$ axis until the effective action breaks down at $A \sim p/g$. This effect of unstable growth of the gauge fields toward an Abelian configuration is called `Abelianization'.



\bibliography{review-MSS}

\end{document}